\documentclass{aastex631}

\usepackage{indentfirst}
\usepackage{xcolor,graphicx,float}
\usepackage{subfigure}
\usepackage{color}
\usepackage{multirow}
\usepackage{booktabs}
\usepackage[figuresleft]{rotating}
\usepackage{subfigure}

\shorttitle{A Study of the Spectral properties of Gamma-Ray Bursts with the Precursors and Main bursts}
\shortauthors{Hui-ying Deng et al.}

\graphicspath{{./}{figures/}}

\begin{document}

\title{A Study of the Spectral properties of Gamma-Ray Bursts with the Precursors and Main bursts}

\author{Hui-Ying Deng}
\affiliation{College of Physics and Electronics information, Yunnan Normal University, Kunming 650500, People's Republic of China\\}
\author{Zhao-Yang Peng}
\affiliation{College of Physics and Electronics information, Yunnan Normal University, Kunming 650500, People's Republic of China; pengzhaoyang412@163.com\\}
\author{Jia-Ming Chen}
\affiliation{Department of Astronomy, School of Physics and Astronomy, Yunnan University, Kunming, Yunnan 650091, People’s Republic of China \\}
\author{Yue Yin}
\affiliation{Department of Physics, Liupanshui Normal College, Liupanshui 553004, People's Republic of China\\}
\author{Ting Li}
\affiliation{Division of Assets, Yunnan Normal University, Kunming 650500, People's Republic of China}

\begin{abstract}
  
There is no consensus yet on whether the precursor and the main burst of gamma-ray bursts (GRBs) have the same origin, and their jet composition is still unclear. In order to further investigate this issue, we systematically search 21 Fermi GRBs with both precursor and main burst for spectral analysis. We first perform Bayesian time-resolved spectral analysis and find that almost all the precursors and the main bursts (94.4$\%$) exhibit thermal components, and the vast majority of them have low-energy spectral index ($\alpha$) (72.2$\%$) that exceed the limit of synchrotron radiation. We then analyse the evolution and correlation of the spectral parameters and find that approximately half of the $\alpha$ (50$\%$) of the precursors and the main bursts evolve in a similar pattern, while peak energy ($E_{p}$) (55.6$\%$) behave similarly, and their evolution is mainly characterized by flux tracking; for the $\alpha-F$ (the flux) relation, more than half of the precursors and the main bursts (61.1$\%$) exhibit roughly similar patterns; the $E_{p}-F$ relation in both the precursor and main burst (100$\%$) exhibits a positive correlation of at least moderate strength. Next, we constrain the outflow properties of the precursors and the main bursts and find that most of them exhibit typical properties of photosphere radiation. Finally, we compare the time-integrated spectra of the precursors and the main bursts and find that nearly all of them are located in similar regions of the Amati relation and follow the Yonetoku relation. Therefore, we conclude that main bursts are continuations of precursors and they may share a common physical origin.

\end{abstract}


\section{Introduction\label{Sections1}}
Gamma-ray bursts (GRBs) represent one of the most intense explosive events in the universe. The energy released during these phenomena can be equivalent to the total energy emitted by thousands of suns within a matter of seconds. According to the classification of $T_{90}$, bursts with a duration longer than $2$ seconds are classified as long GRBs (LGRBs), while bursts with a duration shorter than $2$ seconds are classified as short GRBs (SGRBs) \citep{1993ApJ...413L.101K}. GRBs sometimes exhibit faint emission events prior to the main radiation, and this phenomenon is referred to as a precursor \citep{1974ApJ...194L..19M}.

\cite{1995ApJ...452..145K} first defined the precursor as having a peak intensity slightly lower and distant from the main burst, with a separation phase having an intensity comparable to the background and a duration not shorter than the main burst phase. Based on this criterion, precursors were selected by the naked eye, and durations were defined using signal-to-noise ratios, while spectral properties were quantified using hardness ratios. They found that these GRBs with precursors and other GRBs without precursors had the same spatial distribution; the duration of the precursors correlated with the duration of the main bursts. In addition, they found no other significant connection between the precursors and the main bursts, suggesting that the precursors and the main bursts may be independent of each other. \cite{2005MNRAS.357..722L} selected a bright, long-duration BATSE burst sample and set the precursor criteria: first, the event must be detected before the trigger; second, its flux should decrease prior to the trigger, and it searches for precursors 200 seconds before the GRB is triggered. They concluded that there was no correlation between the precursor properties and those of the main bursts. Moreover, the spectra of most precursors were typically non-thermal power-law spectra. The spectra of these long-delay and non-thermal origin precursors were challenging to explain using existing progenitor models. \cite{2015MNRAS.448.2624C} searched for 2710 long-duration bursts from three instruments, namely BATSE, Swift-BAT, and Fermi-GBM. They adopted the basic precursor definition, which is similar to previous definitions, as an emission preceding the episode with the highest peak intensity (main event), separated from it by a quiescent period with no detectable gamma-ray flux. They obtain the energy in time-frequency bins using an algorithm developed for gravitational-wave data analysis (Q pipeline; \cite{2004CQGra..21S1809C}). In this method, time-frequency bins are tiled using bisquare windows with overlapping Gaussian-enveloped sinusoids. The signal is whitened using linear prediction, after which a high pass filter is applied. To ensure comparability between bins, the energy density in each bin is normalized to account for the varied window size and the tile overlapping. They found no correlation in the temporal properties between the main bursts and the precursors. \cite{2018NatAs...2...69Z} reported on an exceptionally bright burst, GRB 160625B, which exhibited three distinct radiation events separated by two long quiescent periods (a short precursor, an extremely bright main burst, and an extended radiation event). Through time-resolved spectral analysis of the precursor and the main burst, they found that the precursor and the main burst displayed distinctly different spectral properties, with the precursor demonstrating a thermal spectral component, while the main burst exhibited a non-thermal spectral component. This transition serves as a clear indication of the change in jet composition from a fireball to a Poynting-flux-dominated jet. \cite{2019ApJS..242...16L} further analysed GRB 160625B and determined that its precursor and main burst are respectively dominated by thermal and non-thermal components, consistent with the findings of \cite{2018NatAs...2...69Z}. This further suggests that the origins of the main burst and the precursor may be distinct. \cite{2019ApJ...884...25Z} identified 18 short bursts with precursors from 660 SGRBs observed by Fermi and Swift. They found that most of the precursors and main bursts exhibit non-thermal emission properties, they still identified some differences between the precursors and the main bursts. \cite{2020PhRvD.102j3014C} found 217 bursts exhibiting precursors through the analysis of 11 years of Fermi/GBM data. They found that the duration of quiescent interval was a bimodal normal distribution, suggesting the existence of two distinct progenitor stars for these GRBs.

However, \cite{2008ApJ...685L..19B} provided a simple precursors definition: their peak flux is lower than that of the main event, separated by a quiescent period. This period may not exceed the duration of the main burst and is not necessarily before the triggering event. They analysed the spectra of the precursors with known redshifts and compared them with the time-integrated spectra of prompt emission, although no correlation was found between the two slopes, there were no systematic differences observed in spectral hardness or softness. Additionally, precursors exhibit considerable energy levels, slightly below those of entire bursts within the 15–150 keV range. Furthermore, these properties were found to be independent of the quiescent period. Their results suggested that the precursors were phenomena closely associated with the main burst. Additionally, they examined whether precursors contained a thermal component. Potentially due to a low signal-to-noise ratio, they did not identify any distinct thermal radiation component. Later, \cite{2009A&A...505..569B} found that $12.5\%$ of the 2704 observed BATSE bursts exhibited one or more precursors. Through time-resolved spectral analysis, they found striking similarities in the spectral characteristics of precursors and main bursts, and they concluded that the fireball model mechanism of the precursors and the main bursts was supported. \cite{2010ApJ...723.1711T} performed a precursor search on a sample of short bursts observed by Swift and there were no strict constraints on whether the instrument was triggered or not and on the interval between the precursor and the main event. After analysing the time characteristics of the precursors and the main bursts, no significant differences are found between the two, as well as between SGRBs with and without precursors. \cite{2014ApJ...789..145H} used Bayesian algorithms to search for precursors of GRBs observed by Swift BAT, without the requirement of a quiescent period for precursors. They further indicated through spectral analysis that the origin of the precursor is consistent with that of the main burst. \cite{2015PhDT.......100Z} investigated the characteristics of Fermi GRBs with precursors. They selected precursors by comprehensively considering the methods defined by previous researchers and classified them into three categories for separate studies. The three categories of precursors include Type I, where the precursor is much dimmer than the dominant emission and preceded by a well-defined quiescent period; Type II, similar to Type I but the quiescent period is not well-defined; and Type III, where the precursor is dimmer than the main emission but not by much, with the background-subtracted precursor peak flux being more than a third but less than the background-subtracted dominant emission peak flux. They compared distributions of temporal and spectral parameters and found no statistically significant differences between the precursors and main emissions, indicating they originate from the same source. \cite{2021ApJS..252...16L} investigated SGRBs data observed by Swift/BAT, focusing on examining short burst events that simultaneously have precursors, main bursts, and extended emissions. They found a correlation between their peak fluxes, supporting that these three events originate from similar central engine activities. \cite{2022ApJ...928..152L} analysed 52 LGRBs with precursors selected from the third Swift BAT catalog. They discovered that both the temporal characteristics of the precursors and the main bursts follow relationship between peak time ($\tau_{p}$) and pulse width ($\omega$), suggesting that the precursors and the main bursts may have a common physical origin. \cite{10.1088/1674-4527/ad0497} compared the power-law relationship between the pulse width and energy of the precursor and the main bursts, revealing that they may share a common physical origin.

Some theoretical explanations have been proposed to investigate the physical origins of precursors, which can be categorized as follows: (1) When the fireball becomes transparent, the transient radiation produced by the photosphere is released, referred to as photospheric precursors \citep{2000ApJ...543L.129L}; (2) \cite{2001ApJ...550..410M} proposed the concept of shock breakout precursor. For LGRBs, the central engine is surrounded by the stellar envelope of the progenitor star, and the jet must penetrate this envelope to be observed for radiation. During this process, the interactions heat the material immediately ahead of the jet. When this heated material breaks through this envelope, it releases thermal emission in the form of a shock breakout. For SGRBs, a similar scenario can occur if the central engine releases a dense wind before emitting a jet and the central engine must be a magnetar rather than a black hole. Furthermore, a second precursor with increased energy is generated after the interaction of high-energy particles and thermal photons within the jet; (3) In fallback precursors model, the central engine initiates an initially weak jet that successfully penetrates the stellar envelope. However, interactions slow down the jet, causing some of its material to fall back and be accreted by the central engine. This process powers the emergence of a second, stronger jet. The first jet produces the precursor while the stronger jet produces the main emission \citep{2007ApJ...670.1247W}; (4) Before the merger, electromagnetic signals can arise from the interaction between the magnetospheres of two neutron stars (NSs), potentially serving as precursor emission \citep{2012ApJ...757L...3L,2013PhRvL.111f1105P,2018ApJ...868...19W,2020ApJ...902L..42W}; (5)Prior the merger, potential crust cracking in one or both NSs could give rise to precursor radiation \citep{2012PhRvL.108a1102T,2020PhRvD.101h3002S}. Based on the cumulative progress in the research mentioned above, there is currently no unified consensus on whether the precursors and main bursts, whether they are long or short burst, share a common origin. There are divergent theoretical explanations regarding their physical origins. Furthermore, few studies have compared the precursors of long and short bursts with the spectral characteristics of the main burst. Therefore, we analyse the precursors and main bursts of long and short bursts from the perspective of time-resolved spectra and time-integrated spectra. We not only compare their spectral components and spectral characteristics using the best-fit model, but also, based on previous researcher theoretical model, constrain the outflow properties of the precursors and main bursts. This comparison aims to investigate whether they share a common origin and provides clues for explaining their physical origins. 

This paper is organized as follows. Section \ref{Sections2} describes the selection of the sample and analysis methods. The models used in this study and the criteria for selecting the best model are presented in Section \ref{Sections3}. Section \ref{Sections4} illustrates the analysis results of spectrum parameters. Section \ref{Sections5} characterizes the radiation parameters of the photosphere. Section \ref{Sections6} shows the Amati relation and the Yonetoku relation, and in Sections \ref{Sections7} and \ref{Sections8} the discussion and conclusions are given. Throughout the article, the consistent cosmological parameters with values are \(H_0 = 67.4 \, \text{km s}^{-1} \, \text{Mpc}^{-1}\), \(\Omega_M = 0.315\), and \(\Omega_\Lambda = 0.685\), following the concordance cosmology \citep{2020A&A...641A...6P}.

\section{Sample Selection and Analysis Methods \label{Sections2}}
The data in this study are sourced from the Fermi satellite, equipped with two instruments: the Gamma-ray Burst Monitor (GBM) and the Large Area Telescope (LAT). The GBM comprises $14$ detectors, each with $128$ energy channels, including $12$ NaI detectors covering an effective energy range of $8 keV-1 MeV$, and two BGO detectors covering an effective energy range of $200 keV-40 MeV$. The observational data from GBM are stored in three file formats: CTIME files, CSPEC files, and TTE files. Among these, TTE data consume more memory. Typically, only data recorded within 30 seconds before and 300 seconds after the trigger are used. In comparison to the first two file types, TTE files have the highest time and energy resolution, making them suitable for analyzing the time-resolved spectra of GRBs. Therefore, TTE data are used for time-resolved spectral analysis in this study. The TTE data and standard response files are provided by the GBM team. We select data from all NaI detectors (usually one to three) triggered by GBM and the brightest BGO detector. 

In this study, a similar precursor definition to \cite{2008ApJ...685L..19B} is adopted: the peak intensity of the precursor is lower than the subsequent main burst, and there is a separation period between the precursor and the main burst. This separation period is not necessarily longer than the duration of the main burst, and it does not necessarily occur before the triggering event. We perform a detailed time-resolved spectral analysis of the sample and select relatively brighter bursts, which have precursors with fluence greater than $3.5 \times 10^{-7}$erg/cm$^2$. We obtain $21$ GRBs from GBM, comprising $16$ long bursts and $5$ short bursts. However, three short bursts (GRB130310840, GRB100717372 and GRB081216531) do not meet the fluence criterion, so we will only be used in the time-integrated analysis. The remaining two short bursts (GRB 180703B and GRB 140209A) will be used in both time-resolved  and time-integrated analysis.

In this paper, we use the Multi-Mission Maximum Likelihood Framework (3ML; \cite{2015arXiv150708343V}) for Bayesian analysis, serving as the primary tool to conduct time-resolved spectral analyses on both precursors and main bursts \citep{2018ApJ...857..120Y}. Performing time-resolved spectral analysis and effectively binning the data is crucial in this investigation. \cite{2014MNRAS.445.2589B} studied various methods for GRB spectral time binning. Bayesian Blocks (\cite{2013ApJ...764..167S}) are the effective methods for time binning, characterized by the following features: (1) each time interval conforms to a constant Poisson rate; (2) the algorithm is used to subdivide the light curve of GRBs for the selection of time bins; (3) the selection of time bins reflects the genuine variability in the data; (4) it has variable width and variable signal-to-noise ratio. However, the Bayesian Blocks method does not guarantee that each time bin contains enough photons to perform accurate spectral fitting. Traditional signal-to-noise ratio methods ensure that there are enough photons for spectral fitting, but they can sometimes disrupt the physical structure. Both methods have their advantages and disadvantages. Therefore, we combine the strengths of both methods to choose the time bin. First, we use the Bayesian Blocks method to divide the data into time bins, and then calculate the statistical significance S (an appropriate measure of signal-to-noise ratio) for each individual time bin. Due to the significant difference in peak flux between precursors and main bursts, we attempt to use selection criteria for S that may not necessarily be identical but should at least be close. The precursor selection criterion is $S \geq 15$, and the main burst criterion is $S \geq 20$ \citep{2018ApJS..236...17V}. Since there are relatively few precursors satisfying $S \geq 15$, we choose precursors with $S \geq 15$ and main bursts with $S \geq 20$ as gold sample. For silver sample, we choose precursors with $S \geq 5$ and main bursts with $S \geq 20$. Similarly, for copper sample, we choose precursors with $S \geq 2$ and main bursts with $S \geq 20$. The classification is presented in Table \ref{tab:thermal component}. Among them, there are 7 GRB candidate sources in the gold sample, 10 GRB candidate sources in the silver sample, and 1 GRB candidate source in the copper sample. Simultaneously, we apply the Bayesian Blocks method with a false alarm probability $p = 0.01$ to rebin the TTE light curves of one of the brightest NaI detectors for each burst, with other triggered detectors following the same time bin information.

\section{Spectral Models and Selection of the Best Model \label{Sections3}}

To investigate the spectral components of precursors and main bursts, we employ three empirical spectral models commonly used in the literature. The Band spectral component, Band function (Band) \citep{1993ApJ...413..281B}, which is written as 

\begin{equation}
N_{\text{Band}}(E) = A\left\{ \begin{array}{ll}
\left( \frac{E}{100\text{keV}} \right)^\alpha \exp \left( -\frac{E}{E_0} \right), & E < (\alpha - \beta)E_0 \\
\left( \frac{(\alpha - \beta)E_0}{100\text{keV}} \right)^{\alpha - \beta} \exp(\beta - \alpha) \left( \frac{E}{100\text{keV}} \right)^\beta, & E > (\alpha - \beta)E_0
\end{array} \right.
\end{equation}

where
\begin{equation}
{E_p} = \left( {2 + \alpha } \right){E_0},
\end{equation}
where \(N(E)\) is the photon flux (ph $cm^{-2} keV^{-1} s^{-1}$), \(A\) represents the normalization constant of the spectrum, $\alpha$ is the low-energy spectral index, $\beta$ is the high-energy spectral index, ${E_0}$ is the break energy, and $E_{p}$ is the peak energy in units of keV in the observed $\nu {F_\nu }$ spectrum.
The cutoff power-law function (CPL) is written as follows, corresponding to the first portion of the Band function:
\begin{equation}
{N_{CPL}}\left( E \right) = A{\left( {\frac{E}{{100keV}}} \right)^\alpha }\exp \left( { - \frac{E}{{{E_0}}}} \right),
\end{equation}
where A is the normalization factor at 100 keV in units of photons ${s^{ - 1}}c{m^{ - 2}}ke{V^{ - 1}}$, $\alpha$ is the low-energy spectral index and ${E_0}$ is the break energy in keV. 
Some GRBs have additional thermal components, which are generally fitted by Planck blackbody (BB) function. Planck function, given by:

\begin{equation}
N_{B B}(E)=A \frac{E^2}{\exp [E / k T]-1},
\end{equation}
where E is the photon energy, A is the normalization constant for energy at 1 keV, \(kT\) is the energy corresponding to the blackbody temperature, with units of keV, and \(k\) is the Boltzmann constant.

In our analysis, we initially fit the time-resolved spectra of each GRB using the Band and CPL models and screen the fitting results to identify the best-fit model (best model). Subsequently, we incorporate BB components separately into the Band and CPL models, resulting in Band+BB and CPL+BB models. Through further data analysis, we identify the best model with a BB component (best model+BB). Finally, by comparing the best model and best model+BB, we determine the presence of a thermal component. In this paper, we introduced the Deviance Information Criterion (DIC) \citep{2019ApJS..245....7L,2021ApJ...920...53C} to assess the quality of the fitting models. The expression for DIC is given by:
\begin{equation}
DIC = -2 \log \left[ p(\text{data} \mid \hat{\theta}) \right] + 2 p_{\text{DIC}},
\end{equation}
here, $\hat{\theta}$ represents the posterior mean of the parameters, $p_{DIC}$ is the effective number of parameters. Different models are fitted to the same data points, and a smaller DIC value indicates a better model. The difference in DIC values between two models, denoted as (\(\Delta\)DIC = DIC\(_{j}\) - DIC\(_{i}\)), is used to assess the model's goodness of fit. The criterion \citep{2018ApJ...866...13H} is as follows: (a) the range of $\Delta$DIC is 0-2: the goodness of fit between model j and model i is indistinguishable; (b) the range of  $\Delta$DIC is 2-6: there is positive evidence supporting model i; (c) the range of  $\Delta$DIC is 6-10: there is strong evidence supporting model i; (d) $\Delta$ $DIC > 10$: there is very strong evidence supporting model i.

\section{Analysis Results of Spectrum Parameters \label{Sections4}}
\subsection{Spectral Component Analysis}
\begin{figure}[htb]
\centering
{
\includegraphics[scale=0.5]{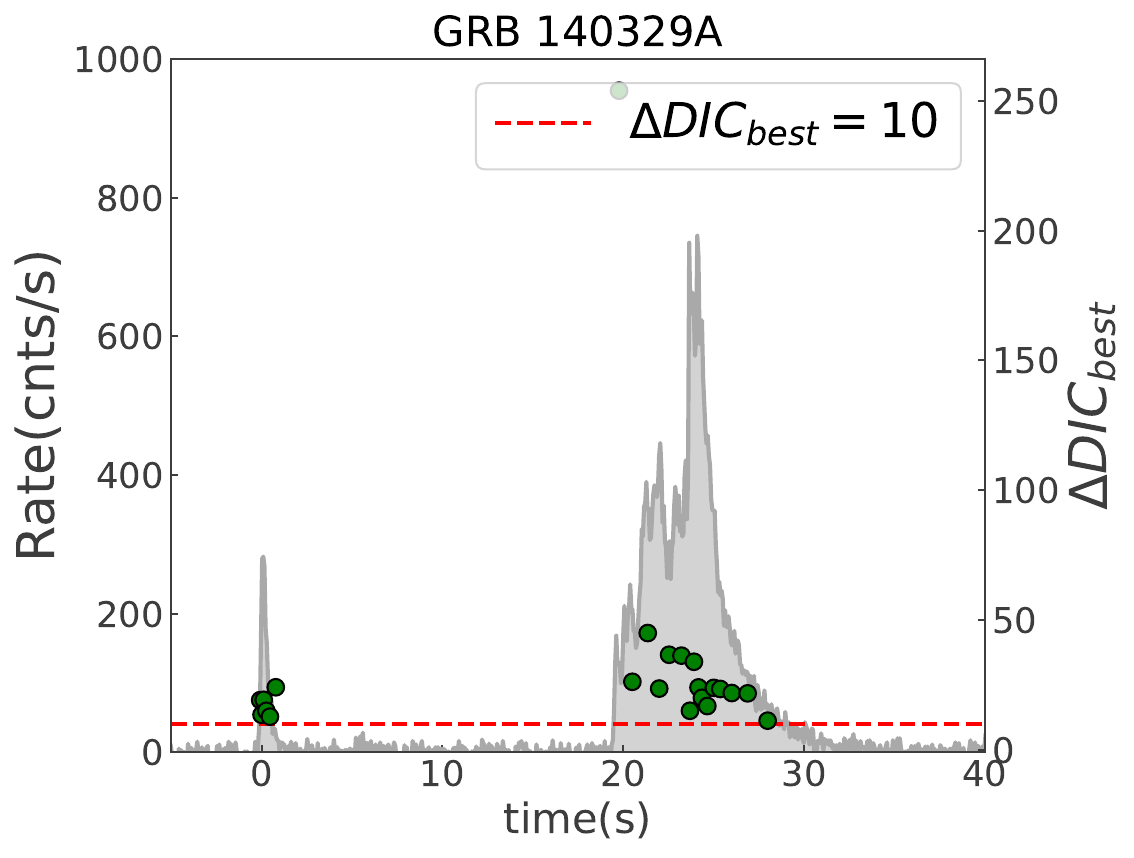}}
\hspace{0in}    
 \caption{The evolution of  $\Delta DIC_{best}$ over time  for GRB 140329A. The gray shading is the light curve, and the red dotted line indicates $\Delta DIC_{best}$ =10.}
   \label{Fig1} 
   \end{figure}

Taking the precursor of GRB 140329A as an example, this phase is divided into 6 time bins, as illustrated in Figure~\ref{Fig1}. The results of our spectral fitting, along with the goodness of fit values (PGSTAT/dof), are presented in Table \ref{table1}. We initially employ the Band and CPL models to fit the spectrum. As shown in Figure~\ref{Fig1} and Table \ref{table1}, for the first time-resolved spectrum of the precursor (-0.12s to -0.03s), the DIC for the Band model is 95.24 and the DIC for the CPL model is 100.21. $\Delta DIC =DIC_{Band} - {DIC}_{CPL} = -4.97$, indicating positive evidence that the Band model is the better fit for the first time-resolved spectrum of the precursor. Subsequently, by introducing the thermal component, we fit the spectra using Band+BB and CPL+BB models. The DIC for Band+BB is 80.97, and for CPL+BB, it is 76.08. $\Delta$DIC = $DIC_{Band+BB}$ -${DIC}_{CPL+BB}$=$4.89$, providing positive evidence in support of the CPL+BB model. Finally, by comparing the DIC values of the two best-fit models, we can determine whether there is a thermal component. The difference in DIC values is $\Delta DIC_{best} = DIC_{best2} - DIC_{best1} = 19.16$, where $DIC_{best2}$ and $DIC_{best1}$ refers to the $DIC_{best}$ value of the best-fit model without and with thermal component, respectively. This indicates strong evidence supporting the CPL + BB model, suggesting the presence of a significant thermal component in this time bin. As shown in Figure~\ref{Fig1} and Table \ref{tab:thermal component} for GRB 140329A, the $\Delta DIC_{best}$ for all time bins in both the precursor and the main burst exceeds 10. Therefore, both the precursor and the main burst exhibit clear evidence of thermal components. Applying the aforementioned method to the remaining $17$ GRBs, the results of thermal component analysis are illustrated in Table \ref{tab:thermal component} and presented in Figure \ref{FigA1} in the appendix. Among them, there is strong evidence indicating that the gold sample exhibit a fundamental similarity in the spectral resolution of precursors and main bursts, both of which contain thermal components. The silver sample is roughly similar to the gold one, except for the precursors of GRB 130815660. Additionally, the copper sample also exhibits similarities with the gold sample. Furthermore, the precursors of the gold sample in Table \ref{tab:thermal component} and Figure \ref{FigA1} not only include spectra with $S\geq15$, but also spectra with $15 \geq S \geq 5$. This indicates that the $15 \geq S \geq 5$ spectra of the gold sample's precursors also exhibit significant thermal components. Hence, this further illustrates the presence of thermal components in the precursors.
  Therefore, almost all precursors and main bursts contain evident thermal components, accounting for 94.4$\%$ (17/18) of the total sample.

\begin{table}
\begin{longtable}{ccccccccc}
\caption{Spectral Fitting Result of the Precursor of the GRB 140329A }
\label{table1}\\
\hline
\hline
$t_{start}-t_{end}$ & $S$ & Model & $\alpha$ & $\beta$ & $E_p$ & kT & DIC& pgstat/dof\\
 s& & & & &$keV$ &$keV$& &\\
\hline
-0.12$-$-0.03 & 7 & band & $-1.24^{+0.29}_{-0.15}$ & $-1.94^{+0.09}_{-0.6}$ & $176.19^{+308.06}_{-52.79}$ & ... & 95.24 & 0.24 \\
 & 7 & cpl & $-1.27^{+0.17}_{-0.22}$ & ... & $268.91^{+343.71}_{-63.97}$ & ... & 100.21 & 0.24 \\
 & 7 & band+bb & $-1.24^{+0.29}_{-0.16}$ & $-1.96^{+0.12}_{-0.57}$ & $117.71^{+216.75}_{-25.84}$ & $102.59^{+6.54}_{-65.95}$ & 80.97 & 0.24 \\
 & 7 & cpl+bb & $-1.27^{+0.27}_{-0.25}$ & ... & $100.41^{+428.7}_{-0.95}$ & $117.49^{+3.65}_{-78.27}$ & 76.08 & 0.24 \\
-0.03$-$0.03 & 15 & band & $-0.5^{+0.23}_{-0.16}$ & $-2.49^{+0.25}_{-0.39}$ & $217.57^{+37.67}_{-37.73}$ & ... & 58.84 & 0.11 \\
 & 15 & cpl & $-0.54^{+0.2}_{-0.15}$ & ... & $238.67^{+37.46}_{-30.06}$ & ... & 55.12 & 0.11 \\
 & 15 & band+bb & $-0.54^{+0.23}_{-0.16}$ & $-2.43^{+0.19}_{-0.4}$ & $214.92^{+26.71}_{-48.96}$ & $71.32^{+27.24}_{-27.88}$ & 45.43 & 0.11 \\
 & 15 & cpl+bb & $-0.55^{+0.21}_{-0.15}$ & ... & $235.29^{+27.57}_{-54.76}$ & $57.41^{+50.13}_{-16.65}$ & 41.58 & 0.11 \\
0.03$-$0.21 & 40 & band & $-0.26^{+0.1}_{-0.11}$ & $-2.57^{+0.15}_{-0.3}$ & $218.68^{+20.08}_{-16.09}$ & ... & 639.33 & 1.72 \\
 & 40 & cpl & $-0.4^{+0.08}_{-0.07}$ & ... & $257.73^{+14.19}_{-14.3}$ & ... & 638.94 & 1.73 \\
 & 40 & band+bb & $-0.26^{+0.14}_{-0.09}$ & $-2.52^{+0.11}_{-0.33}$ & $209.84^{+21.47}_{-23.36}$ & $90.72^{+14.75}_{-49.52}$ & 625.03 & 1.73 \\
 & 40 & cpl+bb & $-0.24^{+0.07}_{-0.2}$ & ... & $197.9^{+61.92}_{-12.08}$ & $153.81^{+7.83}_{-107.3}$ & 619.67 & 1.72 \\
0.21$-$0.32 & 23 & band & $-0.12^{+0.15}_{-0.18}$ & $-2.69^{+0.15}_{-0.24}$ & $109.14^{+9.55}_{-8.58}$ & ... & 268.01 & 0.68 \\
 & 23 & cpl & $-0.29^{+0.18}_{-0.13}$ & ... & $125.47^{+8.53}_{-9.15}$ & ... & 266.74 & 0.68 \\
 & 23 & band+bb & $-0.16^{+0.28}_{-0.12}$ & $-2.74^{+0.16}_{-0.29}$ & $110.55^{+5.47}_{-12.63}$ & $64.92^{+10.92}_{-34.44}$ & 255.12 & 0.68 \\
 & 23 & cpl+bb & $-0.23^{+0.22}_{-0.14}$ & ... & $117.14^{+11.02}_{-12.88}$ & $90.11^{+11.18}_{-47.93}$ & 251.7 & 0.68 \\
0.32$-$0.6 & 19 & band & $-0.91^{+0.25}_{-0.18}$ & $-2.2^{+0.15}_{-0.34}$ & $104.75^{+27.85}_{-20.63}$ & ... & 731.71 & 2.02 \\
 & 19 & cpl & $-1.05^{+0.13}_{-0.15}$ & ... & $133.55^{+27.92}_{-14.99}$ & ... & 739.11 & 2.02 \\
 & 19 & band+bb & $-0.93^{+0.29}_{-0.16}$ & $-2.24^{+0.15}_{-0.35}$ & $102.07^{+20.67}_{-23.05}$ & $60.94^{+9.4}_{-33.08}$ & 718.91 & 2.03 \\
 & 19 & cpl+bb & $-0.9^{+0.13}_{-0.24}$ & ... & $97.57^{+43.98}_{-9.95}$ & $108.98^{+6.5}_{-72.67}$ & 719.39 & 2.03 \\
0.6$-$0.97 & 7 & band & $-0.85^{+0.44}_{-0.34}$ & $-1.87^{+0.02}_{-0.51}$ & $62.21^{+82.51}_{-13.37}$ & ... & 761.66 & 2.21 \\
 & 7 & cpl & $-1.24^{+0.3}_{-0.2}$ & ... & $166.38^{+168.5}_{-50.6}$ & ... & 798.77 & 2.21 \\
 & 7 & band+bb & $-0.9^{+0.53}_{-0.12}$ & $-2.09^{+0.16}_{-0.44}$ & $52.91^{+27.83}_{-11.99}$ & $82.82^{+3.31}_{-40.15}$ & 775.97 & 2.22 \\
 & 7 & cpl+bb & $-0.83^{+0.3}_{-0.45}$ & ... & $57.54^{+108.04}_{-11.99}$ & $90.51^{+5.43}_{-48.82}$ & 737.55 & 2.21 \\
\hline
\end{longtable}
\end{table}

\begin{deluxetable*}{cccccccc}
\tablenum{2}
\tablecaption{Results of GRBs with Candidate Precursors and Thermal Component Analysis Results\label{tab:thermal component}}
\tablewidth{0pt}
\tablehead{
\colhead{GRB(sample)} & \colhead{z} & \colhead{Detector} & \colhead{$T_{90}$}  &
 \colhead{time period} & \colhead{time period} & \colhead{thermal component } & \colhead{thermal component } \\
\colhead{} & \colhead{} & \colhead{} & \colhead{(s)} &
\colhead{precursors(s)} & \colhead{main bursts(s)} & \colhead{precursor} & \colhead{main bursts}
}
\
\startdata
LGRBs& & & & & & \\
\hline
GRB 130427A(G)&0.34&n6,n9,na,b1&138.24&$-0.1\sim2.5$&$3\sim12.5$&100\%(11/11)&100\%(43/43)\\
GRB130720582(G)&…&n9, na,nb,b1&199.17&$-5\sim50$&$90\sim186$&100\%(12/12)&100\%(17/17)\\
GRB130815660(S)&…&n3,n4,n5,b0&37.89&$-1\sim6.5$&$30\sim45$&33.3\%(1/3)&100\%(6/6)\\
GRB 140329A(G)&…&n8,nb,b1&21.25&$-1\sim3$&$17\sim32$&100\% (6/6)&100\% (16/16)\\
GRB 150330A(G)&0.939&n1,n2,n5,b0&153.86&$-1\sim12$&$123\sim155$&100\%(7/7)&100\% (28/28)\\
GRB 151227B(S)&…&n1,n2,n5,b0&42.75&$-2\sim5$&$20\sim46$&100\%(5/5)&100\%(22/22)\\
GRB 160225B(S)&…&n7,n8,nb,b1&64.26&$-5\sim13$&$45\sim47$&100\% (4/4)&100\%(4/4)\\
GRB 160509A(S)&1.17&n0,n1,n3,b0&369.67&$-1.5\sim6$&$7\sim30$&100\%(5/5)&100\% (22/22)\\
GRB 160625B(G)&1.406&n6,n7,n9,b1&453.39&$-1\sim3$&$185\sim225$&100\% (6/6)&100\% (46/46)\\
GRB 160821A(S)&…&n6,n7,n9,b1&43.0&$-5\sim50$&$115\sim163$&80.0\%(4/5)&100\% (38/38)\\
GRB 180416A(S)&…&n0,n1,n3,b0&103.43&$-5\sim10$&$25\sim125$&100\% (4/4)&100\% (11/11)\\
GRB 180728A(S)&0.117&n3,n6,n7,b1&6.4&$-2\sim5$&$9\sim20$&100\%(3/3)&100\%(19/19)\\
GRB 190829A(S)&0.0785&n6,n7,n9,b1&59.39& $-1\sim10$&$47\sim65$&100\%(4/4)& 85.7\%(6/7)\\
GRB 210801A(S)&…&n9,na,nb,b1&13.82&$-2\sim5$&$6\sim16$&100\%(3/3)&100\%(9/9)\\
GRB 211211A(S)&0.0763&n2,na,b0&34.31&$-0.5\sim0.25$&$1\sim12$&100\% (3/3)&98.4\%(61/62)\\
GRB 230307A(G)&0.065&na,b1&34.56&$-0.65\sim0.4$&$0.7\sim2.67$and $7.22\sim18.3$&100\% (3/3)&100\% (73/73)\\
\hline
SGRBs& & & & & & \\
\hline
GRB 180703B(G)&…&n0,n1,n3,b0&1.54&$-0.9\sim0.6$&$0.9\sim2$&100\%(5/5)&100\%(6/6)\\
GRB 140209A(C)&…&na,n9,b1&1.41&$-0.5\sim1$&$1.1\sim4.1$&100\%(3/3)&100\% (7/7)\\
GRB130310840&…&n9,na,nb,b1&16.0&$-0.2\sim1$&$4\sim6.5$&…&…\\
GRB100717372&…&n7,n8.nb,b1&5.952&$-1\sim0$&$3\sim5$&…&…\\
GRB081216531&…&n7,n8,nb,b1&0.768&$-0.14\sim0.01$&$0.5\sim1.10$&…&…\\
\enddata
\tablecomments{The main burst of GRB 230307A was excluded due to the bad time interval (time-tagged event, TTE: T0 + [3.00, 7.00] s) of GBM caused by pulse pileup \cite{2023GCN.33551....1D}. The redshift of the sample can be found at this website, please see \url{https://gcn.gsfc.nasa.gov/selected.html}. Other than that, the redshift of GRB 150330A can be found at \citep{2022RAA....22f5002L}. In the first column, ``G'' represents gold sample, ``S'' represents silver sample, and ``C'' represents copper sample. Furthermore, the last two columns indicate the percentage of time slices with the presence of a thermal component.}
\end{deluxetable*}

\subsection{Parameter Evolution}
To further compare the spectral properties of precursors and main bursts, we also examine the evolution of their parameters. Early studies have indicated several modes of spectral parameter evolution: the ``hard-to-soft" (h.t.s), the``flux tracking" (f.t) , the ``soft-to-hard" (s.t.h), and other chaotic evolution patterns. If both peak energy ($E_{p}$) and low-energy spectral index ($\alpha$) exhibit ``flux tracking" behavior, it is classified as a ``double tracking" mode of spectral evolution \citep{2019ApJ...884..109L}. We conduct a detailed analysis of the spectral parameters of precursors and main bursts for 18 GRBs, and the results are presented in Table \ref{tab:evolutionary pattern and correlation coefficients}.

\begin{deluxetable*}{ccccccccccc}
\tablenum{3}
\tablecaption{Evolutionary Pattern and Correlation Coefficients of the Precursor and Main Burst\label{tab:evolutionary pattern and correlation coefficients}}
\tablewidth{0pt}
\tablehead{
\colhead{GRB(sample)} & \colhead{$\alpha$} & \colhead{$E_{p}$} & \colhead{$\alpha$}  &
 \colhead{$E_{p}$} & \colhead{$\alpha-F$} & \colhead{$E_{p}-F$ } & \colhead{$E_{p}-\alpha$ } & \colhead{$\alpha-F$ } & \colhead{$E_{p}-F$} & \colhead{$E_{p}-\alpha$ }\\
\colhead{} & \colhead{p} & \colhead{p} & \colhead{m} &
\colhead{m} & \colhead{p} & \colhead{p} & \colhead{p} &
\colhead{m} & \colhead{m} & \colhead{m}
}
\
\decimalcolnumbers
\startdata
LGRBs& & & & & & & & & & \\
 \hline
GRB 130427A(G)&f.t&f.t&f.t&f.t&0.63&0.80&0.72&0.71&0.91&0.55\\
GRB130720582(G)&f.t&f.t&f.t&f.t&0.12&0.64&-0.24&0.27&0.82&-0.11\\
GRB130815660(S)&f.t&h.t.s&f.t&h.t.s&1.00&0.50&0.50&1.00&0.94&0.94\\
GRB 140329A(G)&f.t&f.t&f.t&f.t&0.40&0.80&0.20&0.51&0.85&0.44\\
GRB 150330A(G)&h.t.s.t.f.t&h.t.s.t.f.t&f.t&f.t&0.10& 0.40&0.70& 0.74&0.83&0.69\\
GRB 151227B(S)&h.t.s.t.f.t&f.t&h.t.s.t.f.t&f.t&-0.30&1.00&-0.30&0.17&0.85&0.17\\
GRB 160225B(S)&h.t.s.t.f.t&h.t.s.t.f.t&h.t.s.t.f.t&f.t&-0.20&0.40&0.80&0.20&0.90&0.10\\
GRB 160509A(S)&f.t&f.t&h.t.s.t.f.t&f.t &0.50&1.00&0.50&0.61&0.72&0.47\\
GRB 160625B(G)&f.t&h.t.s.t.f.t&f.t&h.t.s.t.f.t&0.90&0.70&0.60&0.77&0.67&0.29\\
GRB 160821A(S)&f.t&f.t&…&f.t&-0.50&0.70&-0.30&0.62&0.56&-0.06\\
GRB 180416A(S)&f.t&f.t&f.t&f.t&-0.20&1.00&-0.20&0.85&0.88&0.91\\
GRB 180728A(S)&s.t.h&h.t.s&h.t.s&f.t&-1.00&1.00&-1.00&0.53&0.80&2.20\\
GRB 190829A(S)&h.t.s.t.h&f.t&…&h.t.s.t.f.t&0.00&1.00&0.00&-0.07&0.64&-0.17\\
GRB 210801A(S)&h.t.s.t.h&h.t.s&f.t&f.t&-0.50&0.50&0.50&0.80&0.88&0.78\\
GRB 211211A(S)&h.t.s&h.t.s&f.t&f.t&1.00&1.00&1.00&0.32&0.82&-0.14\\
GRB 230307A(G)&h.t.s&h.t.s&f.t&f.t&1.00&1.00&1.00&0.74&0.82&0.70\\
\hline
SGRBs& & & & & & & & & & \\
 \hline
GRB 180703B(G)&f.t&f.t&f.t&f.t&1.00&0.50&0.50&0.89&0.94&0.71\\
GRB 140209A(C)&h.t.s.t.h&f.t&f.t&h.t.s&-0.50&1.00&-0.50&0.93&0.86&0.86\\
\enddata
\tablecomments{The ``p"  and ``m" denotes precursor and main burst, respectively. The columns from 2 to 5 illustrate the evolution of parameters, while columns 6 to 11 represent correlation coefficients. The first column corresponds to Table 2.}
\end{deluxetable*}

The evolution of $\alpha$ is depicted in Figure \ref{fig B1} in the appendix. The precursors and main bursts of the gold sample both have at least one $\alpha$ exceeding the synchrotron death line ($\alpha=-2/3$, \citep{1998ApJ...506L..23P}); for the silver sample, more than half (60$\%$ = 6/10) exhibit at least one $\alpha$ value that exceeds the limit of synchrotron radiation. Among the silver sample, GRB 180728A and GRB 210801A are the only two bursts that have all of their $\alpha$ values below the synchrotron death line. Bursts where the precursor's $\alpha$ exceeds the synchrotron death line but the main burst does not include: GRB 160821A, GRB 190829A. Bursts where the precursor's $\alpha$ does not exceed the synchrotron death line, but the main burst does include: GRB130815660. Overall, the vast majority of precursors and main bursts exhibit at least one $\alpha$ exceeding the synchrotron death line, accounting for 72.2$\%$ (13/18) of the total samples.

The evolution of $\alpha$ over time is illustrated in Table \ref{tab:evolutionary pattern and correlation coefficients}. Both the precursors and main bursts exhibit identical evolutionary patterns of $\alpha$, accounting for 50$\%$ (9/18) of the total samples. Among them, both the precursors and main bursts of GRB 151227B and GRB 160225B exhibit the hard to soft to flux tracking (h.t.s.t.f.t) pattern, while those of the remaining 7 bursts evolve with the f.t pattern. Overall, the $\alpha$ values of the precursors and main bursts demonstrate evolving f.t patterns, accounting for 50$\%$ (9/18) and 66.7$\%$ (12/18) of the total samples, respectively. The evolution of $E_{p}$ over time is illustrated in Table \ref{tab:evolutionary pattern and correlation coefficients} and in Figure \ref{fig B2} in the appendix. Both the precursors and main bursts show the same evolving patterns of $E_{p}$, accounting for 55.6$\%$ (10/18) of the total samples. Among them, the precursors and main bursts of GRB 130815660 exhibit the h.t.s pattern, while those of GRB 160625B exhibit the h.t.s.t.f.t pattern, with the remaining 8 bursts evolving in the f.t pattern. Overall, the $E_{p}$ values of the precursors and main bursts show evolving f.t patterns, accounting for 55.6$\%$ (10/18) and 77.8$\%$ (14/18) of the total sample, respectively. Therefore, the evolution of both spectral parameters ($\alpha$ and $E_{p}$) in precursors and main bursts primarily follows a similar pattern, with the majority evolving in the f.t pattern. This suggests a potential correlation between the precursors and main bursts. Furthermore, there are bursts with "double-track" spectral evolution mode of precursor and main burst: GRB 130427A, GRB 130720582, GRB 180416A, GRB 140329A, GRB 180703B.

\subsection{Correlation and Distribution of Spectral Parameters}
\begin{figure}[htbp]
\centering
\includegraphics [width=8.5cm,height=4.5cm]{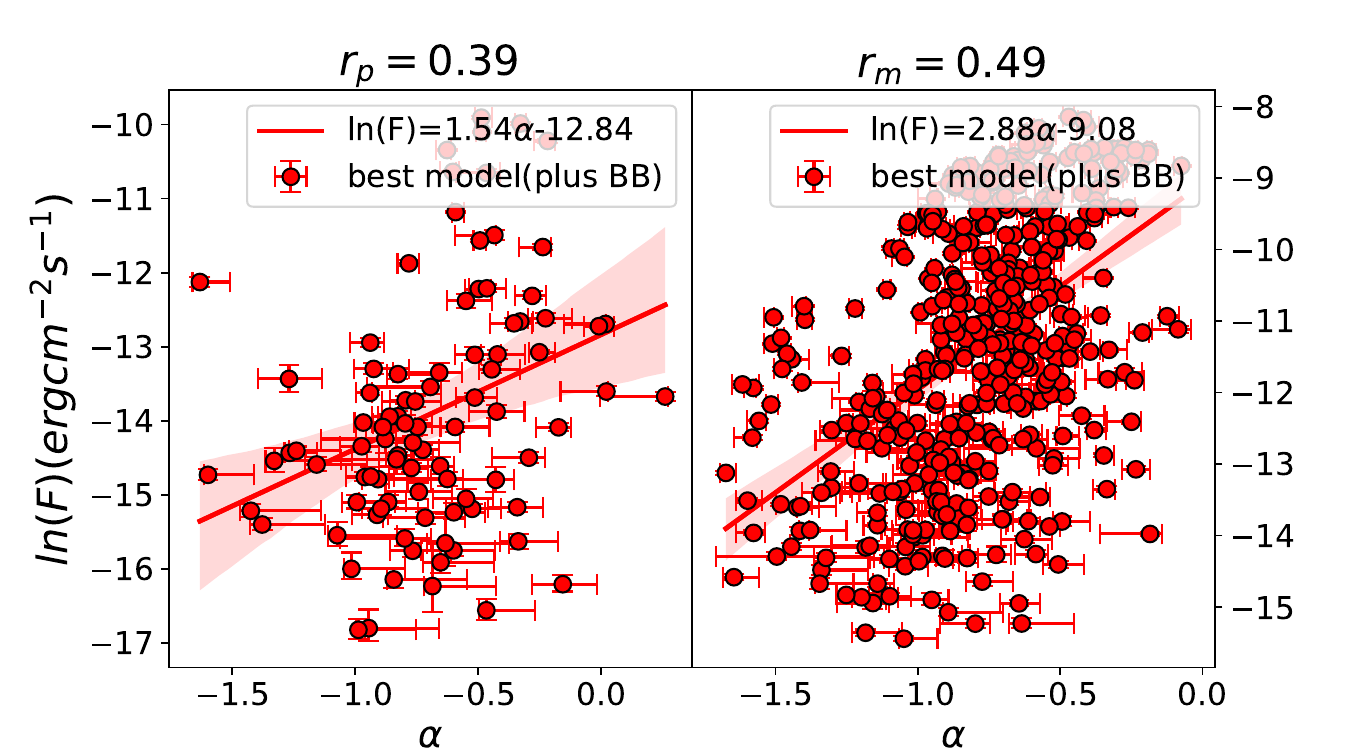}
\includegraphics [width=8.5cm,height=4.5cm]{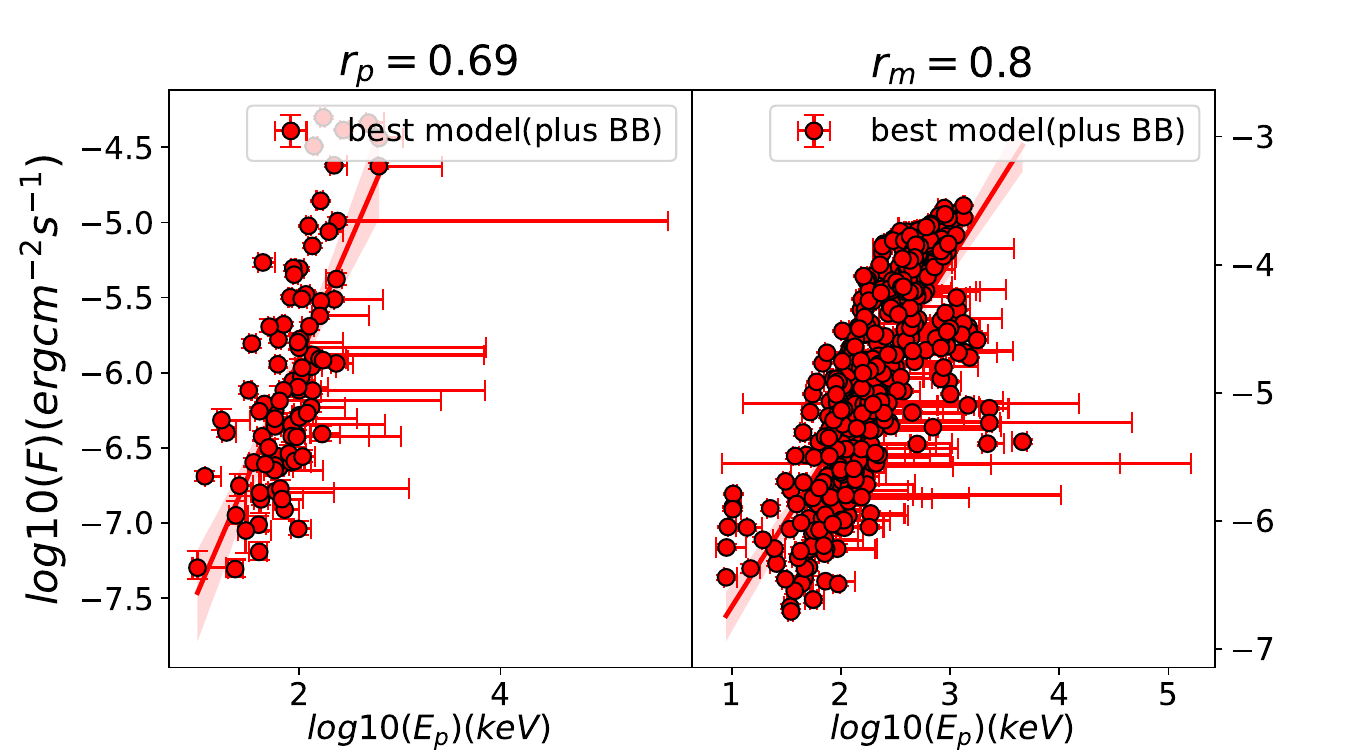}
\includegraphics [width=8.5cm,height=4.5cm]{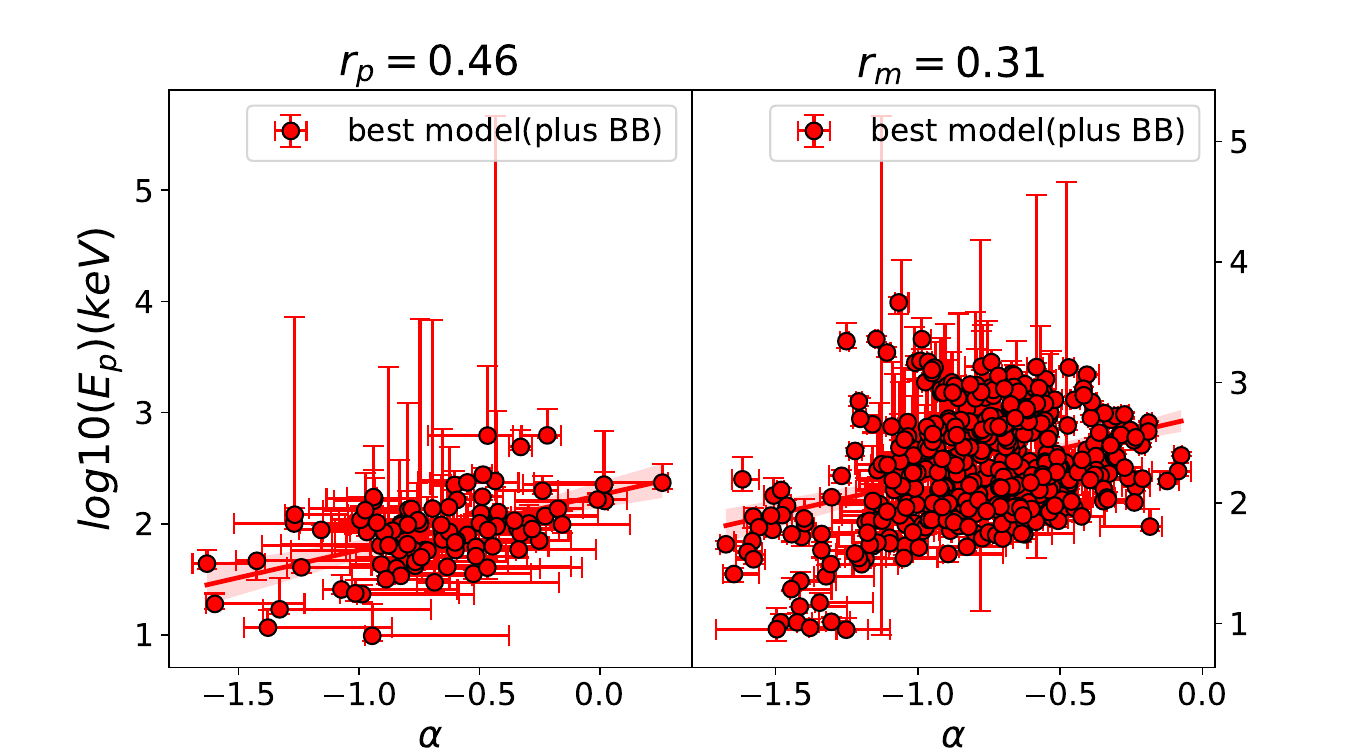}
   \figcaption{The correlations between parameters ($\alpha$, $E_{p}$, $F$) for precursors and main bursts. The left and right panels are the precursors and main bursts, respectively. The red regions represent the confidence regions of the best-fit lines. \label{fig 2}}
\end{figure}

As shown in Table \ref{tab:evolutionary pattern and correlation coefficients}, the correlation among precursor and main burst parameters ($\alpha$, $E_{p}$, $F$ (the flux)) can be observed. The $\alpha–F$ relation depicted in Figure \ref{fig C1} in the appendix reveals that over half of the precursors and main bursts exhibit similar correlations, with 6 in the gold sample and 5 in the silver sample, constituting 61.1$\%$ (11/18) of the total samples. There are bursts where both precursors and main bursts are positively correlated, including GRB 130427A, GRB130815660, GRB 140329A, GRB 160509A, GRB 160625B, GRB 230307A, GRB 180703B, and there is no correlation observed for both precursors and main bursts in GRB130720582, GRB 151227B, GRB 160225B and GRB 190829A. \cite{2019MNRAS.484.1912R} employed the function \(F(\alpha) = Ne^{k\alpha}\) to describe the $\alpha-F$ relation, where $N$ is a normalization constant. The obtained median value for $k$ is approximately $3$. Similarly, we also use this function to fit the relationship between these two parameters, as shown in Figure \ref{fig C1}. We obtain the median values of $k$ for precursors is $0.67$, and for main bursts is $2.98$. The overall sample is displayed in Figure \ref{fig 2}, with correlation coefficients of $0.39$ for precursors and $0.49$ for main bursts.

The $E_{p}- F$ relation is shown in Figure \ref{fig C2} in the appendix. The $E_p-F$ relation in both the precursor and main burst exhibits a positive correlation in all samples. Notably, the precursors and main bursts show at least a moderate positive correlation. The overall sample is displayed in Figure \ref{fig 2}, with correlation coefficients of $0.69$ for precursors and $0.8$ for main bursts.

The $E_{p} - \alpha$ relation is shown in Figure \ref{fig C3} in the appendix. Over half of the precursors and main bursts exhibit similar correlations in the $E_{p} - \alpha$ relation for all samples, constituting 61.1$\%$ (11/18) of the total sample. The bursts demonstrating a positive correlation between \(E_{p}\) and \(\alpha\) for both precursors and main bursts include GRB 130427A, GRB 130815660, GRB 150330A, GRB 160509A, GRB 210801A, GRB 230307A and GRB 180703B. Bursts with no correlation in both are GRB 130720582, GRB 151227B, GRB 160821A and GRB 190829A. The overall sample is shown in Figure \ref{fig 2}, with a correlation coefficient of $0.46$ for precursors and $0.31$ for main bursts.

\subsection{Distributions of Spectral Parameters}
\begin{figure}[htbp]
\centering
\includegraphics [width=12cm,height=6cm]{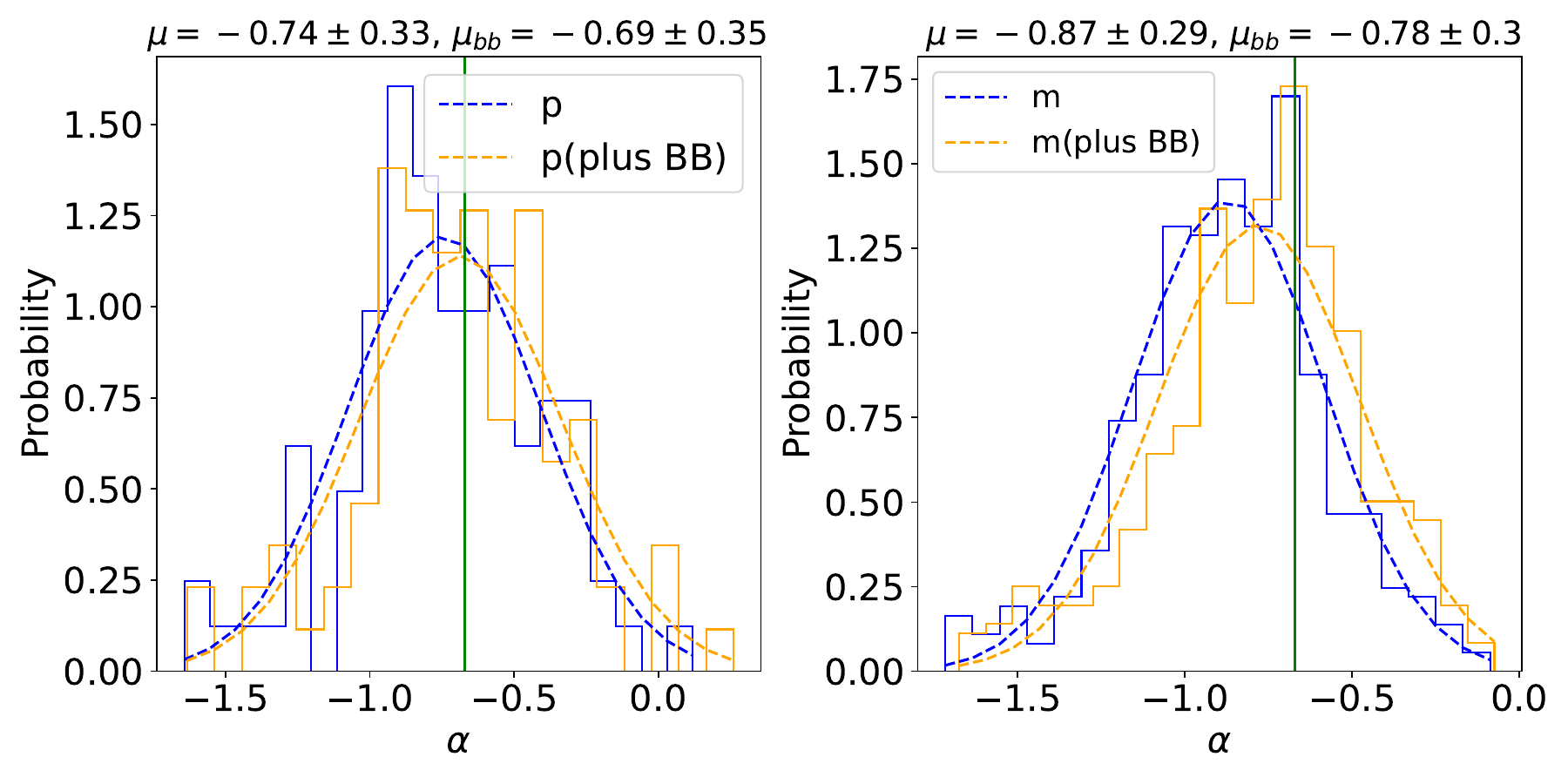}
\includegraphics [width=12cm,height=6cm]{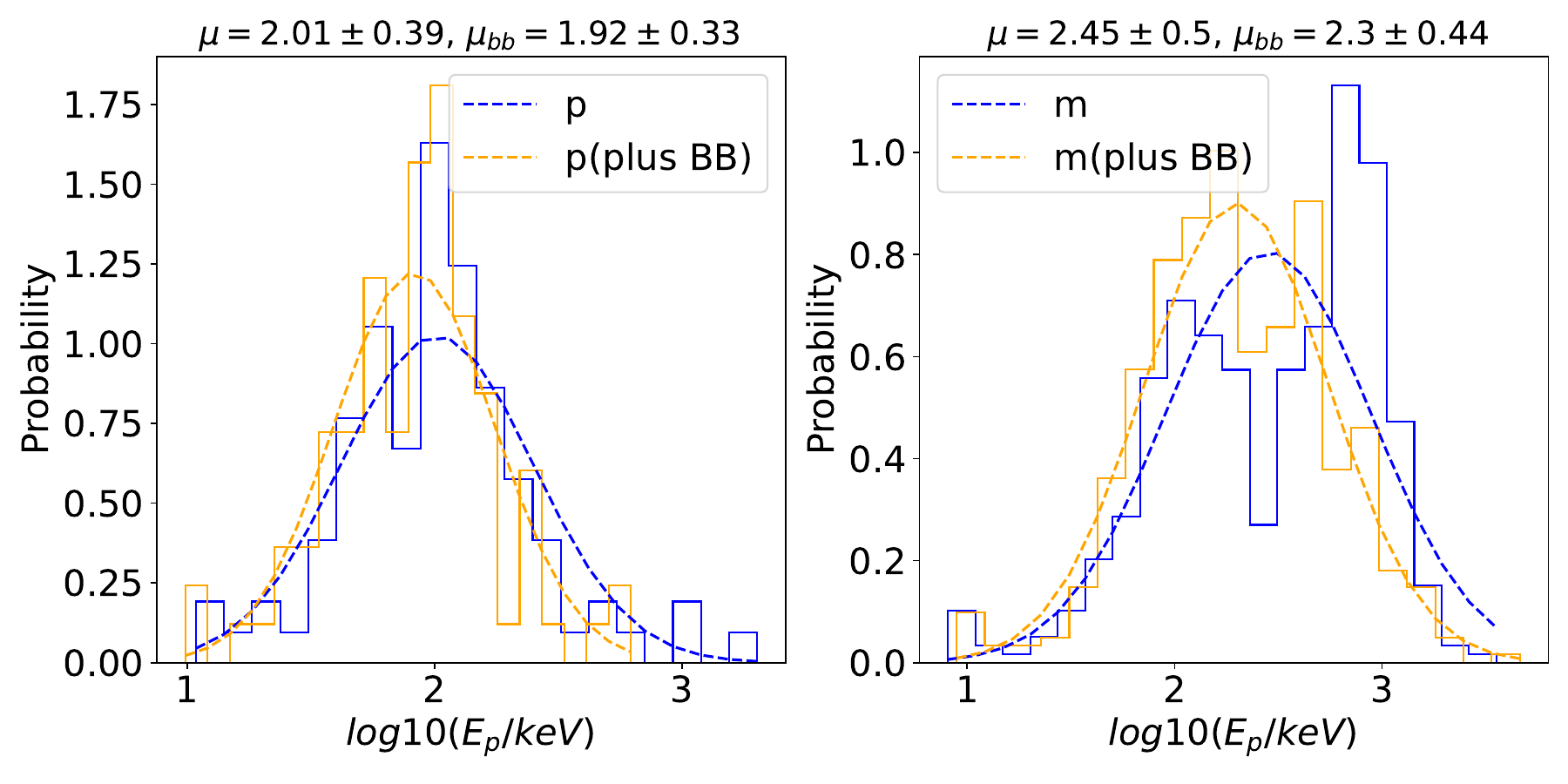}
   \figcaption{The upper panel displays the distribution of $\alpha$, while the lower panel shows the distribution of $E_{p}$. The left and right sides of the figure correspond to the precursors and the main bursts, respectively. The green solid line represents $\alpha=-0.67$. \label{fig 3}}
\end{figure}

In the upper panel of Figure \ref{fig 3}, we obtain the distributions of $\alpha$ and $E_p$ for precursors and main bursts in the overall sample. We fit them with Gaussian functions to derive the corresponding mean and standard deviation values. For precursors, the average value of $\alpha$ is $-0.74 \pm 0.33$, and after adding the BB component, $\alpha = -0.69 \pm 0.35$. In the case of main bursts, $\alpha = -0.87 \pm 0.29$, and after adding the BB component, $\alpha = -0.78 \pm 0.3$. From the statistical results, we observe that the values of $\alpha$ of the precursors are harder than those of the main bursts. Moreover, the proportion of the time-resolved spectra exceeding the ``synchrotron line" is greater for the precursors than for the main bursts. Additionally, we also find that after adding of the BB component in both cases $\alpha$ tends to shift $\alpha$ towards ``harder" values.

In the lower panel of Figure \ref{fig 3}, we present the distribution of the peak energy $E_p$ in the overall sample. Similarly, we use Gaussian functions to fit the distributions of the precursors and the main bursts. For the precursors, we obtain $\log(E_p) = 2.01 \pm 0.39$, and after incorporating the BB component, $\log(E_p) = 1.92 \pm 0.33$. In the case of main bursts, $\log(E_p) = 2.45 \pm 0.5$, and after adding the BB component, $\log(E_p) = 2.3 \pm 0.44$. We note that the $E_p$ values for the majority of the precursors are lower than those of the main bursts, and when both are fitted with the BB component, the $E_p$ values moderately decrease. 

\section{Photosphere radiation parameters\label{Sections5}}
GRB jets primarily undergo two acceleration mechanisms: thermal and magnetic. The former is associated with a hot fireball and progresses rapidly, while the latter is related to outflows dominated by Poynting flux and tends to progress relatively slowly \citep{2015ApJ...801..103G}. In this section, we use empirical relationships to constrain the outflow properties of thermal pulses, aiming to examine the radiative sources of the photosphere for precursors and main bursts. \cite{2007RSPTA.365.1171P} developed a method utilizing thermal radiation to determine the initial size \(r_0\) and Lorentz factor \(\Gamma\) of GRB fireballs. In this section, we employ this method to estimate \(r_0\) and \(\Gamma\) for each time-resolved spectrum. Additionally, it allows us to derive the effective transverse size $\Re$ of the radiative region, the photospheric radius \(r_{\text{ph}}\), and the saturation radius \(r_{\text{s}}\) \citep{2009ApJ...702.1211R}.

\subsection{Parameter $\Re$}
In the spherically symmetric scenario, \cite{2007RSPTA.365.1171P} determined the ratio $\Re$ of the observed quantity \(F_{\text{BB}}\) to the temperature \(T\) (BB temperature), where \(r_{\text{ph}} > r_{\text{s}}\). This ratio can be obtained through the following equation:
\begin{equation}
\Re  = {\left( {\frac{{F_{BB}^{{\rm{ob}}}}}{{\sigma {T^{o{b^4}}}}}} \right)^{1/2}} = \left( {1.06} \right)\frac{{{{\left( {1 + z} \right)}^2}}}{{{d_L}}}\frac{{{r_{ph}}}}{\Gamma },
\end{equation}
here, \(\sigma_{\text{SB}}\) is the Stefan-Boltzmann constant, $z$ is the redshift and $d_{L}$ is the photometric distance. For bursts without detected redshift, we assume an average redshift \(z = 1\) for the calculation \citep{2015ApJ...813..127P}.

The time evolution of $\Re$ of the precursors and the main bursts is illustrated in Figure \ref{fig D1} in the appendix, and their respective averages are presented in Table \ref{tab:mean values}. The average values of $\Re$ for the two are \((5.69 \pm 1.19) \times 10^{-20}\) and \((6.55 \pm 1.04) \times 10^{-20}\) respectively. The average values are completely consistent within the margin of error. Furthermore, the effective transverse size $\Re$ of the precursors and the main emission remains completely consistent with a straight line. Therefore, the $\Re$ of the precursors and the main emission are mostly roughly equivalent.

\subsection{Lorentz factor $\Gamma $}
The Lorentz factor for the gliding phase(${r_{ph}} > {r_s}$) can be given by
\begin{equation}
\Gamma  \propto {\left( {F/\Re } \right)^{1/4}}{Y^{1/4}},
\end{equation}
where $Y$ relates to the radiative efficiency of the burst which is given by
\begin{equation}
Y = \frac{{{L_0}}}{{{L_{obs,\gamma }}}},
\end{equation}
where $L_{0}$ is the total kinetic luminosity and $L_{obs,\gamma}$ is the observed gamma-ray luminosity.

The time evolution of the Lorentz factor \(\Gamma\) is also illustrated in Figure \ref{fig D2} in the appendix, and its average values are presented in Table \ref{tab:mean values}. Including all samples, the average values of $\Gamma$ for the precursor and main burst are  $(288 \pm 35.8) Y^{1/4}$ and $(450 \pm 32.4) Y^{1/4}$, respectively. The Lorentz factor $\Gamma$ values of the main burst are greater than those of the precursor. From Figure \ref{fig D2}, it can be noted that there is generally an increase in the Lorentz factor $\Gamma$ as the burst transitions from the precursor to the main emission.

\subsection{Parameter $r_{0}$, $r_{s}$, $r_{ph}$}
The initial radius $r_{0}$ represents the radius at which the jet begins to accelerate. In cases where $r_{ph} > r_{s}$, the calculation of $r_{0}$ is given by the following formula:
\begin{equation}
{r_0} \propto {\left( {{F_{BB}}/FY} \right)^{3/2}}\Re,
\end{equation}
where the saturation radius $r_{s}$ represents the radius at which the Lorentz factor reaches its maximum, and with $r_{0}$ we can obtain an estimate of the saturation radius $r_{s}$, which is given by the following equation
\begin{equation}
{r_s} = \Gamma {r_0}.
\end{equation}
In the realm of relativistic holonomic motion, the optical depth (\(\tau\)) of the photons traversing a distance \(ds\) is expressed as:
\begin{equation}
\tau  = \int_{{r_{\text{ph}}}}^\infty  {\frac{{n{\sigma _T}}}{{2{\Gamma ^2}}}dr},
\end{equation}
here, \({\sigma _T}\) signifies the Thompson cross-section, \(n\) is the electron number density, and \(ds = \frac{{(1 - \beta \cos \theta )dr}}{{\cos \theta }}\) with \(\theta = 0\). Assuming a constant Lorentz factor, the photosphere radius is determined by the equation (\(\tau = 1\)):
\begin{equation}
{r_{\text{ph}}} = \frac{{{L_0}{\sigma _T}}}{{8\pi {m_p}{c^3}\Gamma _{\text{ph}}^3}},
\end{equation}
where \({L_0}\) represents the total kinetic luminosity, defined as \({L_0} = 4\pi d_L^2Y{F_{\text{tot}}}\), \(d_L\) is the luminosity distance, and \(F_{\text{tot}}\) is the observed \(\gamma\)-ray flux.

The evolution of $r_0$, $r_s$, and $r_{\text{ph}}$ over time for all samples is depicted in Figure \ref{fig D3} in the appendix, with their respective average values summarized in Table \ref{tab:mean values}. The initial radii, $r_0$, of the precursor and the main burst exhibit average values of $(1.22 \pm 0.84) \times 10^7 \text{Y}^{-3/2} \text{cm}$ and $(1.44 \pm 0.57) \times 10^7\text{Y}^{-3/2}\text{cm}$, respectively. The average values of $r_0$ of the precursor and the main burst are comparable. The average values of the precursor and main burst of the saturation radius $r_s$ are $(3.41 \pm 2.24) \times 10^9 \text{Y}^{-5/4}\text{cm}$ and $(6.38 \pm 1.97) \times 10^9 \text{Y}^{-5/4}\text{cm}$, respectively. The average value of the saturation radius $r_s$ of the main burst is larger than that of the precursor. The average values of the precursor and main burst of the photospheric radius $r_{\text{ph}}$ are $(0.65 \pm 0.14) \times 10^{11} \text{Y}^{1/4} \text{cm}$ and $(1.10 \pm 0.17) \times 10^{11} \text{Y}^{1/4} \text{cm}$, respectively. The average value of the photospheric radius $r_{\text{ph}}$ of the main burst is greater compared to the precursor. The three characteristic radii ($r_{0}$, $r_s$, $r_{ph}$) at the end of the precursor and the beginning of the main emission are so close that there seems to be a smooth transitional trend for all samples.  The observational statistical characteristics suggest that the two periods are related.

\begin{deluxetable*}{ccccccccccc}
\tablenum{4}
\tablecaption{Mean Values of Photosphere Radiation Parameters \label{tab:mean values}}
\small %
\tablewidth{0pt}
\tablehead{
\colhead{GRB(sample)} & \colhead{$\Re \times 10^{-20}$} & \colhead{$\Gamma \times 10^{2}$} & \colhead{$r_0 \times 10^{7}$}  &  \colhead{$r_s \times 10^{9}$} & \colhead{$r_{\text{ph}} \times 10^{11}$} & \colhead{$\Re \times 10^{-20}$} & \colhead{$\Gamma \times 10^{-2}$} & \colhead{$r_0 \times 10^{7}$} & \colhead{$r_s \times 10^{9}$} & \colhead{$r_{\text{ph}} \times 10^{11}$}\\
\colhead{} & \colhead{p} & \colhead{p} & \colhead{p(cm)} & \colhead{p(cm)} & \colhead{p(cm)} & \colhead{m} & \colhead{m} & \colhead{m(cm)} & \colhead{m(cm)} & \colhead{m(cm)}}
\startdata
LGRBs& & & & & & & & & & \\
 \hline
GRB 130427A(G)&7.24$\pm$0.95&3.95$\pm$0.35&1.72$\pm$0.49&6.54$\pm$2.07&0.80$\pm$0.07&8.47$\pm$0.73&5.77$\pm$0.29&2.02$\pm$0.22&10.22$\pm$1.05&1.29$\pm$0.09\\
GRB130720582(G)&4.73$\pm$0.24&2.92$\pm$0.13&0.66$\pm$0.21&1.84$\pm$0.57&0.67$\pm$0.02&6.19$\pm$0.54&3.12$\pm$0.09&1.26$\pm$0.28&3.75$\pm$0.77&0.92$\pm$0.07\\
GRB130815660(S)&6.01$\pm$0.07&2.51$\pm$0.23&2.33$\pm$2.00&4.97$\pm$4.11&0.75$\pm$0.06&6.46$\pm$0.70&3.64$\pm$0.30&1.76$\pm$0.70&7.23$\pm$2.96&1.12$\pm$0.03\\
GRB 140329A(G)&7.59$\pm$2.42&4.47$\pm$0.53&0.42$\pm$0.14&2.00$\pm$0.75&1.57$\pm$0.38&6.09$\pm$0.71&5.80$\pm$0.35&2.69$\pm$0.49&14.09$\pm$2.46&1.68$\pm$0.19\\
GRB 150330A(G)&4.81$\pm$1.26&3.97$\pm$0.50&1.48$\pm$0.79&5.00$\pm$2.45&0.85$\pm$0.19&4.93$\pm$0.45&5.08$\pm$0.27&3.22$\pm$0.43&15.45$\pm$2.07&1.16$\pm$0.11\\
GRB 151227B(S)&4.84$\pm$0.27&3.48$\pm$0.29&2.79$\pm$0.96&10.22$\pm$4.15&0.84$\pm$0.11&4.50$\pm$0.43&4.43$\pm$0.22&2.64$\pm$0.51&10.98$\pm$2.13&0.94$\pm$0.09\\
GRB 160225B(S)&5.30$\pm$0.77&2.26$\pm$0.28&3.11$\pm$1.52&6.35$\pm$2.57&0.57$\pm$0.05&8.83$\pm$2.56&2.83$\pm$0.28&3.76$\pm$1.85&9.87$\pm$4.14&1.16$\pm$0.26\\
GRB 160509A(S)&2.88$\pm$0.87&4.29$\pm$0.70&1.85$\pm$1.17&6.78$\pm$4.57&0.51$\pm$0.11&5.24$\pm$0.72&6.30$\pm$0.29&2.62$\pm$0.44&15.77$\pm$2.69&1.61$\pm$0.22\\
GRB 160625B(G)&6.79$\pm$0.66&4.51$\pm$0.21&0.04$\pm$0.01&0.15$\pm$0.05&1.60$\pm$0.19&4.54$\pm$0.55&10.12$\pm$0.64&0.57$\pm$0.12&4.84$\pm$1.07&1.91$\pm$0.21\\
GRB 160821A(S)&5.09$\pm$0.94&2.25$\pm$0.20&2.64$\pm$1.43&6.09$\pm$3.19&0.54$\pm$0.08&9.11$\pm$2.08&6.80$\pm$0.43&0.80$\pm$0.38&3.51$\pm$1.36&2.30$\pm$0.45\\
GRB 180416A(S)&4.25$\pm$0.81&2.99$\pm$0.52&1.19$\pm$0.66&2.77$\pm$1.32&0.57$\pm$0.05&5.45$\pm$0.45&3.28$\pm$0.28&1.47$\pm$0.61&5.74$\pm$2.30&0.85$\pm$0.06\\
GRB 180728A(S)&7.69$\pm$0.22&0.79$\pm$0.12&0.62$\pm$0.16&0.46$\pm$0.04&0.08$\pm$0.01&8.37$\pm$1.58&1.97$\pm$0.13&0.28$\pm$0.10&0.56$\pm$0.19&0.21$\pm$0.04\\
GRB 190829A(S)&5.26$\pm$0.91&0.85$\pm$0.11&0.16$\pm$0.11&0.16$\pm$0.13&0.04$\pm$0.01&7.50$\pm$0.58&1.07$\pm$0.05&0.06$\pm$0.03&0.06$\pm$0.03&0.07$\pm$0.01\\
GRB 210801A(S)&6.13$\pm$1.94&2.39$\pm$0.09&0.28$\pm$0.04&0.66$\pm$0.10&0.74$\pm$0.25&6.22$\pm$0.56&3.69$\pm$0.33&0.25$\pm$0.08&0.86$\pm$0.30&1.08$\pm$0.08\\
GRB 211211A(S)&6.84$\pm$1.65&1.26$\pm$0.13&0.03$\pm$0.02&0.03$\pm$0.02&0.08$\pm$0.03&5.33$\pm$0.38&3.12$\pm$0.17&0.68$\pm$0.09&1.63$\pm$0.18&0.12$\pm$0.01\\
GRB 230307A(G)&8.17$\pm$1.79&2.20$\pm$0.26&0.20$\pm$0.10&0.48$\pm$0.23&0.14$\pm$0.04&7.88$\pm$0.98&3.88$\pm$0.15&0.22$\pm$0.07&0.72$\pm$0.20&0.19$\pm$0.02\\
\hline
SGRBs& & & & & & & & & & \\
 \hline
GRB 140209A(C)&4.08$\pm$1.65&2.61$\pm$0.55&1.74$\pm$0.47&4.09$\pm$0.69&0.47$\pm$0.13&6.36$\pm$0.75&5.06$\pm$0.49&1.09$\pm$0.46&6.51$\pm$2.83&1.56$\pm$0.20\\
GRB 180703B(G)&4.71$\pm$0.56&4.20$\pm$0.38&0.75$\pm$0.61&2.82$\pm$2.16&0.95$\pm$0.08&6.51$\pm$0.70&5.04$\pm$0.45&0.53$\pm$0.34&3.03$\pm$1.96&1.57$\pm$0.11\\
\enddata
\tablecomments{The table denotes ``p" as precursor and ``m" as main burst. The mean values of  photosphere radiation parameters in this table are done under the assumption Y = 1. See the text for details. The first column corresponds to Table 2.}
\end{deluxetable*}

\section{Amati Relation and Yonetoku Relation \label{Sections6}}
The Amati relation, proposed by \cite{2002A&A...390...81A}, is commonly used for the classification of GRBs. In this context, \(E_{p,z}=(1+z)E_p\) represents the rest-frame peak energy, and \(E_{\gamma,iso}\) denotes the isotropic equivalent energy. The expression for \(E_{\gamma,iso}\) is given by:

\begin{equation}
E_{\gamma,iso} = \frac{4\pi d_L^2 kS_{\gamma}}{1+z},
\end{equation}
here, \(d_L\) is the luminosity distance, \(S_{\gamma}\) is the fluence in erg \(\text{cm}^{-2}\), and \(k\) is a correction factor that adjusts the energy range in the observer frame to that in the rest frame. The correction factor \(k\) is expressed as:
\begin{equation}
k = \frac{\int_{\frac{E_1}{1+z}}^{\frac{E_2}{1+z}} EN(E) dE}{\int_{e_1}^{e_2} EN(E) dE}, 
\end{equation}
where \(E_1\) and \(E_2\) correspond to the energy range in the rest frame as 1 keV and \(10^4\) keV, respectively. \(e_1\) and \(e_2\) correspond to the energy range of Fermi-GBM as 8 keV and 40 MeV, respectively.

The Yonetoku relation, proposed by \cite{2004ApJ...609..935Y}, for the peak isotropic luminosity (\(L_{p,iso}\)) can be expressed as:

\begin{equation}
L_{p,iso} = 4\pi d_L^2 F_{\gamma} k,
\end{equation}
here, \(F_{\gamma}\) is the flux in \(\text{erg} \, \text{cm}^{-2} \, \text{s}^{-1}\), which can be obtained from spectral parameters.

\begin{figure}
  \centering
  \includegraphics[width=13cm,height=10cm]{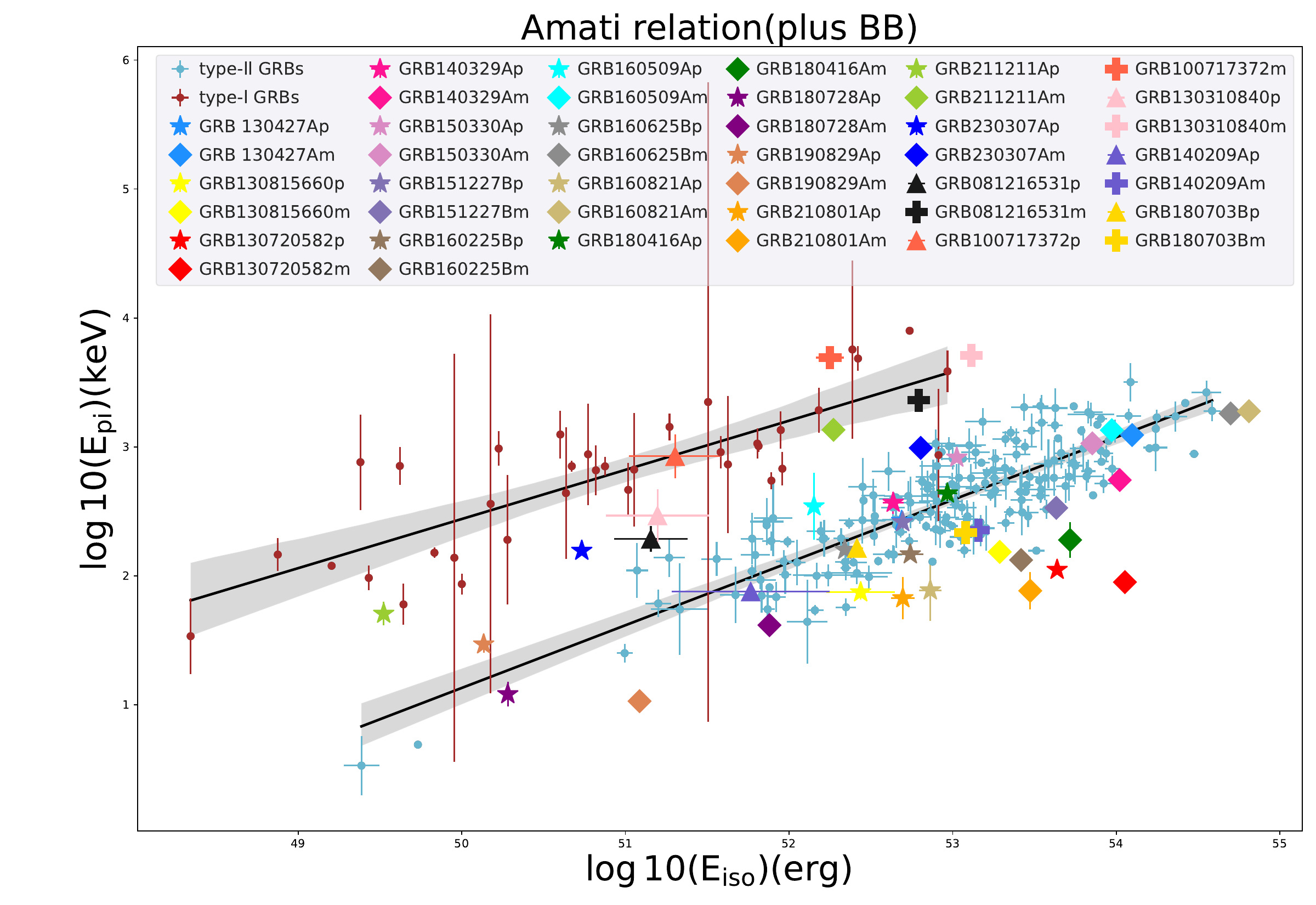}
  \includegraphics[width=13cm,height=10cm]{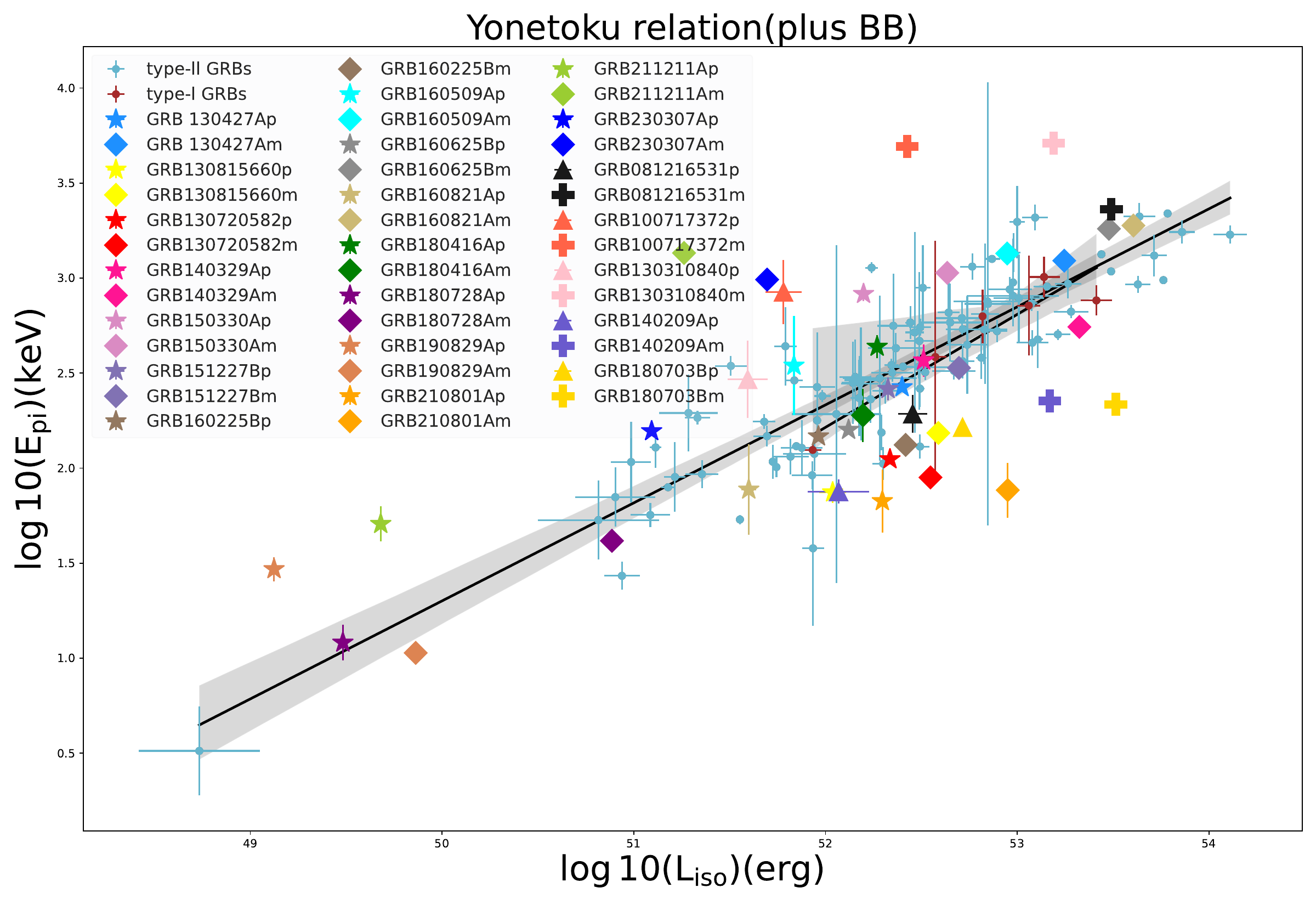}
  \caption{The Amati relation and Yonetoku relation. In the figure, maroon circles represent type-I GRBs, while blue circles represent type-II GRBs. Pentagram symbols indicate precursors of long-duration GRB samples, diamonds represent main bursts of long-duration samples, and triangles represent precursors of short-duration GRB samples. The ``+" sign represents main bursts of short-duration samples. GRBs with the same color indicate the same event. The upper panel corresponds to the Amati relation, while the lower panel represents the Yonetoku relation.} \label{fig 4}
\end{figure}

Known redshifts for GRBs are presented in Table \ref{tab:thermal component}. For GRBs with unknown redshifts, we assume a redshift of 1. The resulting Amati relation is depicted in the upper part of Figure \ref{fig 4}, and the Yonetoku relation is illustrated in the lower part of Figure \ref{fig 4}. Nearly all precursors and main bursts follow the Amati relation and are located in the same region. Additionally, both the precursors and main bursts closely follow the Yonetoku relation. 

\section{Discussion\label{Sections7}}
Performing time-resolved spectral and time-integrated spectral analyses on the precursors and main bursts of 16 LGRBs and 5 SGRBs can help examine whether their origins are the same and provide clues for explaining their physical origins. Due to the scarcity of bright precursors, we further divide the sample into gold, silver, and copper samples, finding that their results are generally similar.

\cite{2010ApJ...709L.172R} found that the fireball characteristics of GRB 090902B can be represented by a multicolor blackbody or Planck function. With the passage of time, the spectrum widens and the photosphere radius increases, suggesting a possible variation in the jet components. In fact, the fitting of a combination of blackbody and non-thermal components has shown good agreement in many bursts \citep{2011ApJ...727L..33G,2012ApJ...757L..31A,2013MNRAS.433.2739I,2015ApJ...801..103G,2016MNRAS.456.2157I}. \cite{2017IAUS..324...74Z} and \cite{2017ApJ...849...71L} both found that both the precursor and main burst of GRB 160625B contain a thermal component. Notably, \cite{2009A&A...505..569B} found that the spectral characteristics of precursors and main bursts are remarkably similar. They supported the fireball model for both precursors and main bursts, implying a common origin \citep{2010ApJ...723.1711T,2014ApJ...789..145H,2021ApJS..252...16L,2022ApJ...928..152L,10.1088/1674-4527/ad0497}. For the samples investigated in this study, except for GRB 130815660, all the precursors and the main bursts exhibit noticeable thermal components. Therefore, we believe there is a close correlation in their origins. However, the precursor of GRB 130815660 may be dominated by non-thermal component, while the main burst may be dominated by a thermal component. In the jet-cocoon model \citep{2005ApJ...629..903L}, there is interaction between the jet and its surrounding cocoon. This interaction can give rise to precursor activity. While thermal radiation exists within the cocoon, the initially ejected jet from the cocoon still contains non-thermal radiation.  Changes in jet bursts and opening angles can affect the long quiescent period \citep{2022ApJ...928..152L}. This can explain why the precursor of GRB130815660 is dominated by non-thermal radiation. Another possibility is that the signal-to-noise ratio is too low to detect the presence of significant thermal components. 

For the two short bursts GRB 180703B and GRB 140209A with \(T_{90} < 2 s\), both the precursor and the main burst contain thermal components. Among them, there is strong evidence supporting the presence of thermal components in all the time-resolved spectra of both the precursor and the main burst of GRB 180703B, with most of the time-resolved spectra having $\alpha$ exceeding the synchrotron death line. Therefore, the precursor of GRB 180703B can be explained by photospheric precursors \citep{2000ApJ...543L.129L,2020ApJ...902L..42W}. However, the number of bright short bursts with precursors is relatively limited, and it is hoped that more samples can be obtained in the future for further study.

For total sample, both the precursors and main bursts exhibit identical evolutionary patterns of $\alpha$ and $E_{p}$, accounting for 50$\%$ (9/18) and 55.6$\%$ (10/18) of the total sample, respectively. Consequently, this indicates a possible correlation between the precursors and main bursts. Additionally, the $\alpha$ values of both the precursors and main bursts exhibit evolving f.t patterns, representing 50$\%$ (9/18) and 66.7$\%$ (12/18) of the total samples, respectively. Similarly, the $E_{p}$ values of the precursors and main bursts also demonstrate evolving f.t patterns, accounting for 55.6$\%$ (10/18) and 77.8$\%$ (14/18) of the total sample, respectively. Thus, the evolution of $\alpha$ and $E_{p}$ in both precursors and main bursts is primarily governed by the f.t patterns, consistent with the findings of \cite{2021ApJS..254...35L}. The evolution of $E_{p}$ from h.t.s may come from the ICMART model, and f.t may come from internal shocks or the photosphere \citep{2012ApJ...756..112L}. If both the precursors and the main bursts are predominantly governed by thermal components, it suggests that flux tracking may be related to the photosphere. If the thermal component is minimal or absent, it may originate from internal shocks.

We examine the correlation between the parameters ($\alpha$, $E_{p}$, $F$) of the precursors and the main bursts. We use \(F(\alpha) = Ne^{k\alpha}\) to describe the $\alpha-F$ relation in all samples. We obtain the median values of \(k\) for precursors and main bursts as 0.67 and 2.98, respectively. The values for main bursts are larger than those for precursors and are closer to the results of \cite{2019MNRAS.484.1912R}. This is attributed to the trend of smaller \(k\) values when \(E_p\) is lower \cite{2019MNRAS.484.1912R}. However, in over half of the bursts (61.1$\%$=11/18), the $\alpha$–$F$ relation between precursors and main bursts is generally consistent, with the majority exhibiting a moderately strong positive correlation. \cite{2019MNRAS.484.1912R} proposed that the observed positive correlations may be indicative of heating in the sub-photospheric layer during outflows with varying entropy. For the $E_p$–$F$ relation, both the precursor and main burst exhibit a positive correlation in total samples, and both show at least a moderately strong positive correlation. Notably, \cite{2014ApJ...785..112D} proposed that in complex photospheric scenarios, the natural emergence of a positive correlation in the $E_p$–$F$ relation can occur. Over half of the precursors and main bursts (61.1$\%$=11/18) share a similar $\alpha$–$E_p$ relation, further emphasizing the close correlation between precursors and main bursts.

\cite{2009ApJ...702.1211R} showed that thermal radiation can be utilized to study the photosphere properties and investigate the physical parameters of GRB fireballs. Overall, the evolutionary trends of $\Re$ in both precursors and main bursts are generally similar, and so is their range. Transitioning from precursors to main bursts, the majority of $\Re$ exhibits an stable trend. The Lorentz factors of all bursts essentially satisfy the condition $10^2 \leq \Gamma Y^{-1/4} \leq 10^3$, which is generally consistent with the findings of \cite{2015ApJ...813..127P}. As the burst transitions from precursor to main emission, there is  typically an increase in the Lorentz factor. The Lorentz factor at the end of the precursor being smaller than at the beginning of the main emission may imply an energy accumulation process following the precursor, leading to the subsequent main burst.

For the vast majority of GRBs (88.9$\%$=16/18), the initial radii of both precursors and main bursts lie within the range of $10^6 \text{cm} \leq r_0 Y^{3/2} \leq 10^9\text{cm}$, which is consistent with the results presented by \cite{2013MNRAS.433.2739I} and \cite{2016MNRAS.456.2157I}. \cite{1986ApJ...308L..43P} and \cite{1986ApJ...308L..47G} proposed an optically thick ``fireball'' made of electron--positron plasma and photons, which gives rise to blackbody radiation from the fireball photosphere at a photospheric radius of $\sim10^{11}-10^{12}$ cm. Based on the average values presented in Table \ref{tab:mean values}, it can be observed that the radii of the photospheres \(r_{\text{ph}}\) for both precursors and main bursts are generally close to this radius in the vast majority of GRBs (77.8$\%$=14/18). These indicate that both the precursor and the main burst are most of them exhibit typical properties of photosphere radiation. Furthermore, the three characteristic radii ($r_0$, $r_s$, $r_{\text{ph}}$) at the end of the precursor and the beginning of the main emission are so similar that a smooth transition may be observed in all samples. Therefore, based on observational statistical characteristics, we speculate that the precursor and the main burst are associated and possibly originate from the same source. However, the definition of a precursor states it as an interval separated from the main emission period. We have yet to determine the reason for the existence of a smooth transitional trend between precursor and main emission.

We also analyse the time-integrated spectra of precursors and main bursts, comparing the Amati relation and the Yonetoku relation. The precursors and main bursts generally follow the Amati relation. Meanwhile, we set the redshift range from 0.01 to 5 and find that the classification of most precursors and main bursts in long and short bursts is consistent, indicating a common physical origin for the majority.
 
Interestingly, GRB 140209A and GRB 180703B (\(T_{90} < 2 s\)), in the redshift range of 0.01 to 5, both the precursor and main burst are consistently classified as long bursts. However, GRB 180703B does not exhibit extended radiation, while GRB 140209A is associated with extended radiation \citep{2020ApJ...902L..42W}. Additionally, for GRBs with $T_{90} > 2 s$, namely GRB130310840 and GRB100717372, in the redshift range from 0.01 to 5, both the precursors and main bursts of these events fall within the classification of short bursts. 

For the Yonetoku relation, for bursts with unknown redshifts, we adopted $z=1$. It can be observed that both the precursor and the main burst generally follow the Yonetoku relation. To gain a deeper understanding of the properties of both precursors and main bursts, searching for additional samples with redshift information is crucial. In addition to spectral analysis, a comparative study from the perspective of light curves is essential to further investigate.

\section{Conclusion\label{Sections8}}
This paper conducts Bayesian time-resolved spectral and time-integrated spectral analyses on precursors and main bursts using the Band (Band+BB) and CPL (CPL+BB) models. Through a comparative analysis, we derive intriguing conclusions regarding the spectral characteristics and photospheric properties of both precursors and main bursts:

1. Our observation indicates that almost all both precursors and main bursts display evident thermal components, constituting 94.4$\%$ of the total sample. Moreover, the majority of GRB precursors and main bursts exhibit $\alpha$ values that exceed the synchrotron radiation limit in all samples. Additionally, the inclusion of a BB component in both precursors and main bursts results in a harder $\alpha$ value. Furthermore, the vast majority of them (72.2$\%$) exhibit at least one low-energy spectral index $\alpha$ that exceed the limit of synchrotron radiation.

2.For all samples, whether it's $E_{p}$ and $\alpha$ values, both precursors and main bursts exhibit similar patterns, accounting for 50$\%$ and 55.6$\%$ of the total samples, respectively. Moreover, The evolution of $\alpha$ and $E_{p}$ for both precursors and main bursts is primarily characterized by a f.t pattern. Hence, this suggest a possible correlation between the precursors and main bursts.

3. Through the analysis of the correlation between spectral parameters in all samples, We find that for the $\alpha - F$ relation, over half of precursors and main bursts ( 61.1$\%$) exhibit similar correlations. For the $E_{p} - F$ relation, all precursors and main bursts show at least a moderately strong  positive correlation. Regarding the $\alpha - E_{p}$ relation, over half of the precursors and main bursts ( 61.1$\%$) exhibit comparable correlation. These suggest a potential common physical origin for these phenomena.

4. After constraining the photospheric radiation parameters ($\Re$, $\Gamma$, $r_{0}$, $r_{s}$ and ${r_{ph}}$) for both precursors and main bursts in all samples, we find the following trends: $\Re$ exhibits an stable tendency from precursors to main bursts; during the transition from precursor to main emission, there is typically a rise in the Lorentz factor; the three characteristic radii ($r_0$, $r_s$, $r_{\text{ph}}$) from the precursor to the main burst may represent a smooth transition. Consequently, this further supports a consistent origin for both precursors and main bursts, essentially indicating that most of them exhibit typical properties of photosphere radiation.

5. We perform time-integrated spectral analysis on both the precursors and main bursts and find that nearly all them are located in the same region of the Amati relation and generally follow the Yonetoku relation.

Therefore, we support that the precursor and the main burst have the same physical origin.

\section*{acknowledgements}
We would like to thank the anonymous referee for constructive suggestions to improve the manuscript.We acknowledge the use of the public data from the Fermi data archives. This work is supported by the National Natural Science Foundation of China(grant 12163007, 11763009), the Key Laboratory of Colleges and Universities in Yunnan Province for High-energy Astrophysics, National Astronomical Observatory Yunnan Normal University Astronomy Science Education Base.

\appendix
\section{The evolution of $\Delta DIC$ and the evolution and correlation of the spectral parameters}
\begin{figure}[htb]
\centering
{
\includegraphics[scale=0.3]{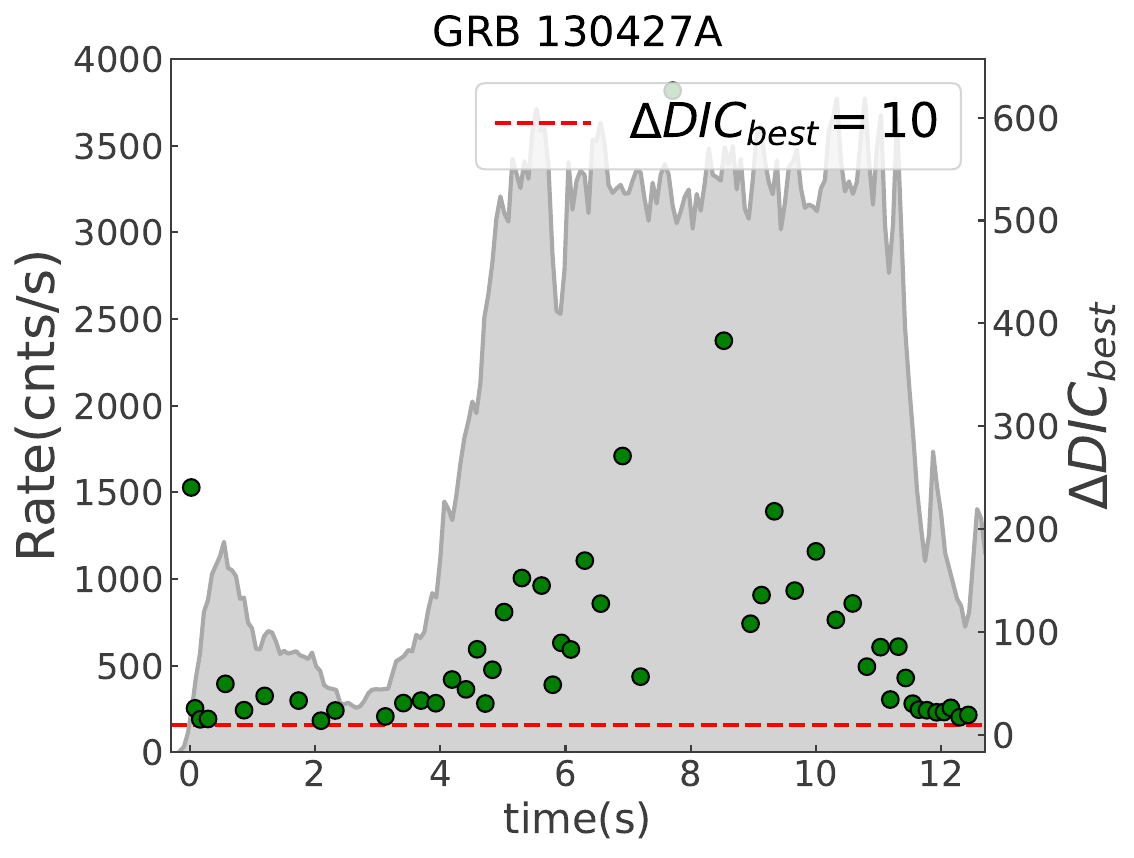}}
\hspace{0in}    
{
\includegraphics[scale=0.3]{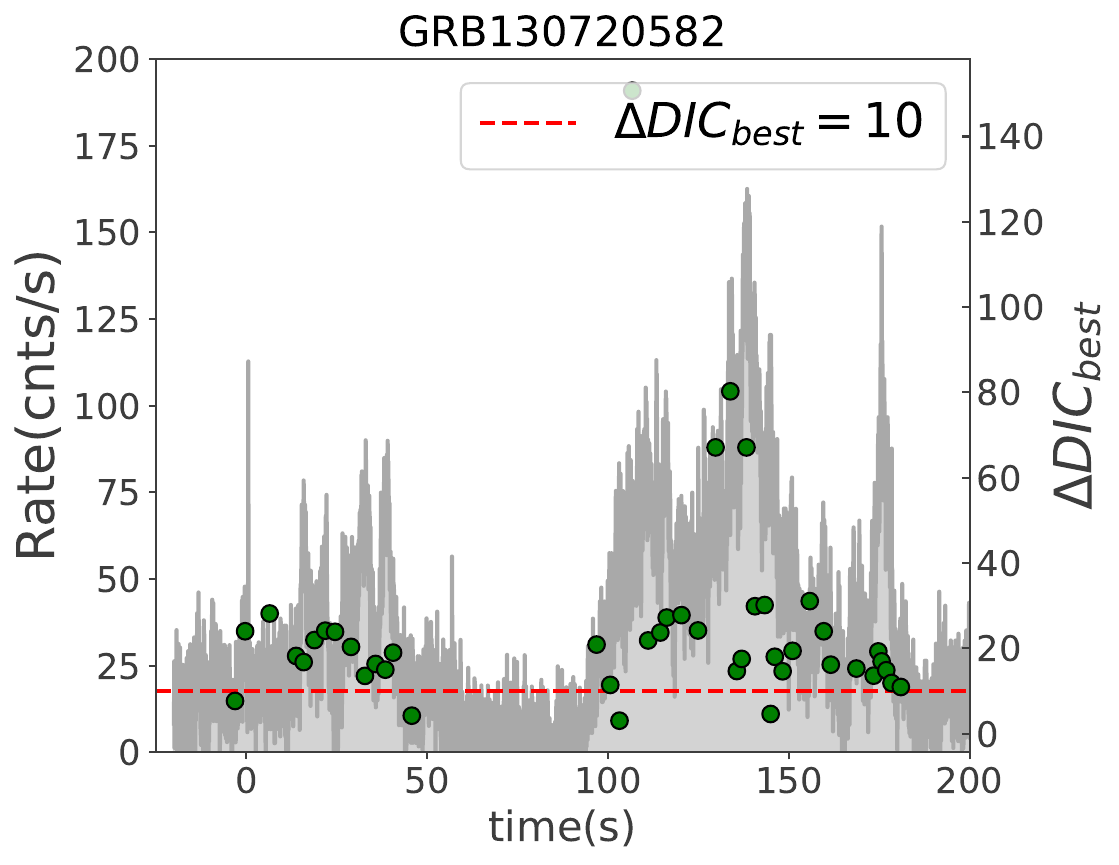}}
\hspace{0in}
{
\includegraphics[scale=0.3]{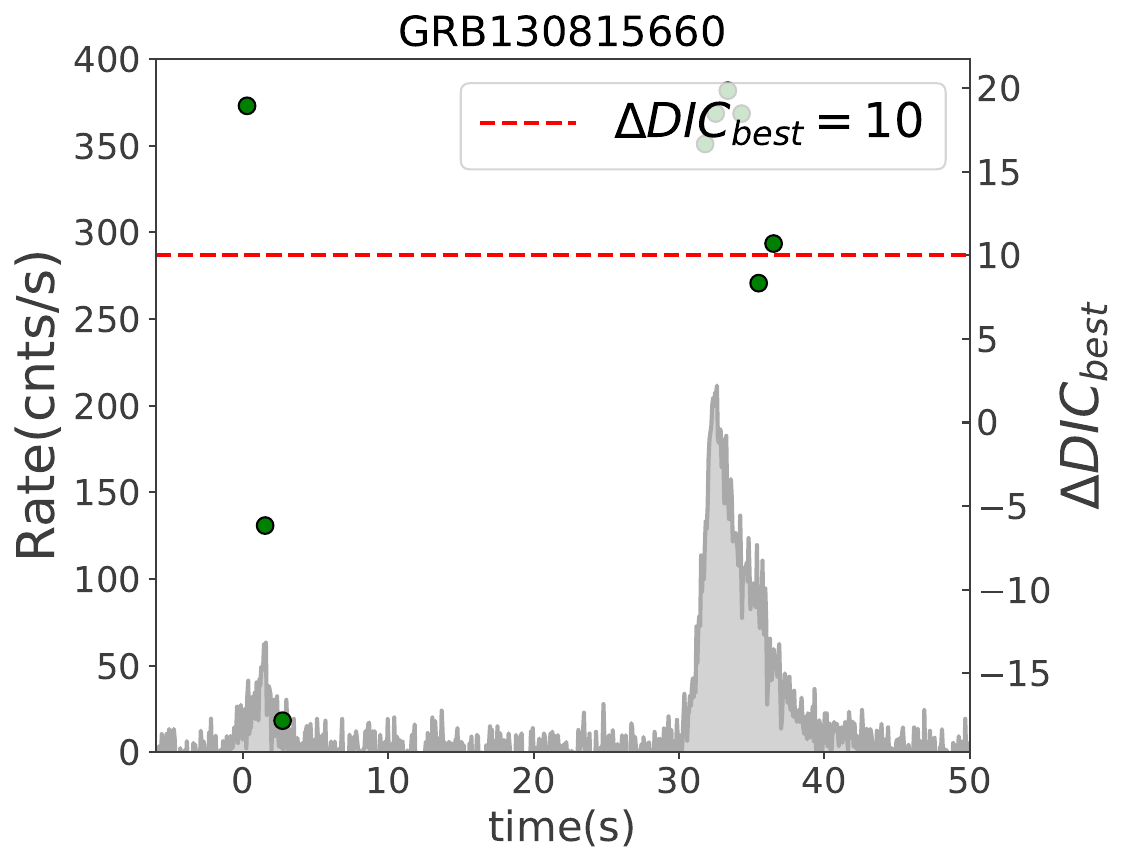}}
\hspace{0in}
{
\includegraphics[scale=0.3]{D4.pdf}}
\hspace{0in}    
{
\includegraphics[scale=0.3]{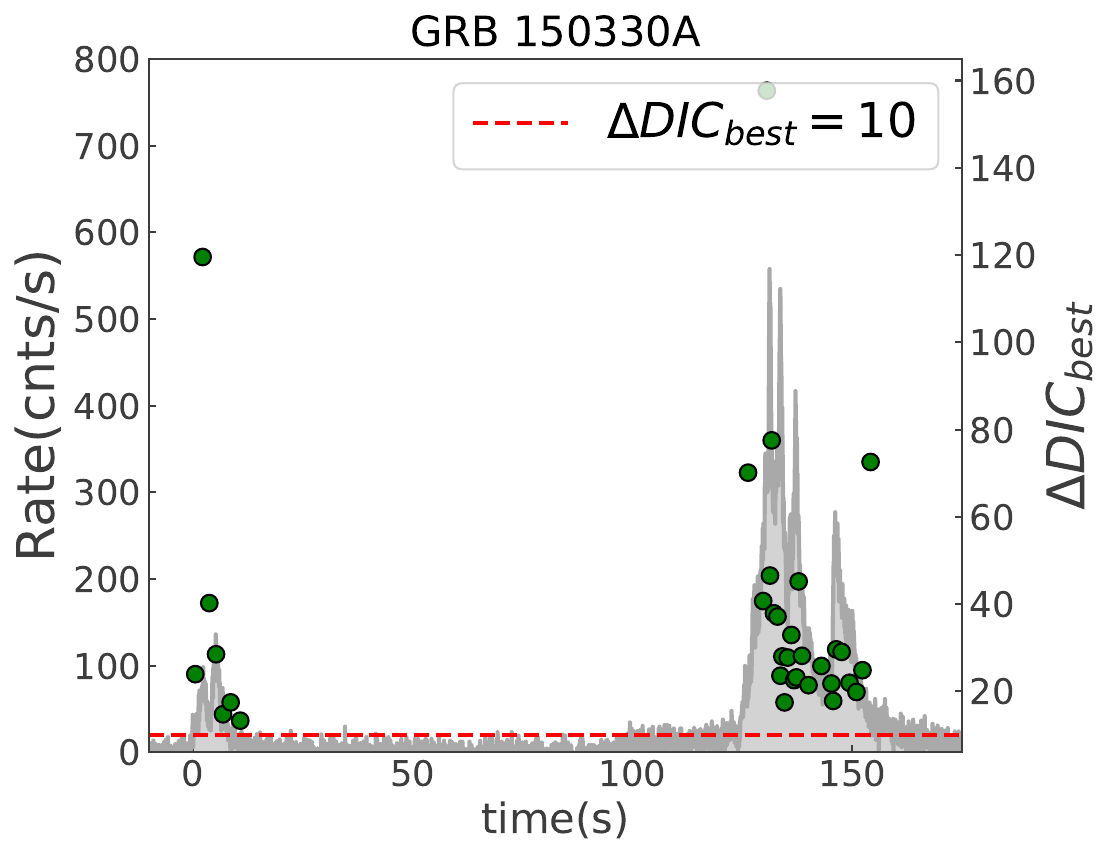}}
\hspace{0in}
{
\includegraphics[scale=0.3]{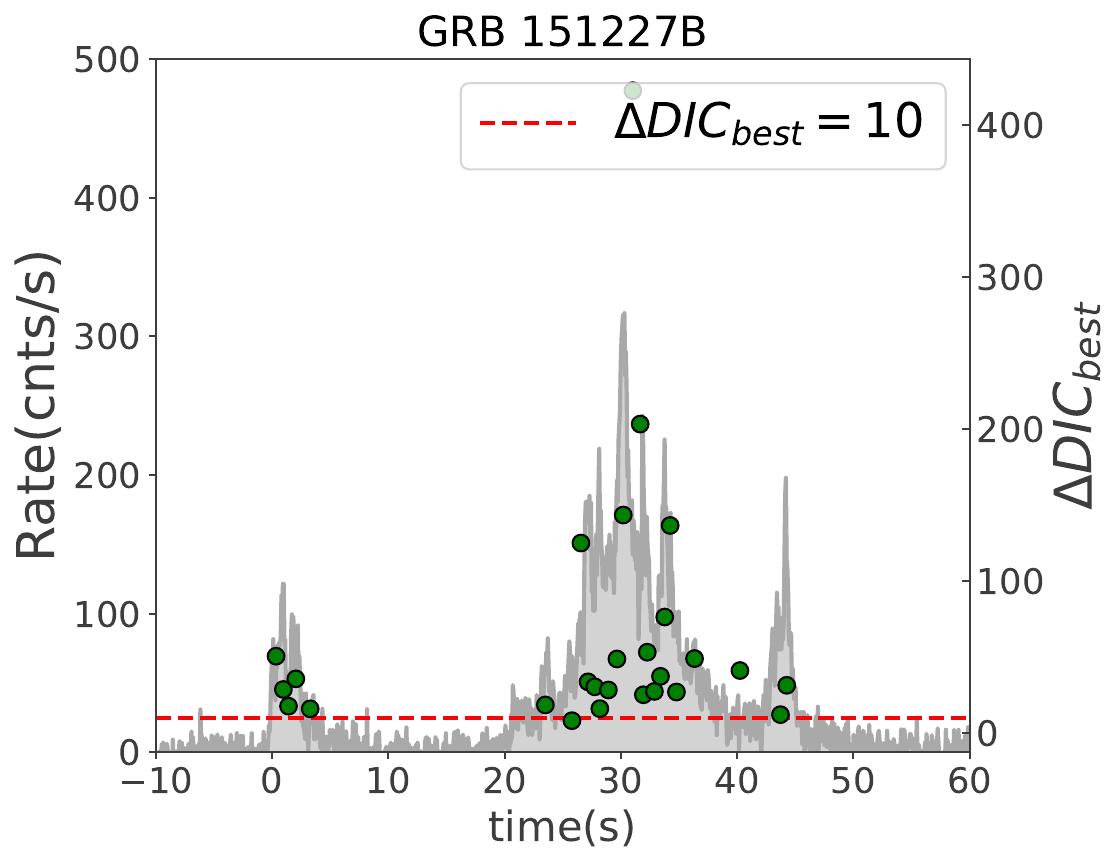}}
\hspace{0in}
{
\includegraphics[scale=0.3]{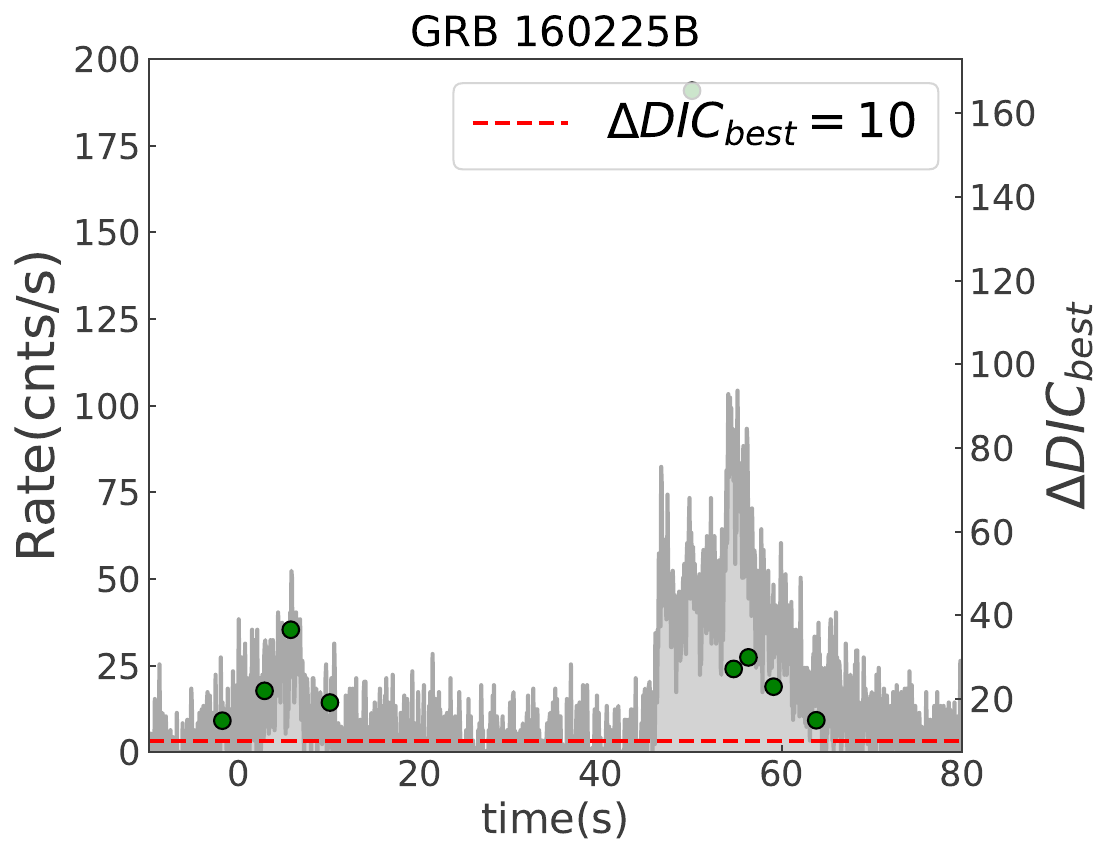}}
\hspace{0in}    
{
\includegraphics[scale=0.3]{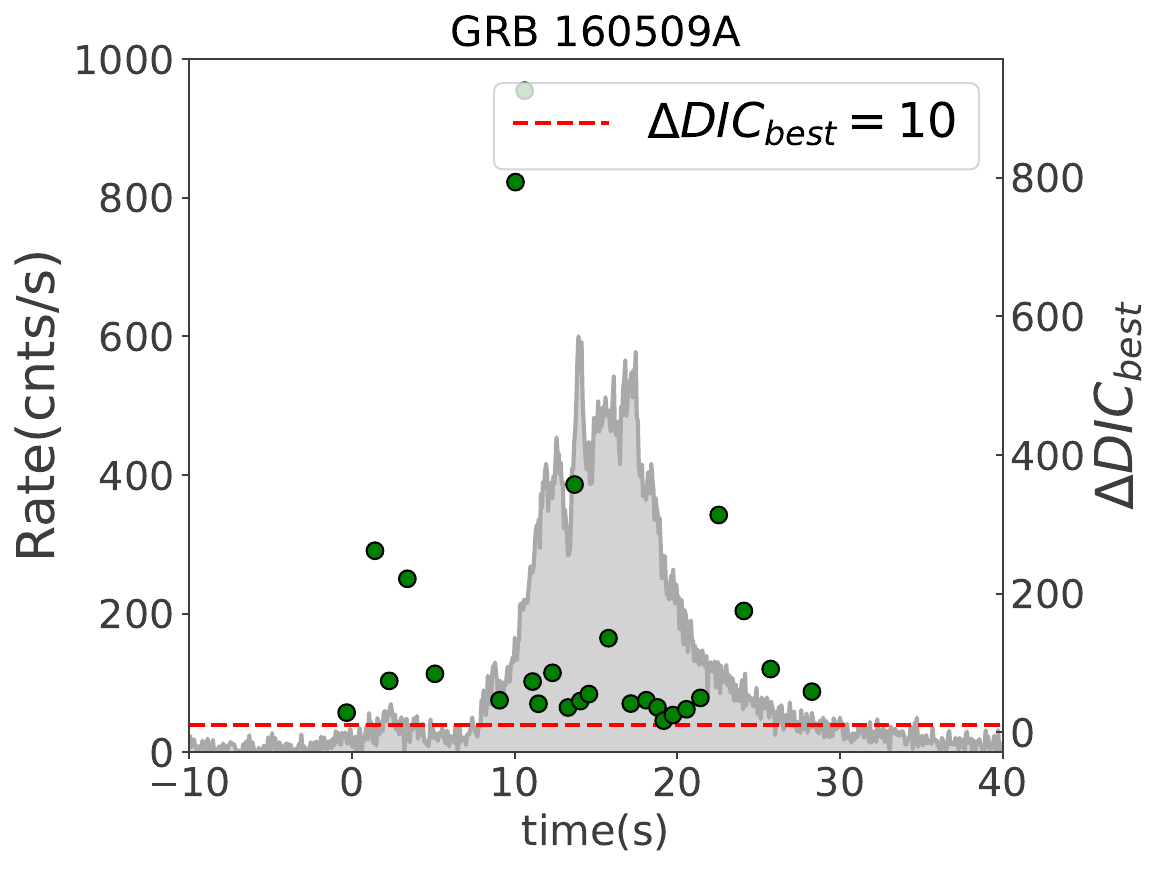}}
\hspace{0in}
{
\includegraphics[scale=0.3]{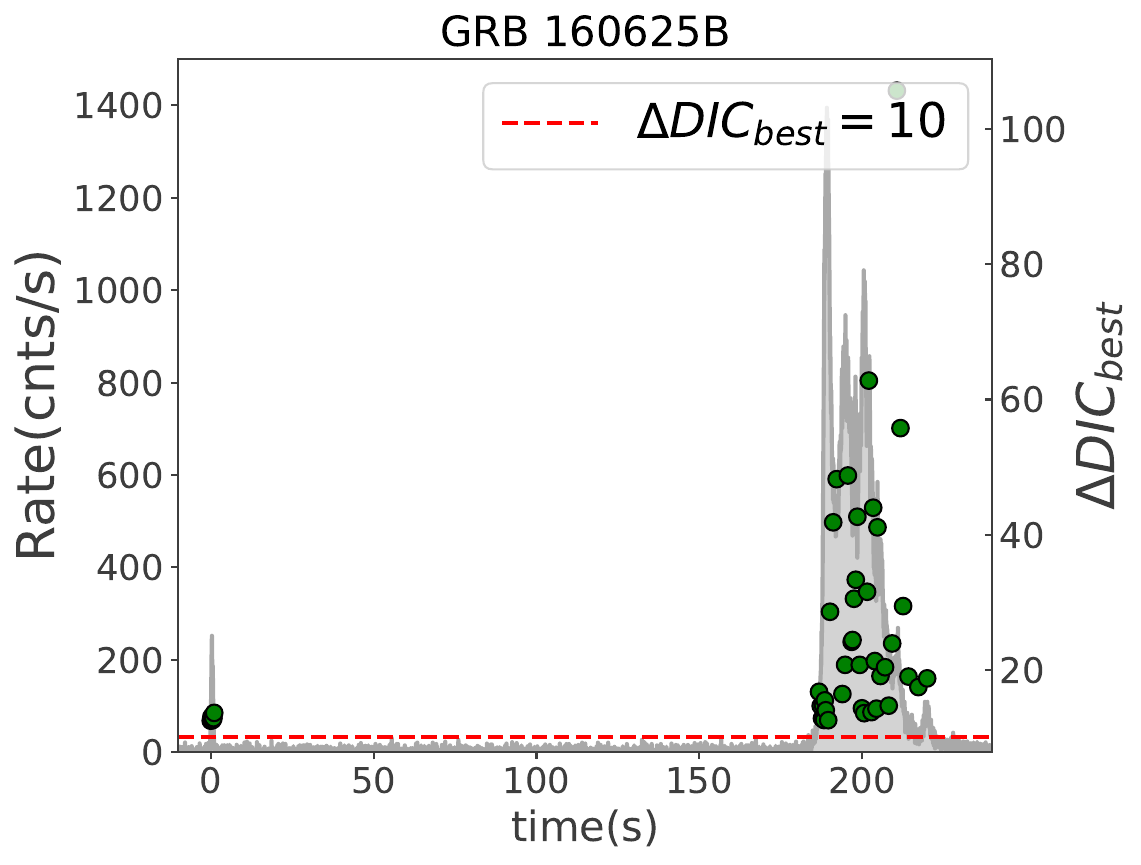}}
\hspace{0in}
{
\includegraphics[scale=0.3]{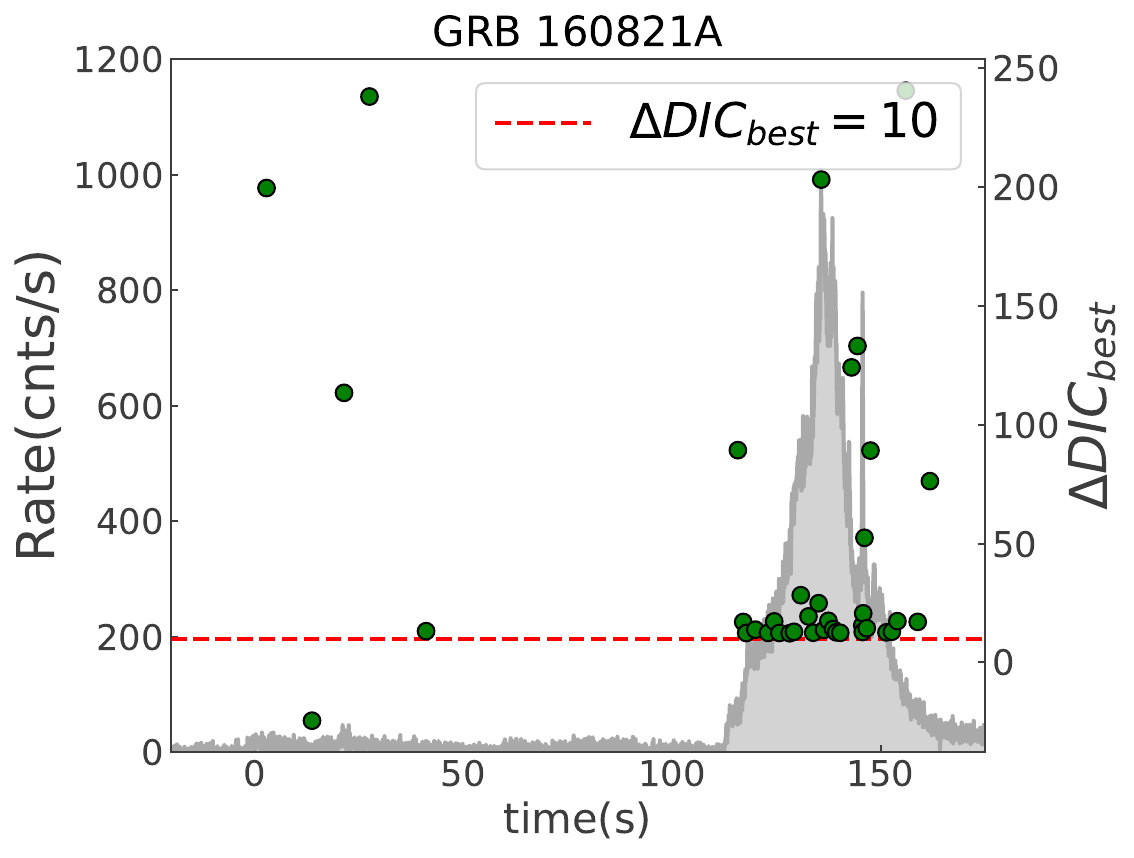}}
\hspace{0in}    
{
\includegraphics[scale=0.3]{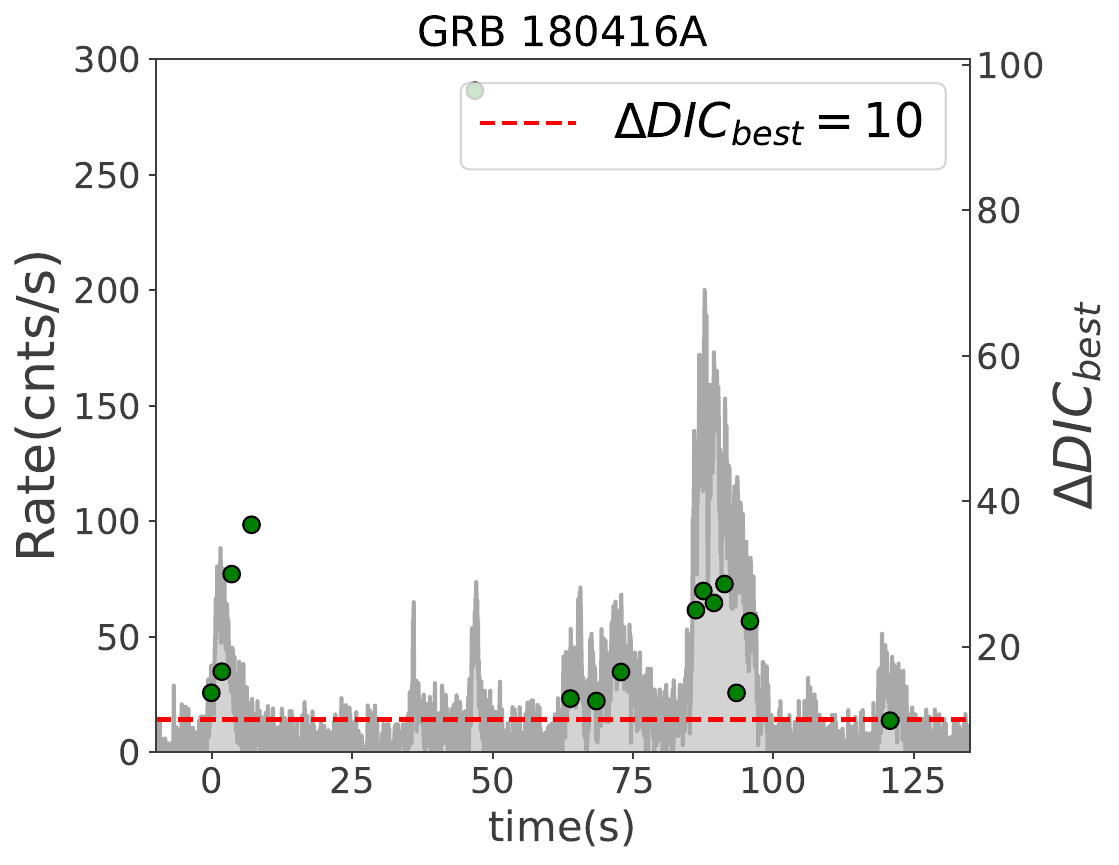}}
\hspace{0in}
{
\includegraphics[scale=0.3]{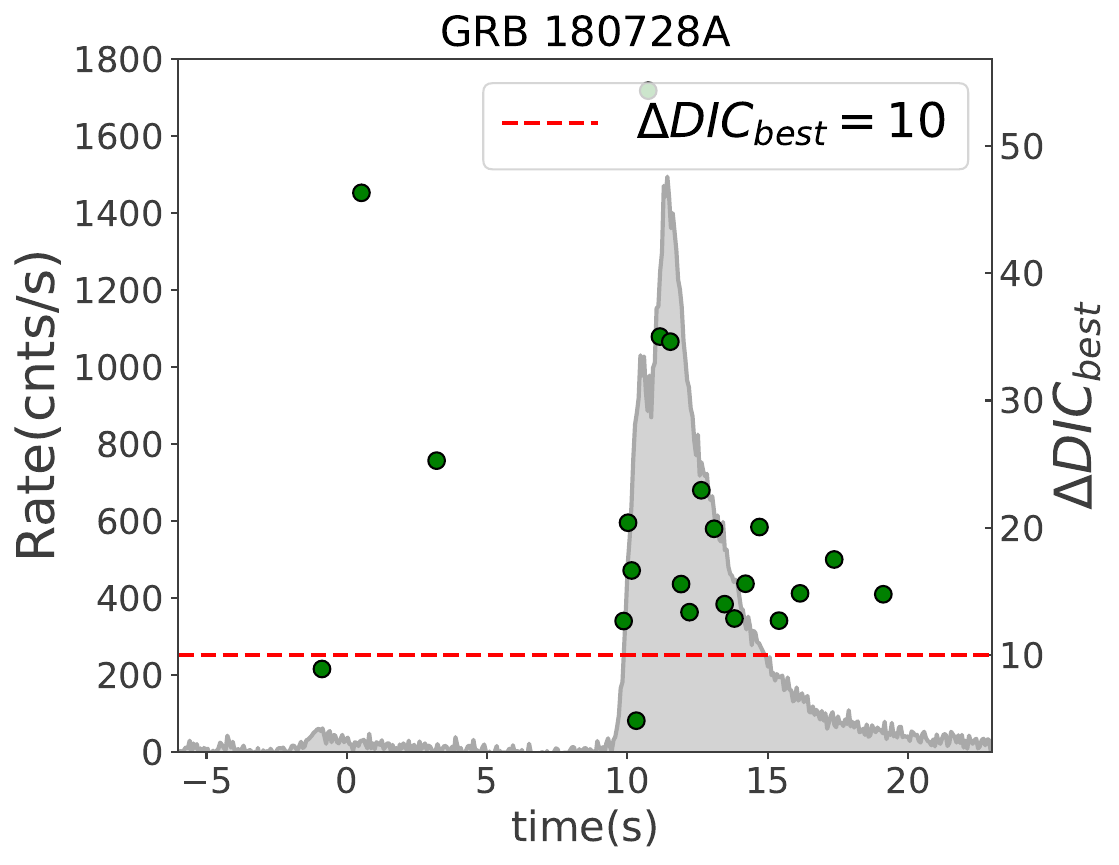}}
\hspace{0in}
{
\includegraphics[scale=0.3]{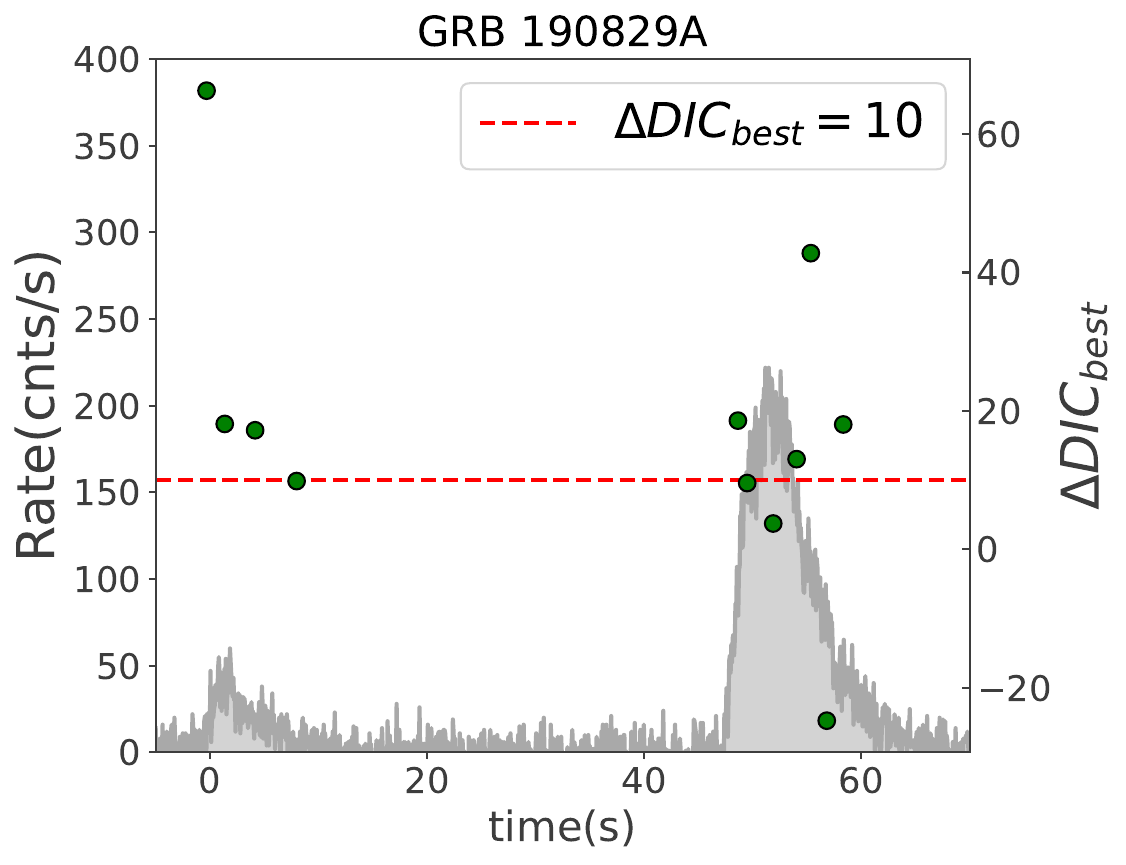}}
\hspace{0in}    
{
\includegraphics[scale=0.3]{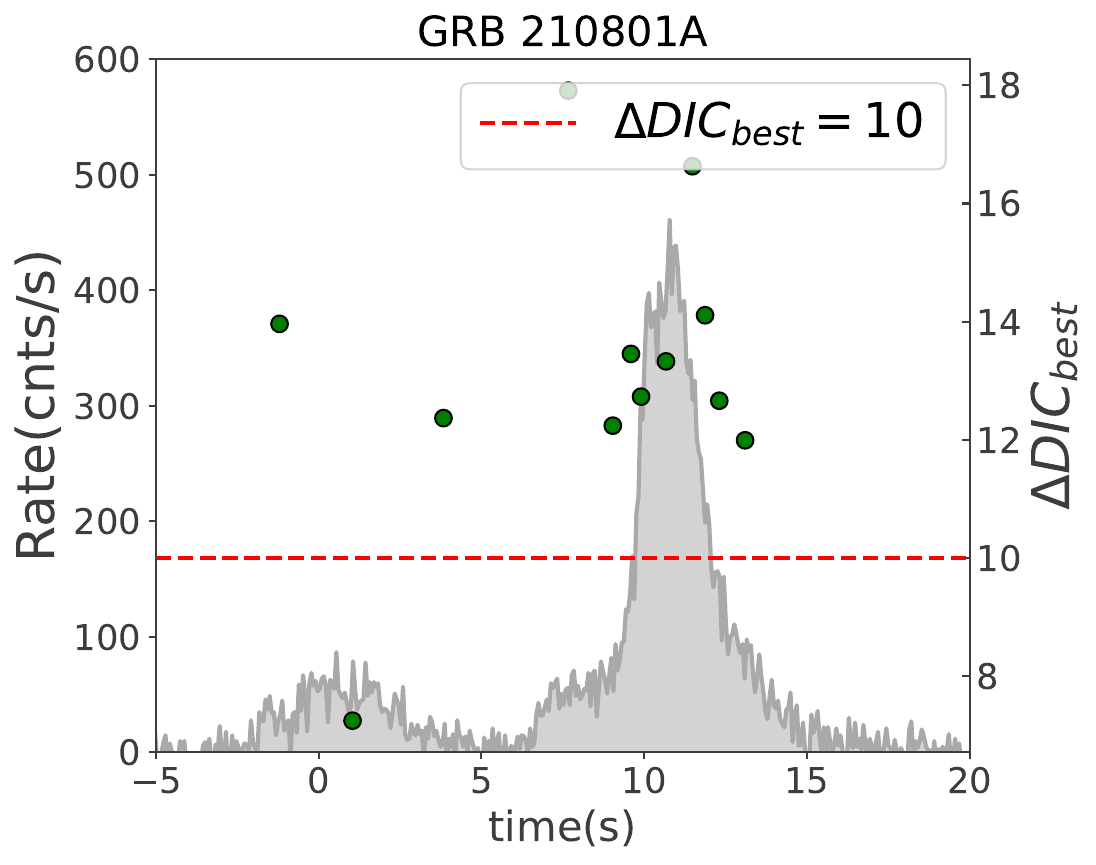}}
\hspace{0in}
{
\includegraphics[scale=0.3]{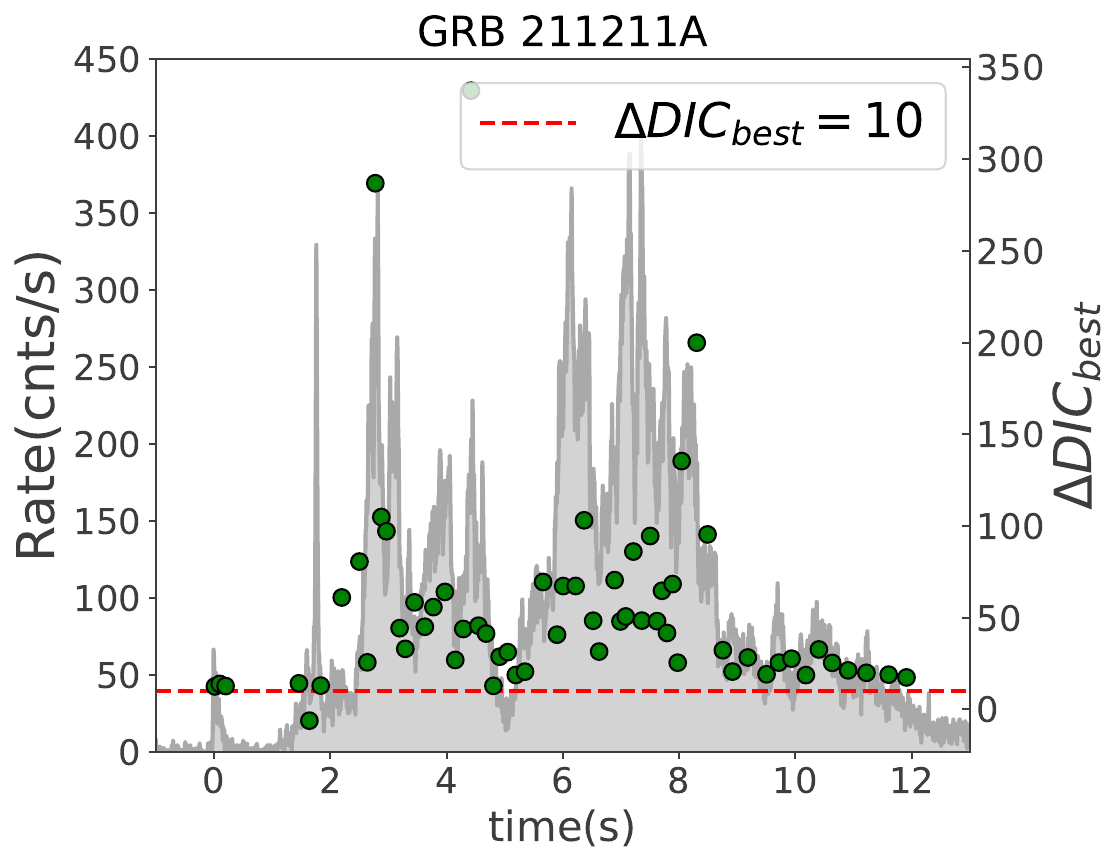}}
\hspace{0in}
\end{figure}
\begin{figure}
\centering
{
\includegraphics[scale=0.3]{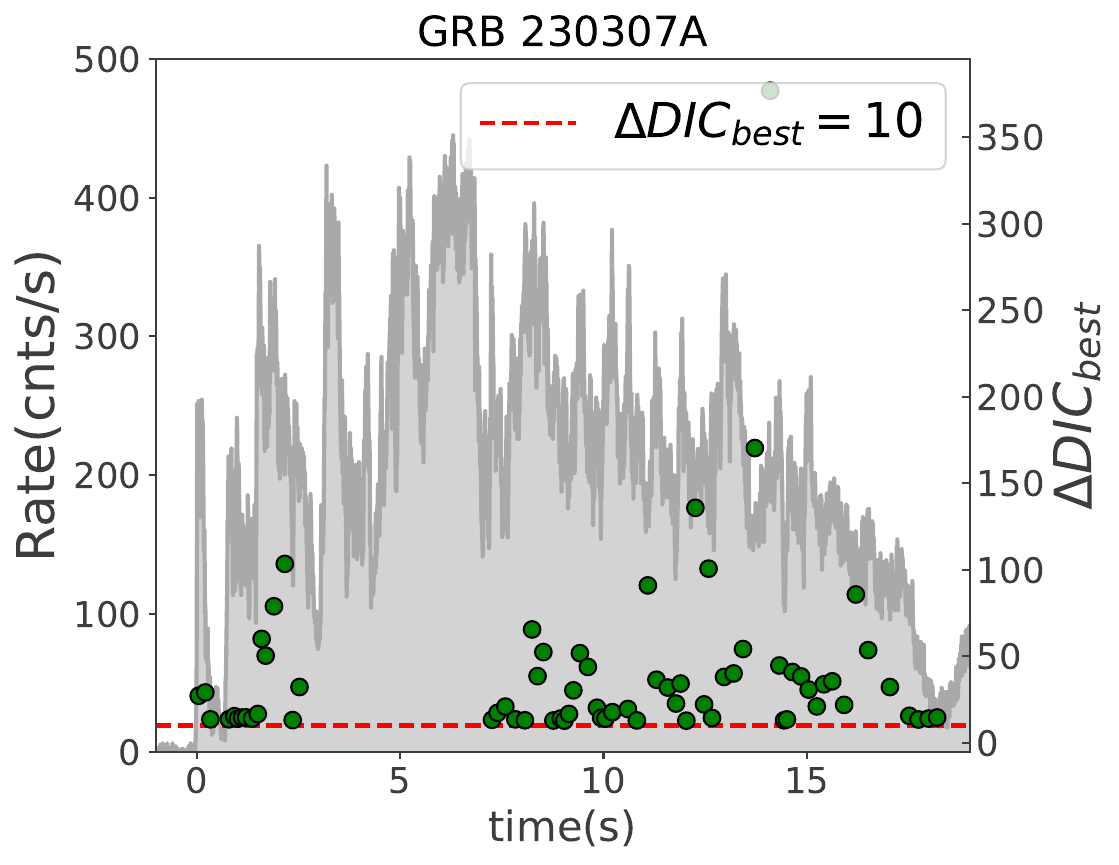}}
\hspace{0in}    
{
\includegraphics[scale=0.3]{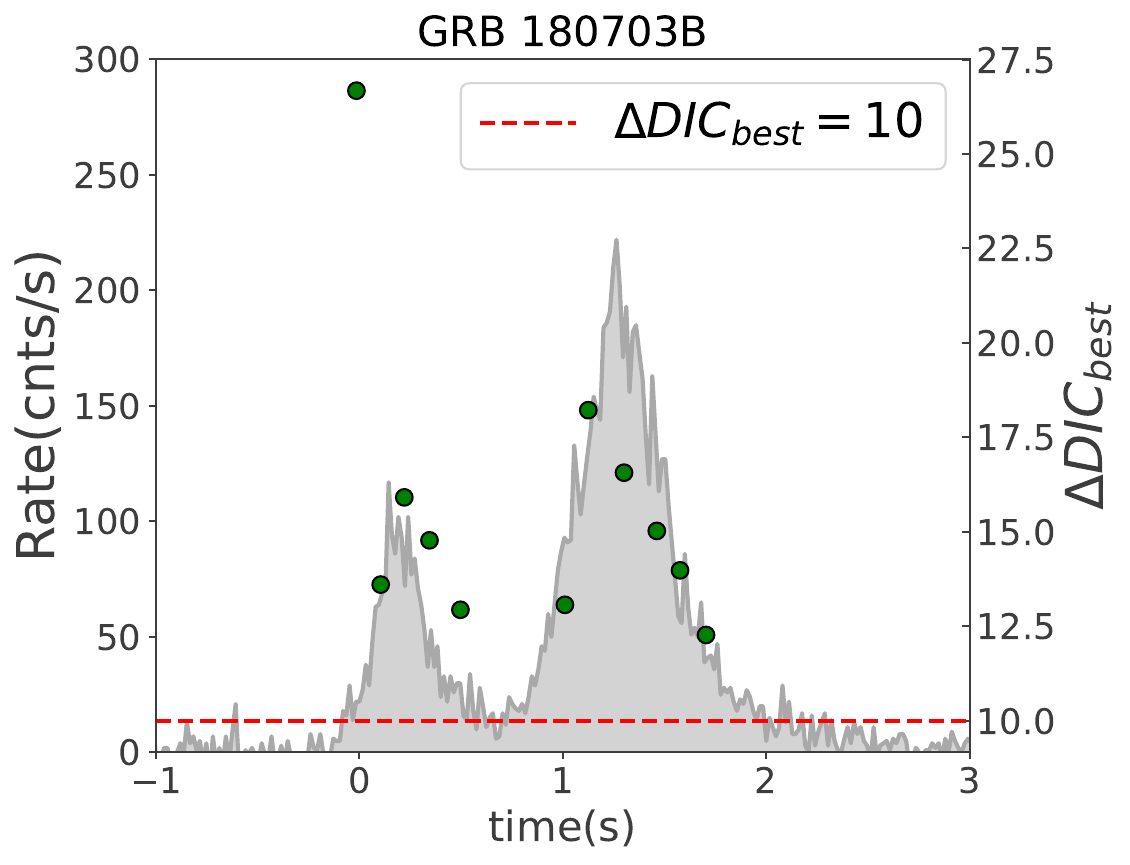}}
\hspace{0in}
{
\includegraphics[scale=0.3]{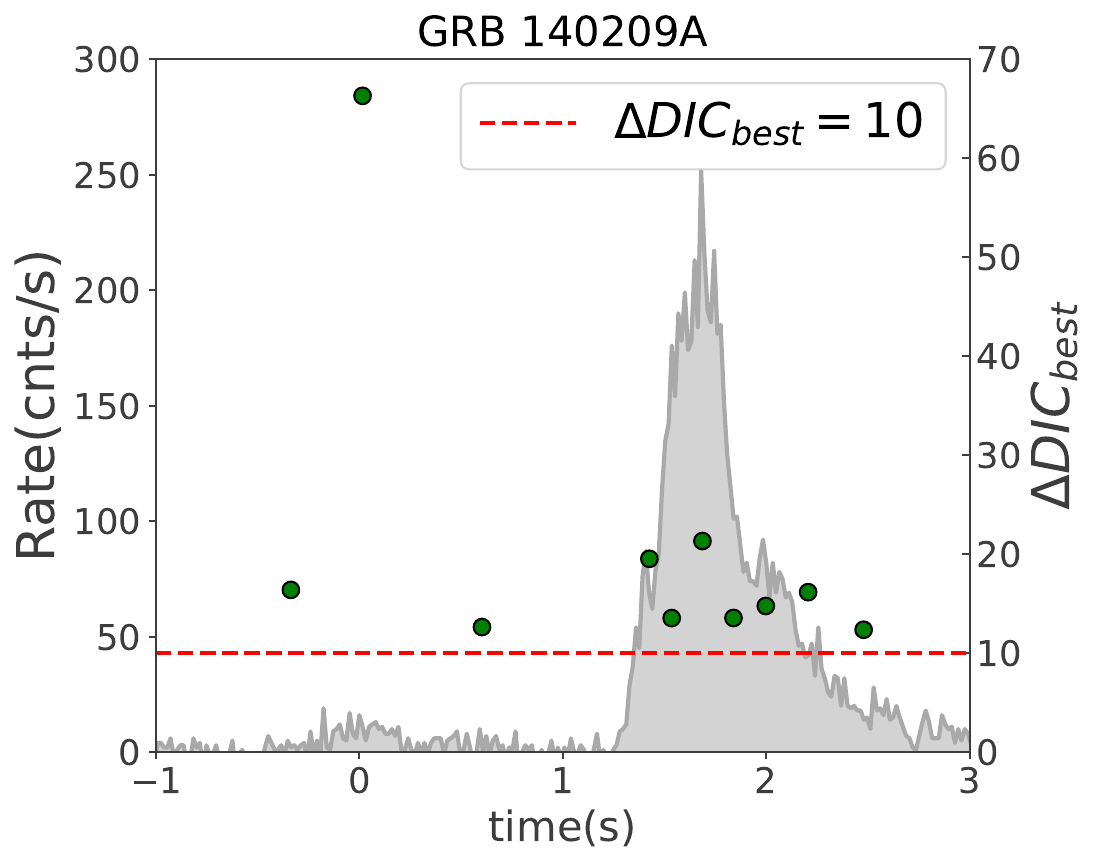}}
\hspace{0in}
 \caption{The evolution of $\Delta DIC_{best}$ over time. The gray shading is the light curve, and the red dotted line indicates $\Delta DIC$ =10. Exceeding the red dashed line indicates compelling evidence of a thermal component and the last two bursts with $T_{90}$ less than 2 seconds.}
   \label{FigA1} 
   \end{figure}

\begin{figure}[htbp]
\centering
\includegraphics [width=8cm,height=5cm]{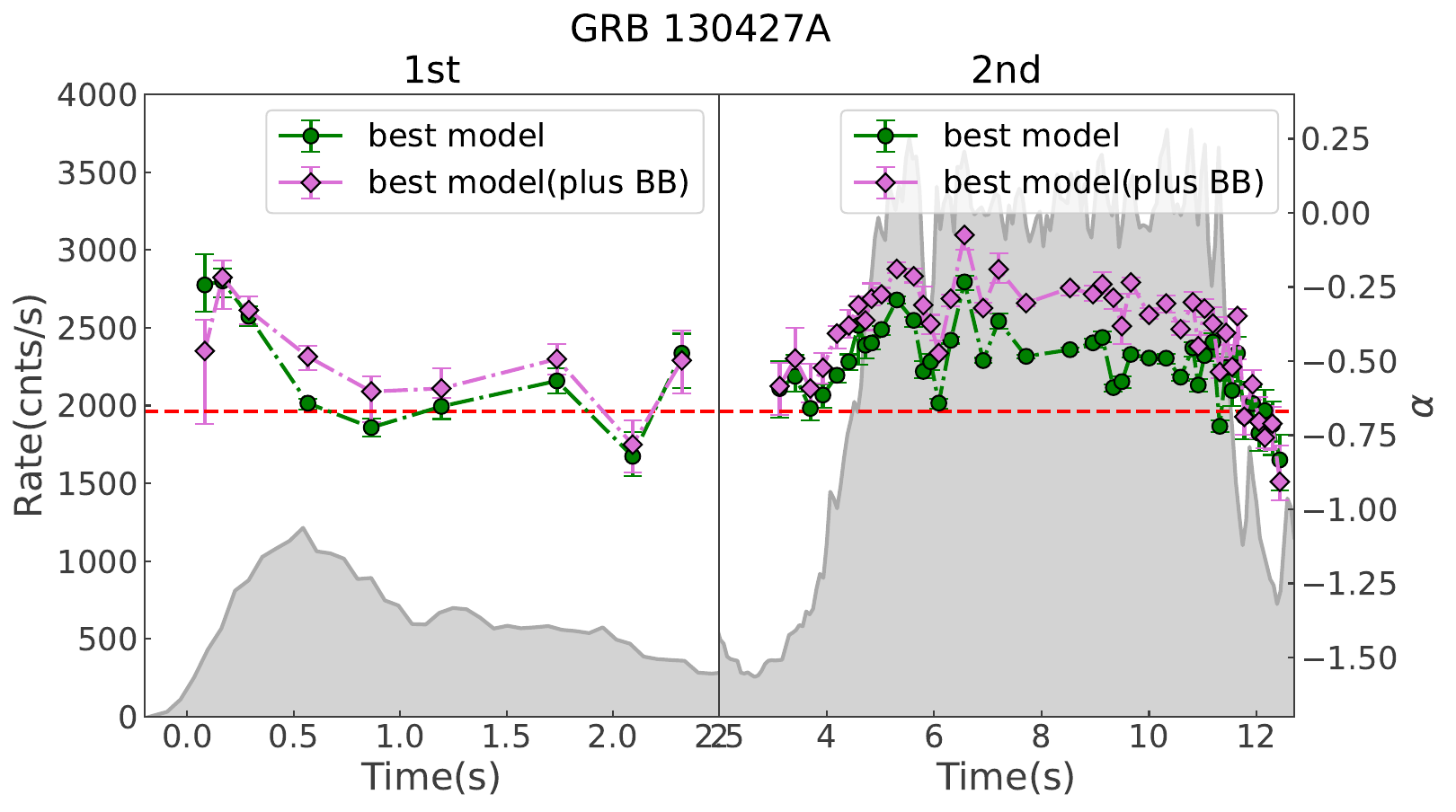}
\includegraphics [width=8cm,height=5cm]{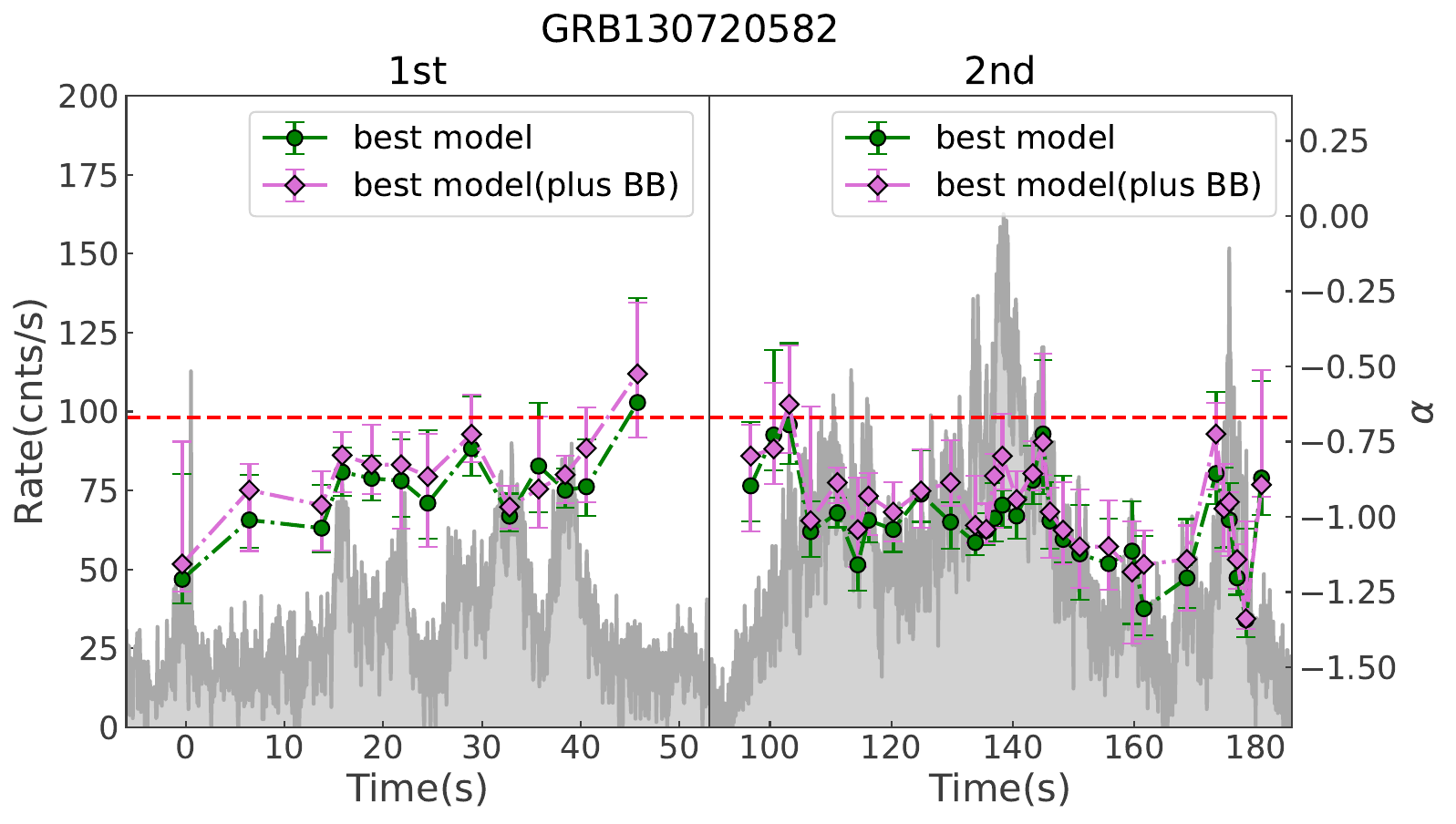}
\includegraphics [width=8cm,height=5cm]{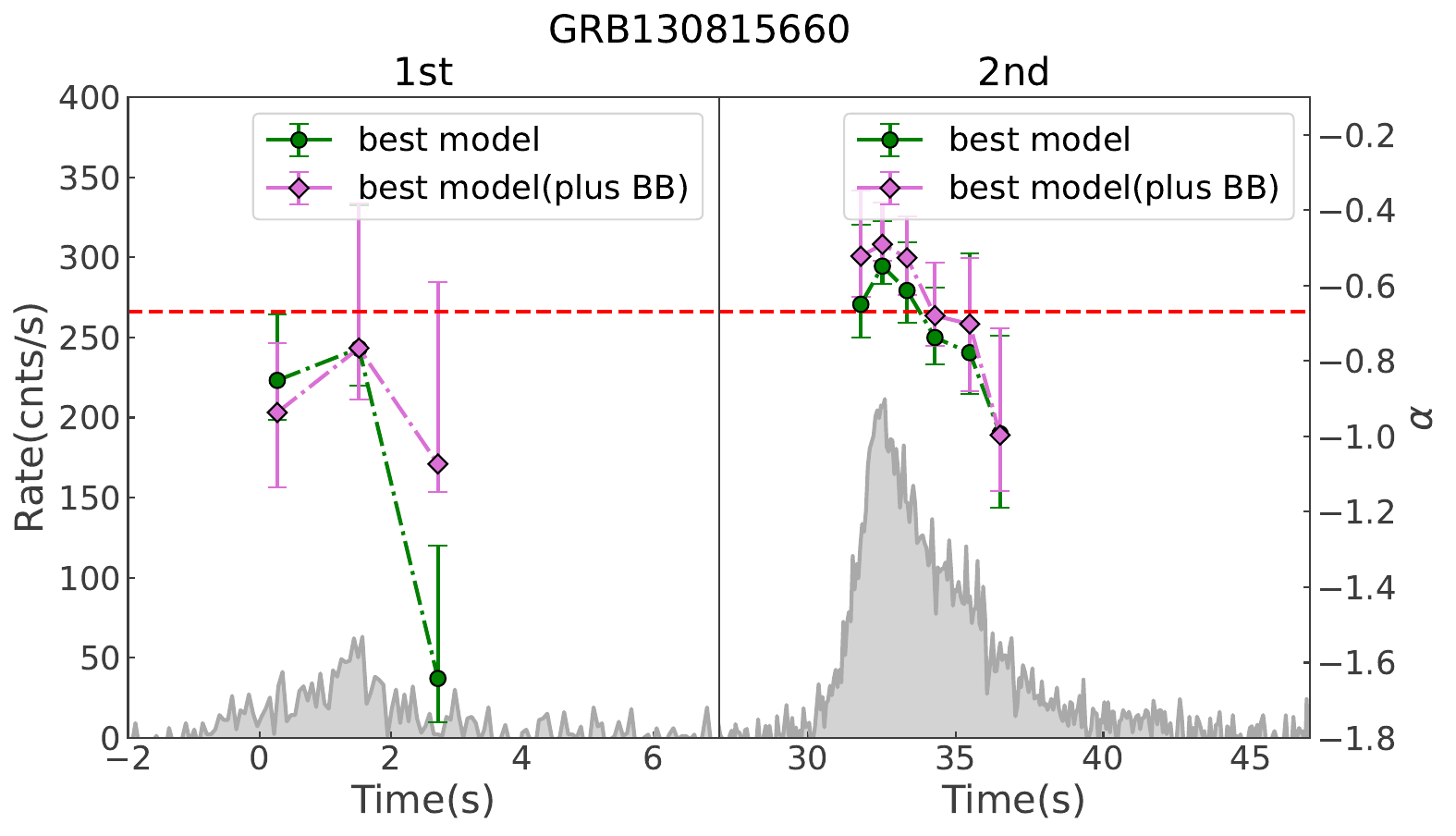}
\includegraphics [width=8cm,height=5cm]{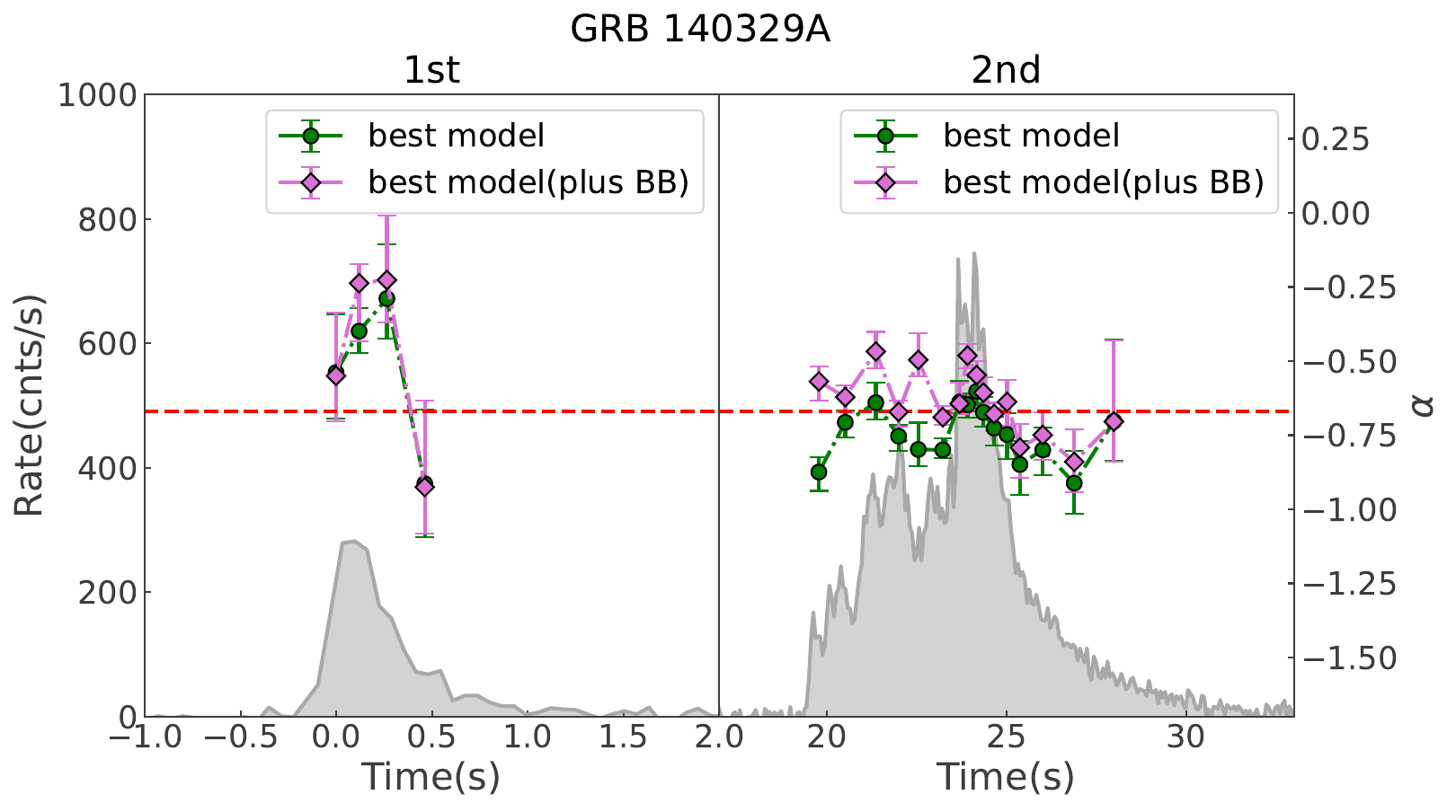}
\includegraphics [width=8cm,height=5cm]{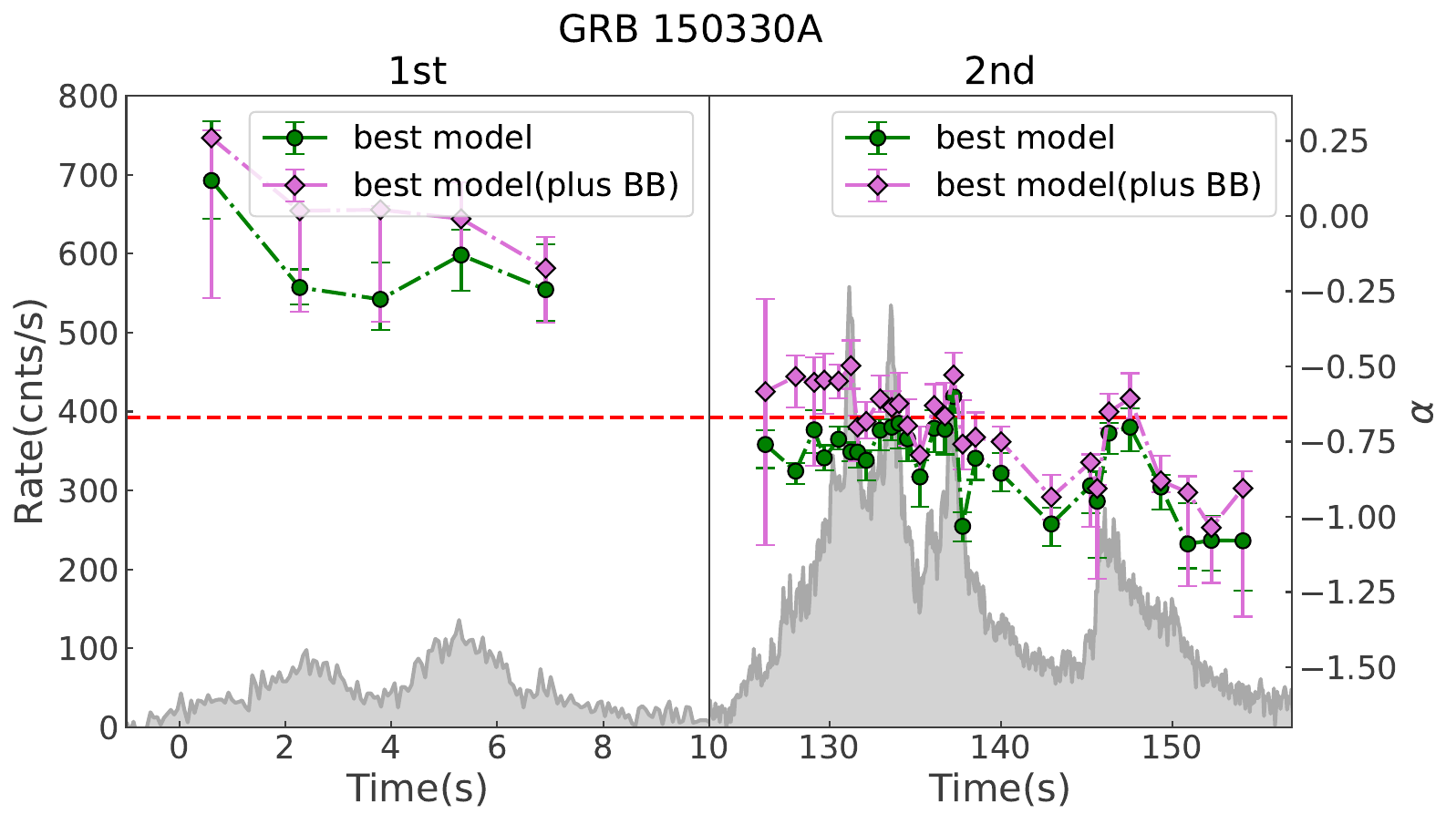}
\includegraphics [width=8cm,height=5cm]{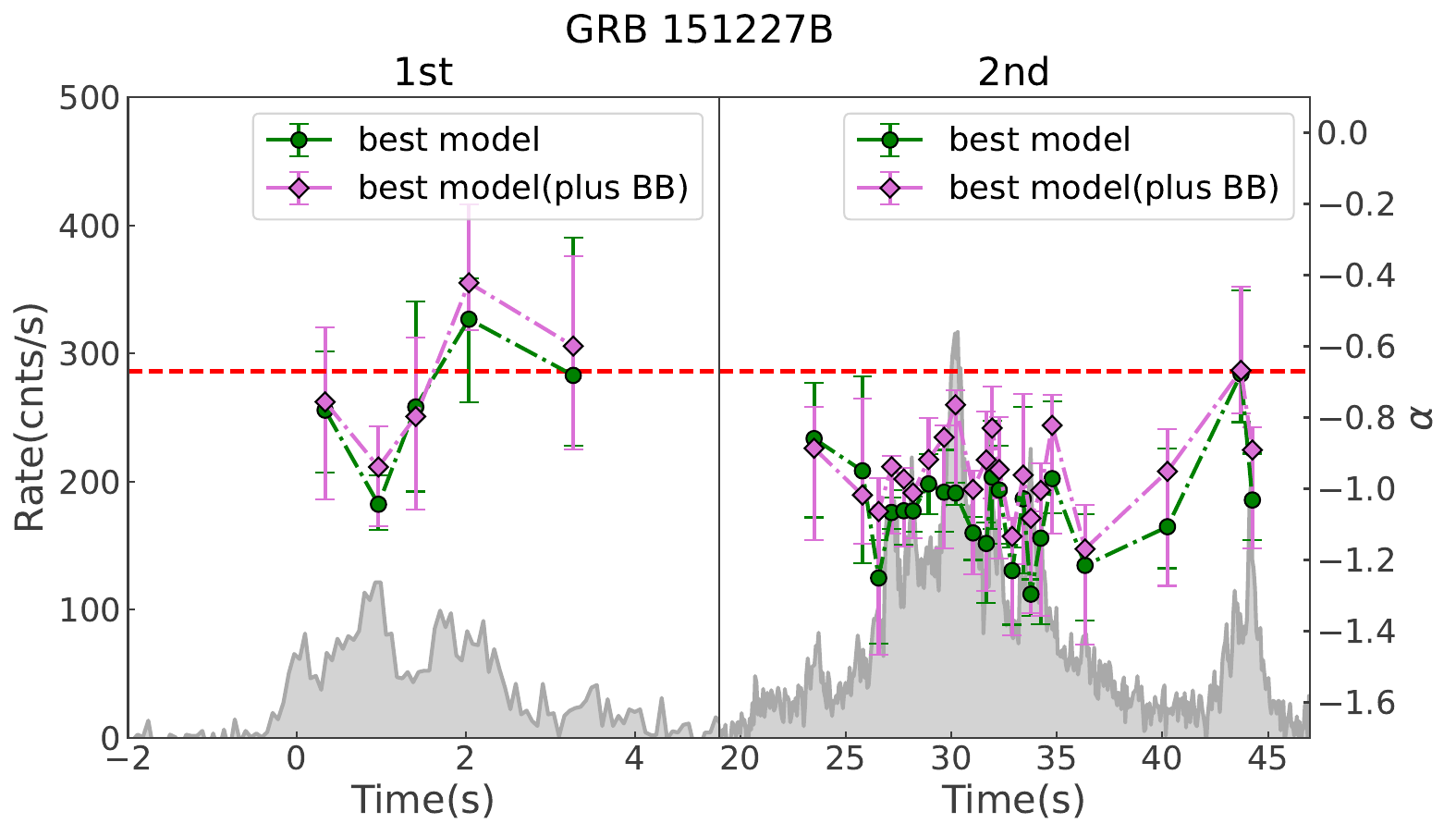}
\end{figure}
\begin{figure}[htbp]
\centering
\includegraphics [width=8cm,height=5cm]{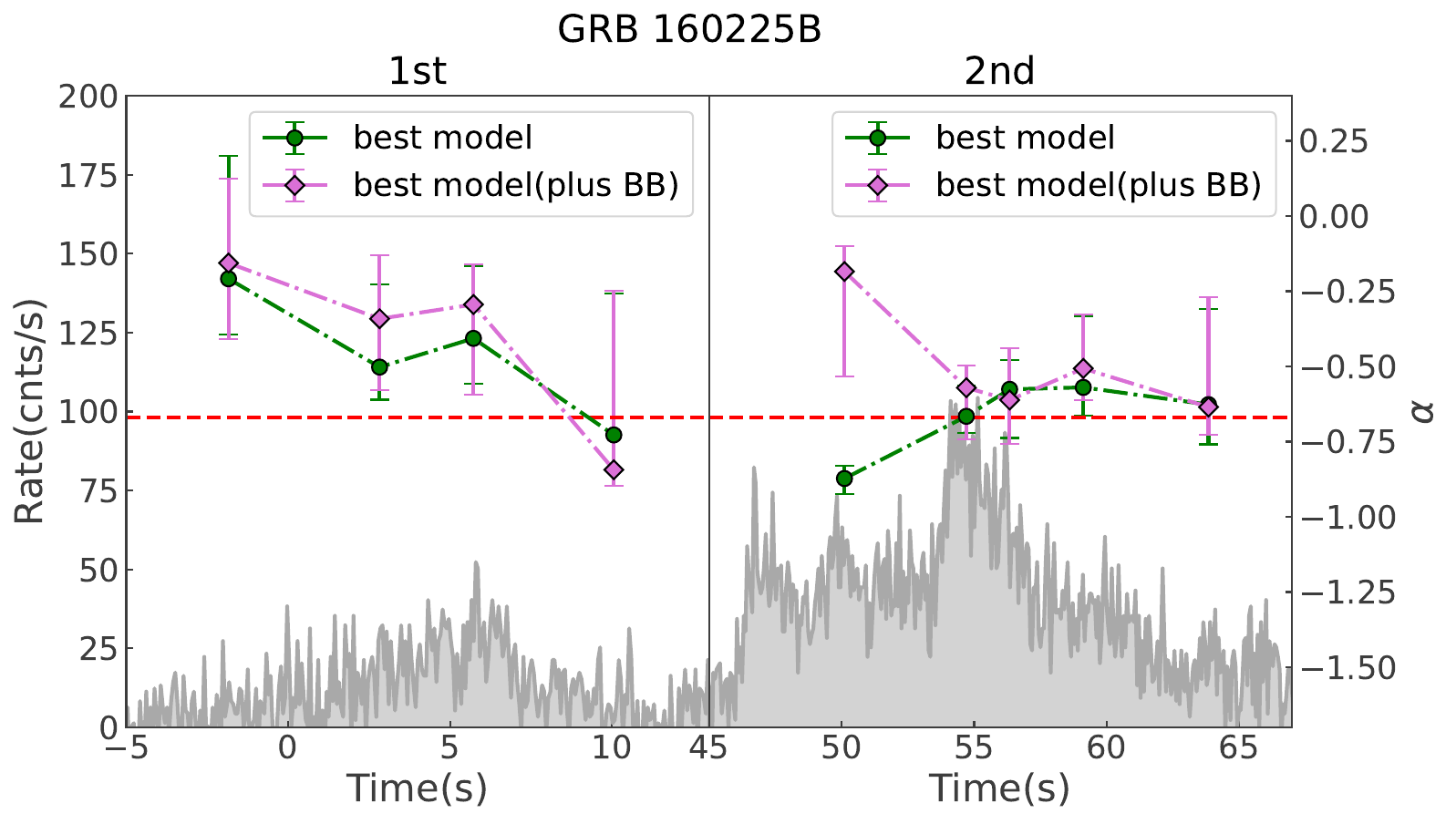}
\includegraphics [width=8cm,height=5cm]{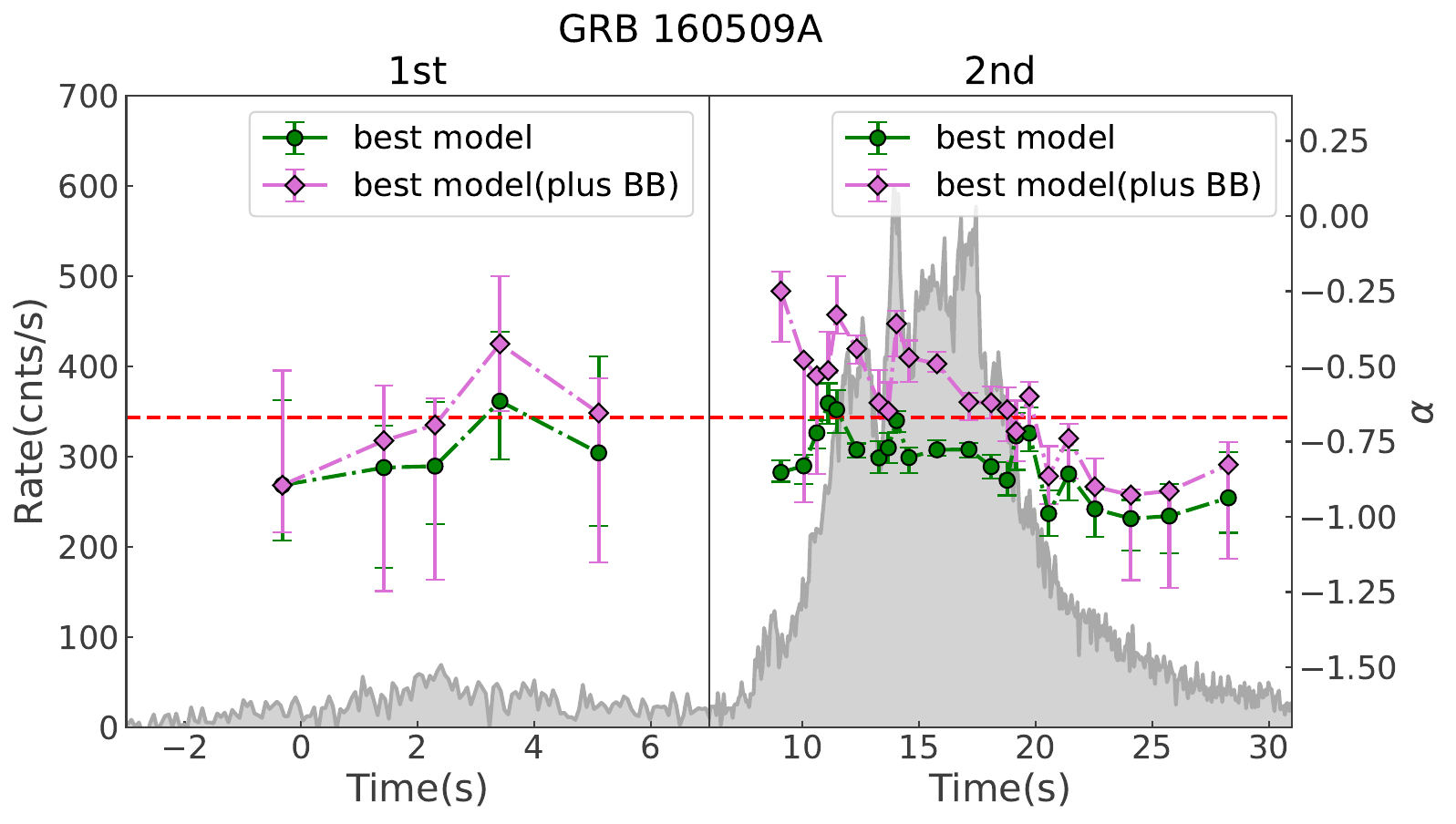}
\includegraphics [width=8cm,height=5cm]{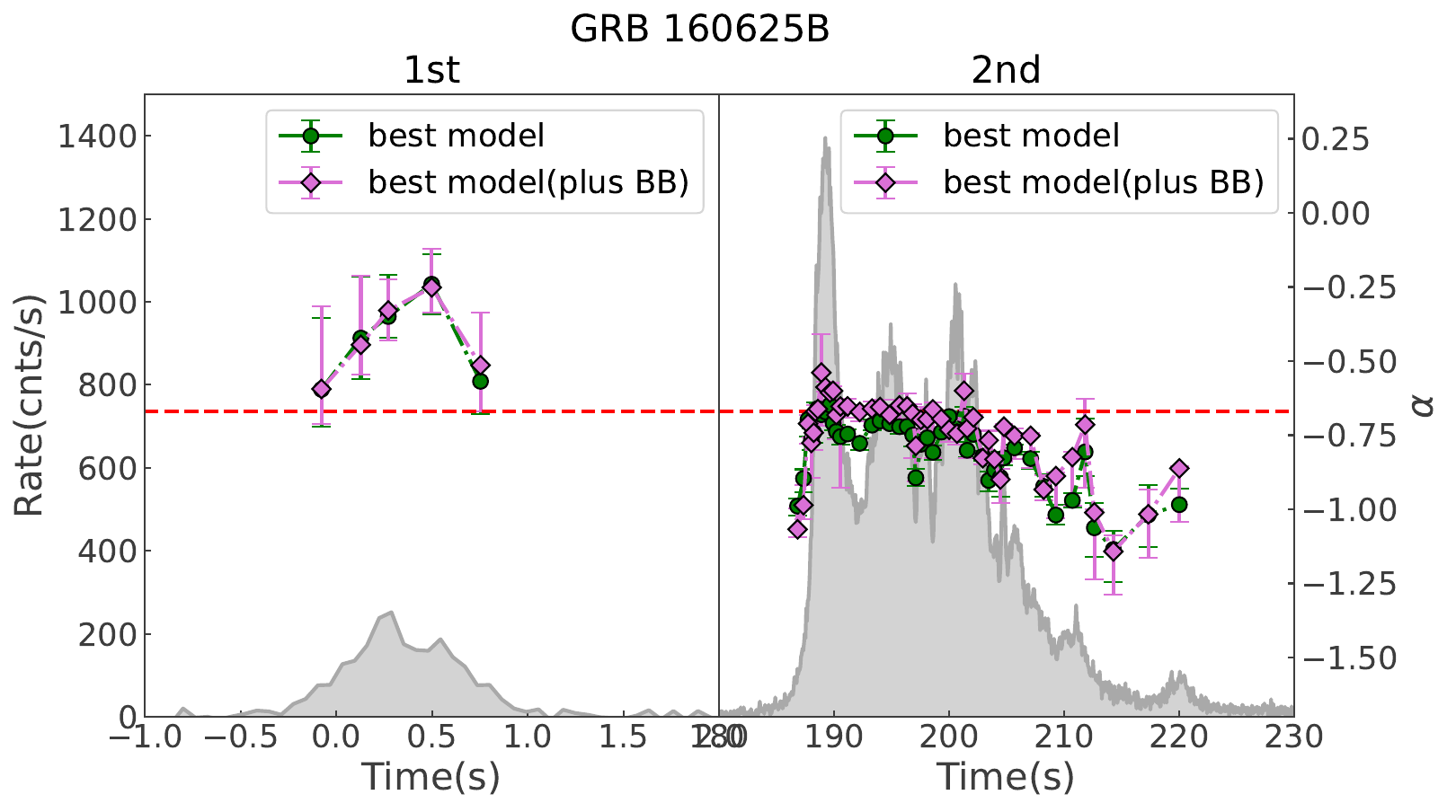}
\includegraphics [width=8cm,height=5cm]{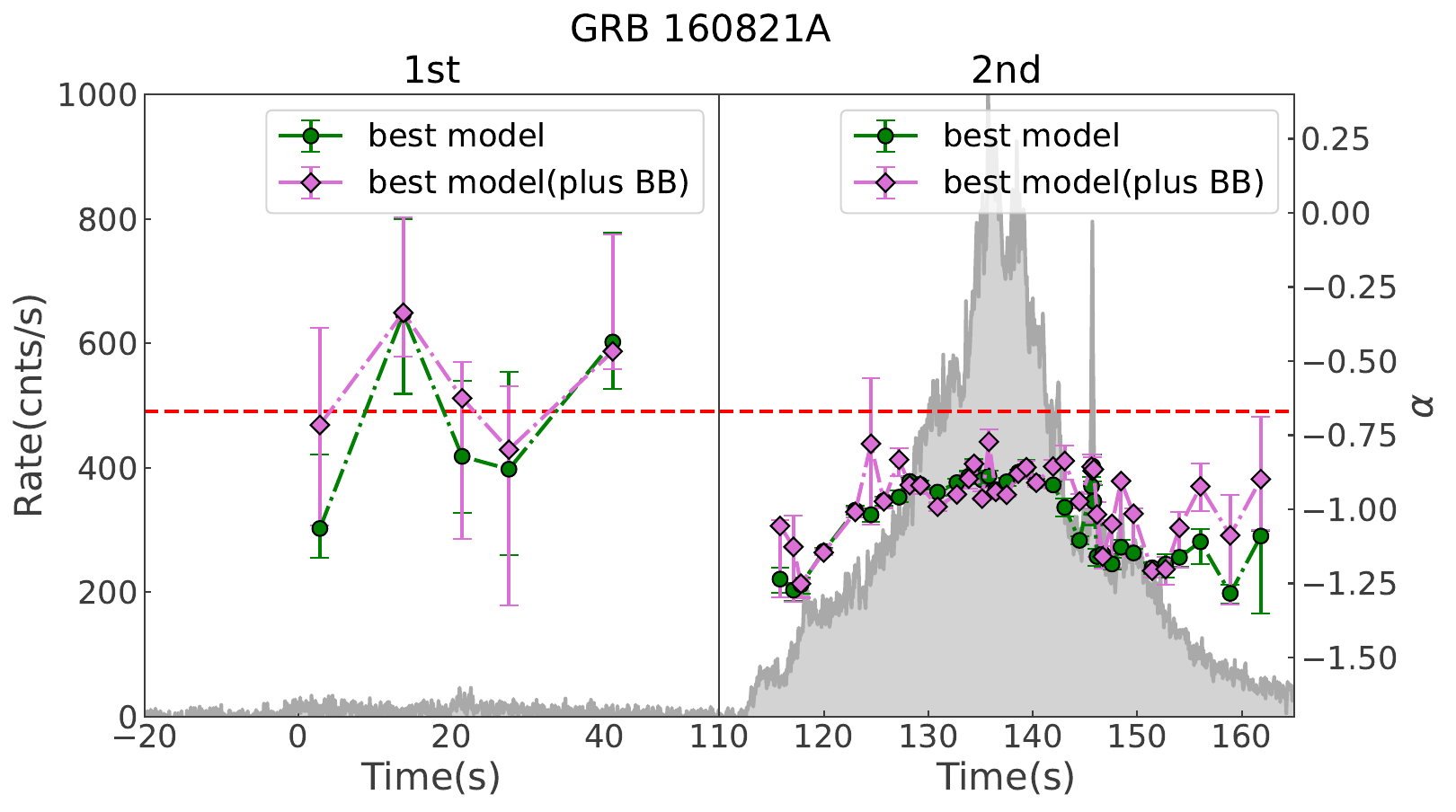}
\includegraphics [width=8cm,height=5cm]{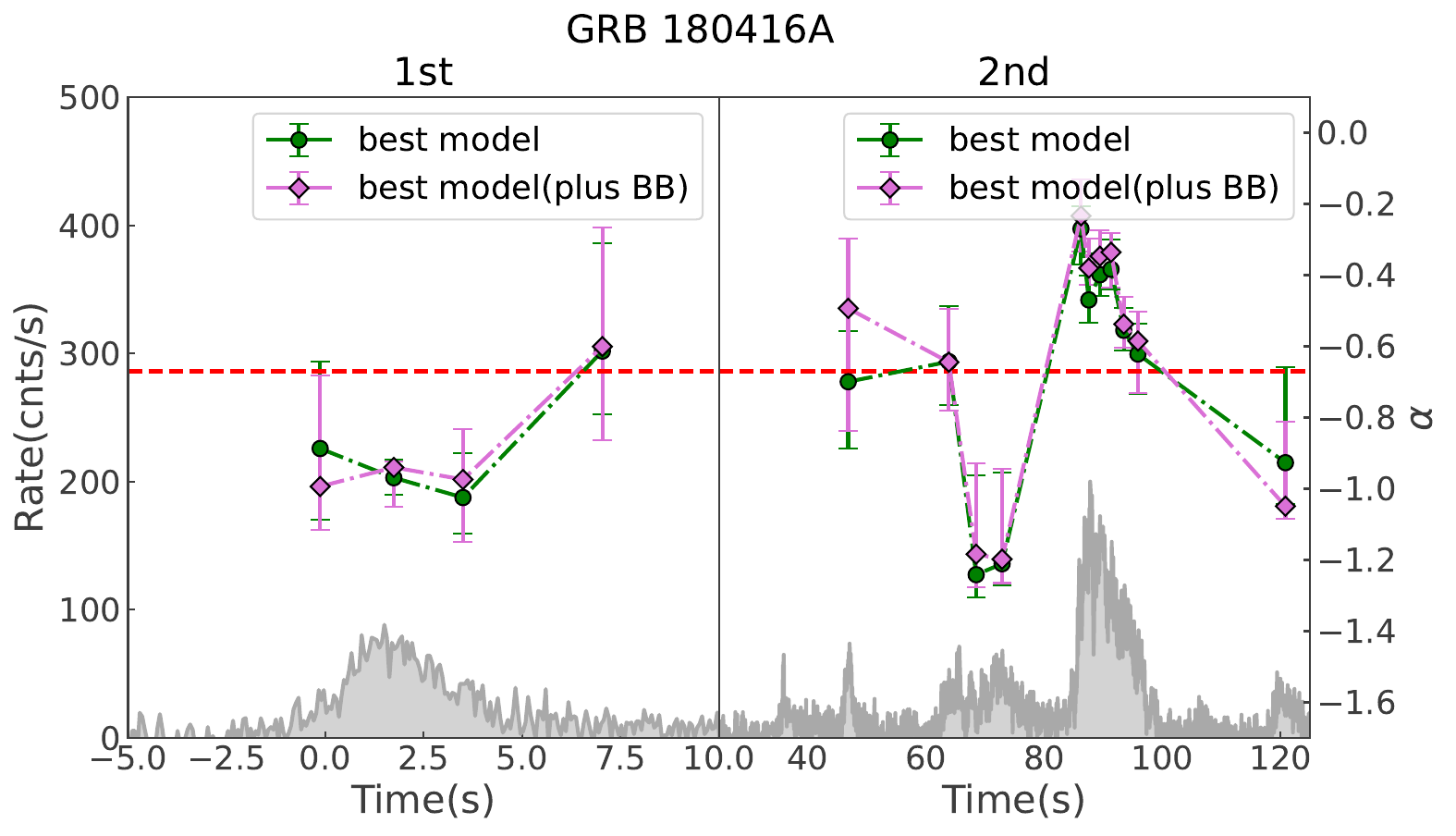}
\includegraphics [width=8cm,height=5cm]{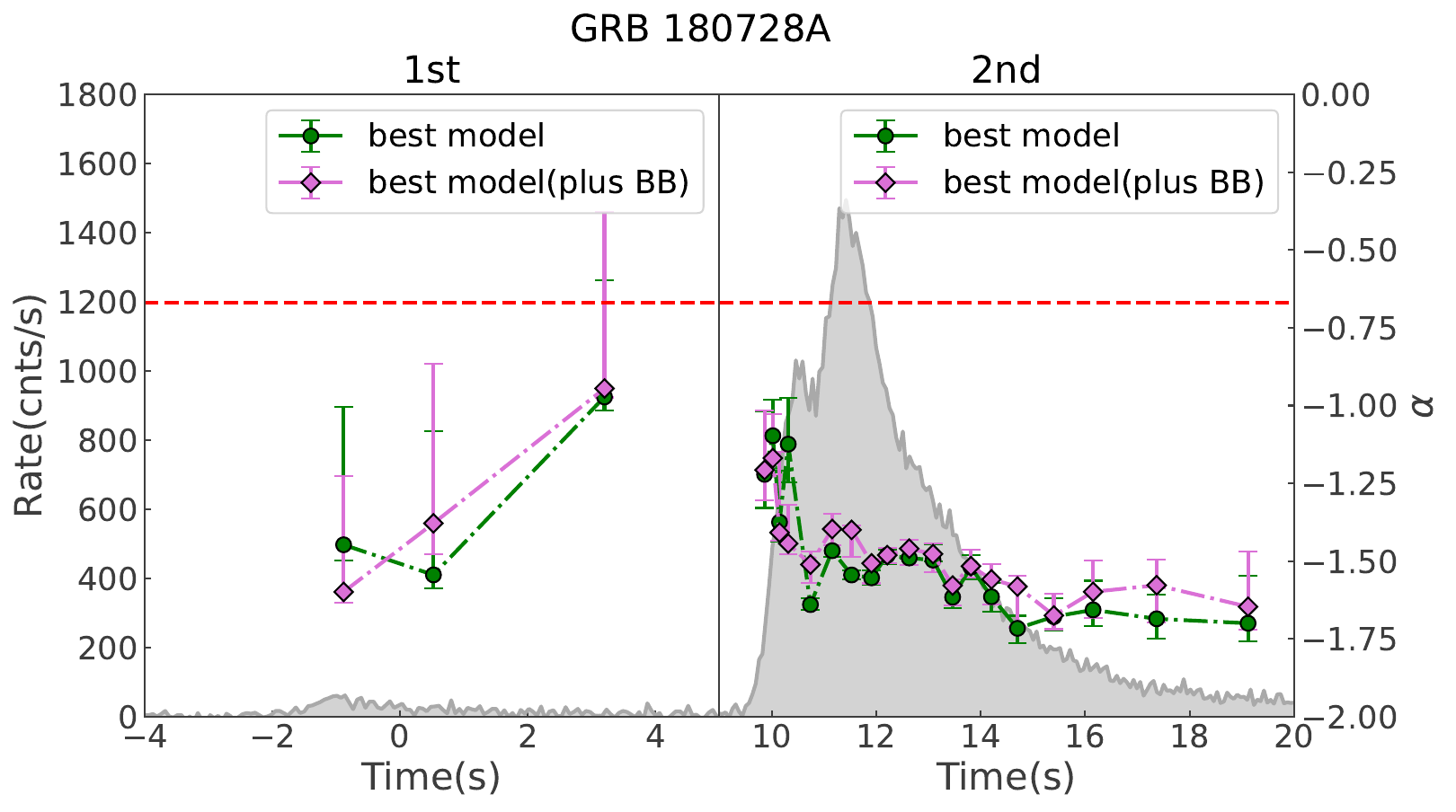}
\includegraphics [width=8cm,height=5cm]{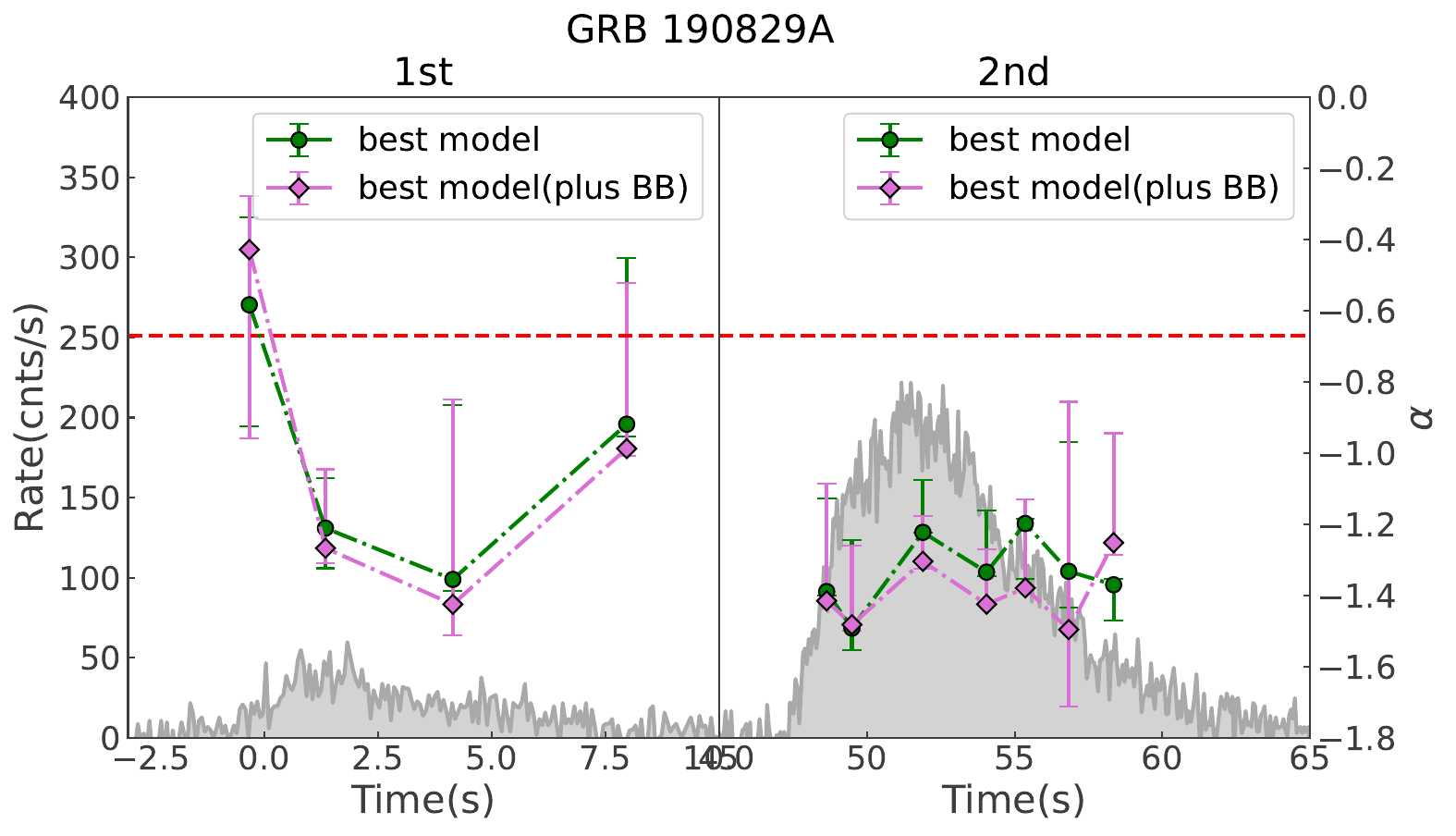}
\includegraphics [width=8cm,height=5cm]{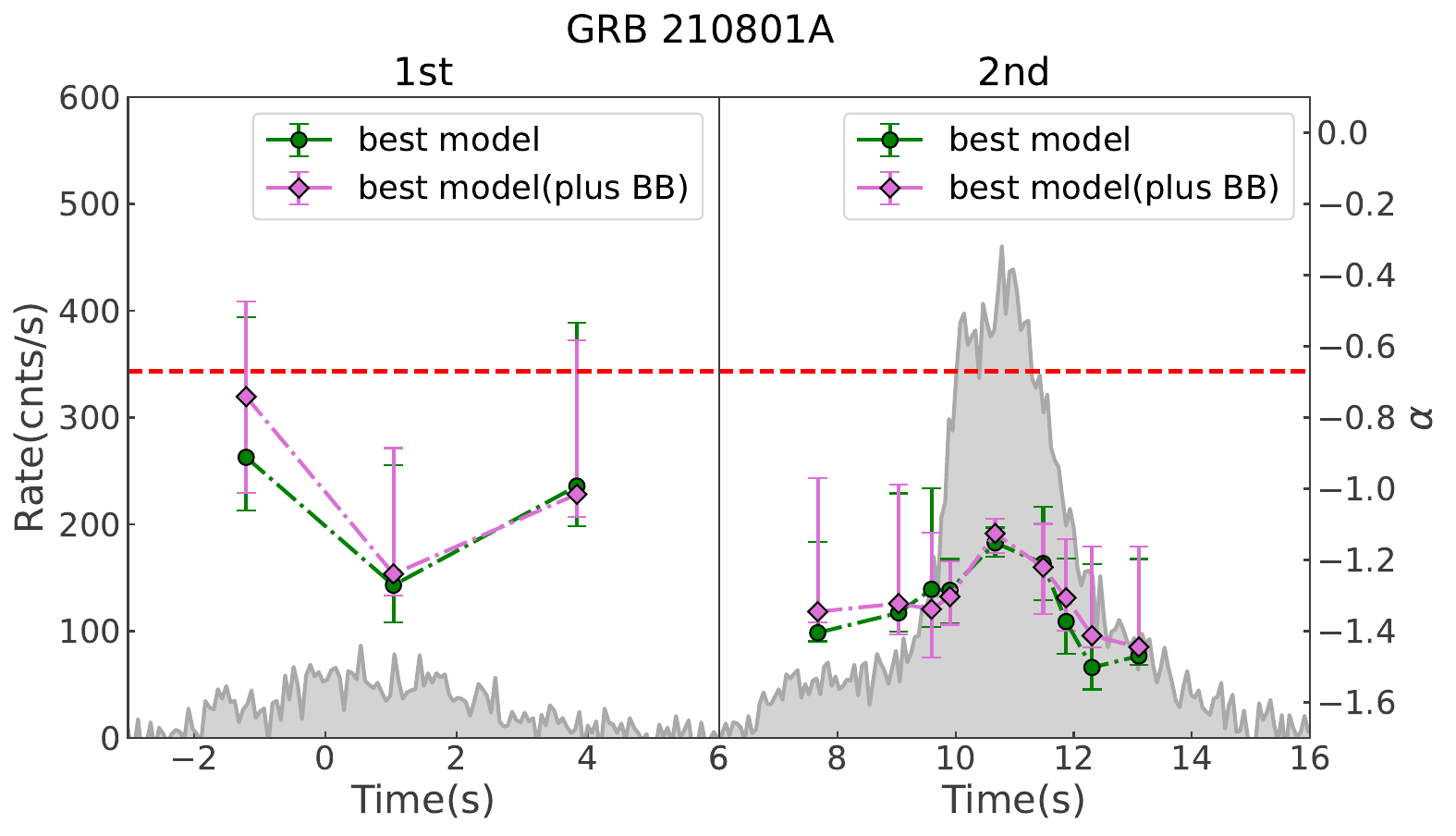}
\includegraphics [width=8cm,height=5cm]{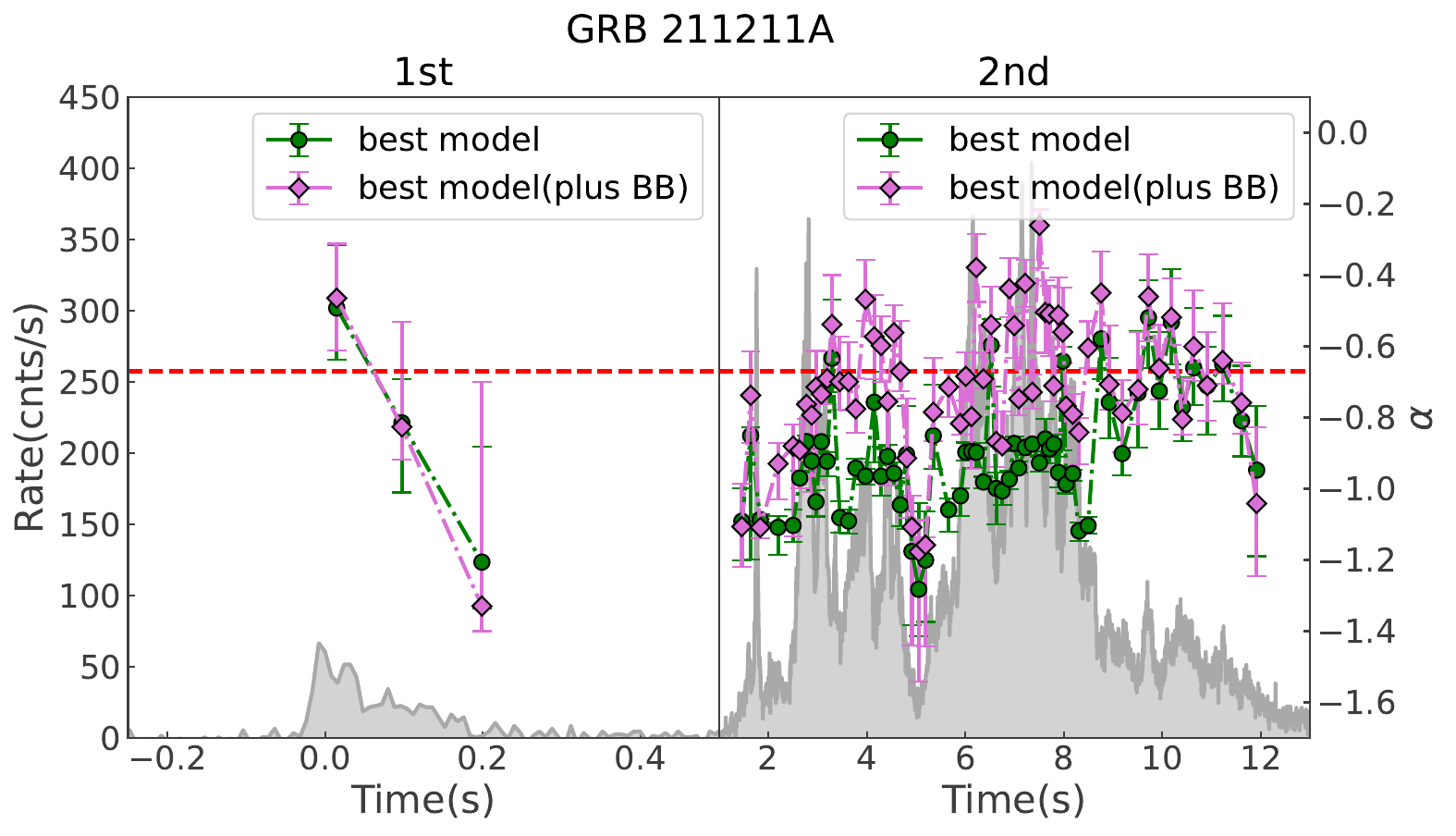}
\includegraphics [width=8cm,height=5cm]{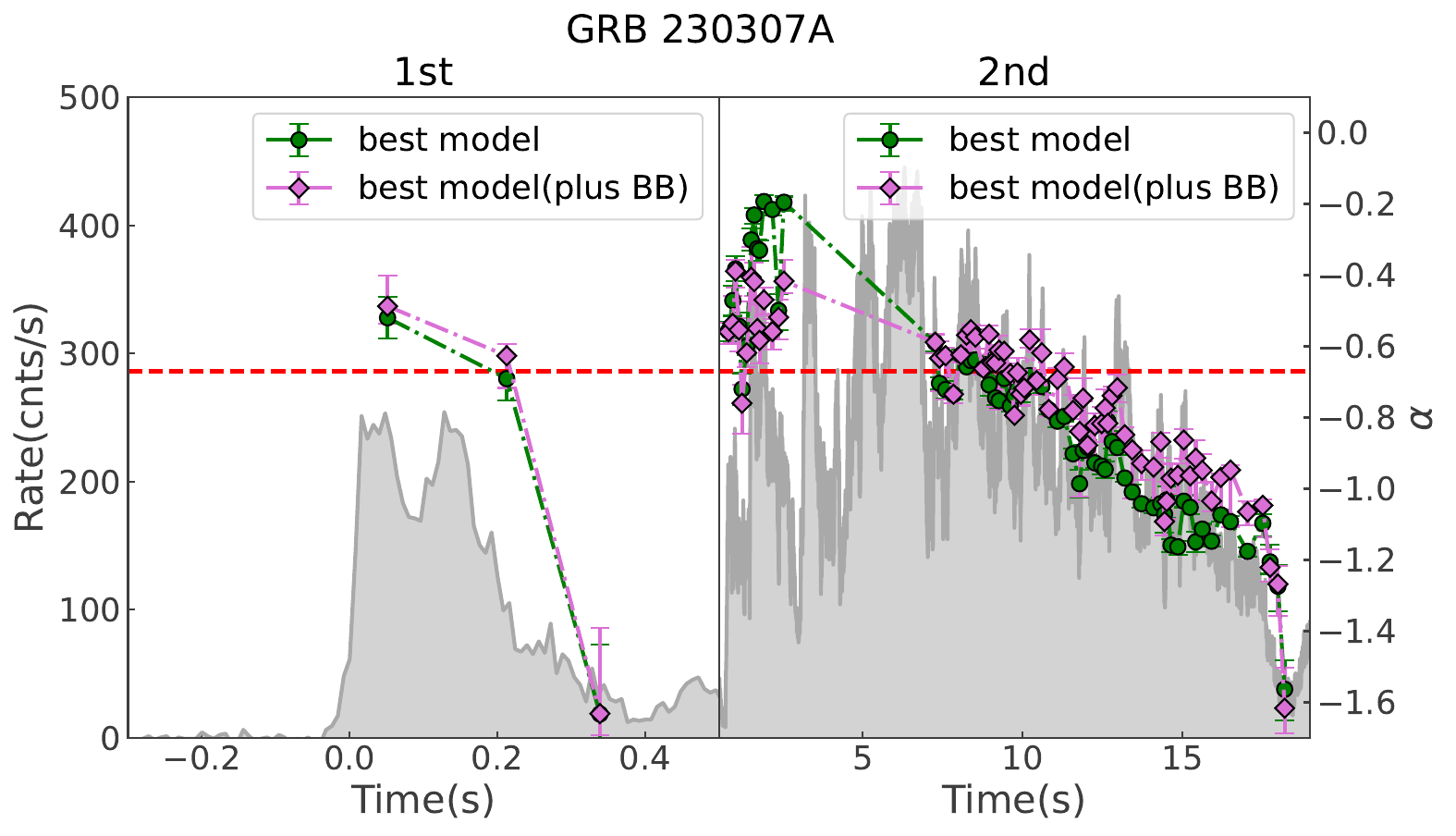}
\end{figure}
\begin{figure}[htbp]
\centering
\includegraphics [width=8cm,height=5cm]{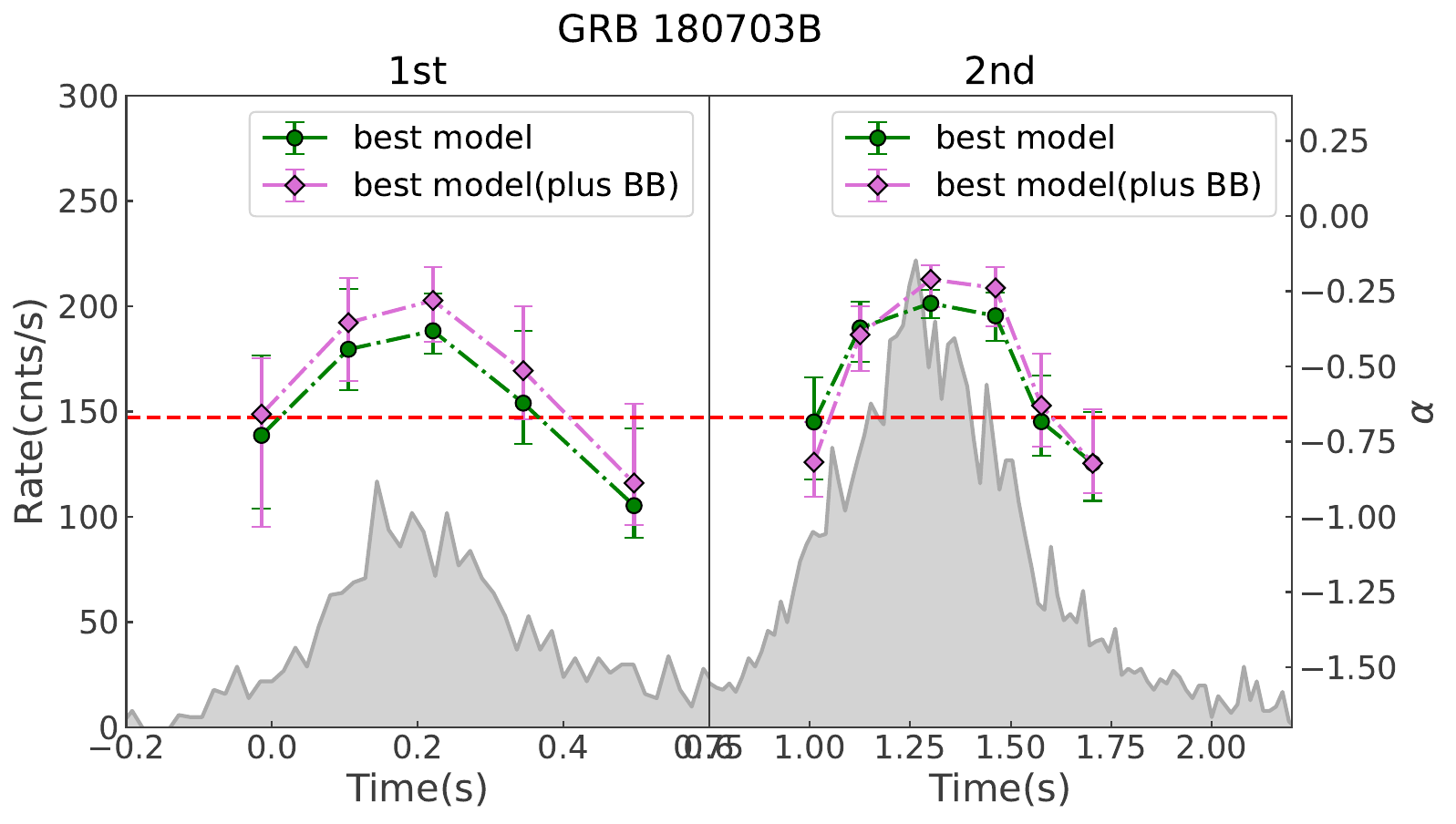}
\includegraphics [width=8cm,height=5cm]{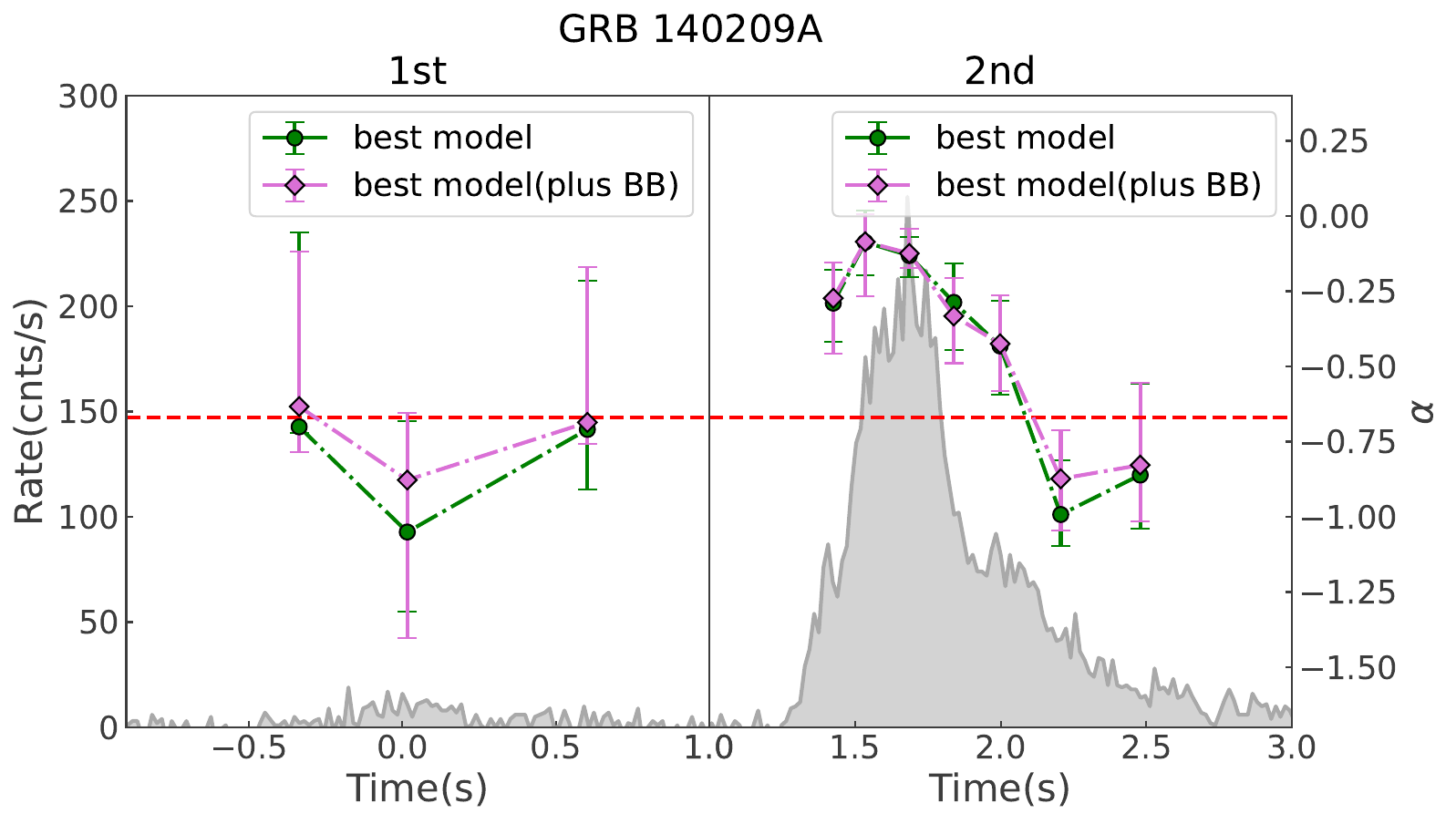}
   \figcaption{Evolution of the spectral parameter $\alpha$ over time fitted with the best model for the GRBs. The green and purple data points represent best model and best model + BB, respectively. ``1st'' denotes precursors, and ``2nd'' denotes main bursts. The red dashed line indicates $\alpha$= -0.67. \label{fig B1}}
\end{figure}
\begin{figure}[htbp]
\centering
\includegraphics [width=8.5cm,height=5cm]{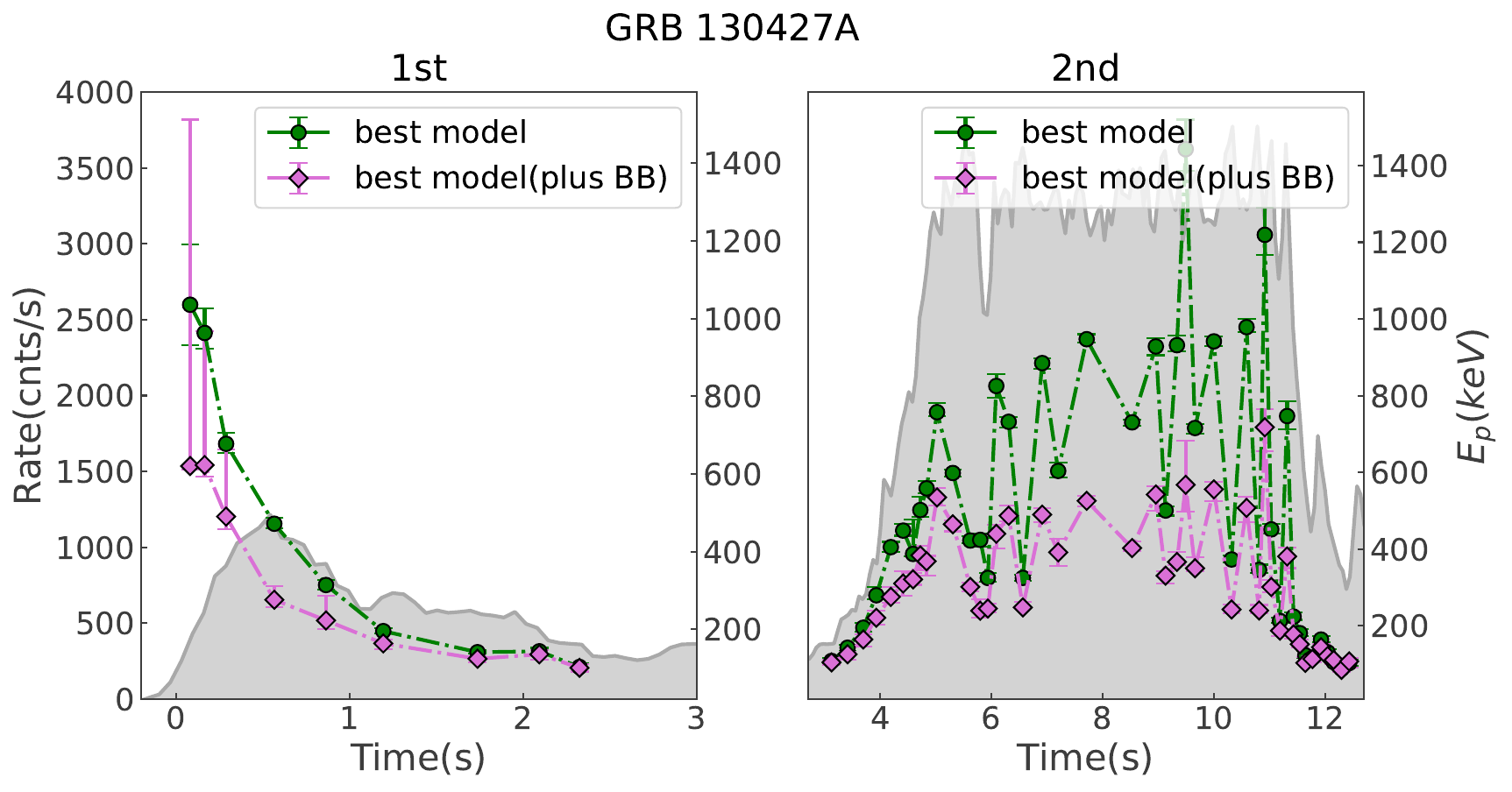}
\includegraphics [width=8.5cm,height=5cm]{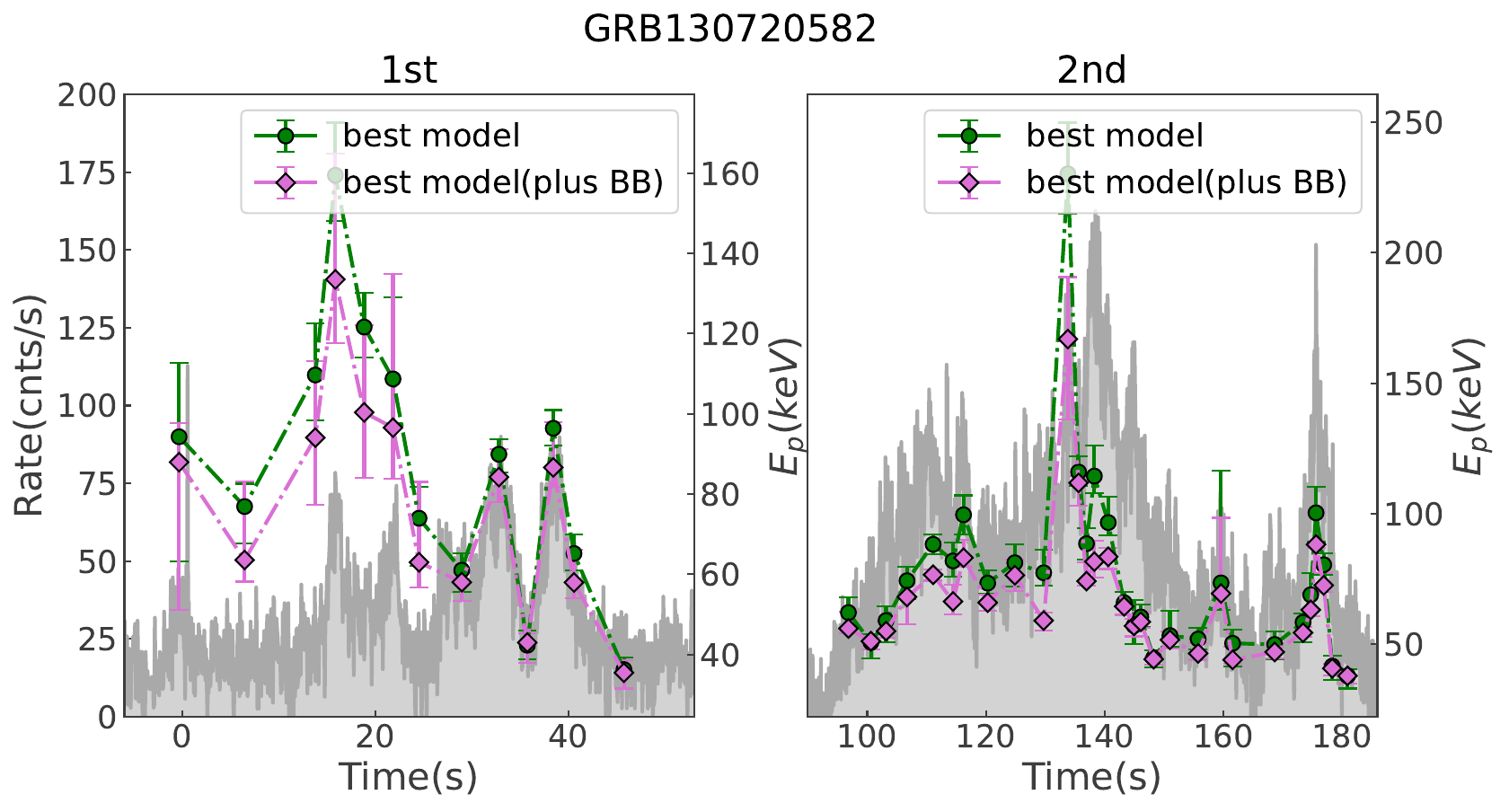}
\includegraphics [width=8.5cm,height=5cm]{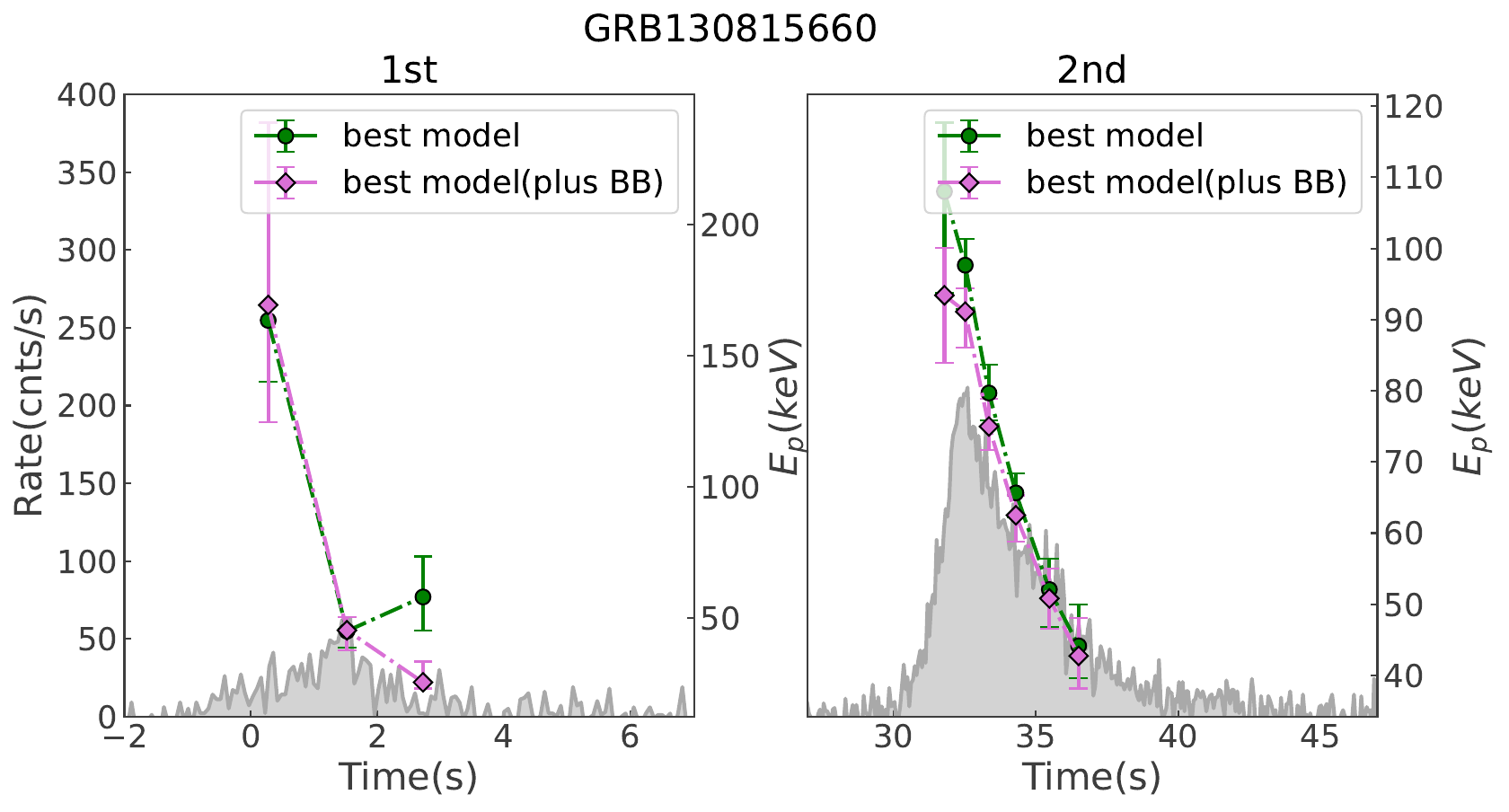}
\includegraphics [width=8.5cm,height=5cm]{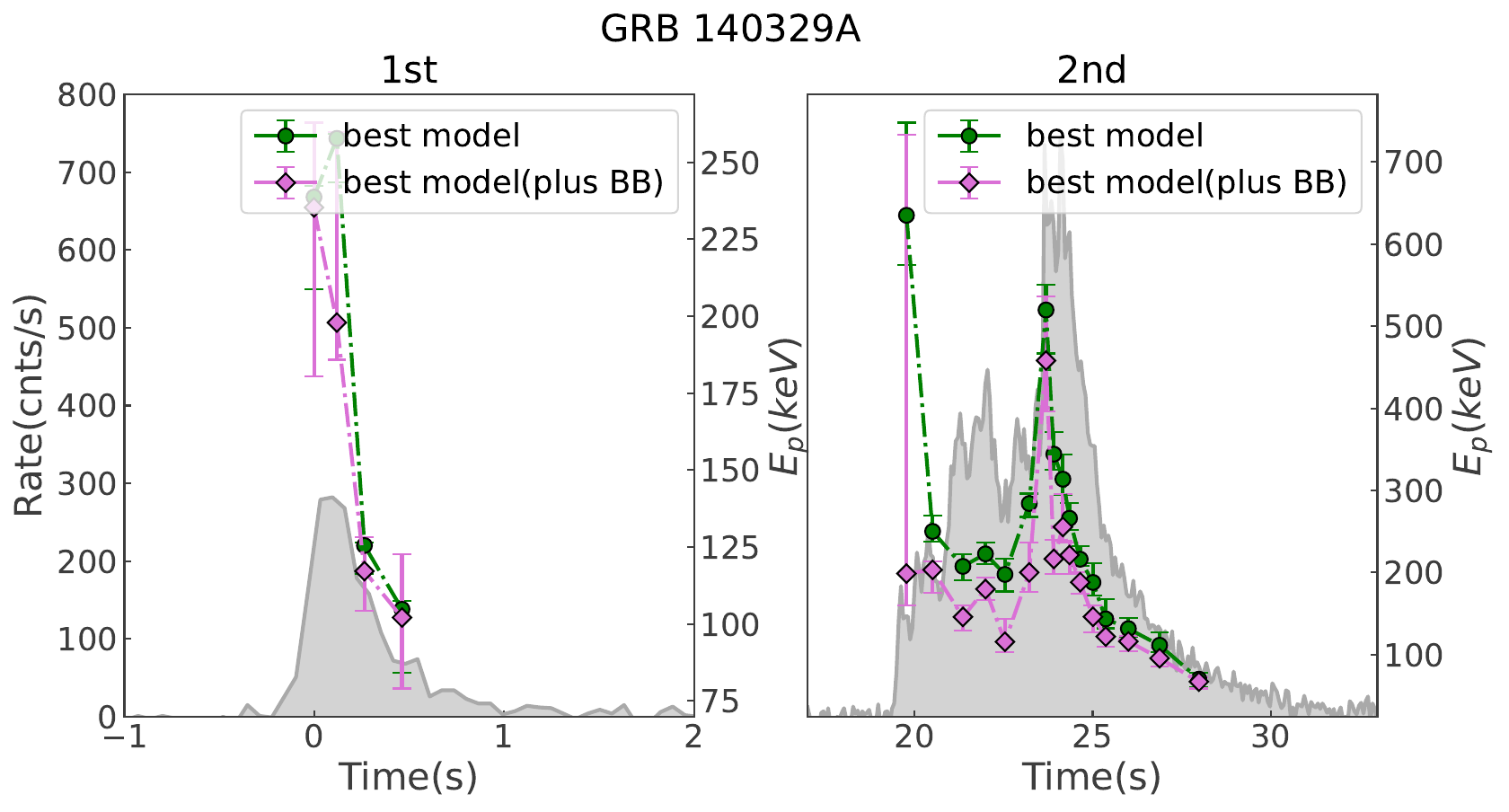}
\includegraphics [width=8.5cm,height=5cm]{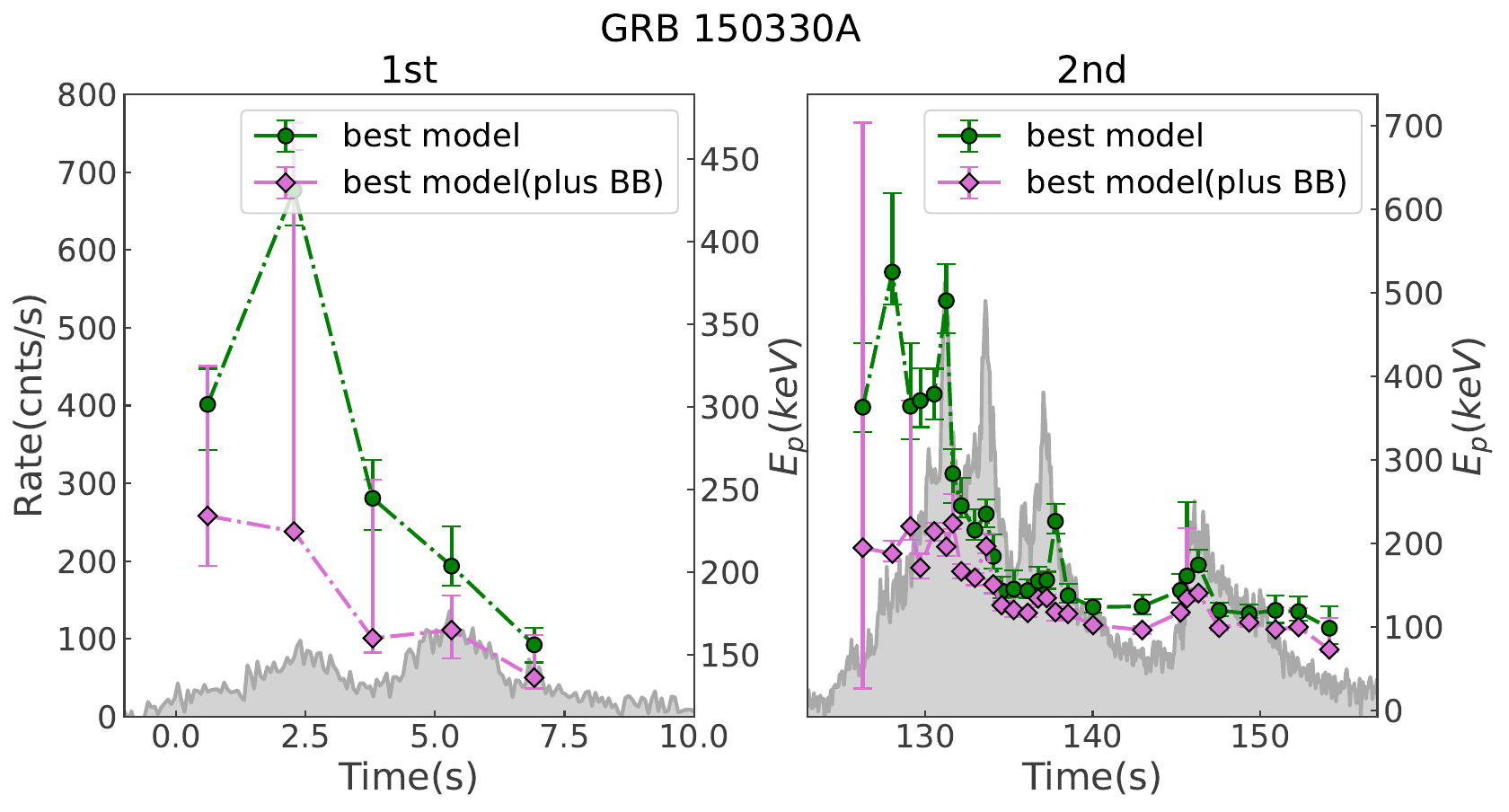}
\includegraphics [width=8.5cm,height=5cm]{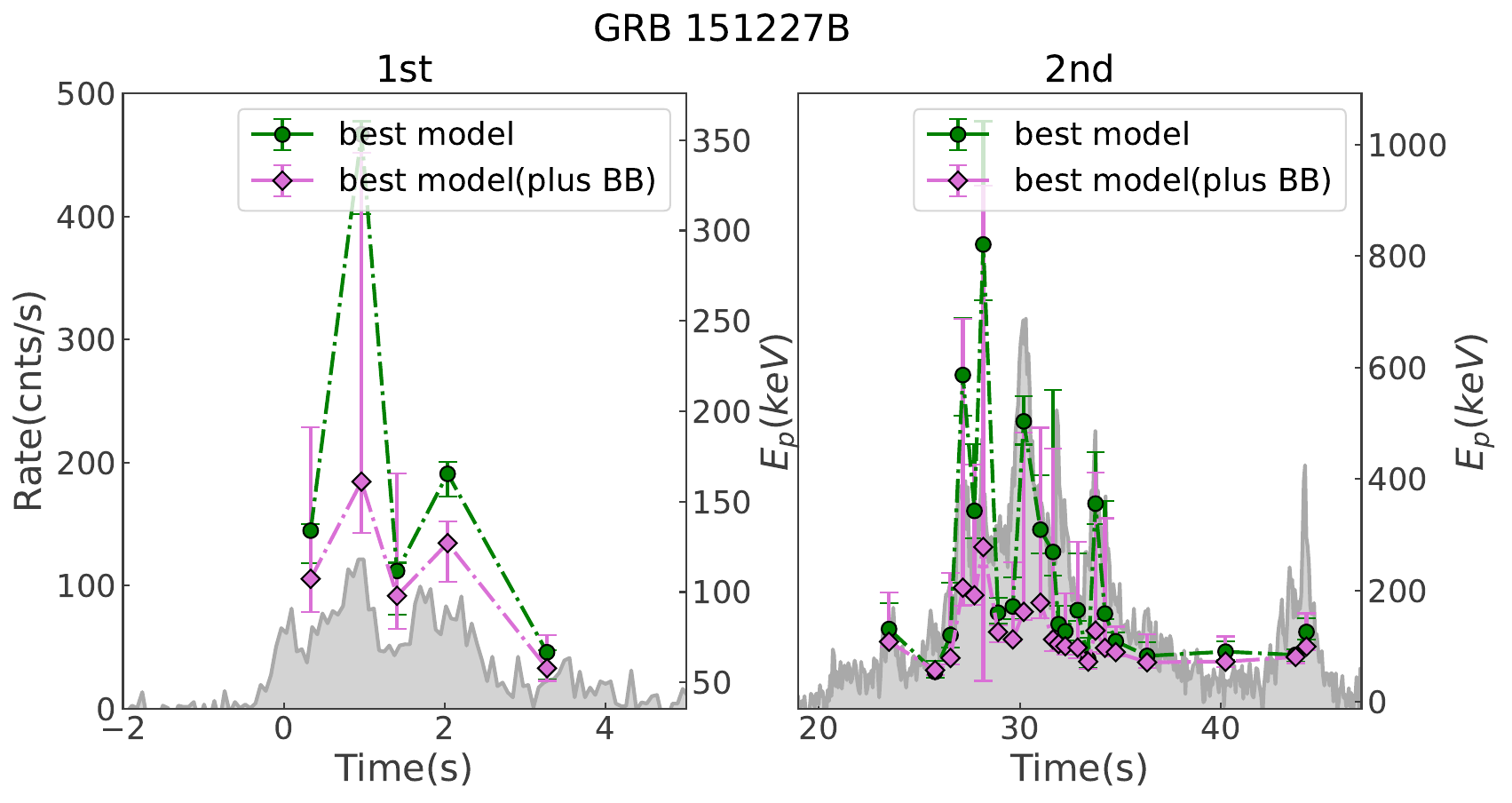}
\end{figure}
\begin{figure}[htbp]
\centering
\includegraphics [width=8.5cm,height=5cm]{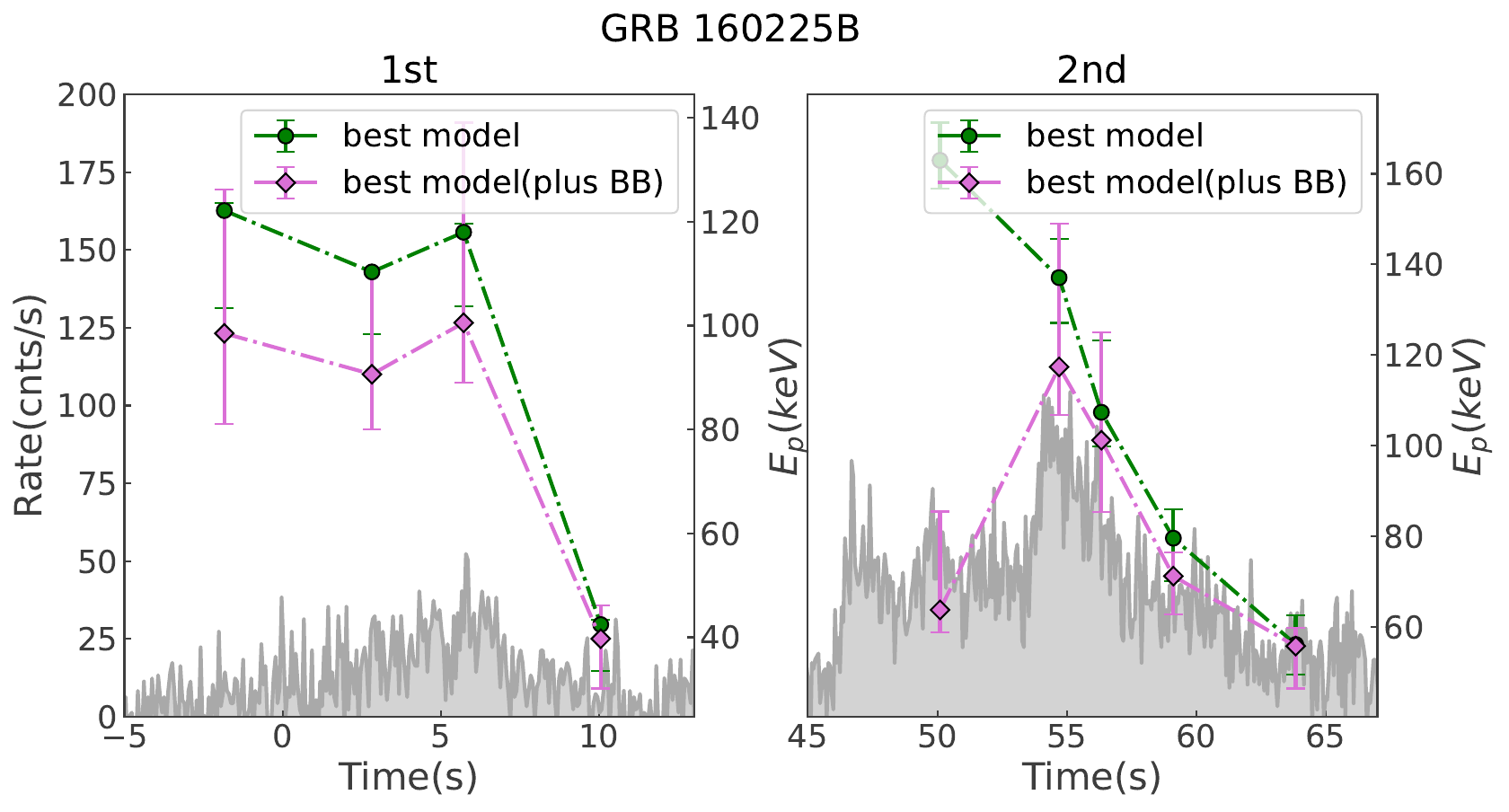}
\includegraphics [width=8.5cm,height=5cm]{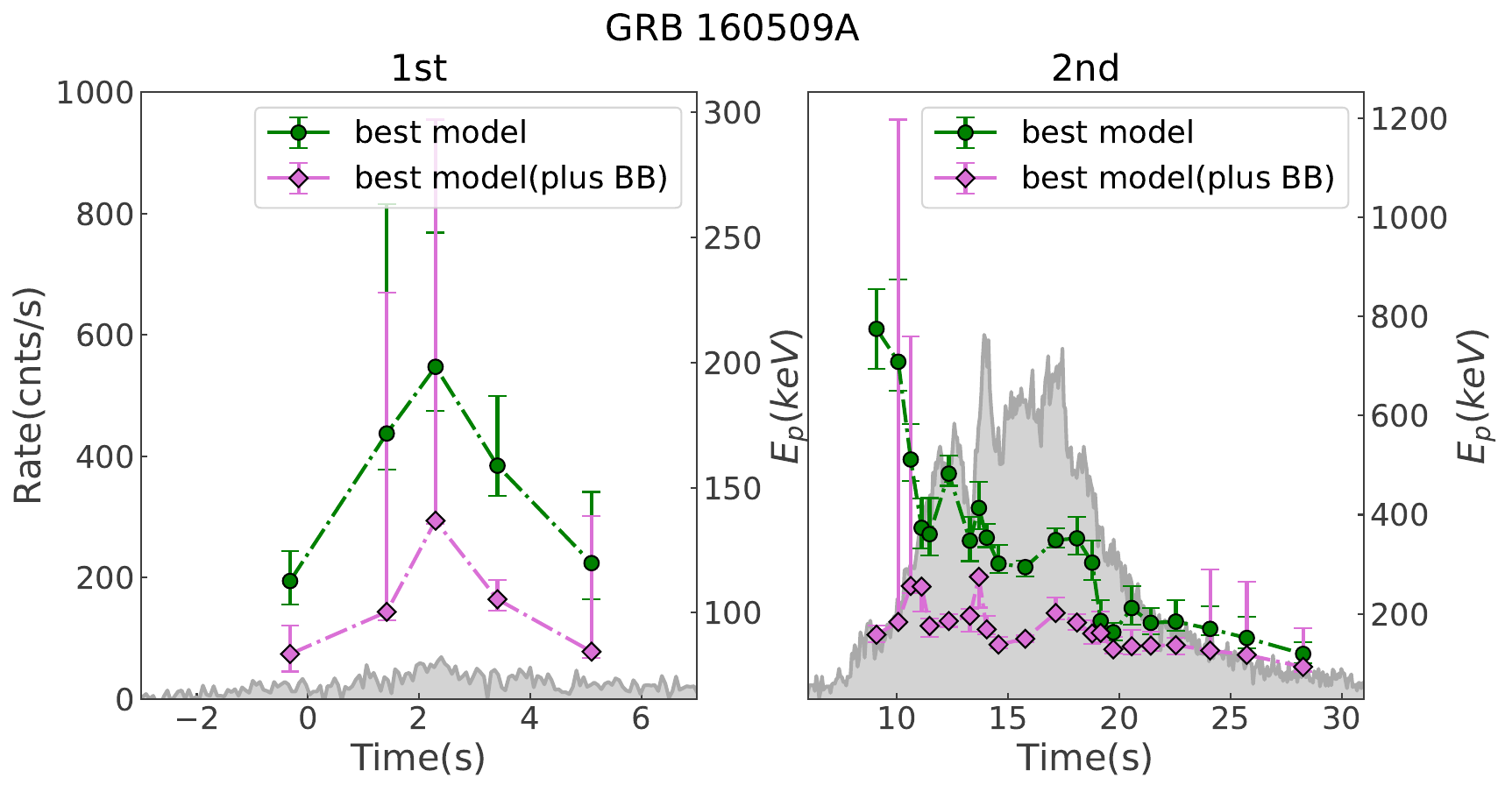}
\includegraphics [width=8.5cm,height=5cm]{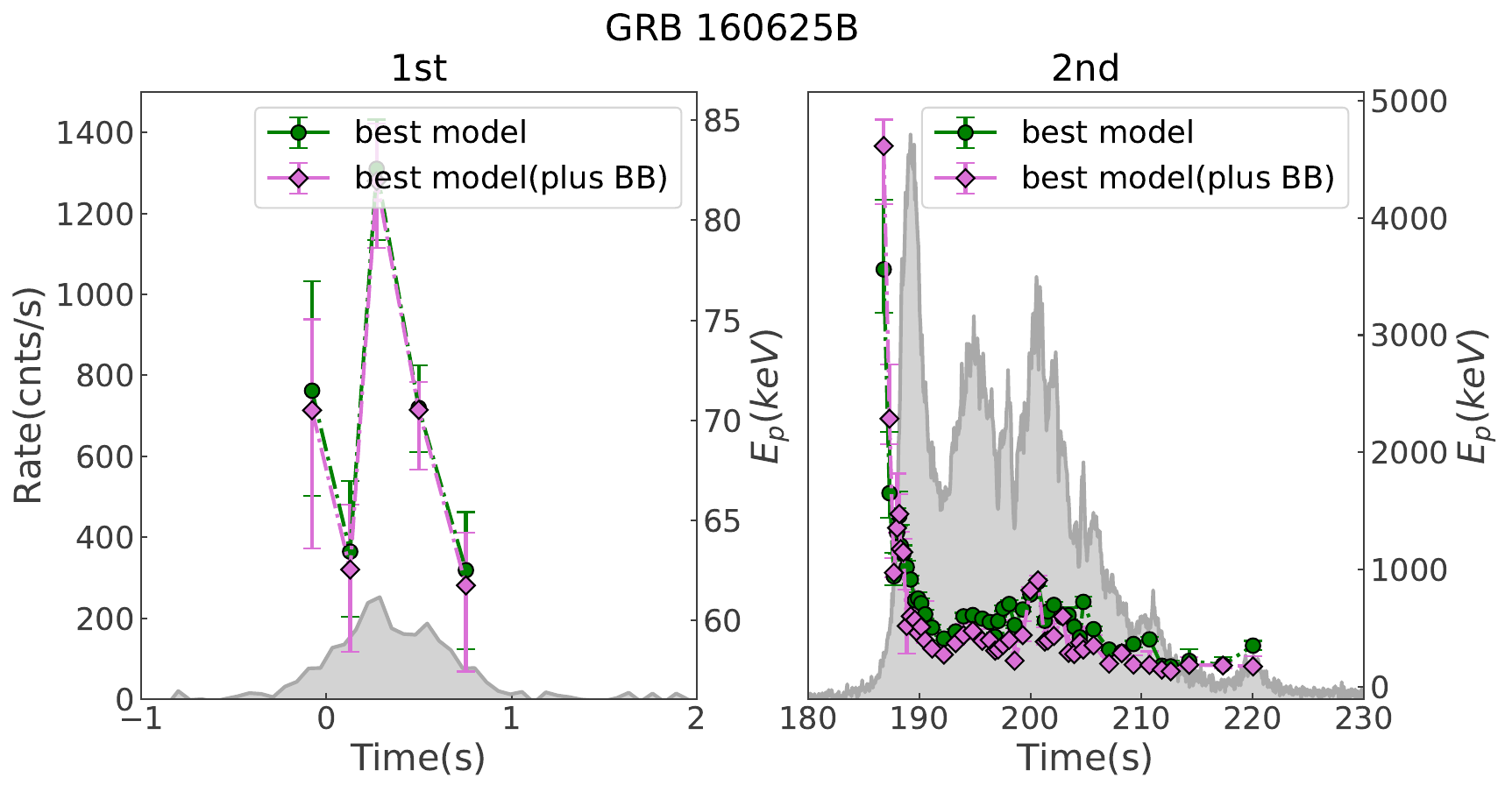}
\includegraphics [width=8.5cm,height=5cm]{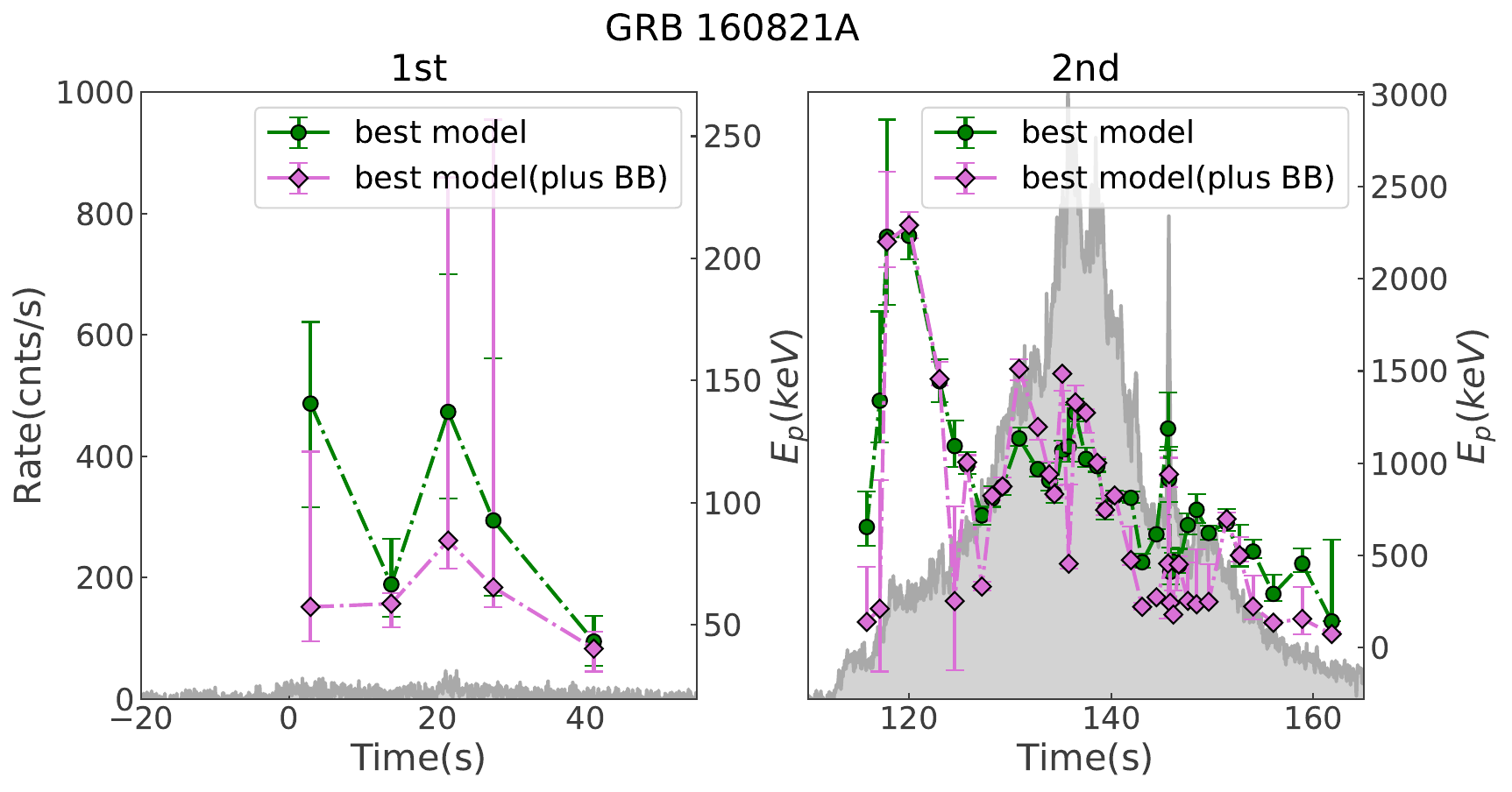}
\includegraphics [width=8.5cm,height=5cm]{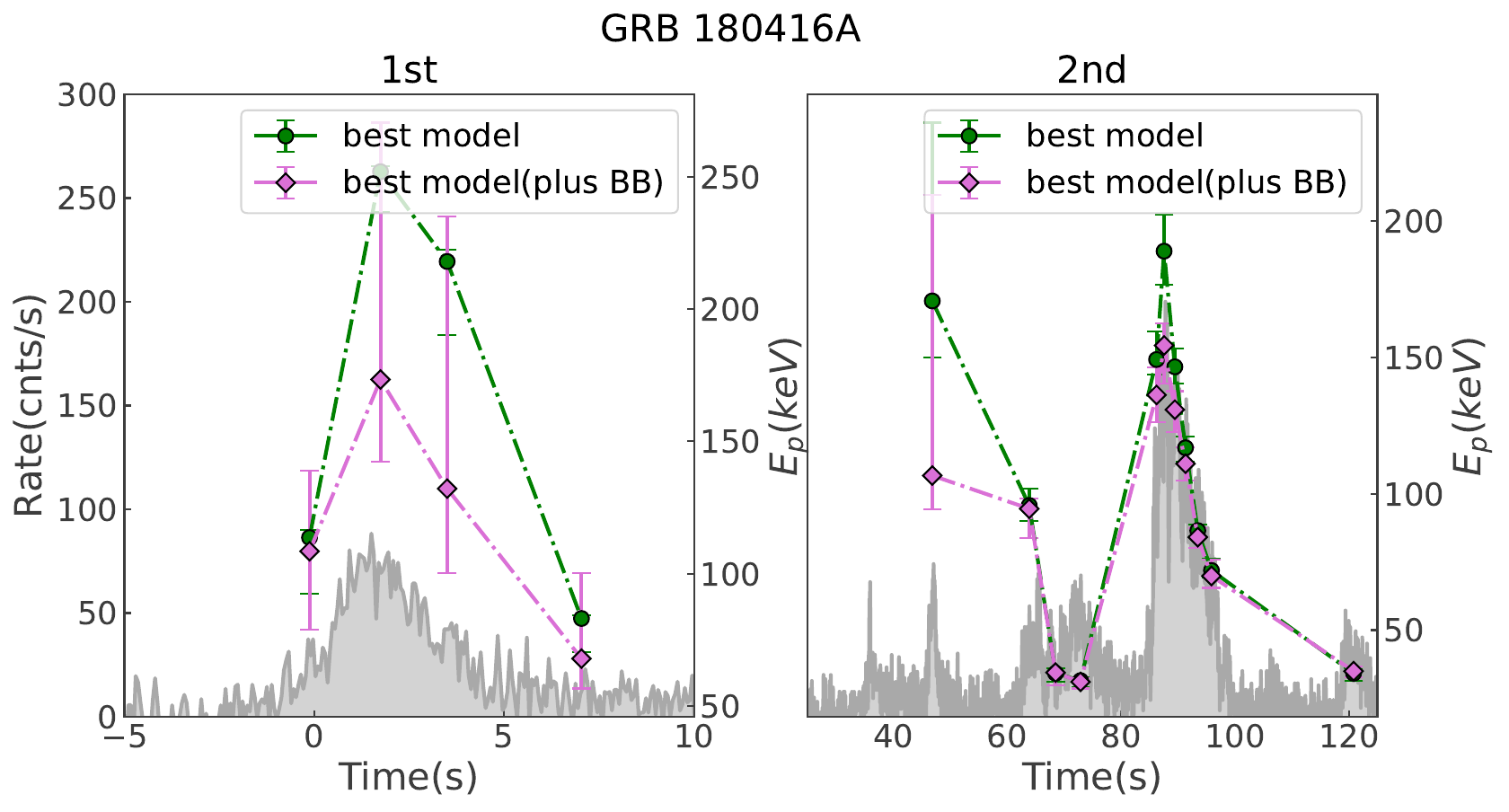}
\includegraphics [width=8.5cm,height=5cm]{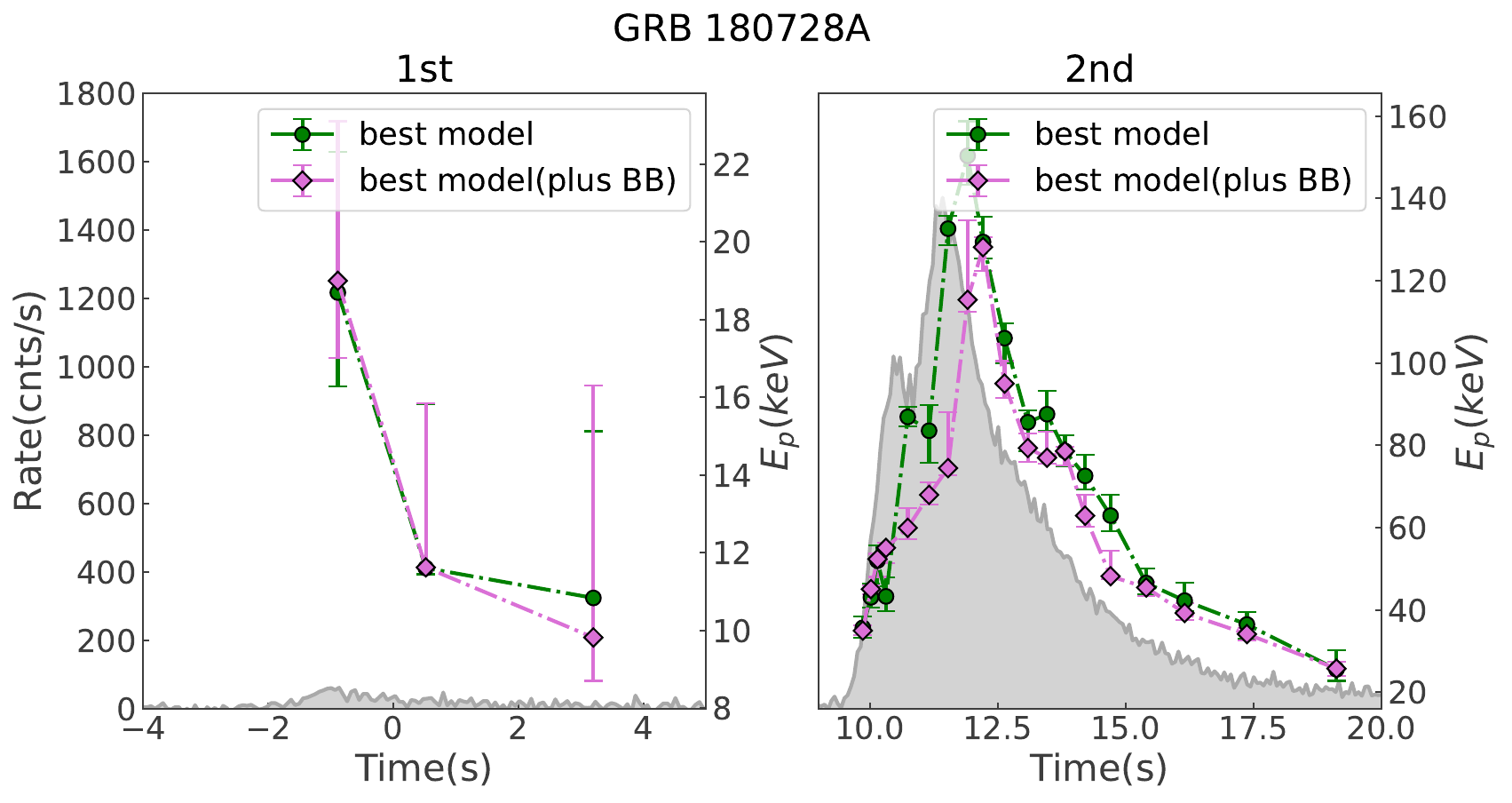}
\includegraphics [width=8.5cm,height=5cm]{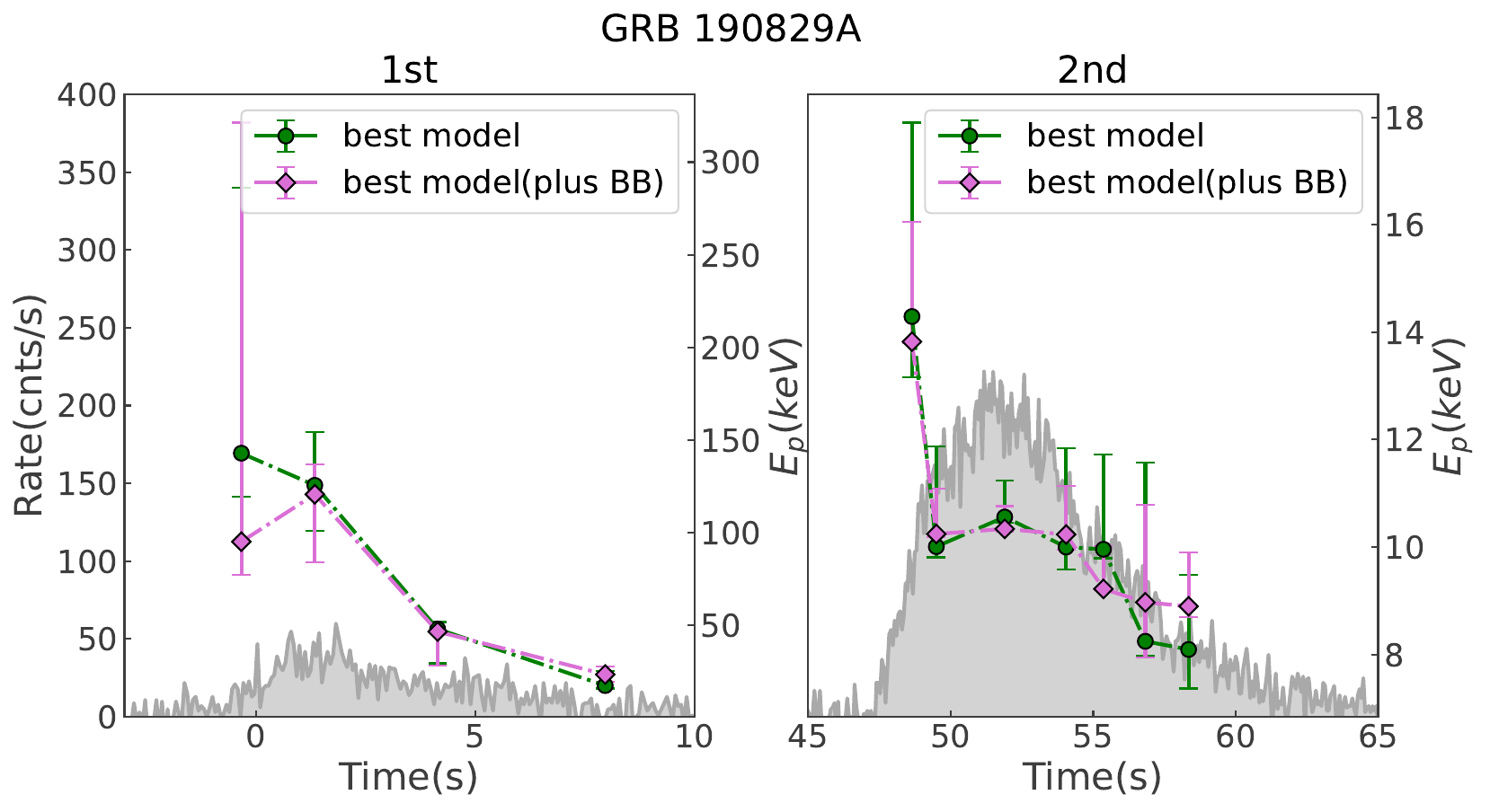}
\includegraphics [width=8.5cm,height=5cm]{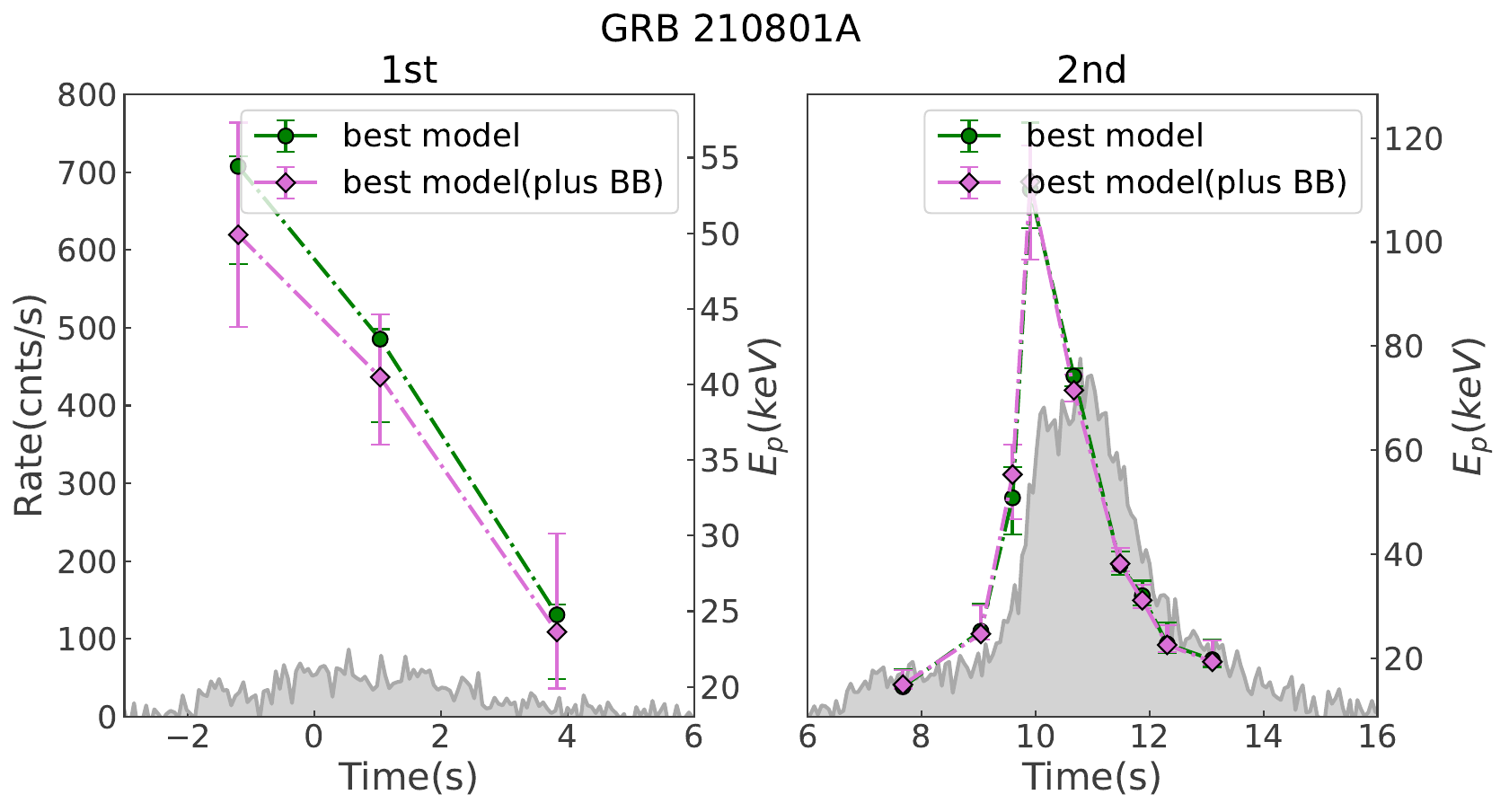}
\end{figure}
\begin{figure}[htbp]
\centering
\includegraphics [width=8.5cm,height=5cm]{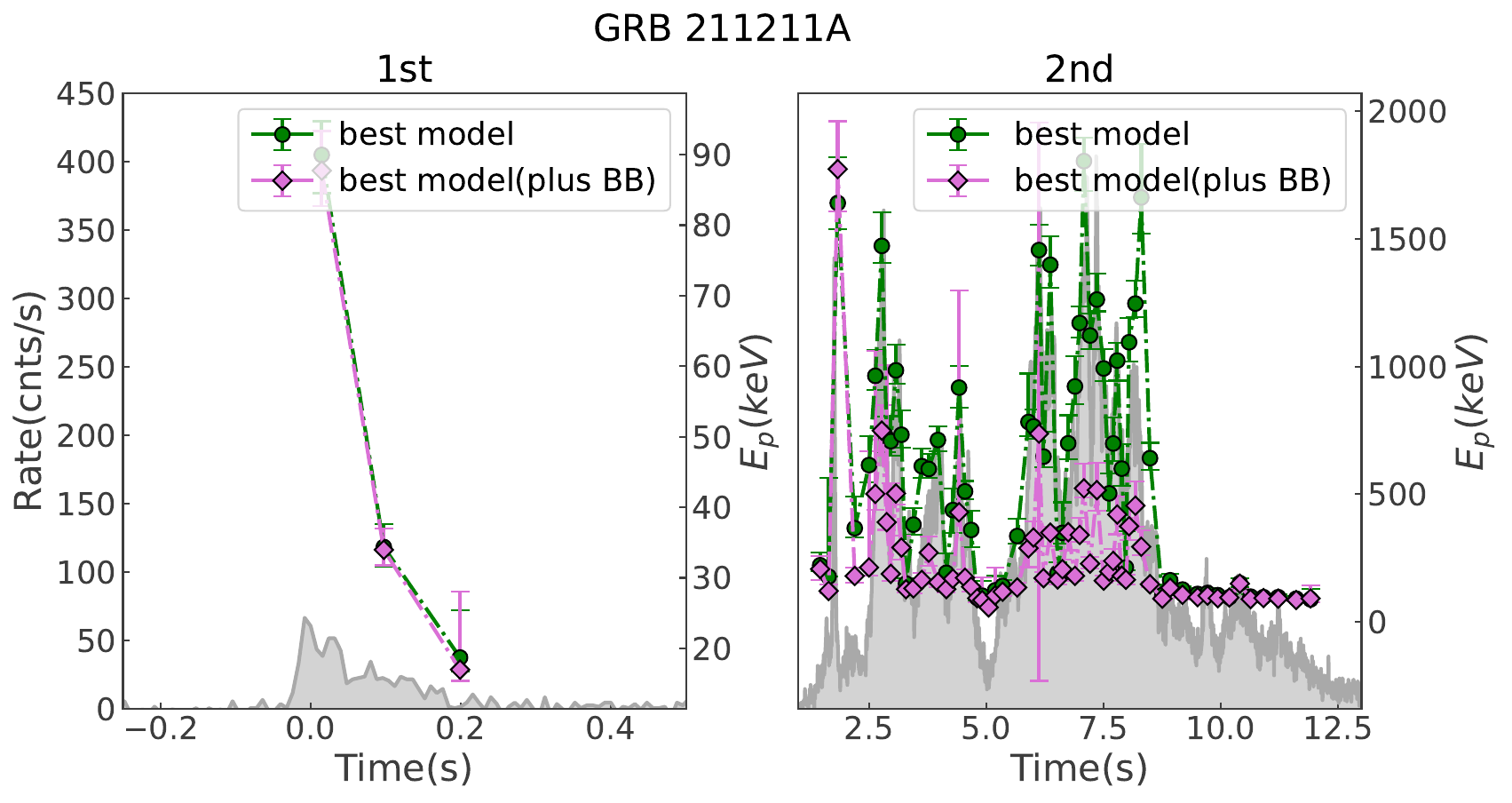}
\includegraphics [width=8.5cm,height=5cm]{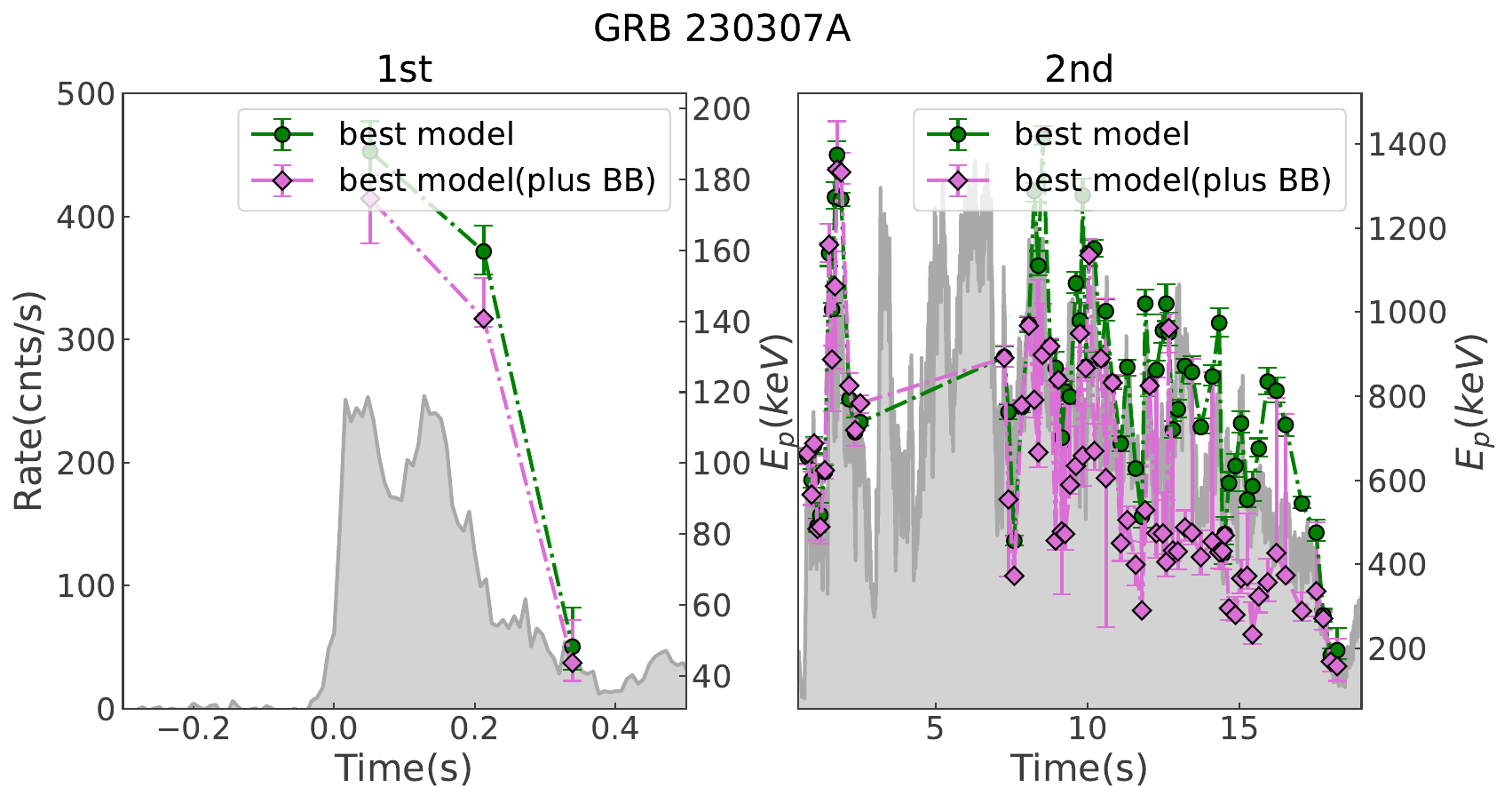}
\includegraphics [width=8.5cm,height=5cm]{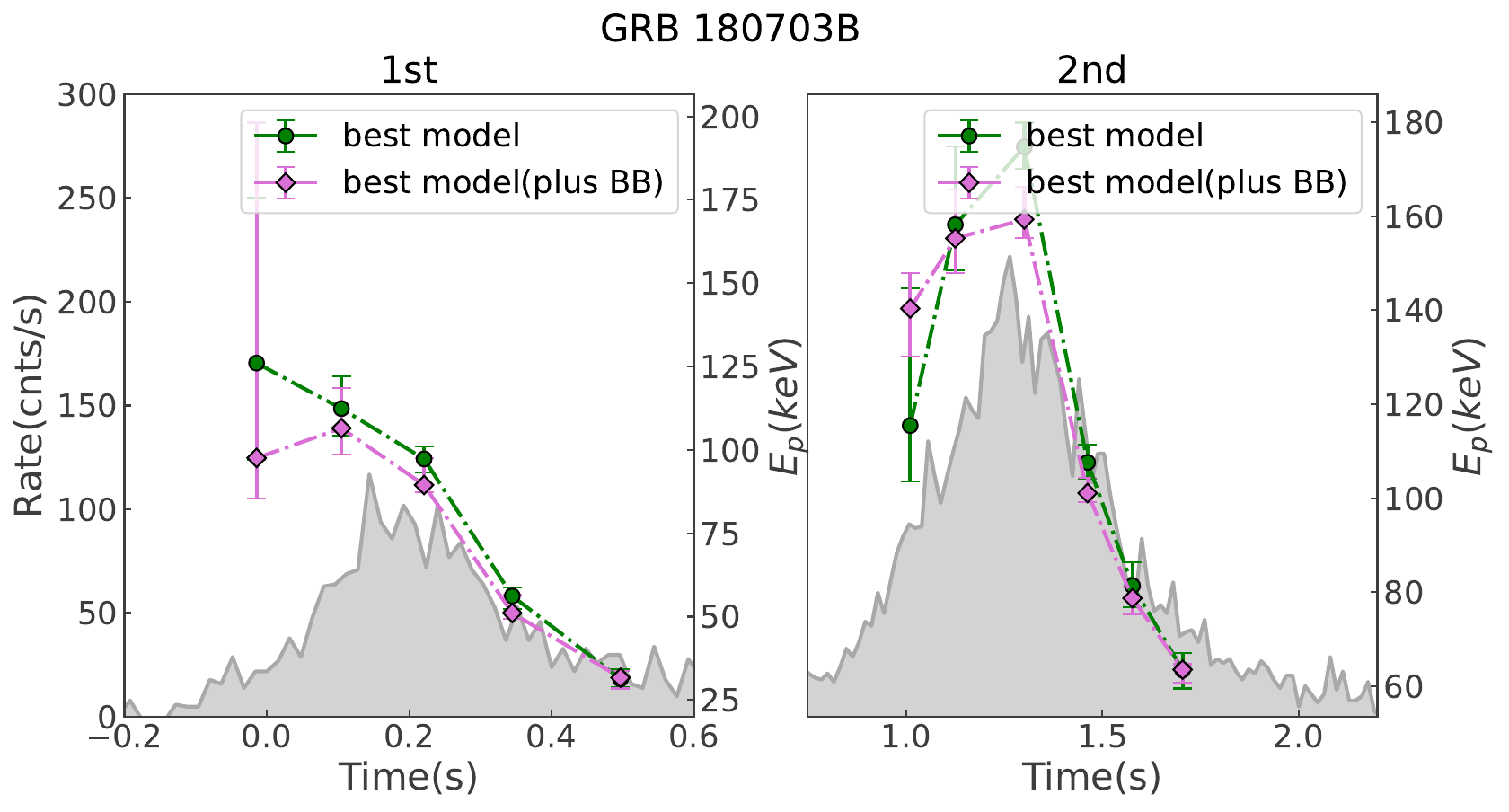}
\includegraphics [width=8.5cm,height=5cm]{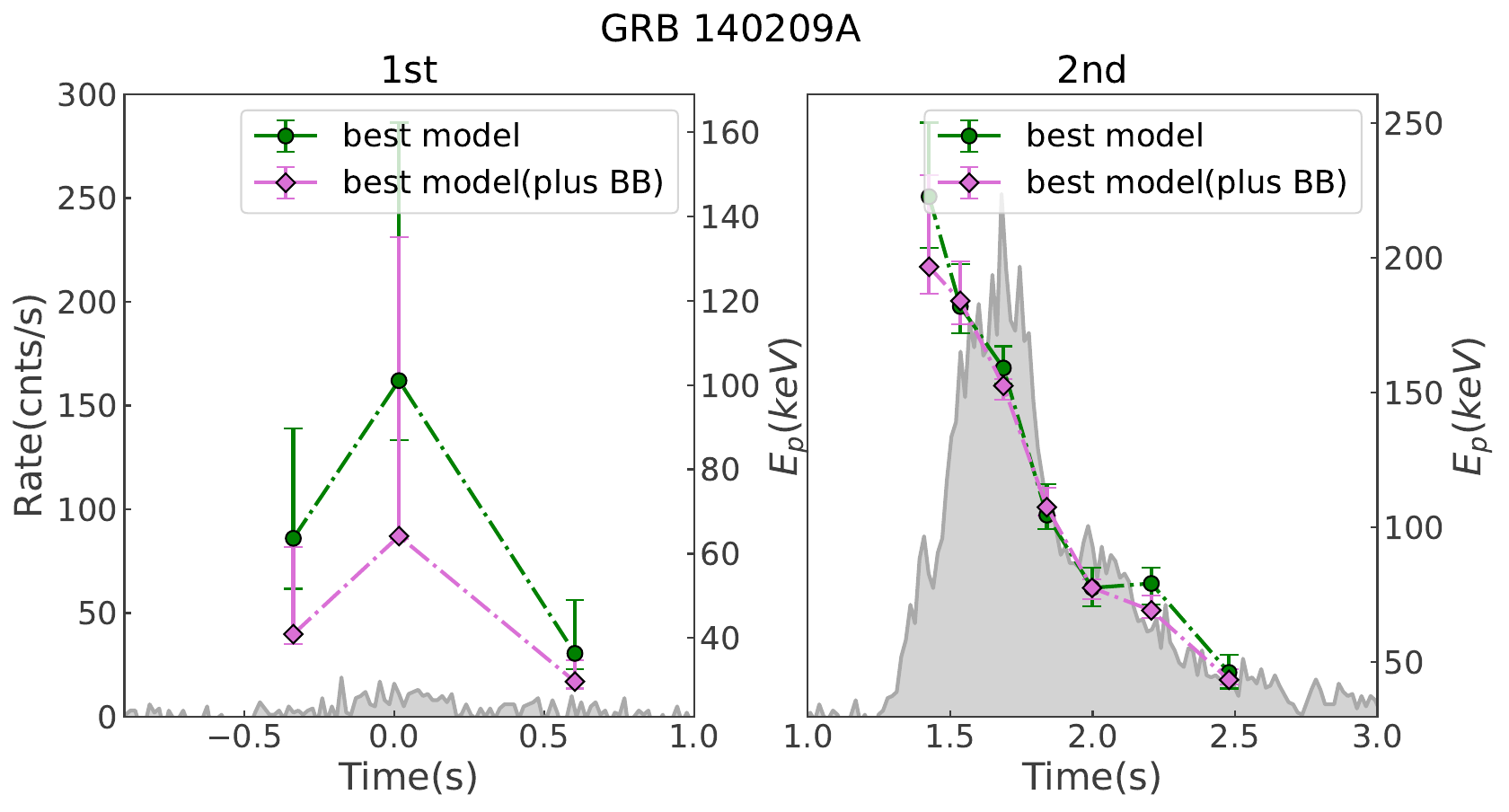}
   \figcaption{Evolution of the spectral parameter $E_{p}$ over time fitted with the best model for the GRBs. All symbols are the same as in Figure \ref{fig B1}. \label{fig B2}}
\end{figure}

\begin{figure}[htbp]
\centering
\includegraphics [width=8.5cm,height=4.5cm]{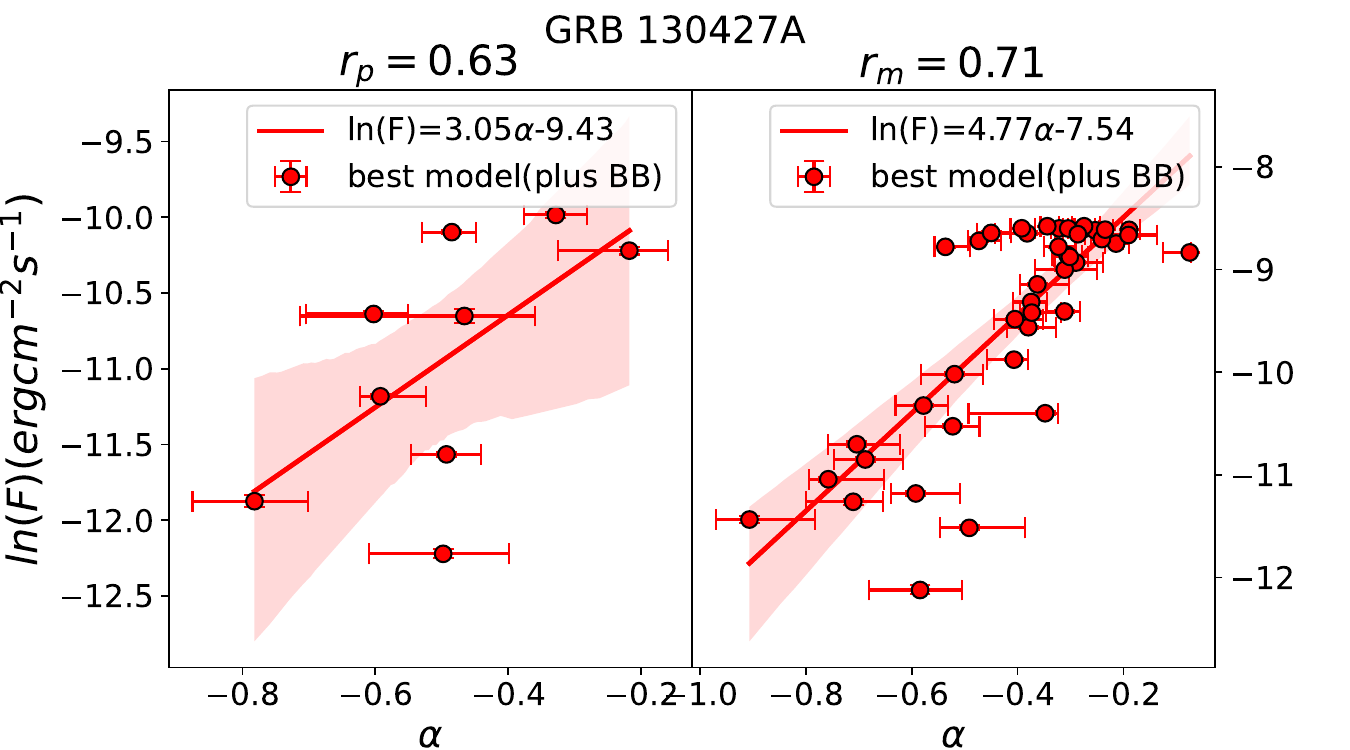}
\includegraphics [width=8.5cm,height=4.5cm]{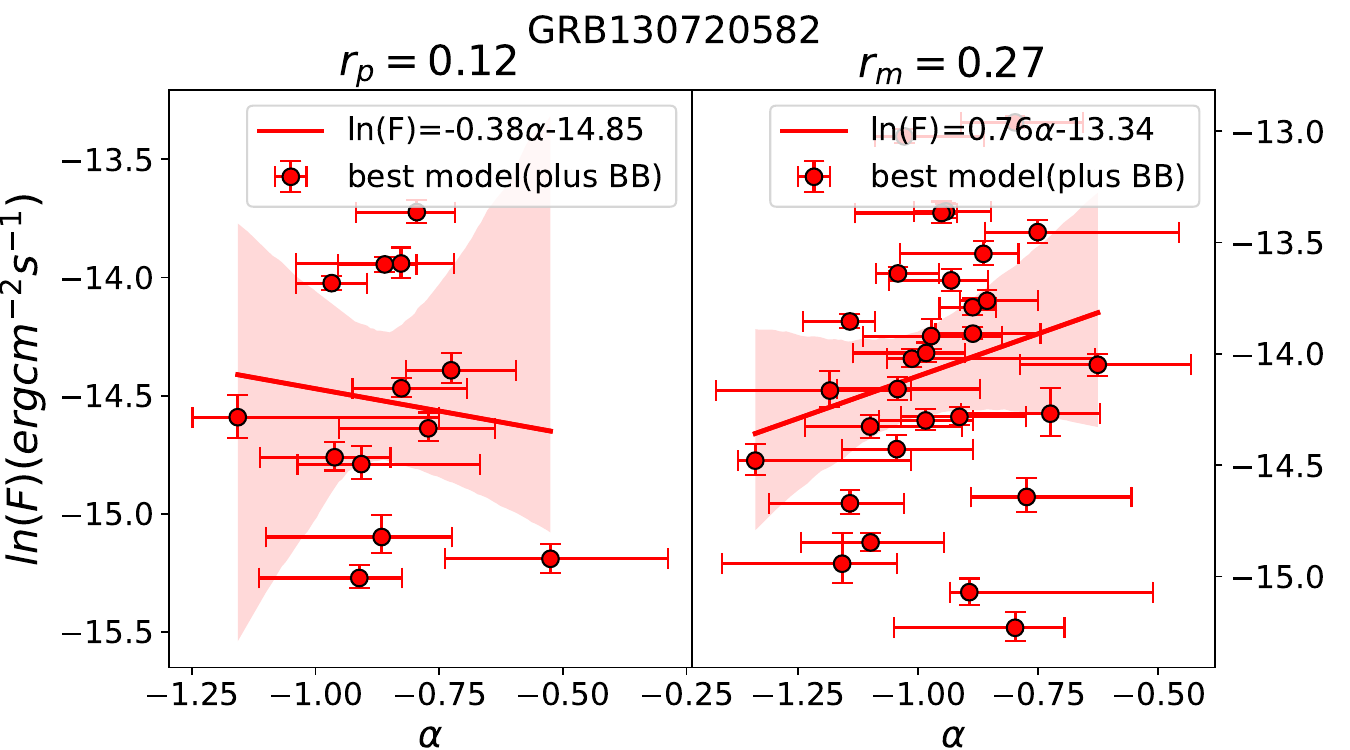}
\includegraphics [width=8.5cm,height=4.5cm]{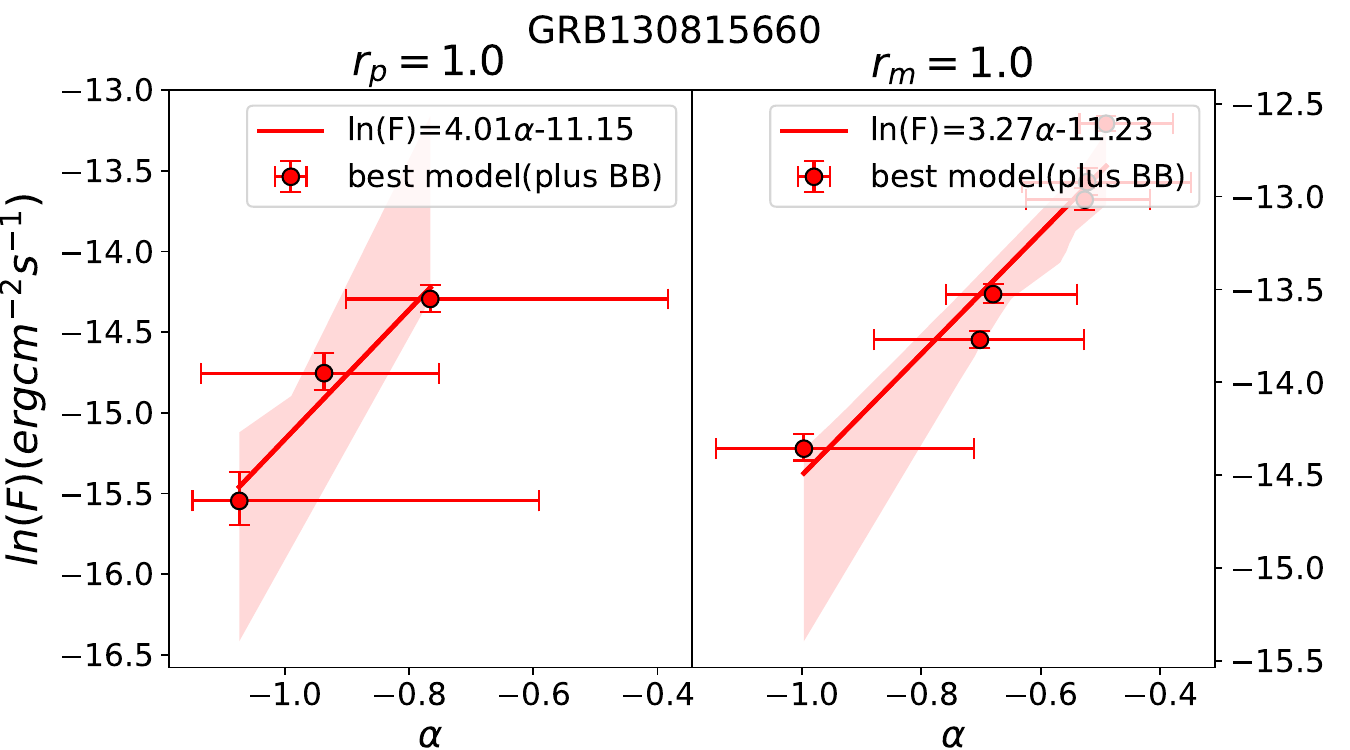}
\includegraphics [width=8.5cm,height=4.5cm]{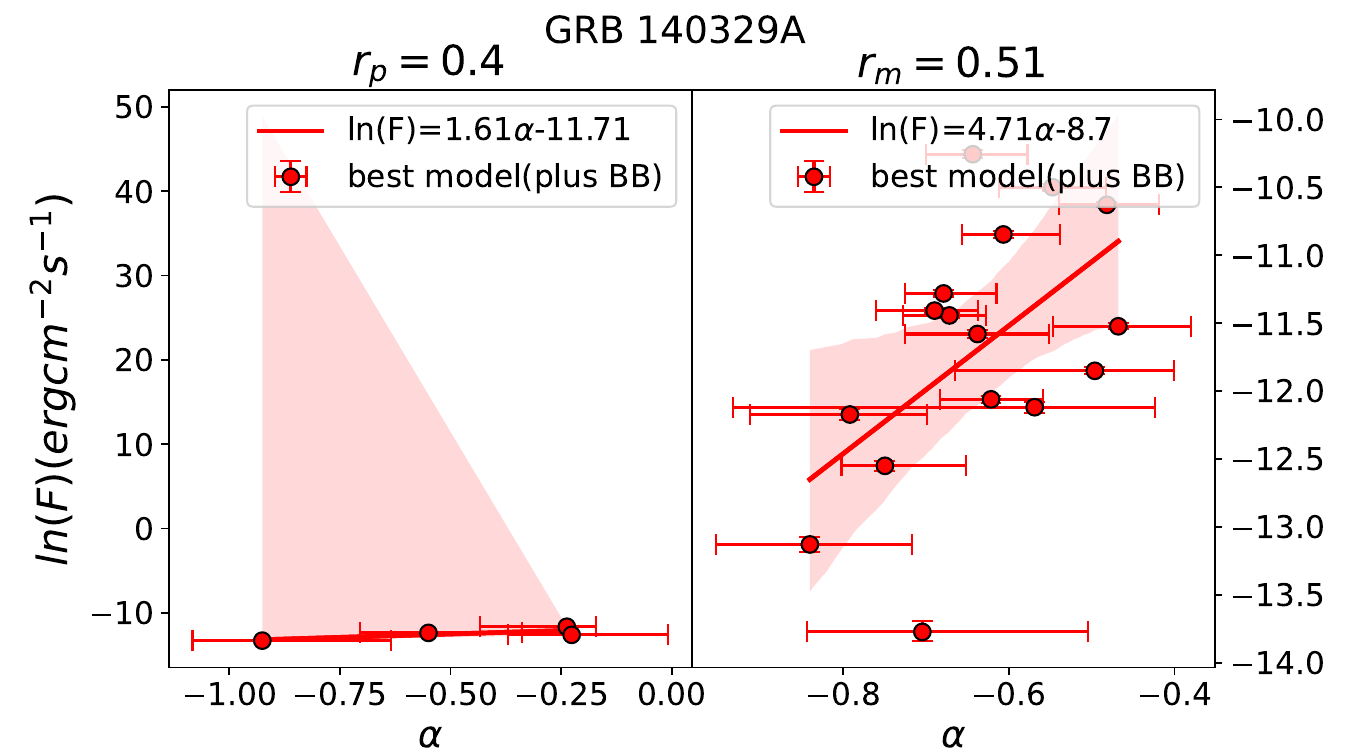}
\includegraphics [width=8.5cm,height=4.5cm]{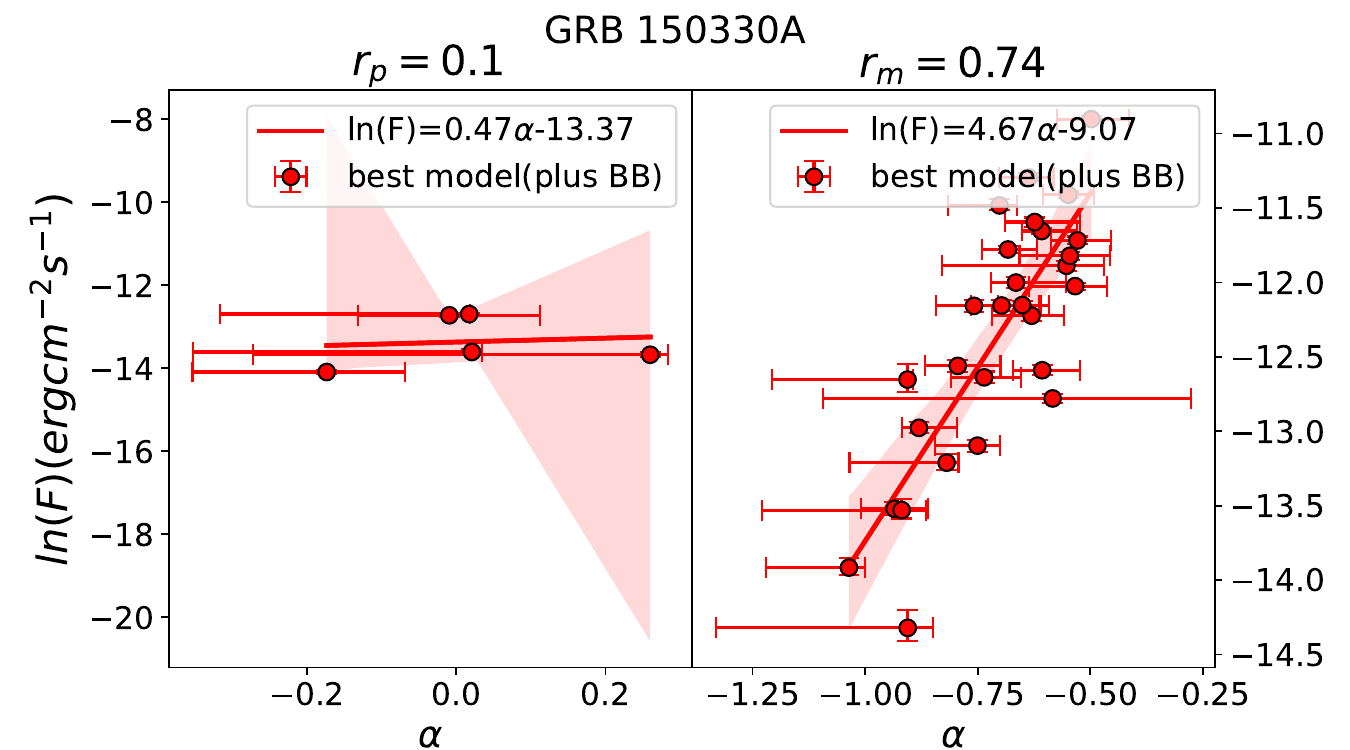}
\includegraphics [width=8.5cm,height=4.5cm]{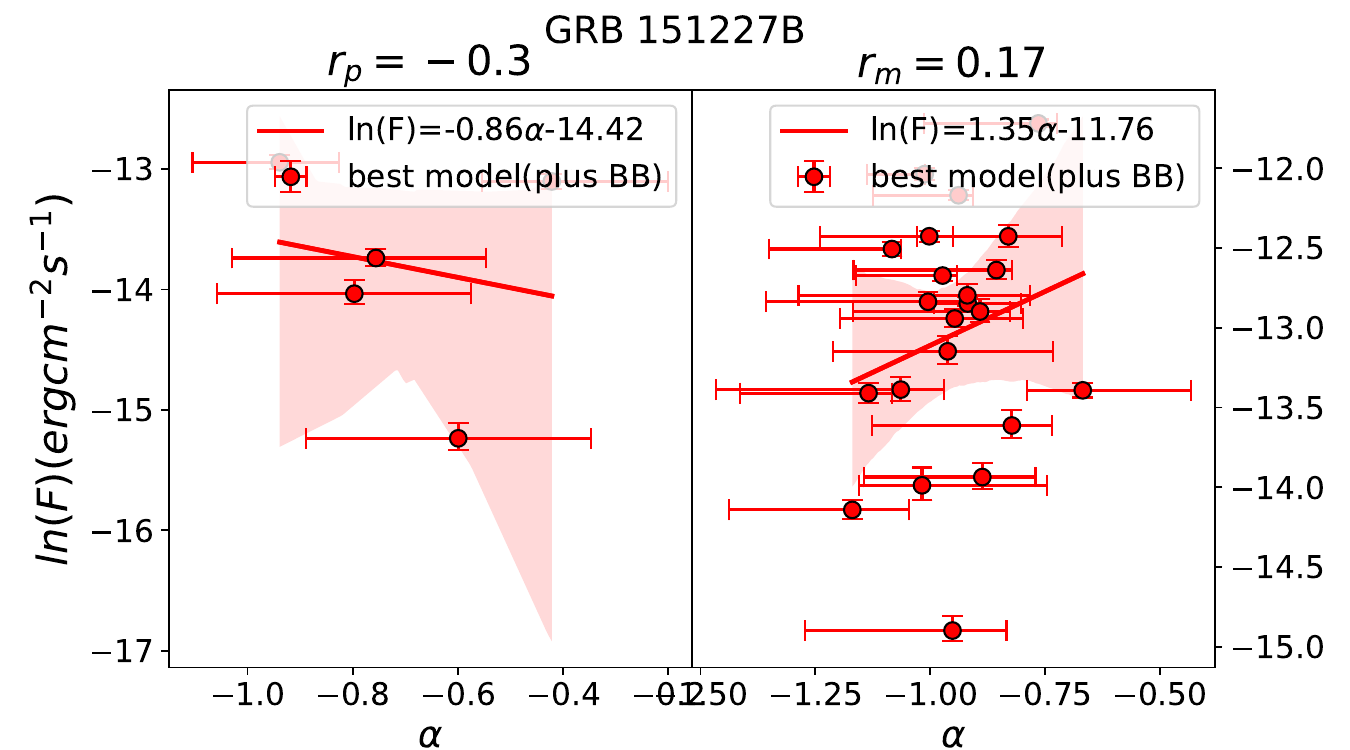}
\includegraphics [width=8.5cm,height=4.5cm]{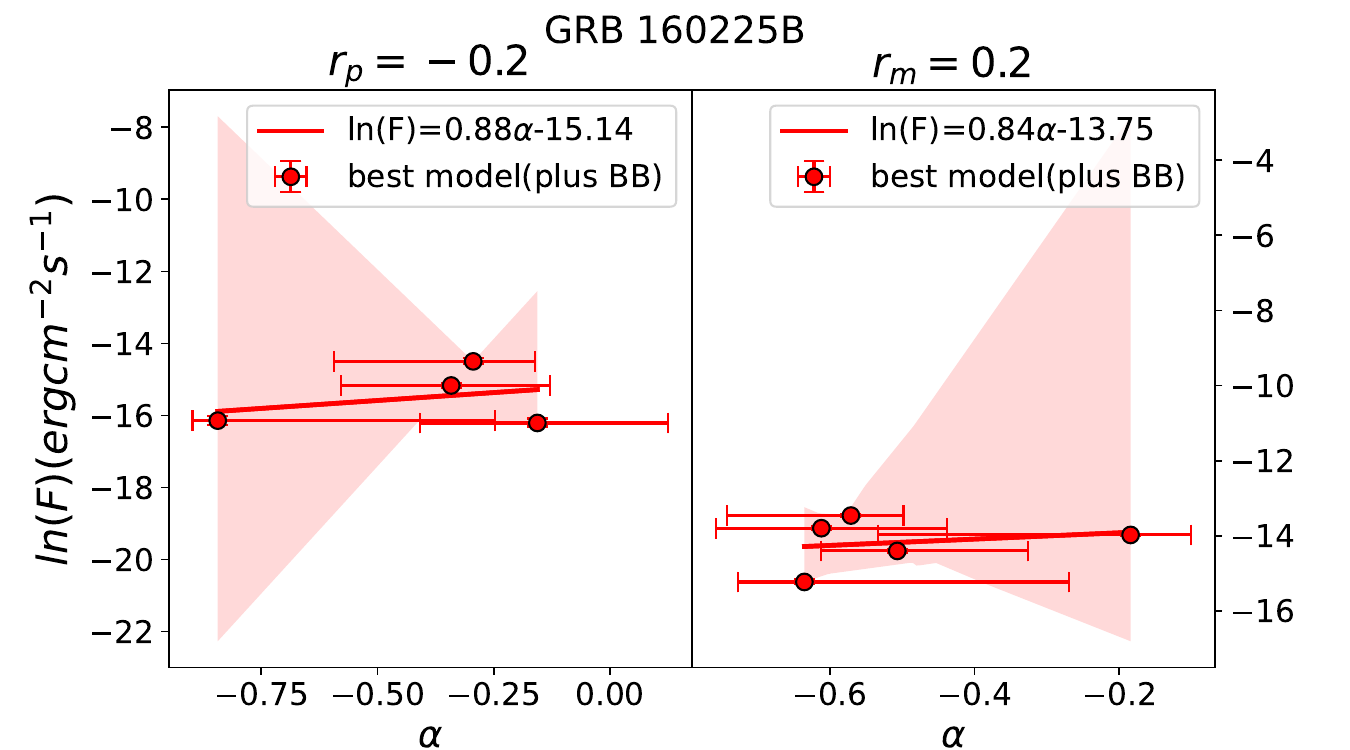}
\includegraphics [width=8.5cm,height=4.5cm]{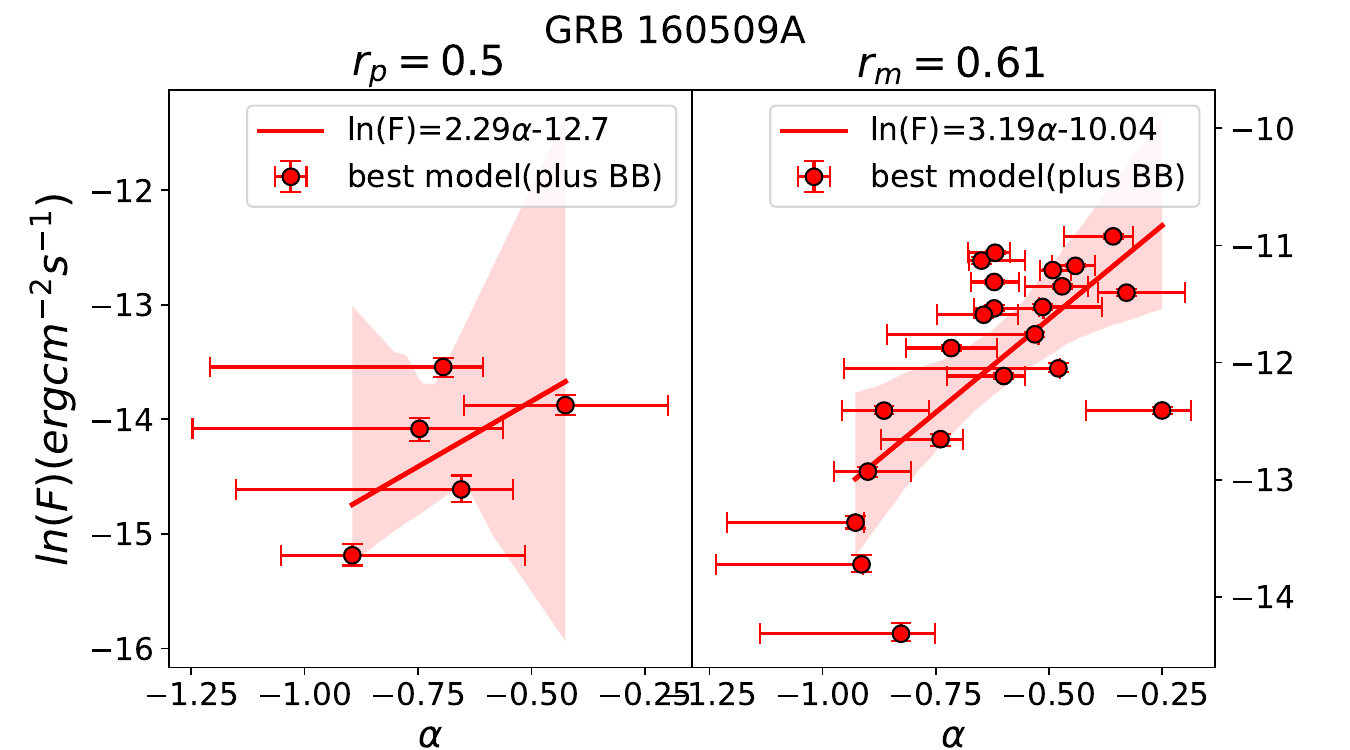}
\includegraphics [width=8.5cm,height=4.5cm]{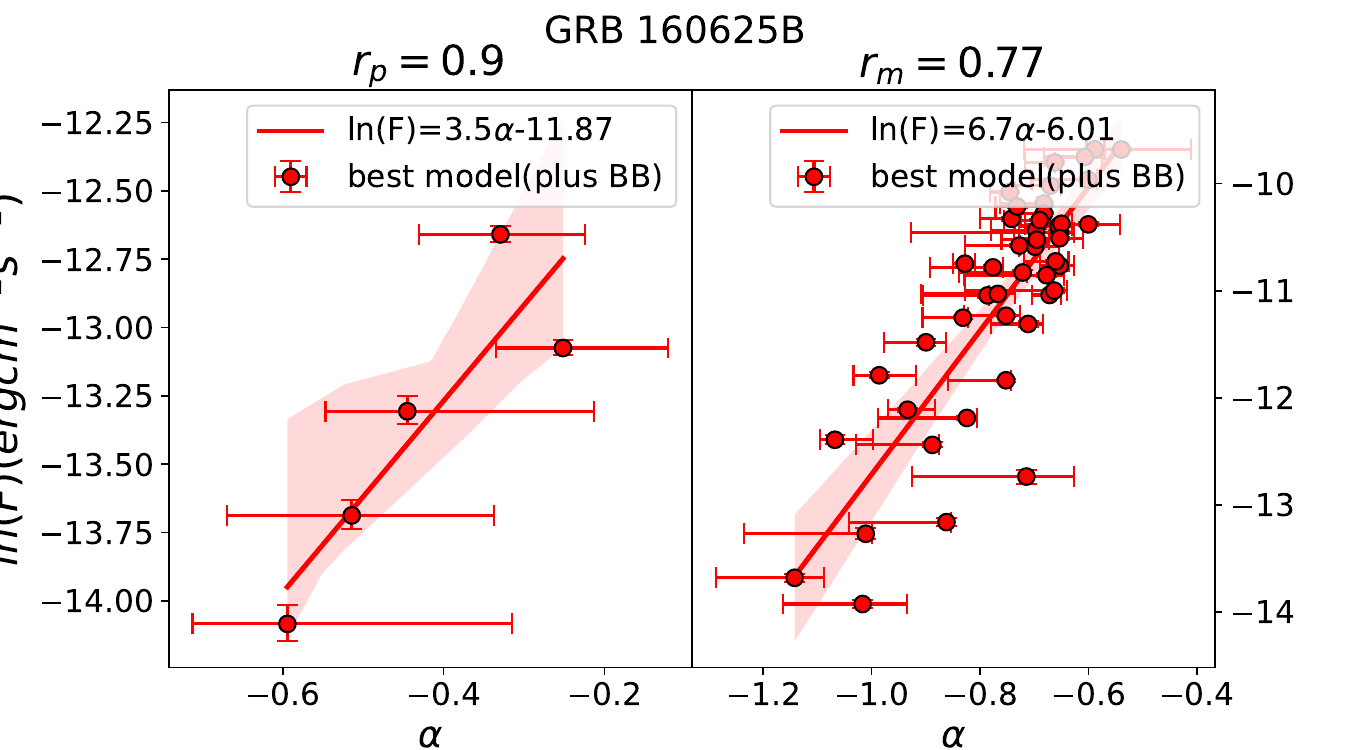}
\includegraphics [width=8.5cm,height=4.5cm]{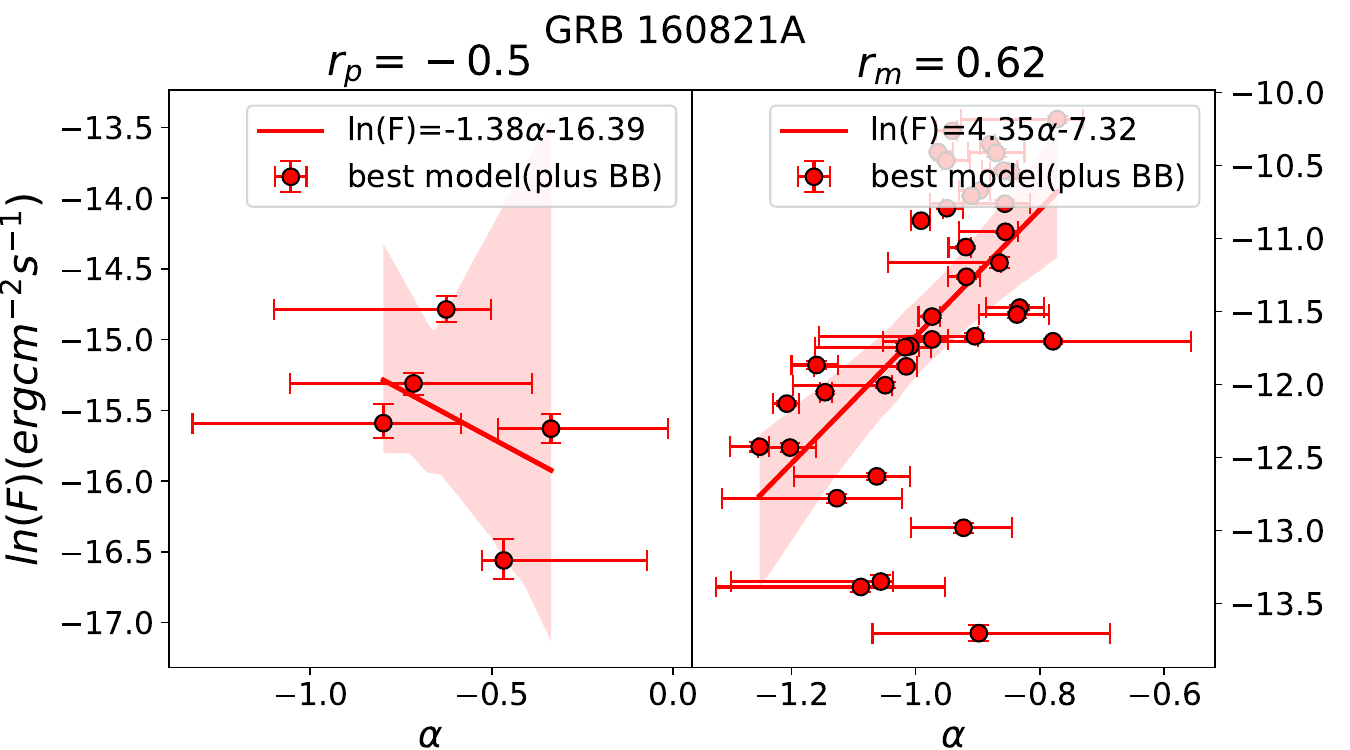}
\end{figure}
\begin{figure}[htbp]
\centering
\includegraphics [width=8.5cm,height=4.5cm]{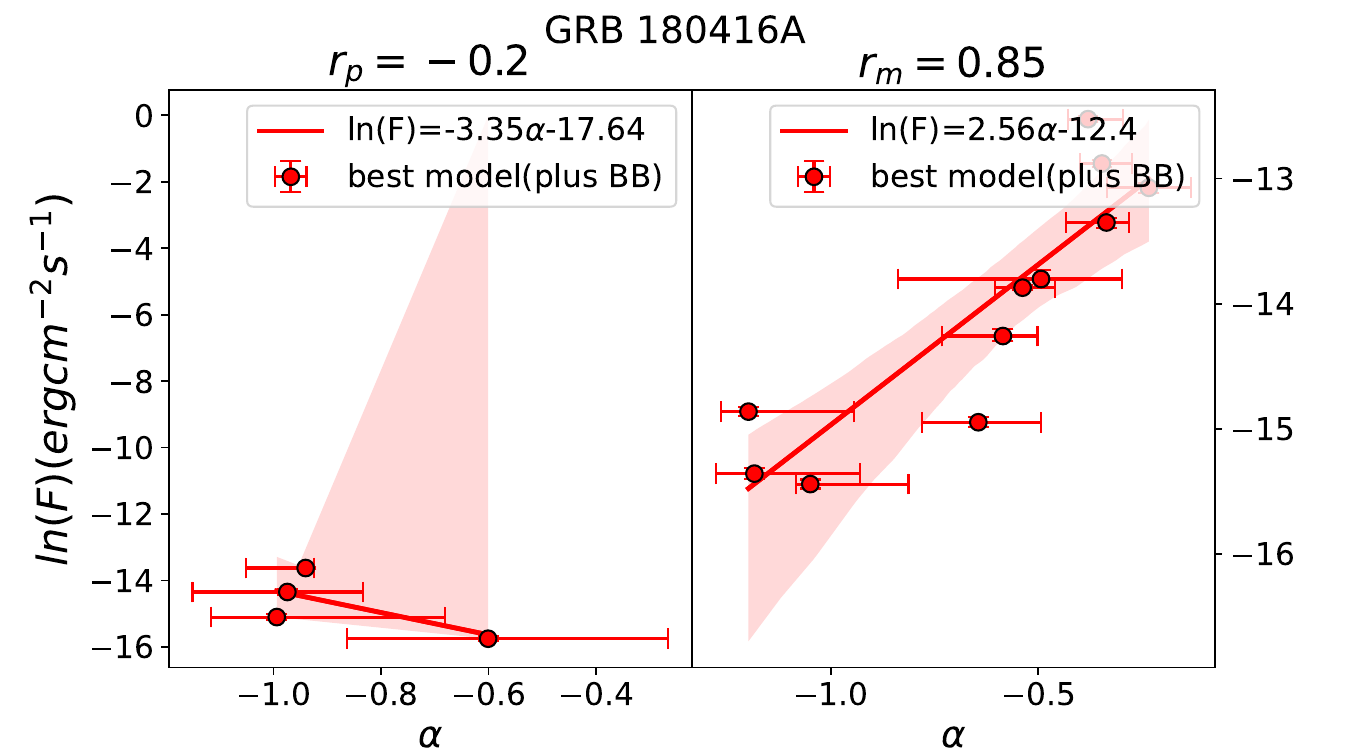}
\includegraphics [width=8.5cm,height=4.5cm]{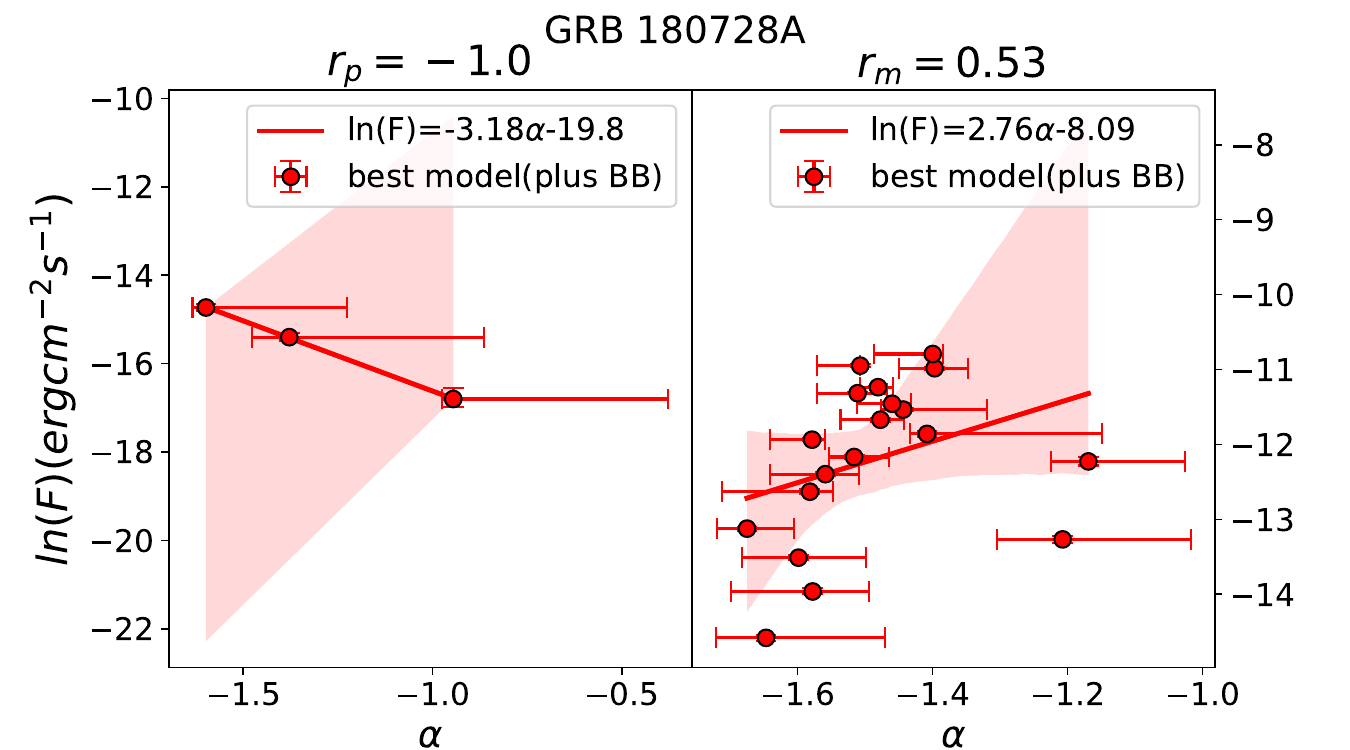}
\includegraphics [width=8.5cm,height=4.5cm]{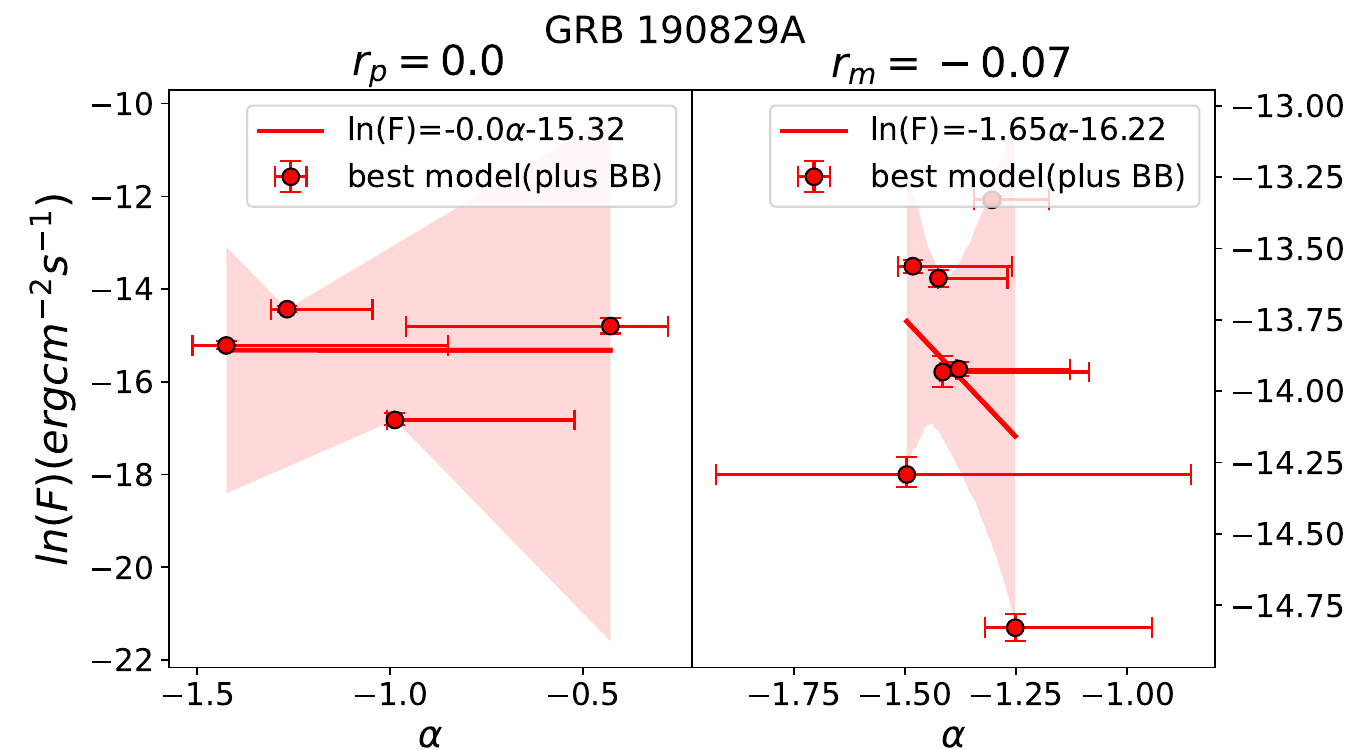}
\includegraphics [width=8.5cm,height=4.5cm]{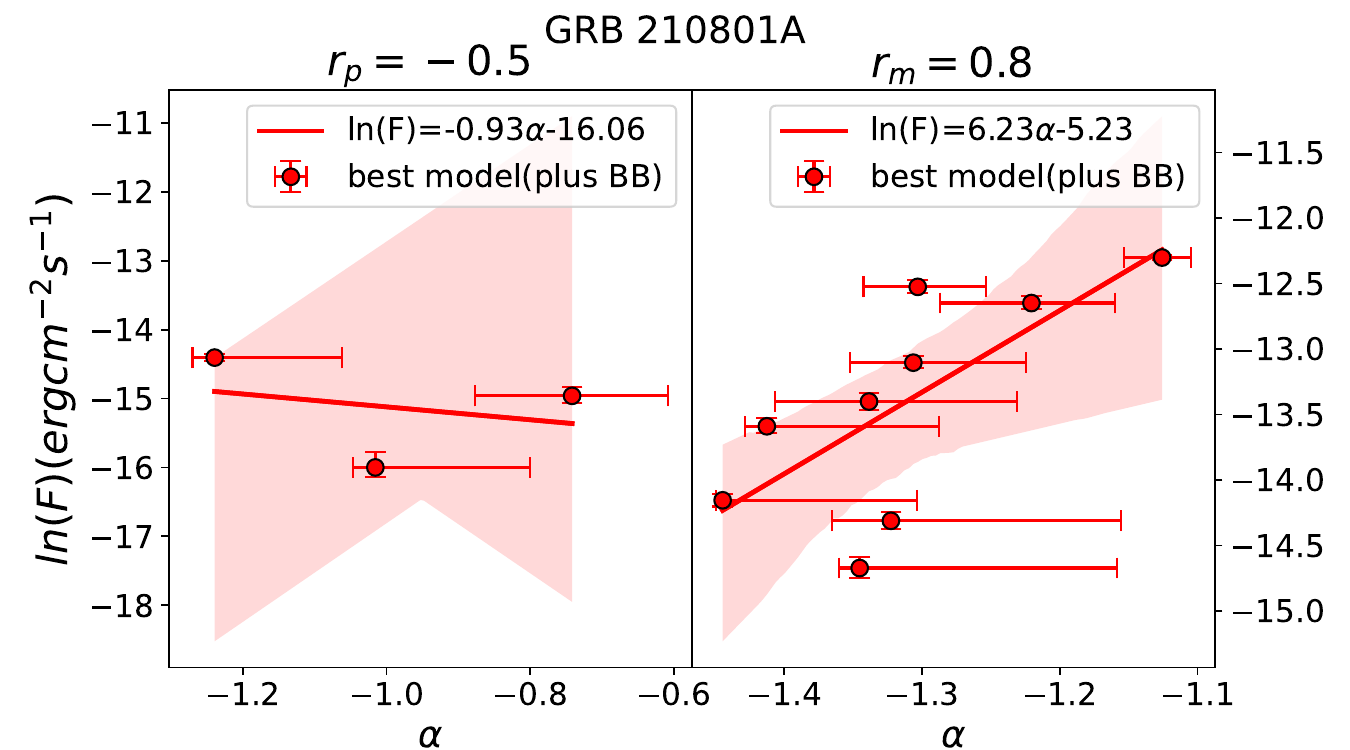}
\includegraphics [width=8.5cm,height=4.5cm]{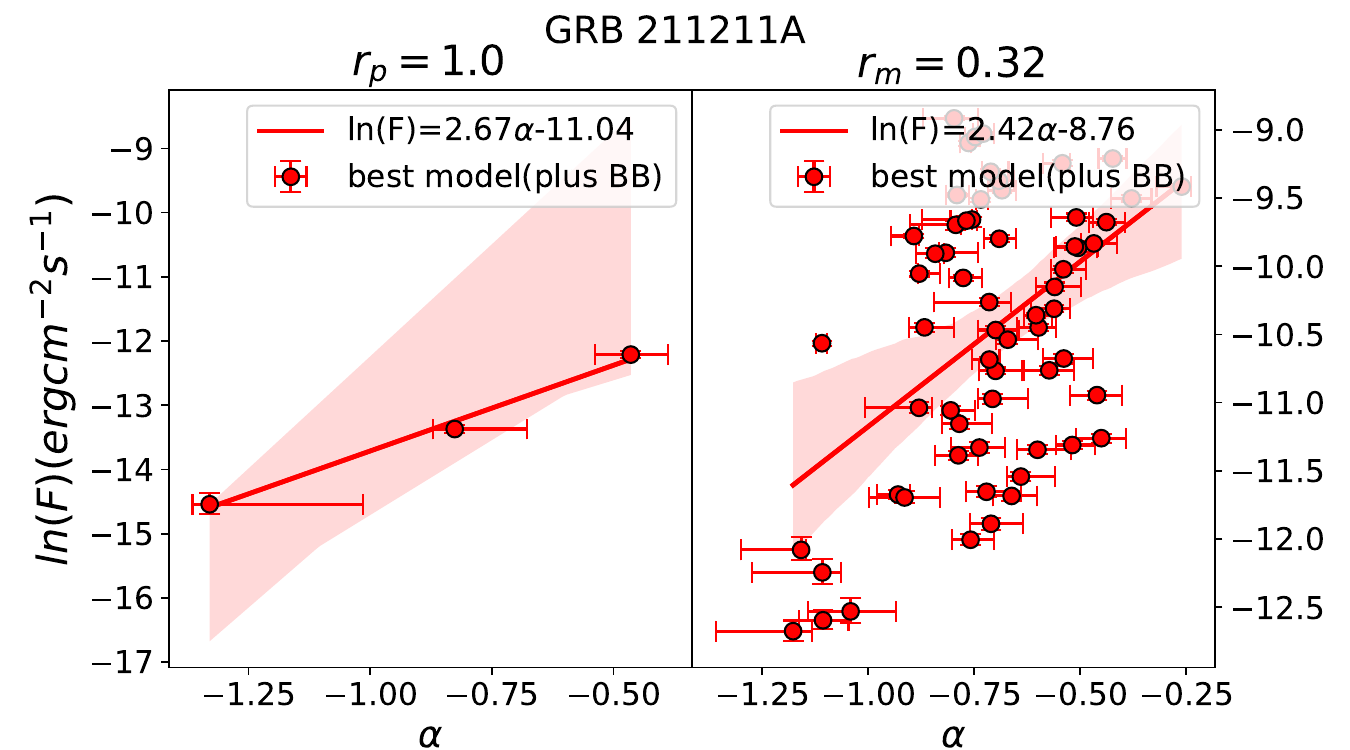}
\includegraphics [width=8.5cm,height=4.5cm]{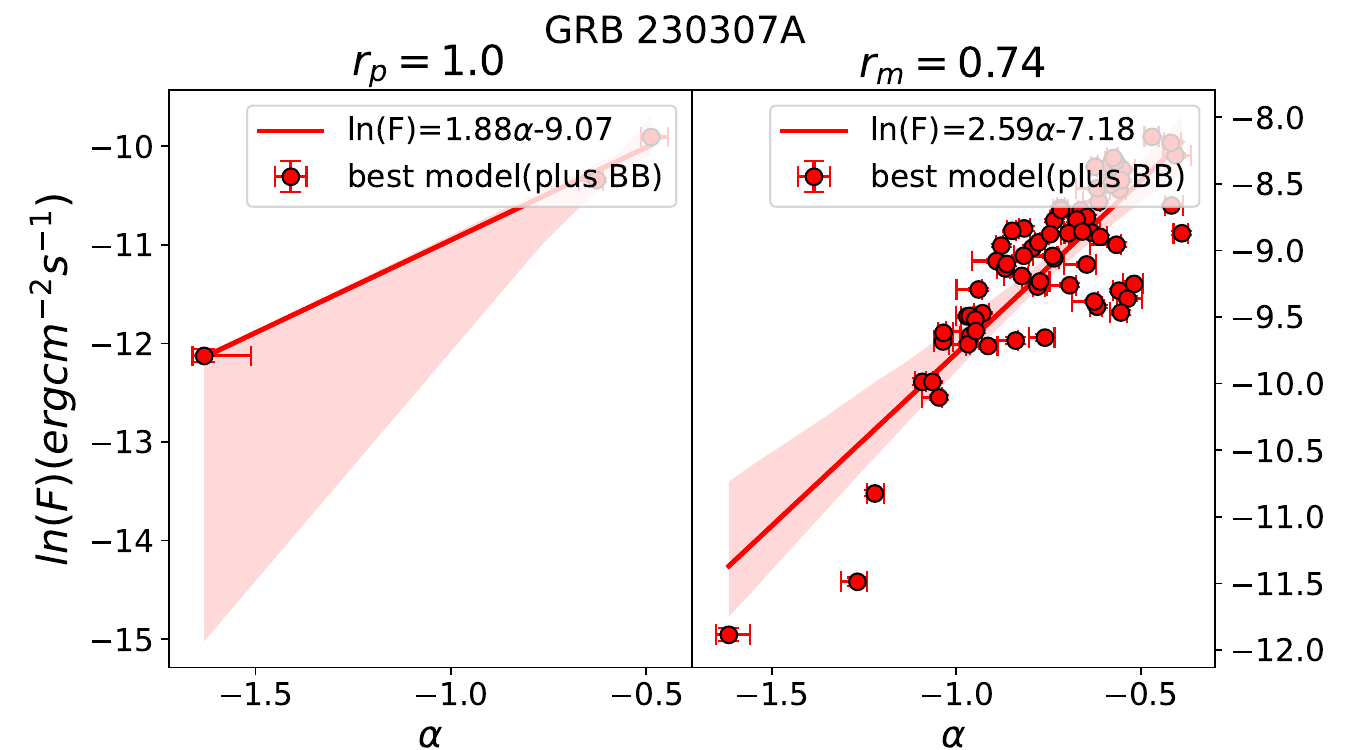}
\includegraphics [width=8.5cm,height=4.5cm]{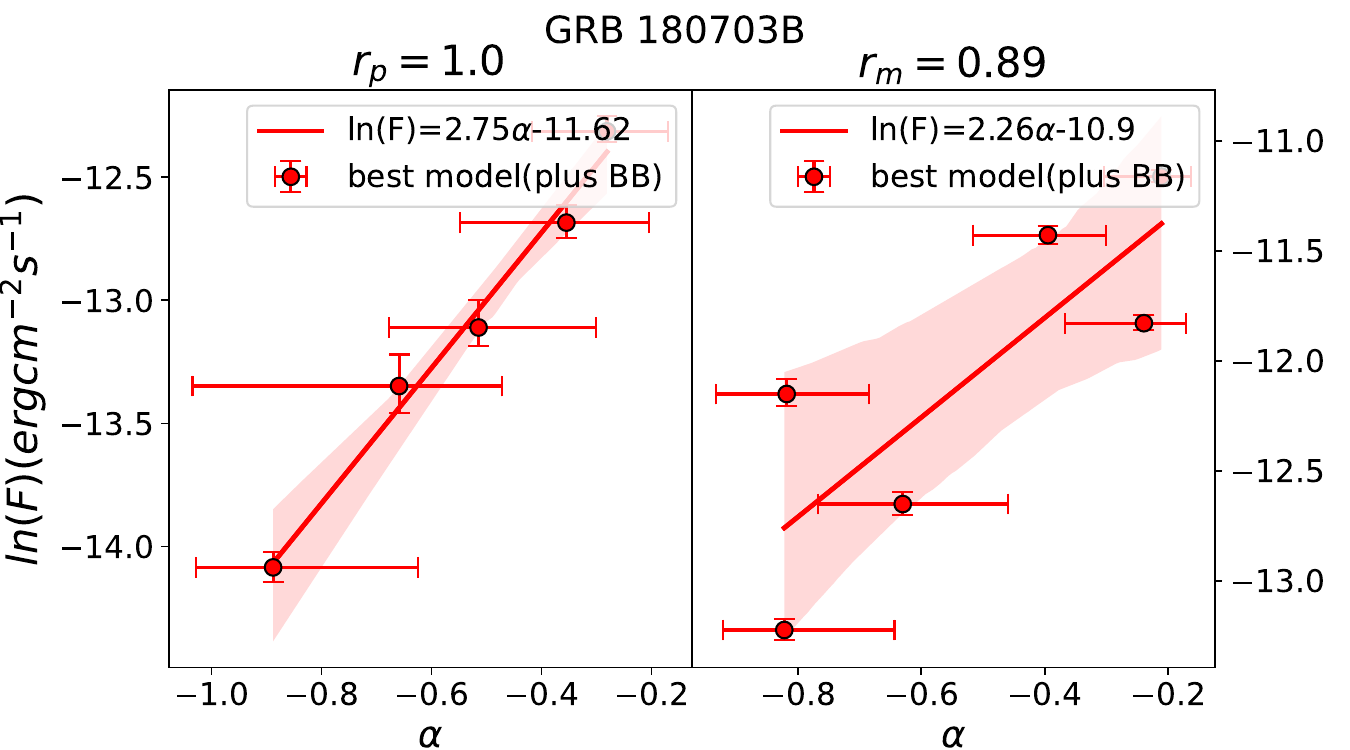}
\includegraphics [width=8.5cm,height=4.5cm]{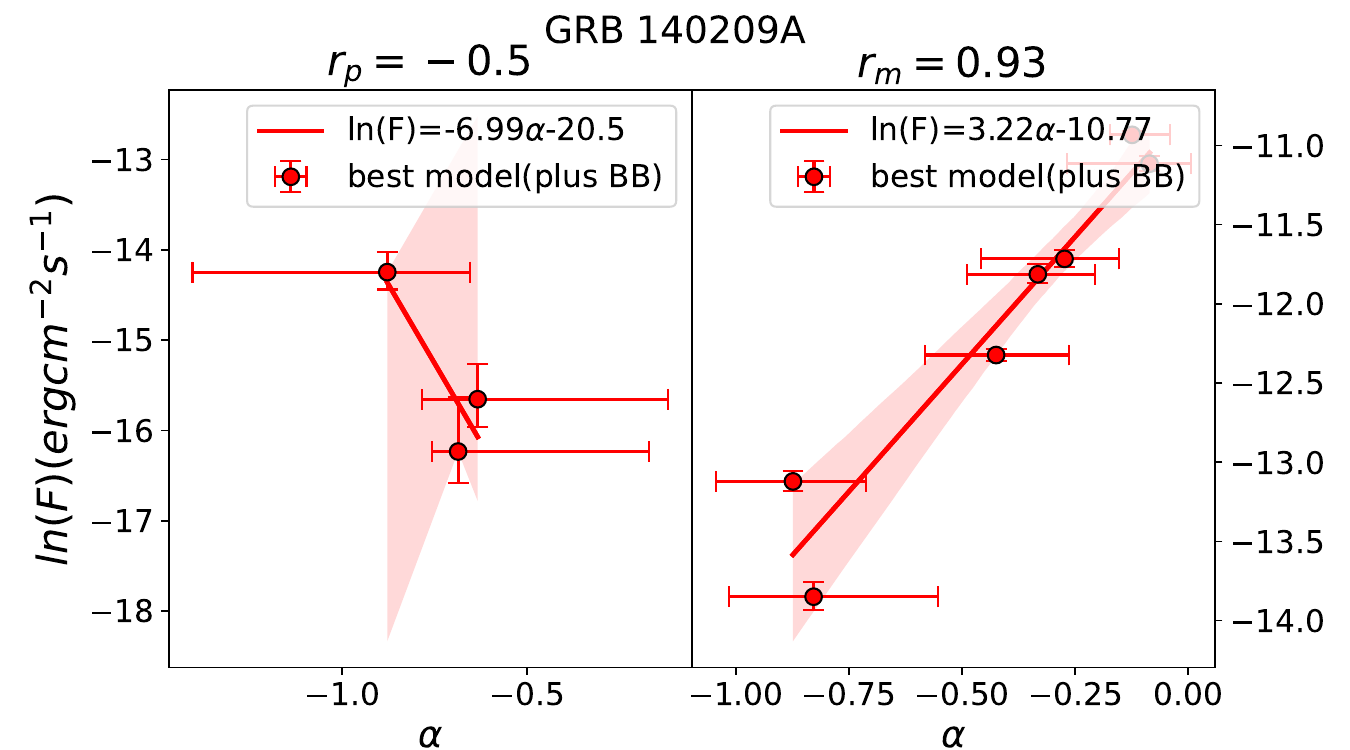}
   \figcaption{Correlations between $F$ and $\alpha$ fitted with the best model. All symbols are the same as in Figure \ref{fig 2}.\label{fig C1}}
\end{figure}

\begin{figure}[htbp]
\centering
\includegraphics [width=8.5cm,height=4.5cm]{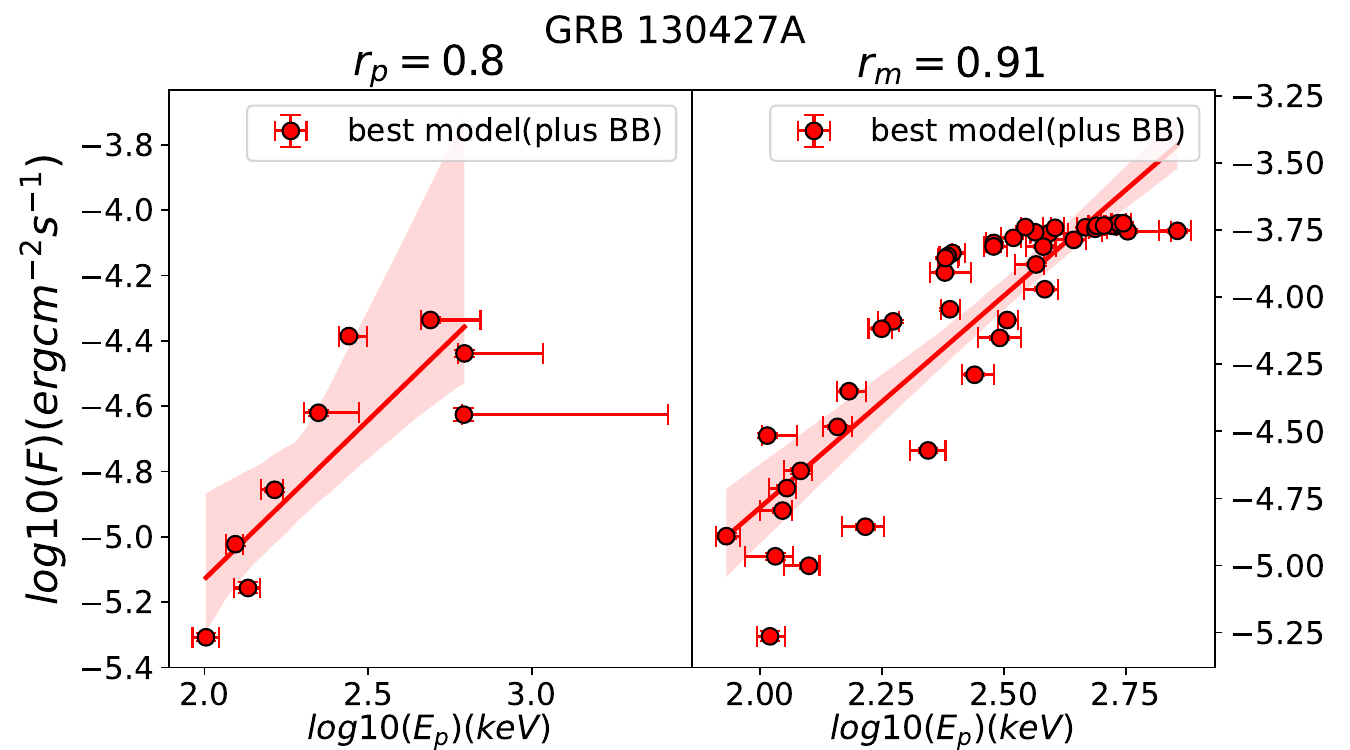}
\includegraphics [width=8.5cm,height=4.5cm]{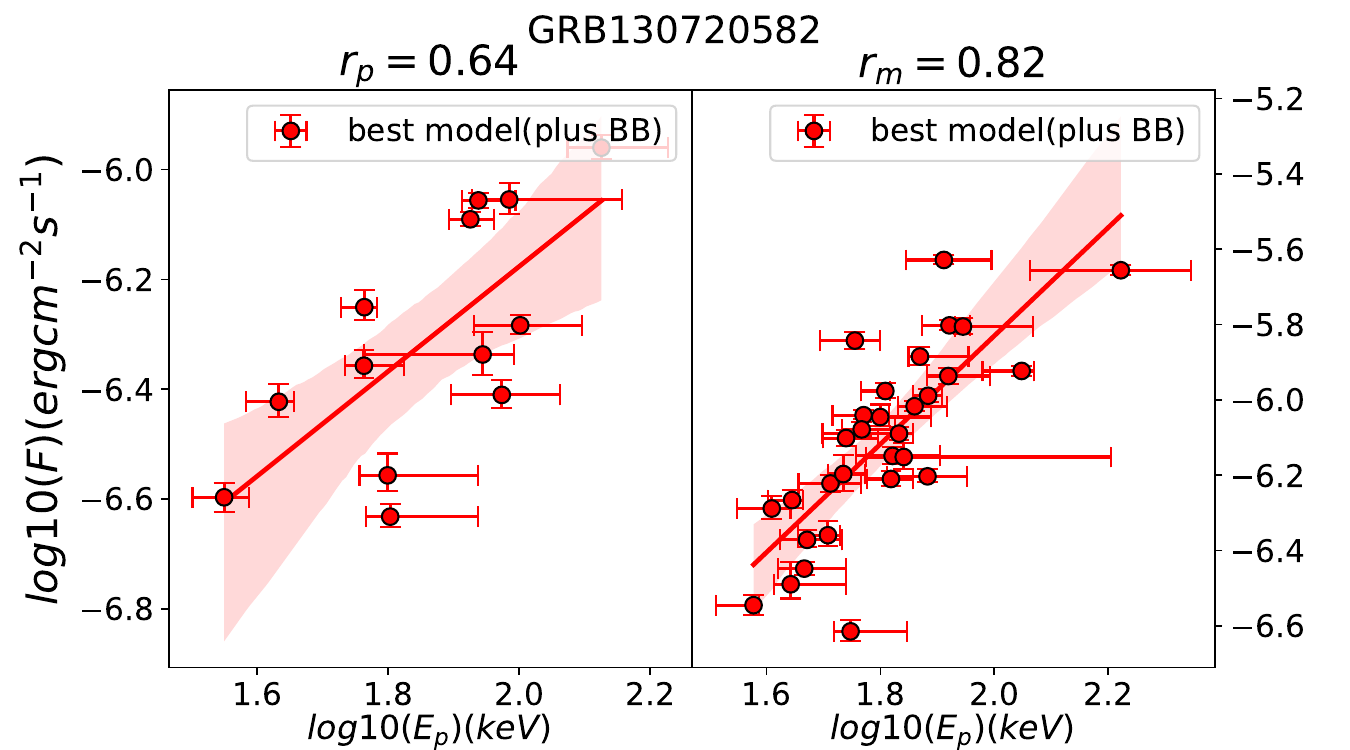}
\includegraphics [width=8.5cm,height=4.5cm]{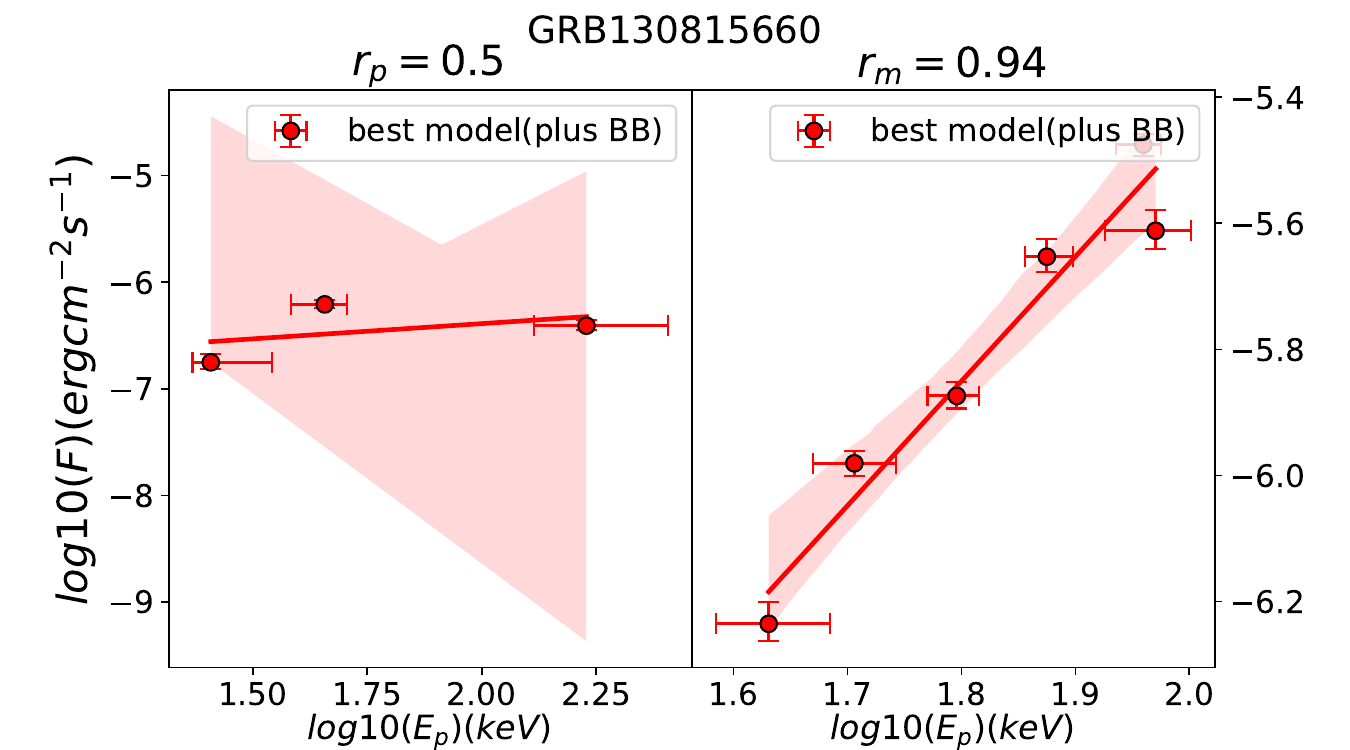}
\includegraphics [width=8.5cm,height=4.5cm]{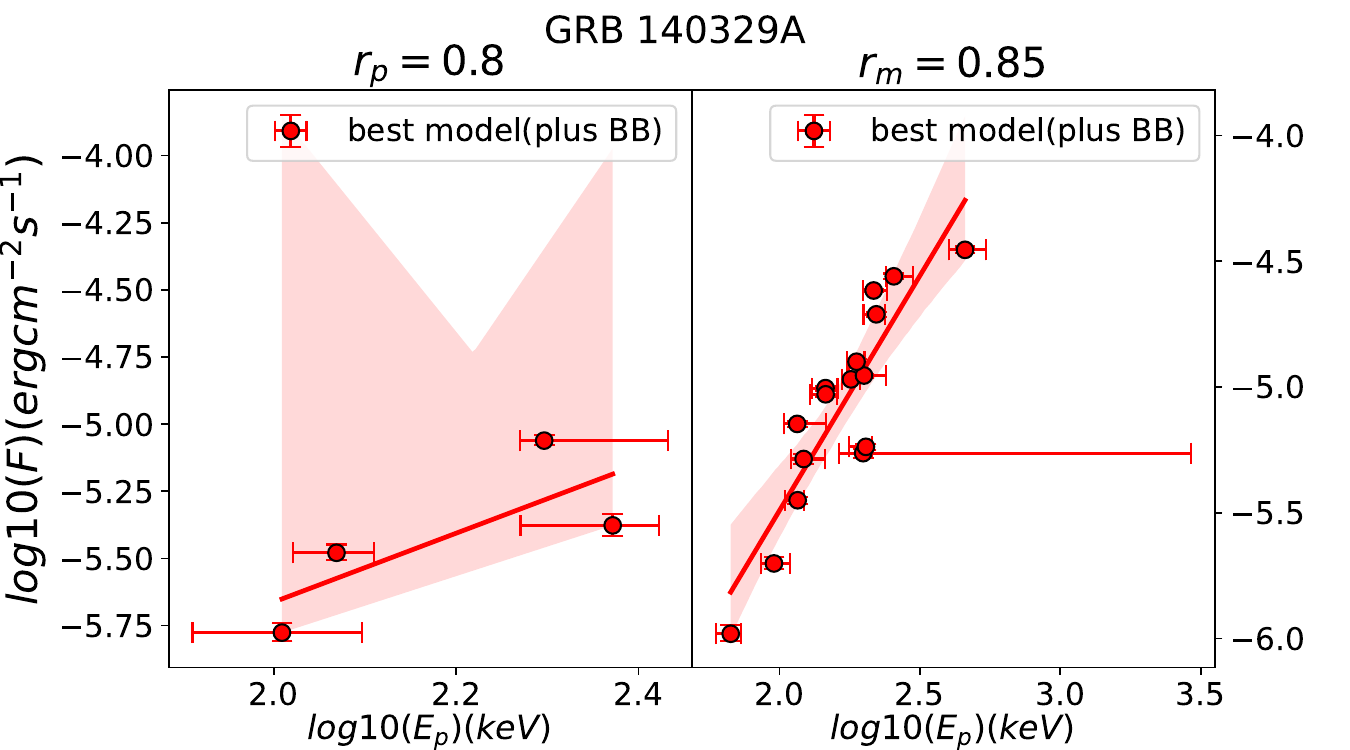}
\includegraphics [width=8.5cm,height=4.5cm]{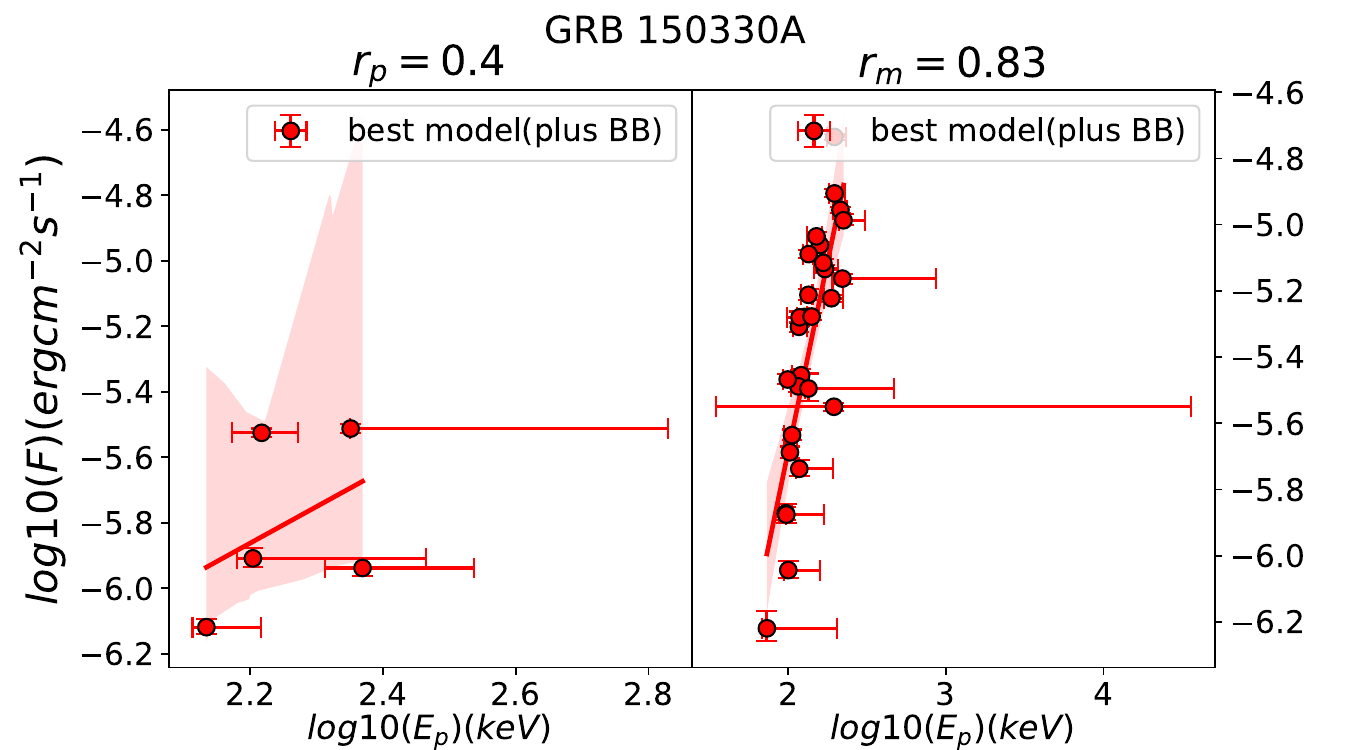}
\includegraphics [width=8.5cm,height=4.5cm]{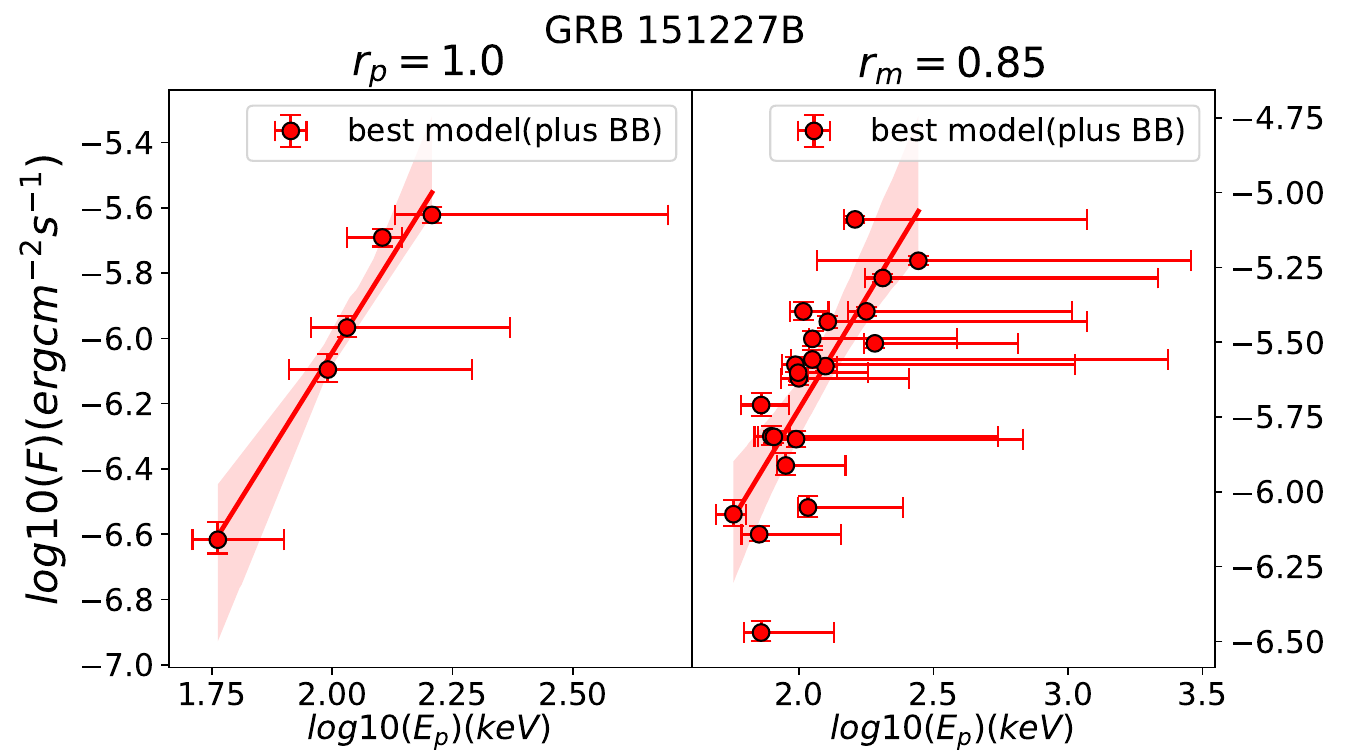}
\includegraphics [width=8.5cm,height=4.5cm]{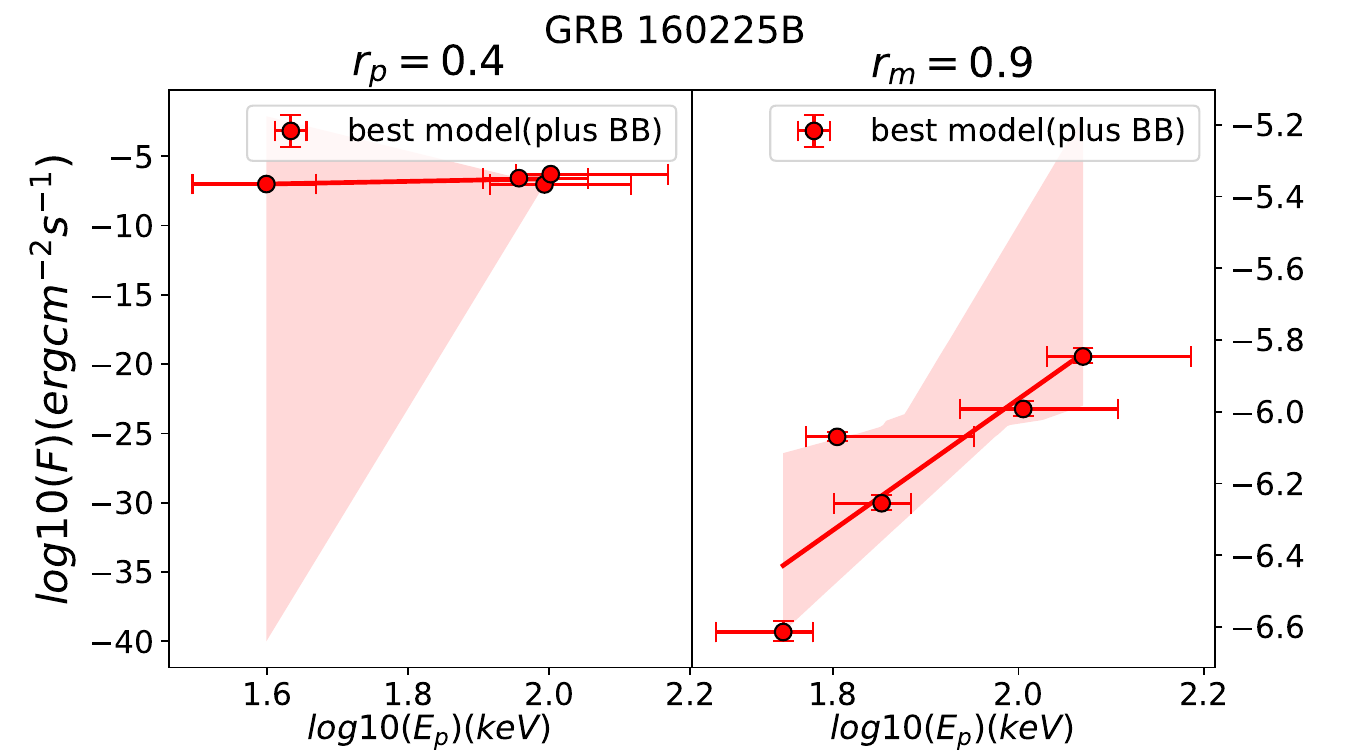}
\includegraphics [width=8.5cm,height=4.5cm]{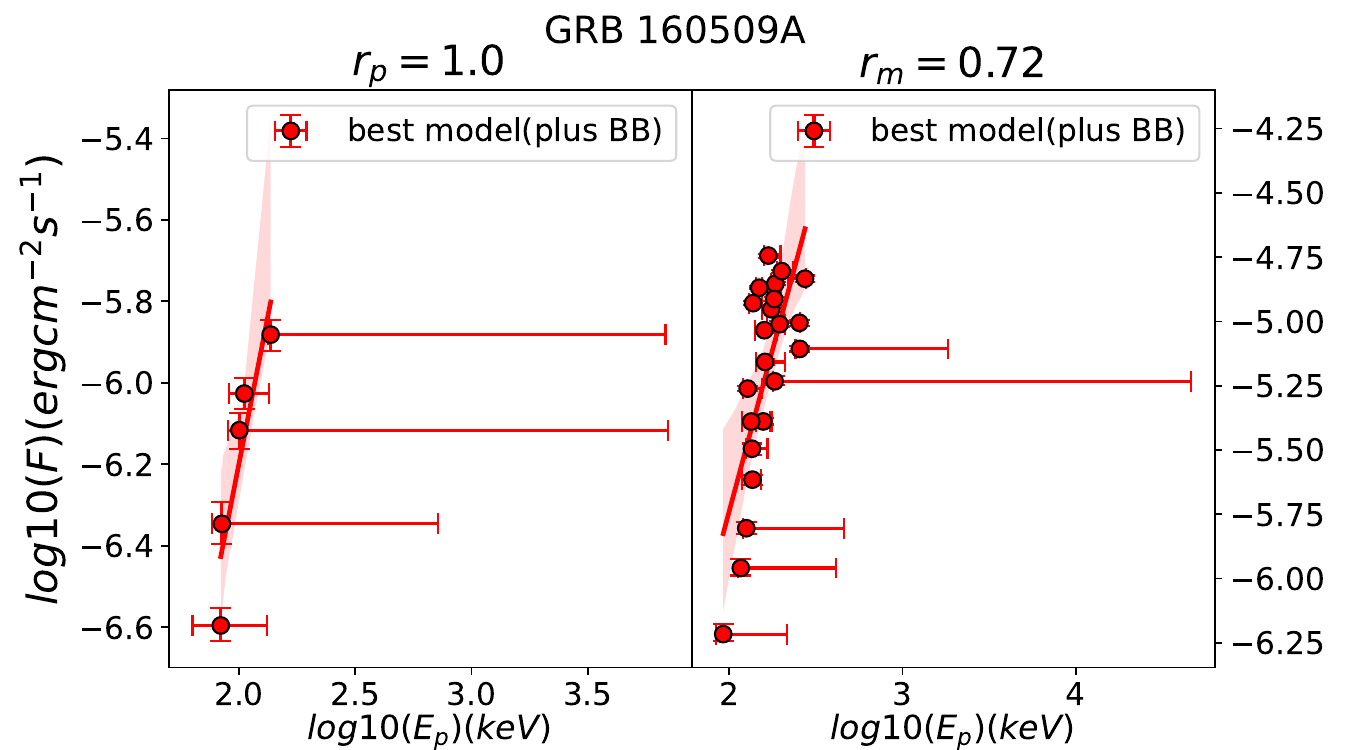}
\includegraphics [width=8.5cm,height=4.5cm]{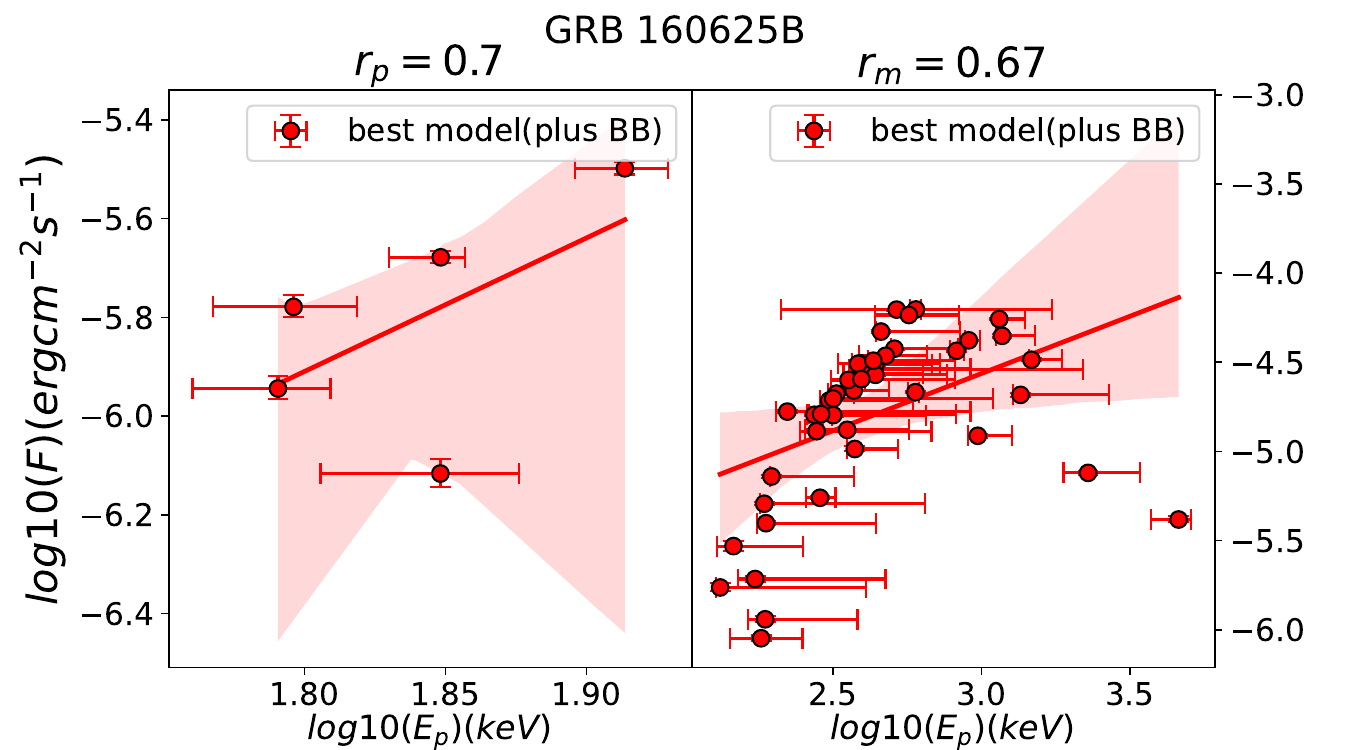}
\includegraphics [width=8.5cm,height=4.5cm]{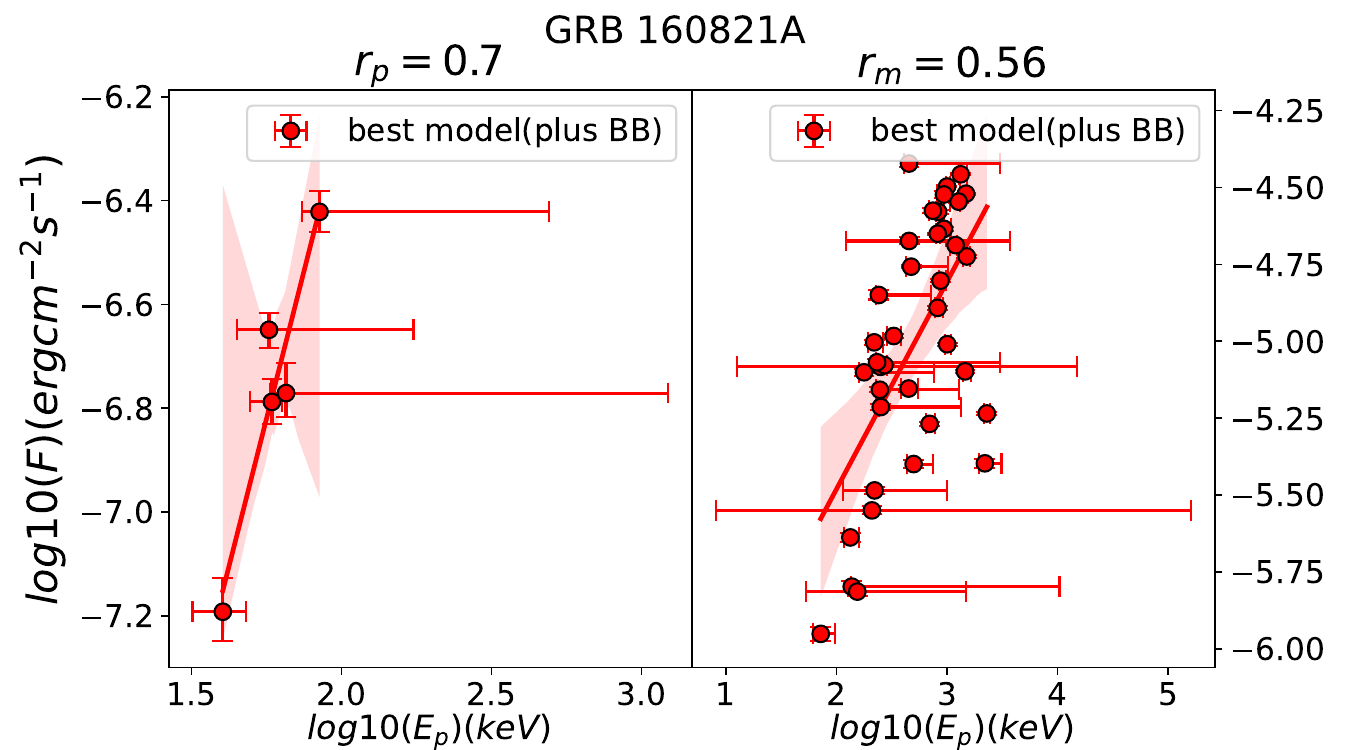}
\end{figure}
\begin{figure}[htbp]
\centering
\includegraphics [width=8.5cm,height=4.5cm]{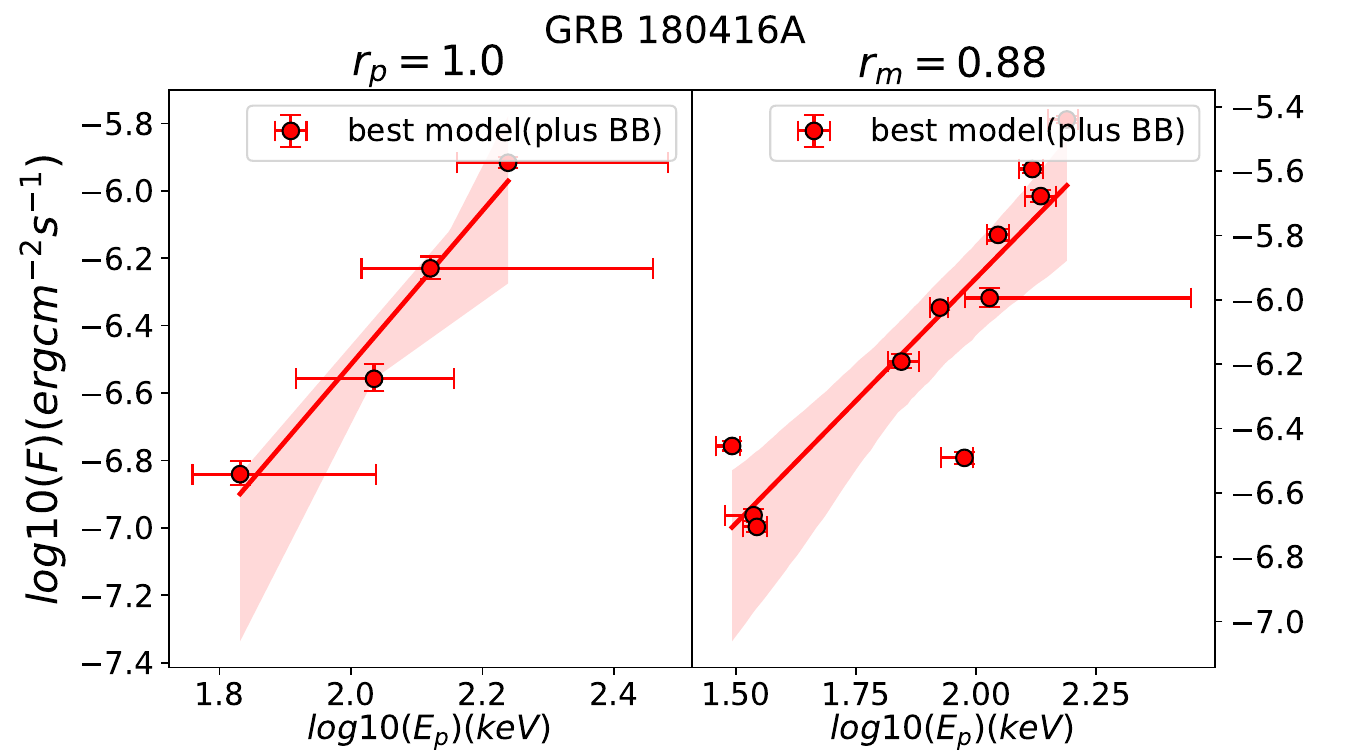}
\includegraphics [width=8.5cm,height=4.5cm]{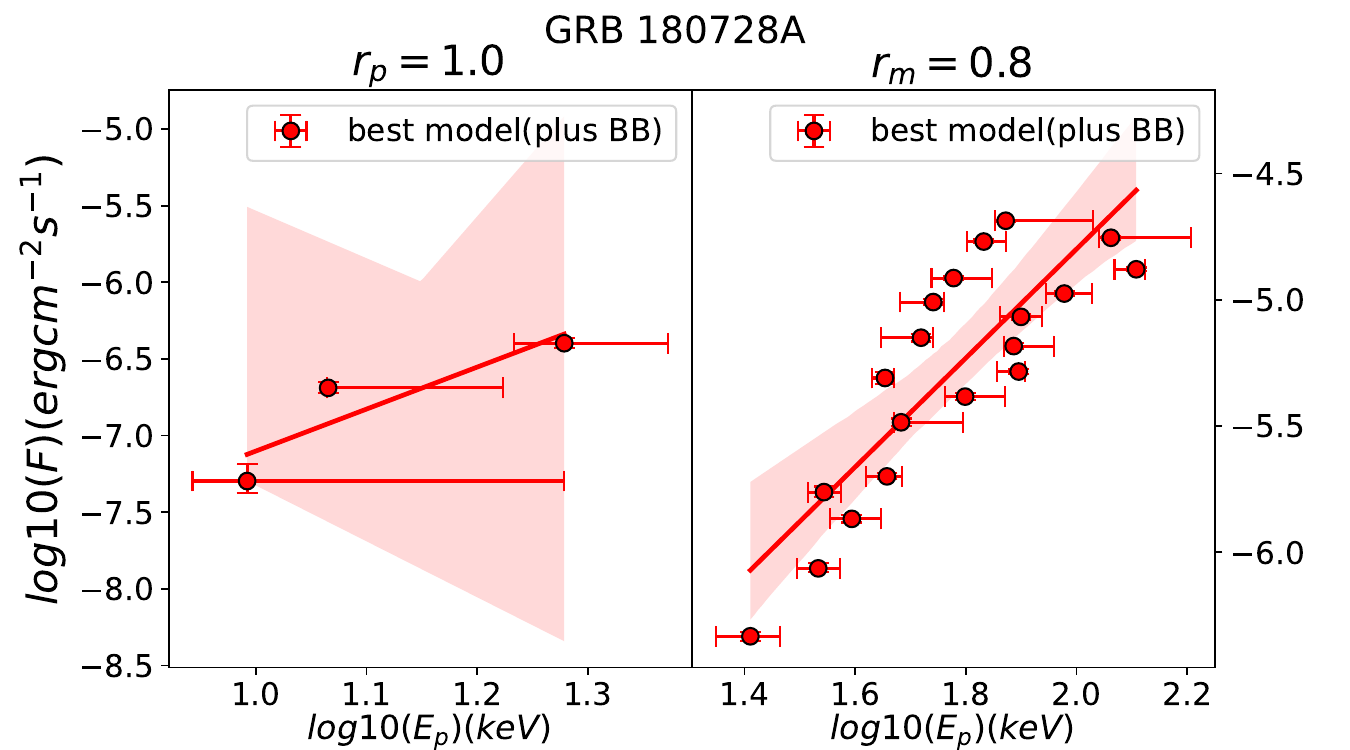}
\includegraphics [width=8.5cm,height=4.5cm]{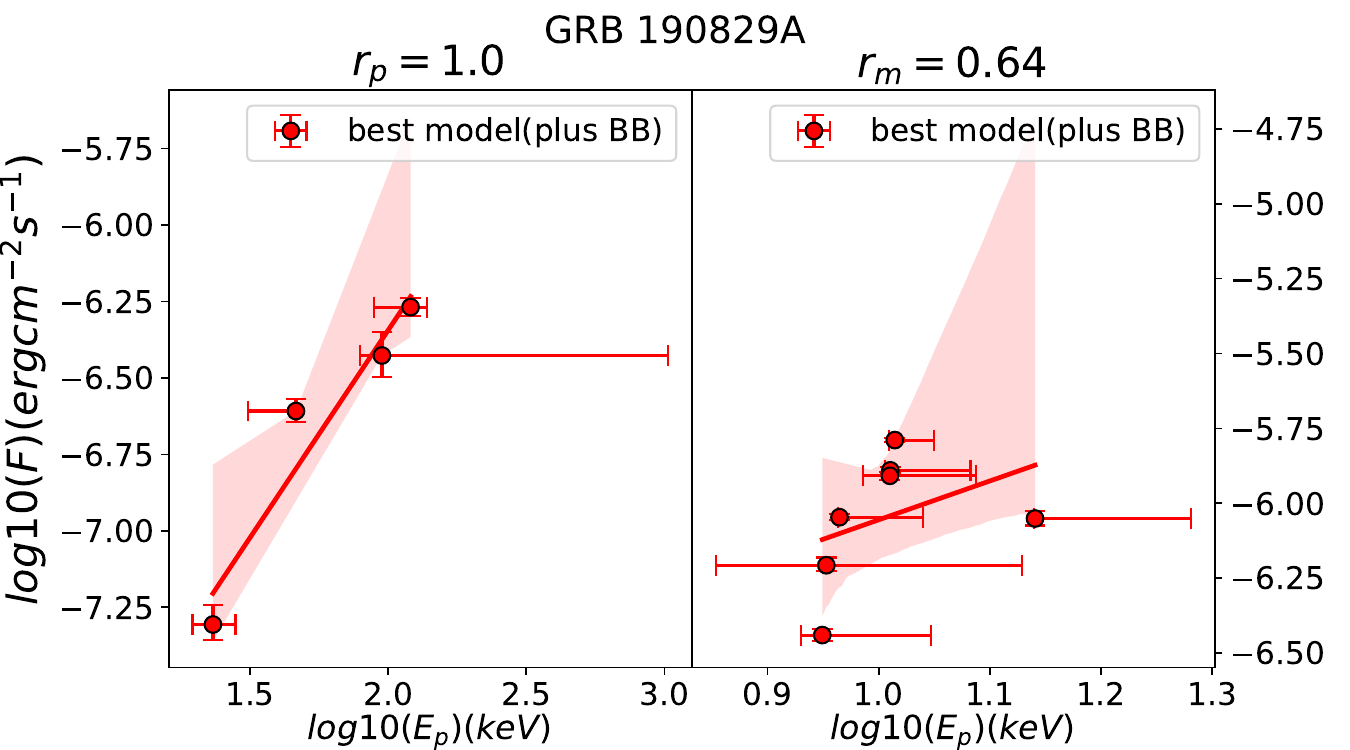}
\includegraphics [width=8.5cm,height=4.5cm]{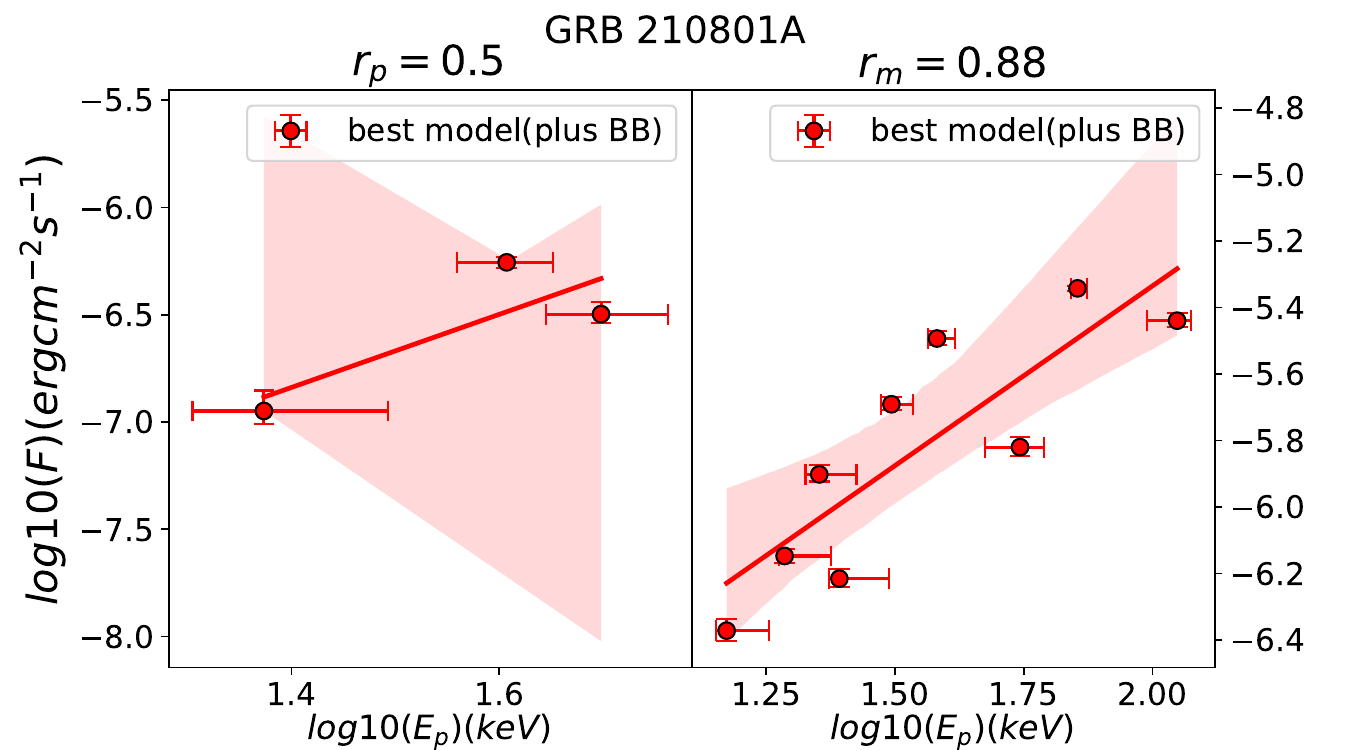}
\includegraphics [width=8.5cm,height=4.5cm]{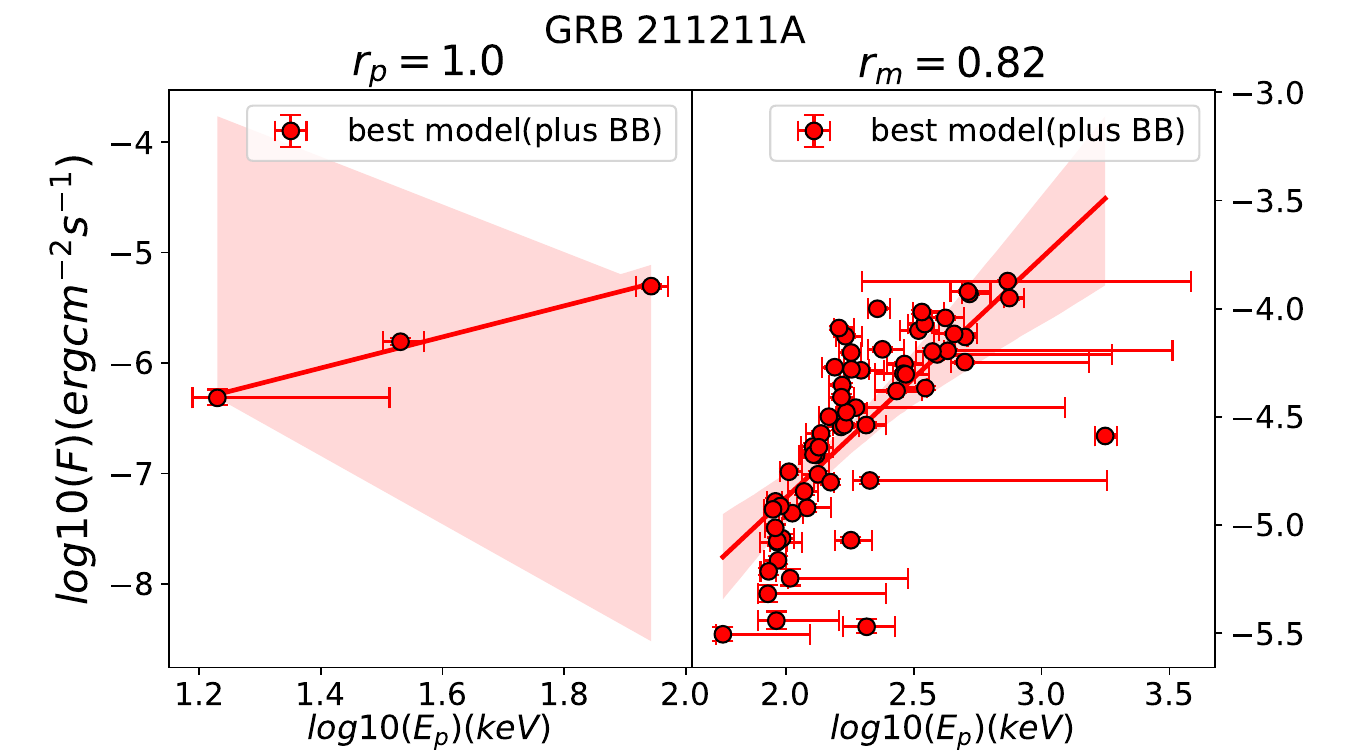}
\includegraphics [width=8.5cm,height=4.5cm]{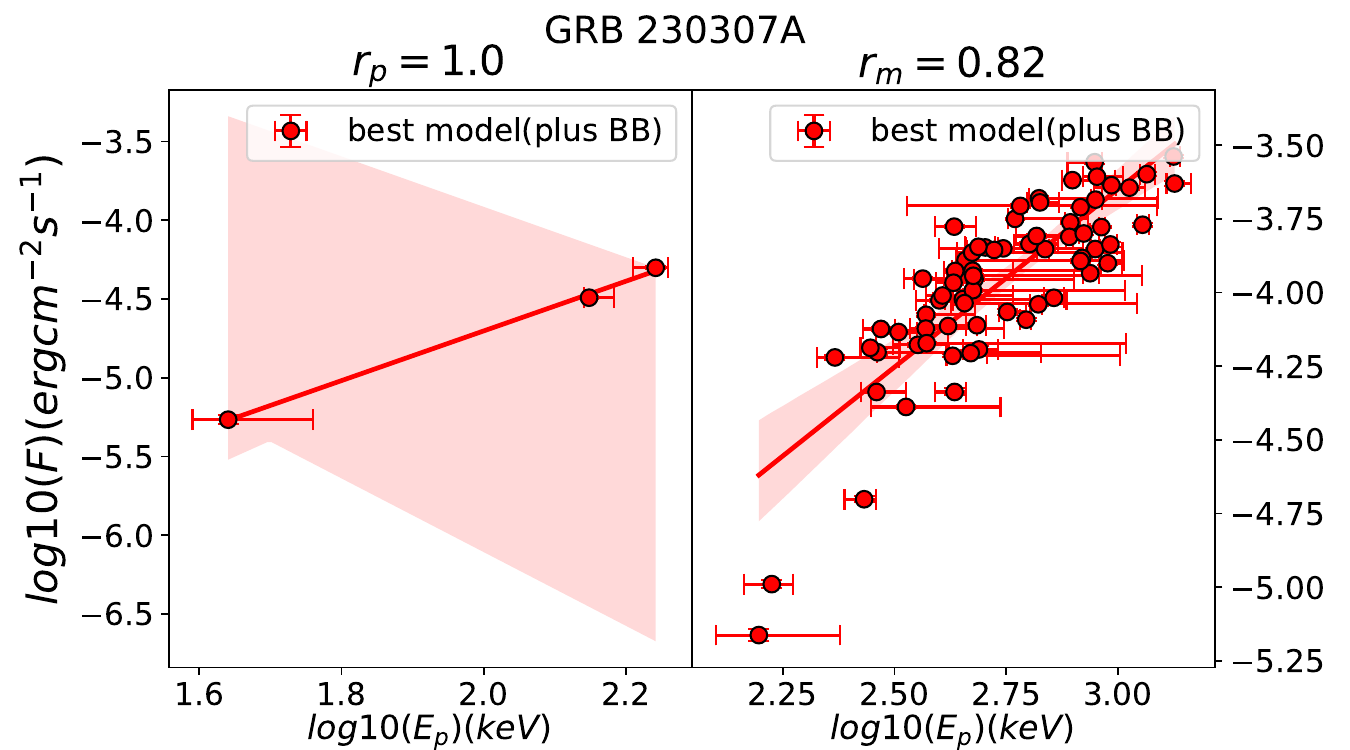}
\includegraphics [width=8.5cm,height=4.5cm]{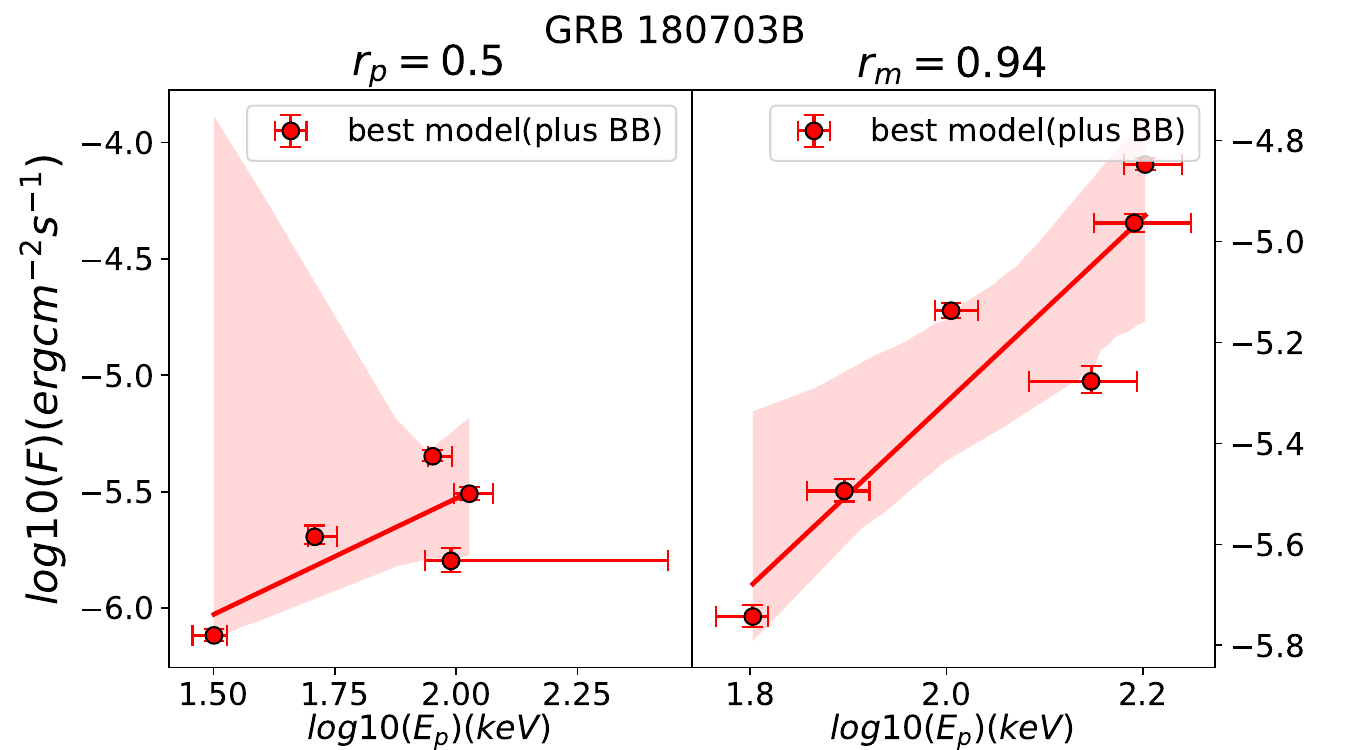}
\includegraphics [width=8.5cm,height=4.5cm]{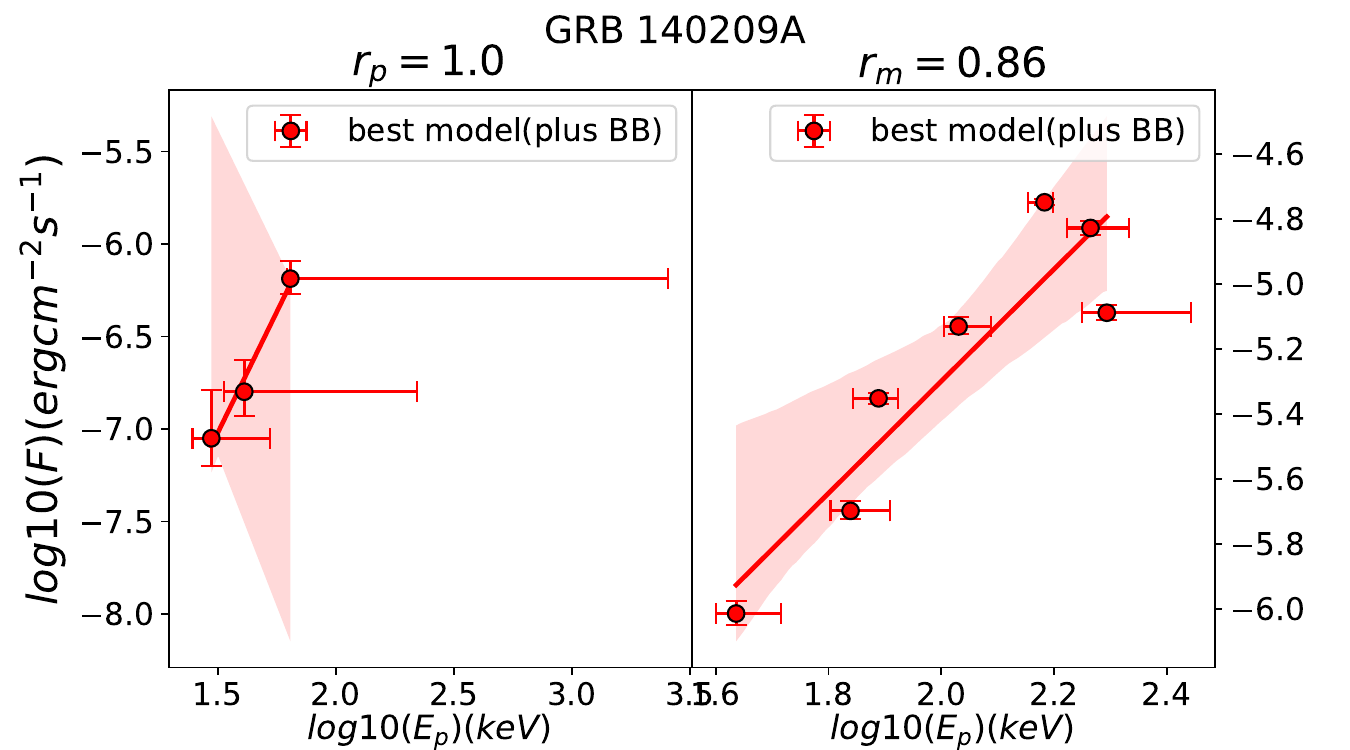}
   \figcaption{Correlations between $F$ and $E_{p}$ fitted with the best model for the precursor and main bursts. All symbols are the same as in Figure \ref{fig 2}. \label{fig C2}}
\end{figure}

\begin{figure}[htbp]
\centering
\includegraphics [width=8.5cm,height=4.5cm]{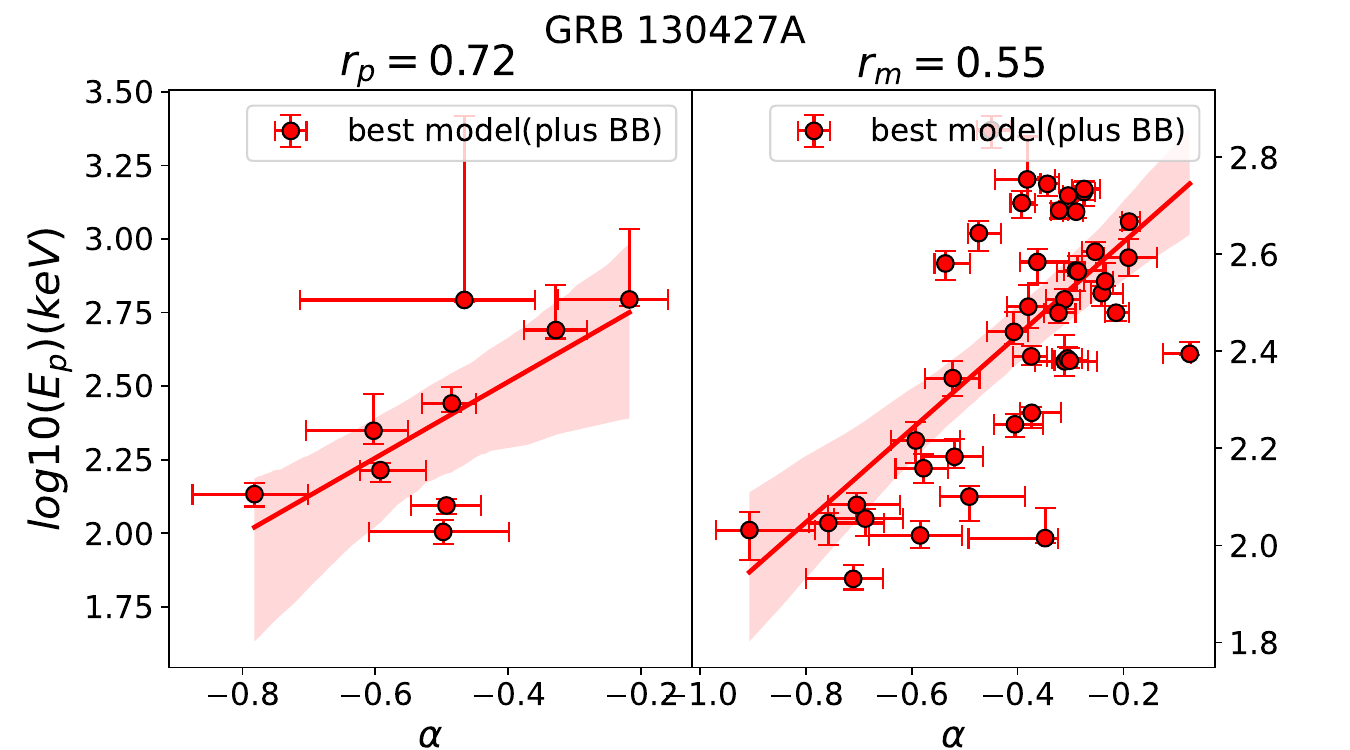}
\includegraphics [width=8.5cm,height=4.5cm]{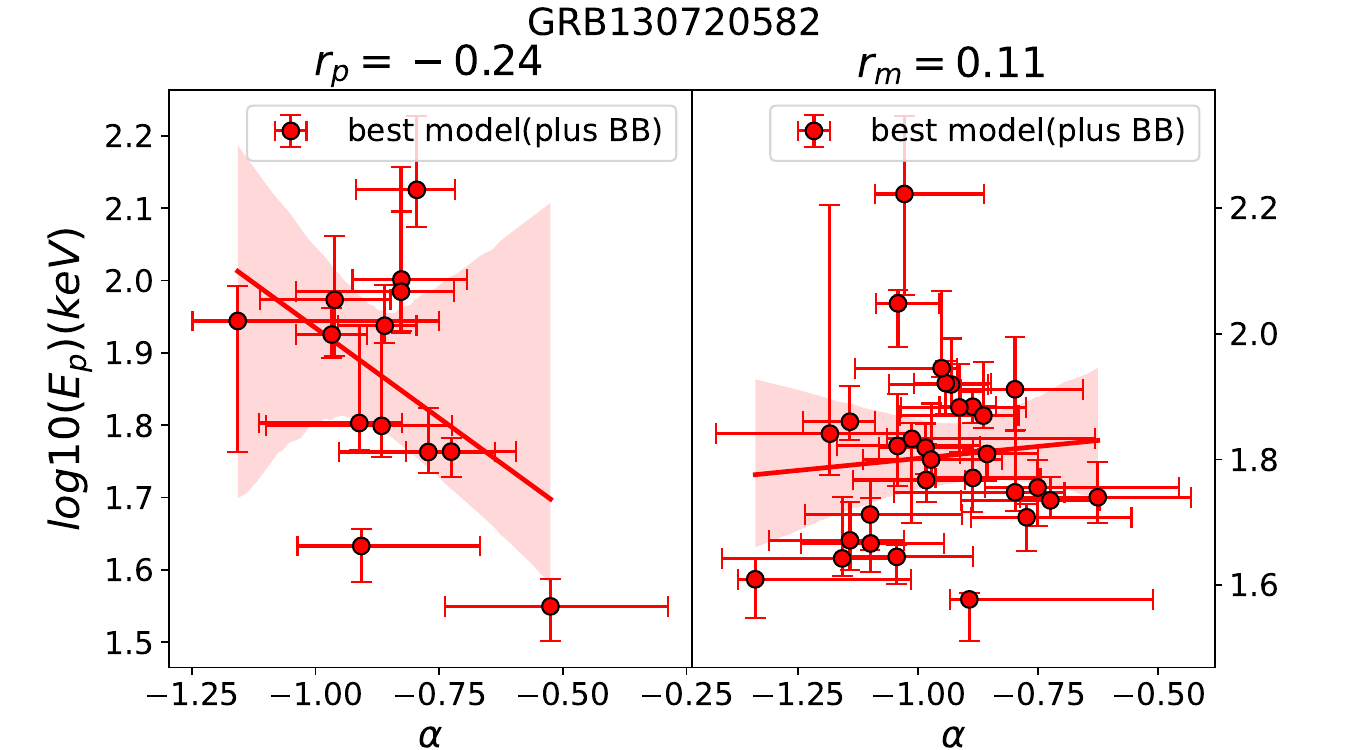}
\includegraphics [width=8.5cm,height=4.5cm]{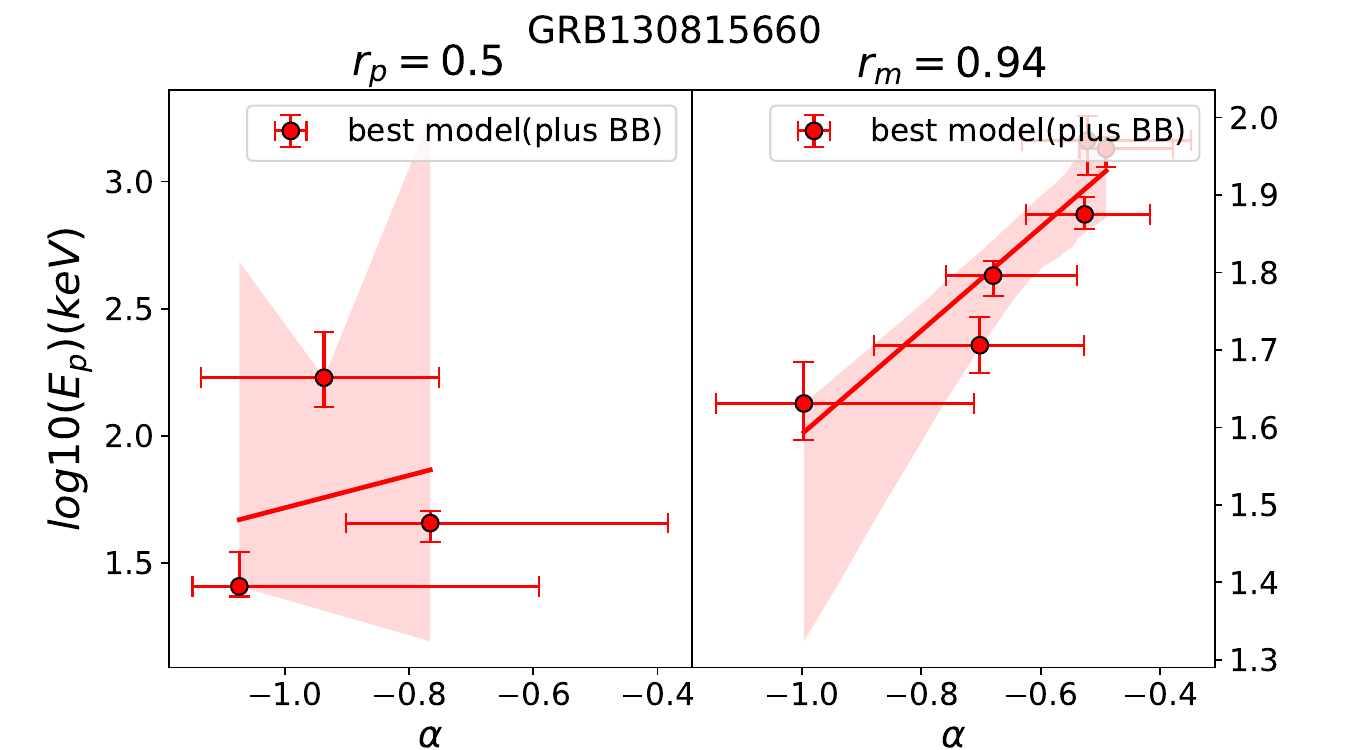}
\includegraphics [width=8.5cm,height=4.5cm]{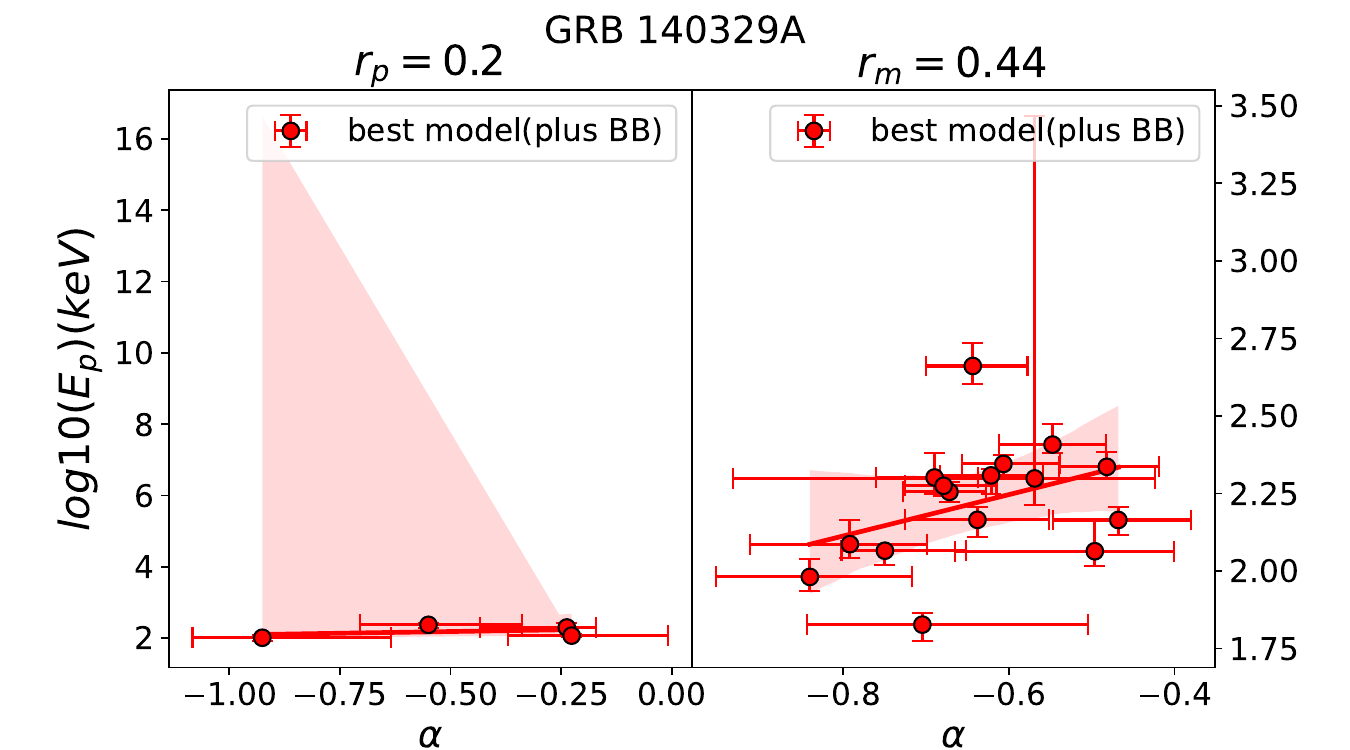}
\includegraphics [width=8.5cm,height=4.5cm]{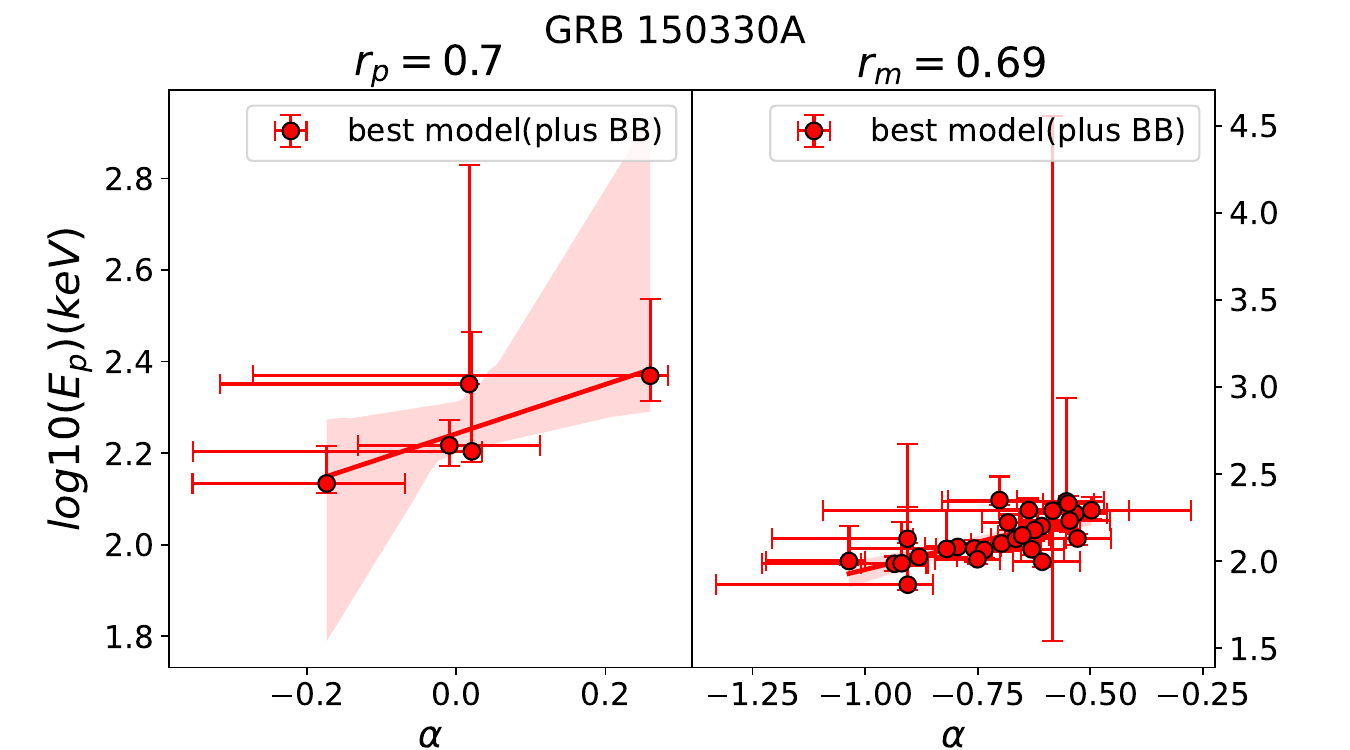}
\includegraphics [width=8.5cm,height=4.5cm]{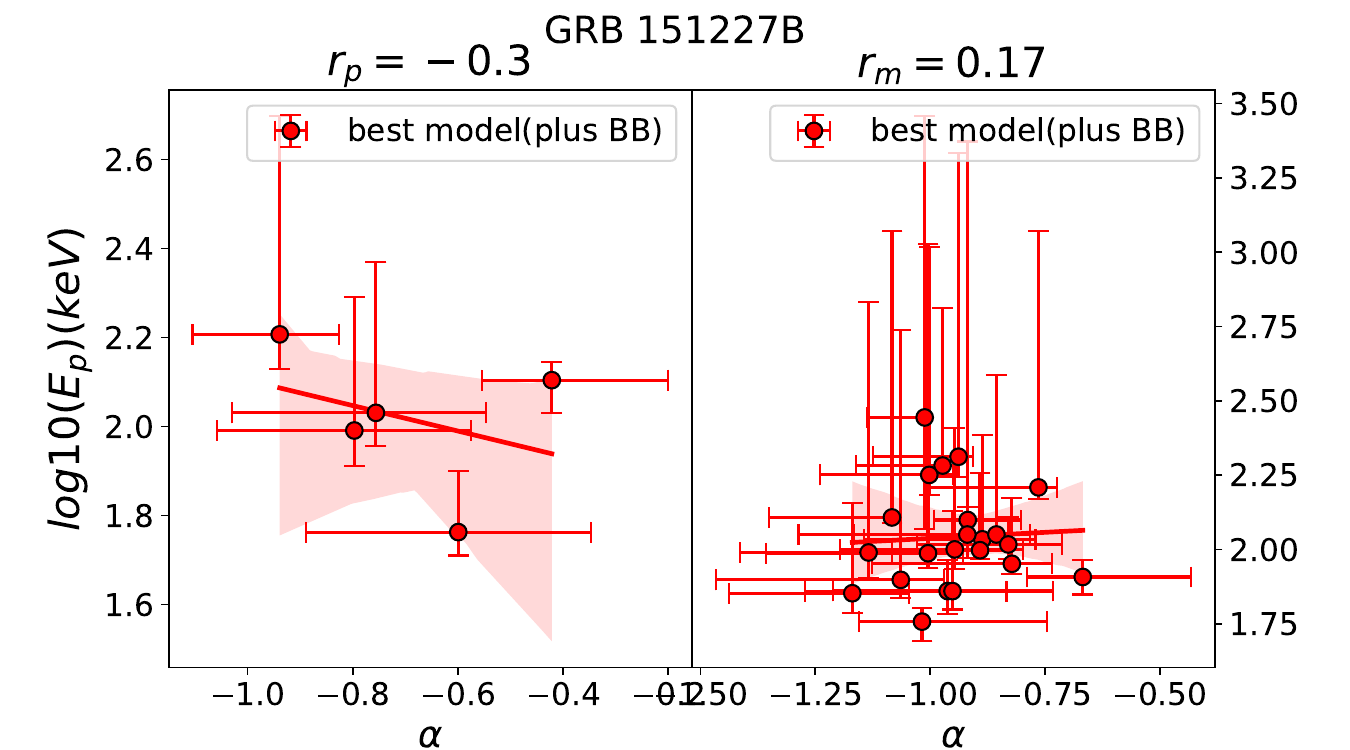}
\includegraphics [width=8.5cm,height=4.5cm]{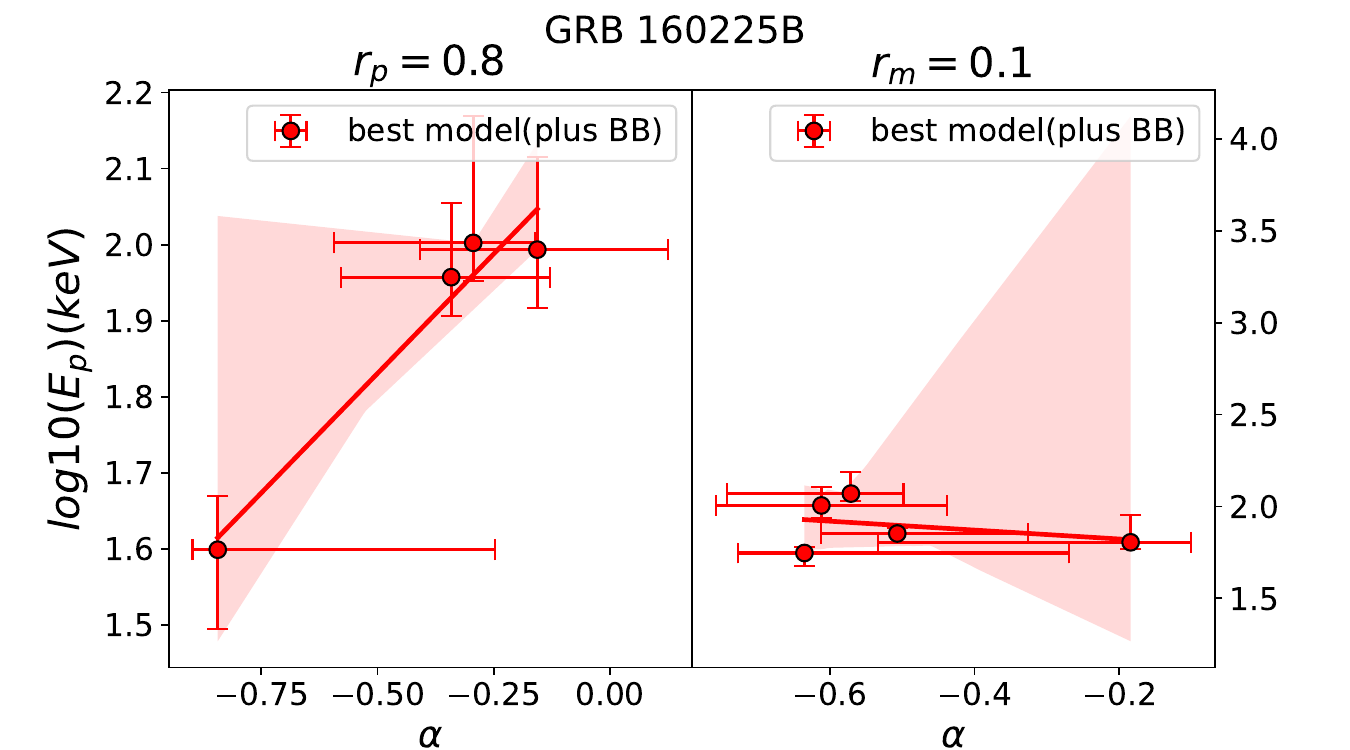}
\includegraphics [width=8.5cm,height=4.5cm]{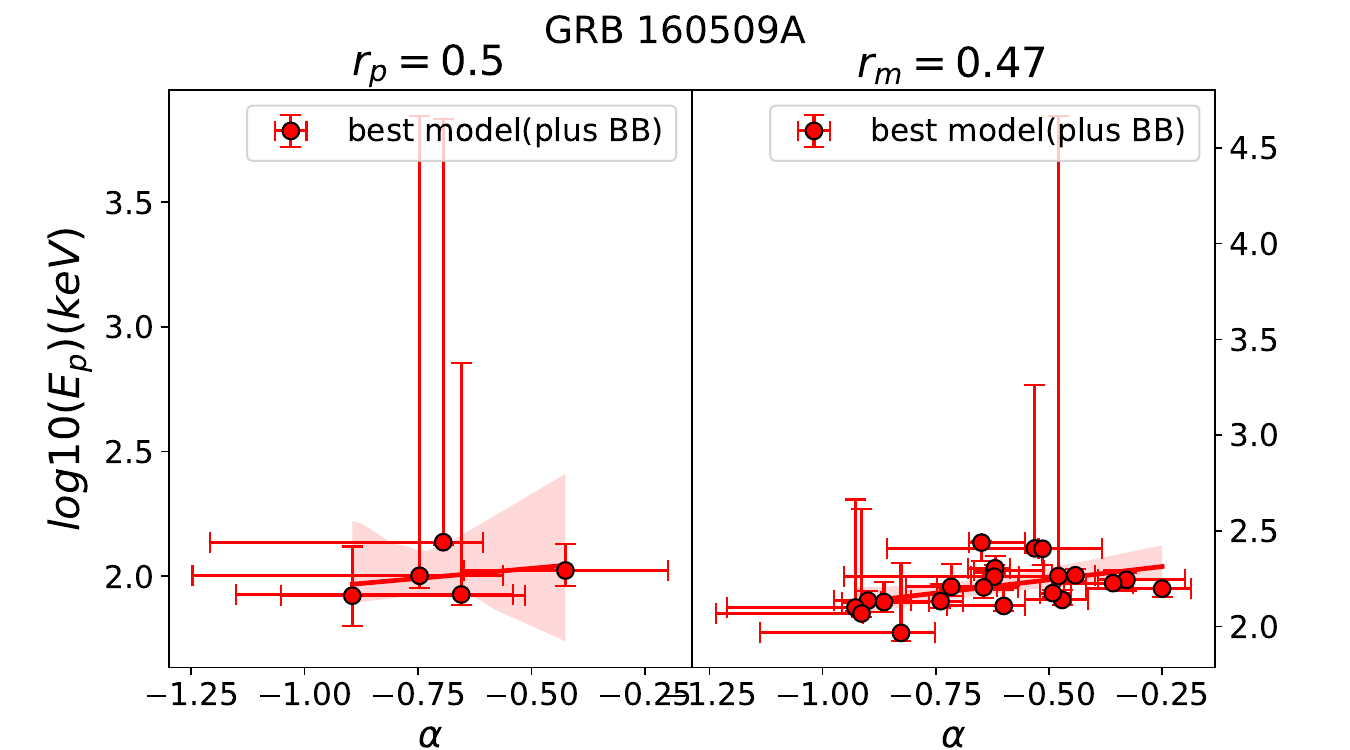}
\includegraphics [width=8.5cm,height=4.5cm]{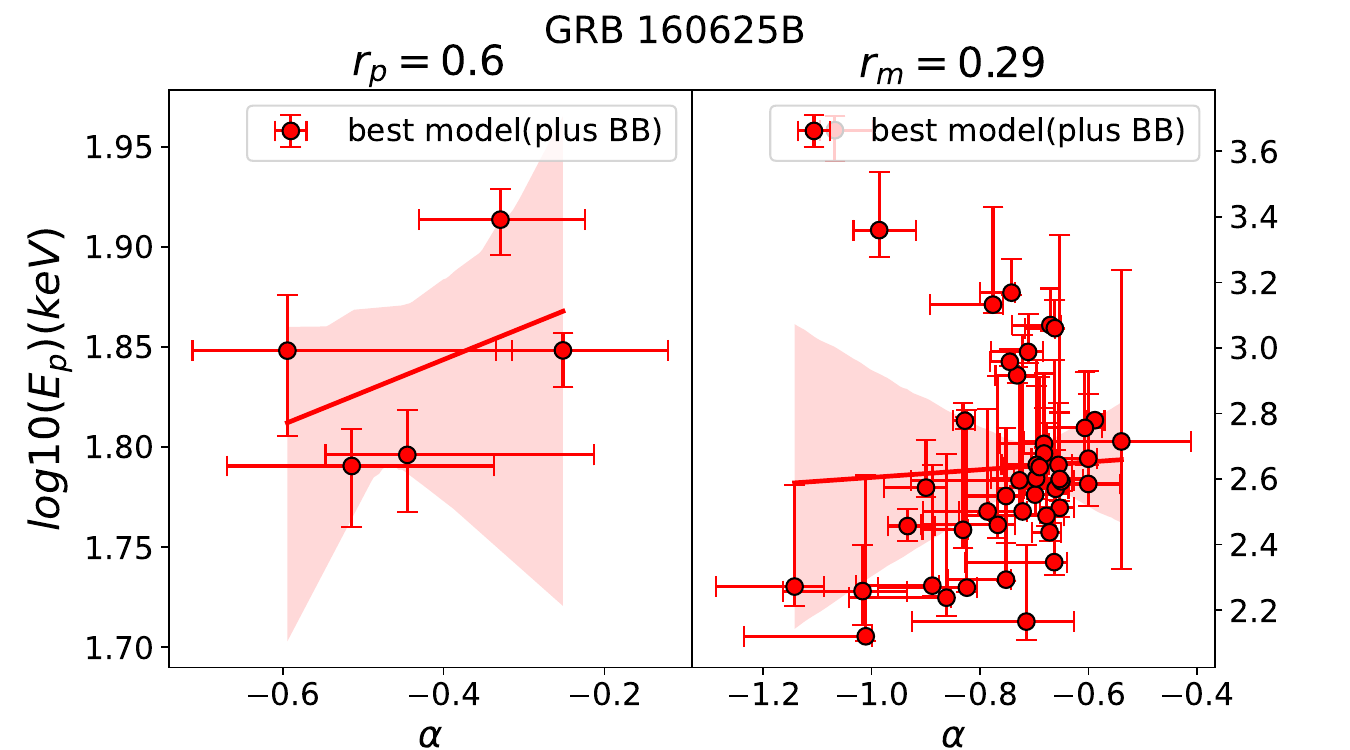}
\includegraphics [width=8.5cm,height=4.5cm]{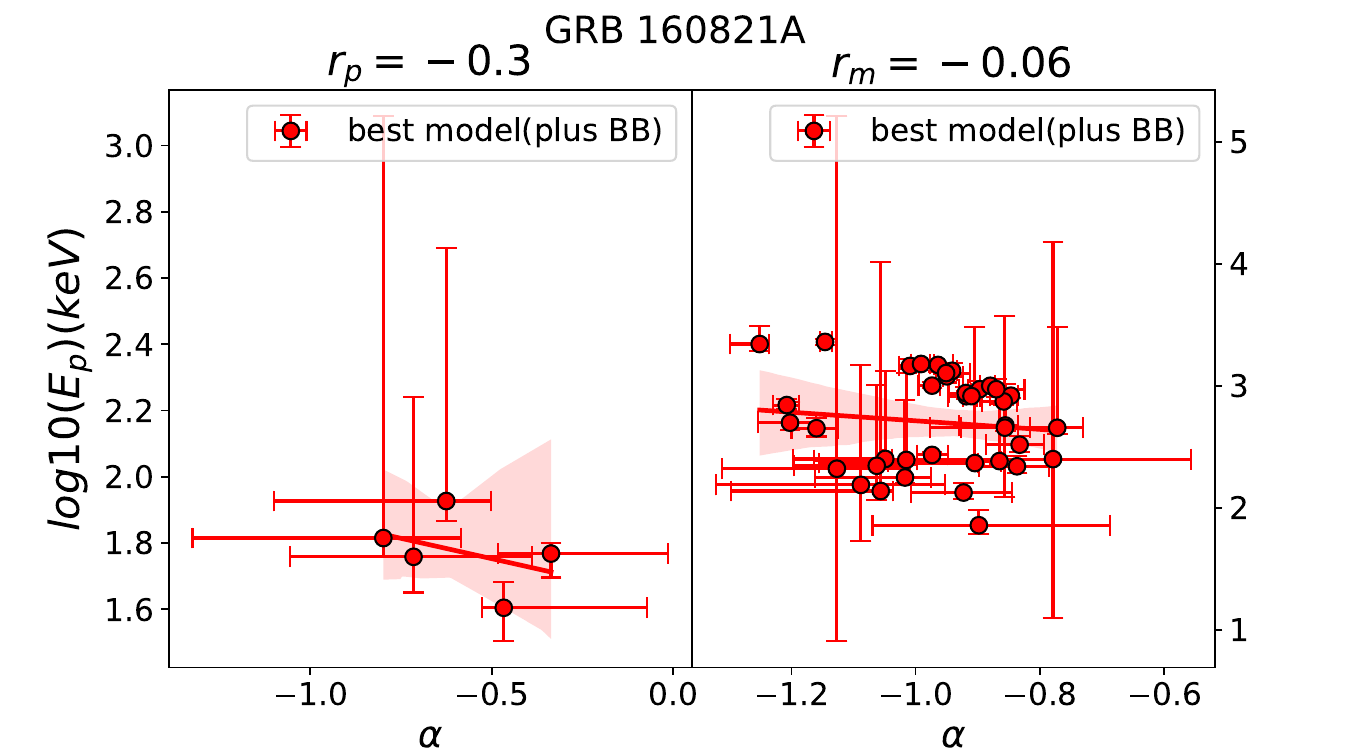}
\end{figure}
\begin{figure}[htbp]
\centering

\includegraphics [width=8.5cm,height=4.5cm]{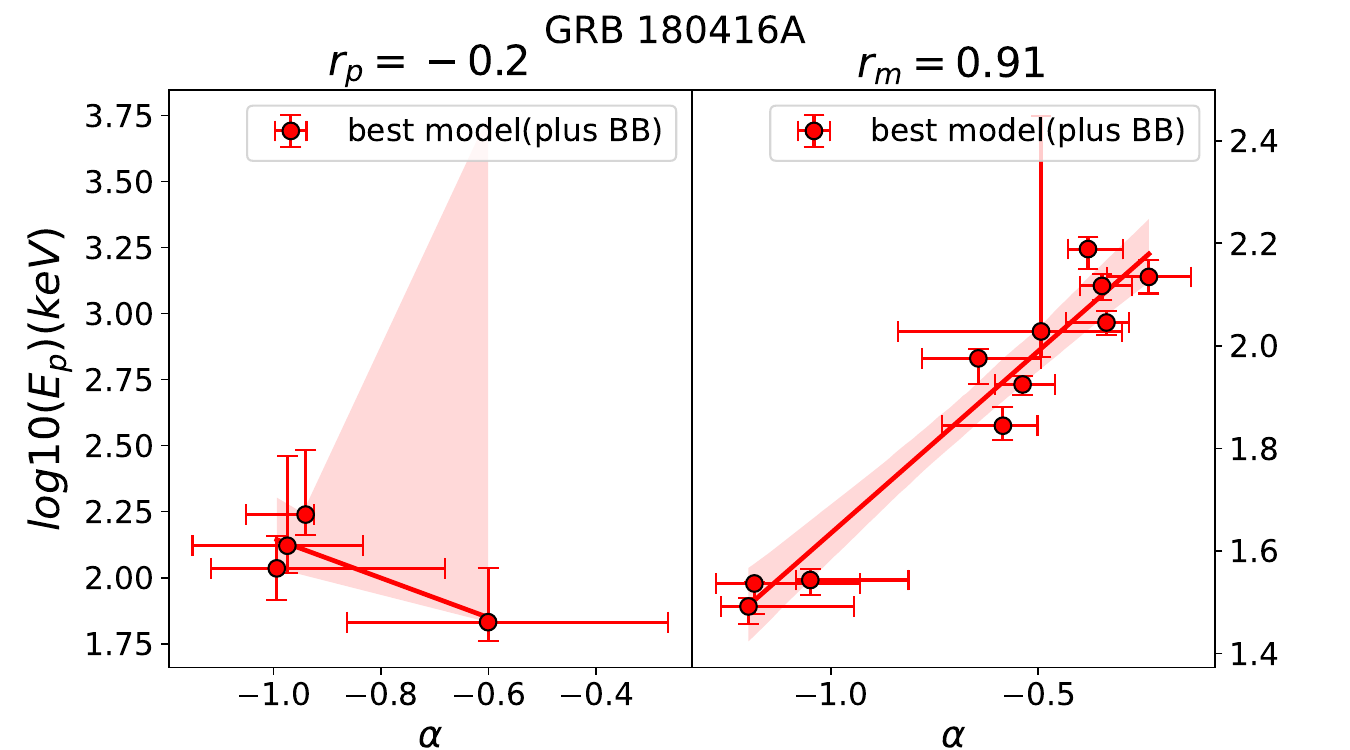}
\includegraphics [width=8.5cm,height=4.5cm]{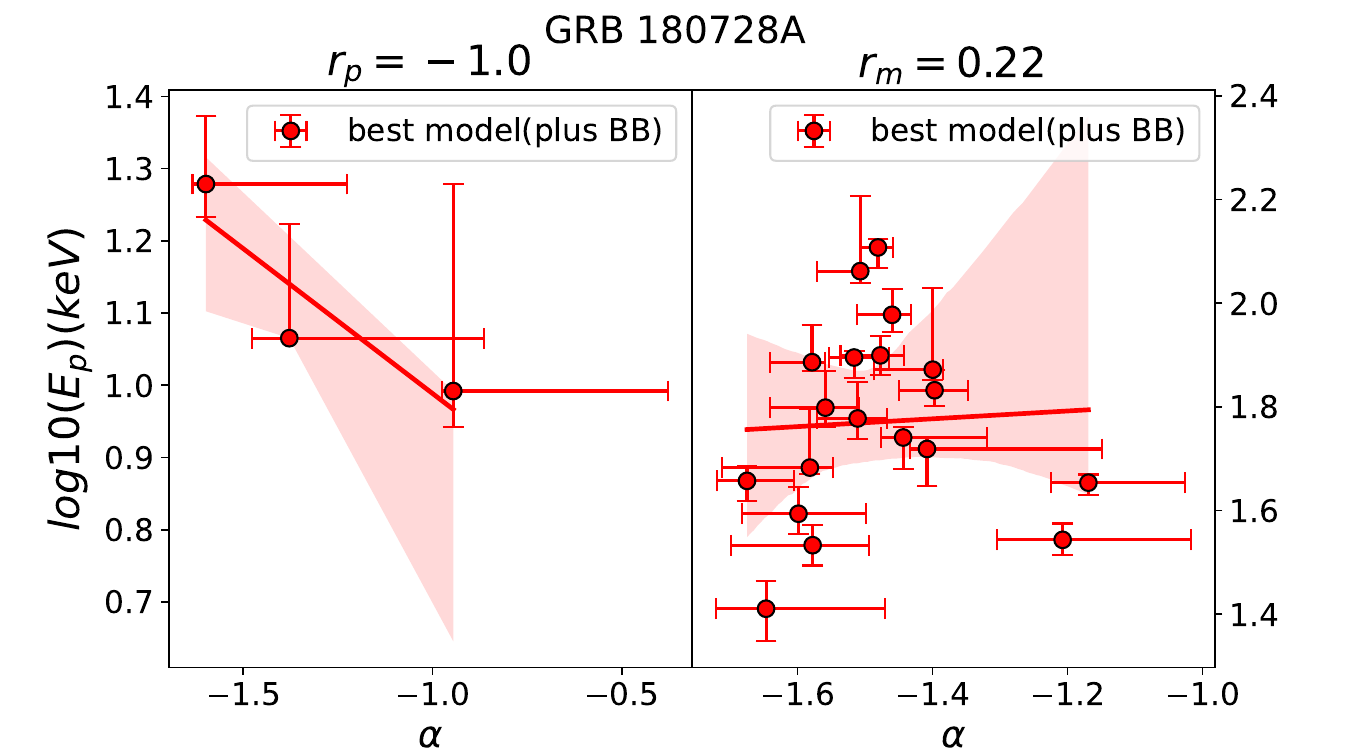}
\includegraphics [width=8.5cm,height=4.5cm]{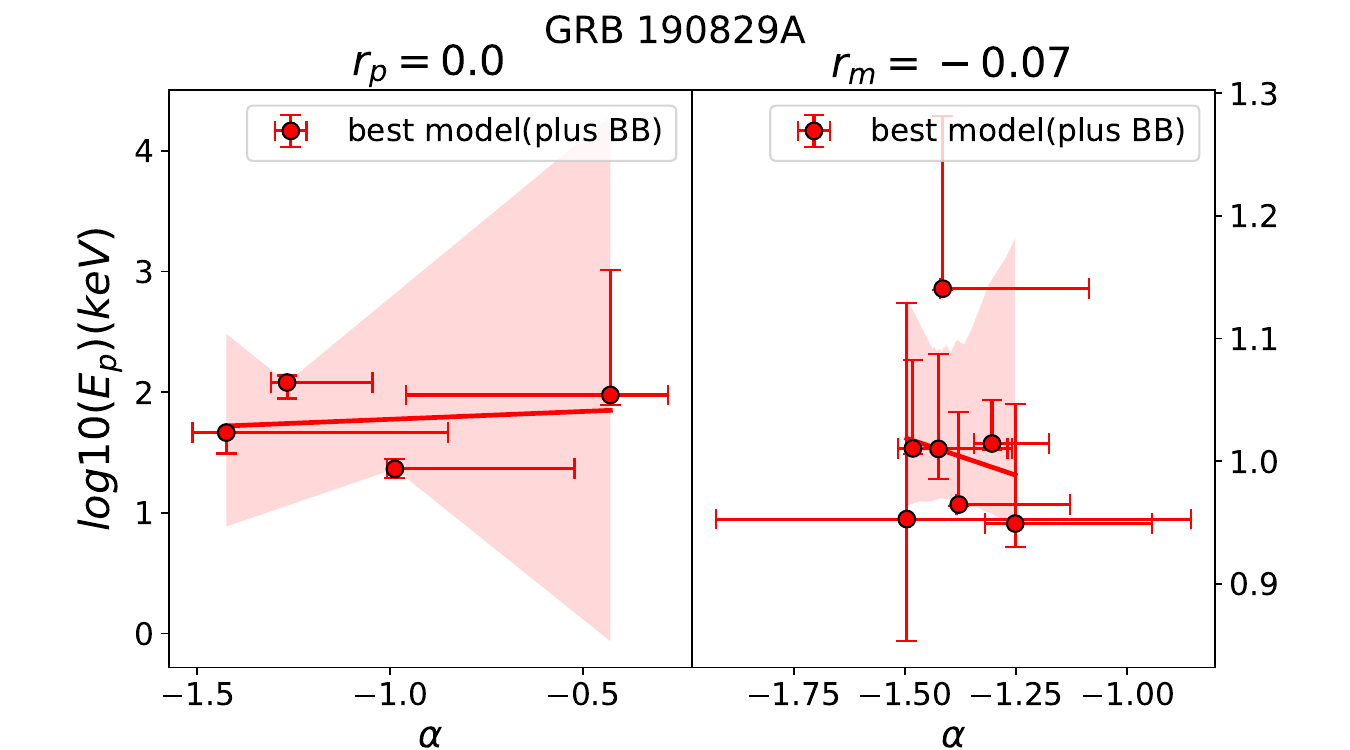}
\includegraphics [width=8.5cm,height=4.5cm]{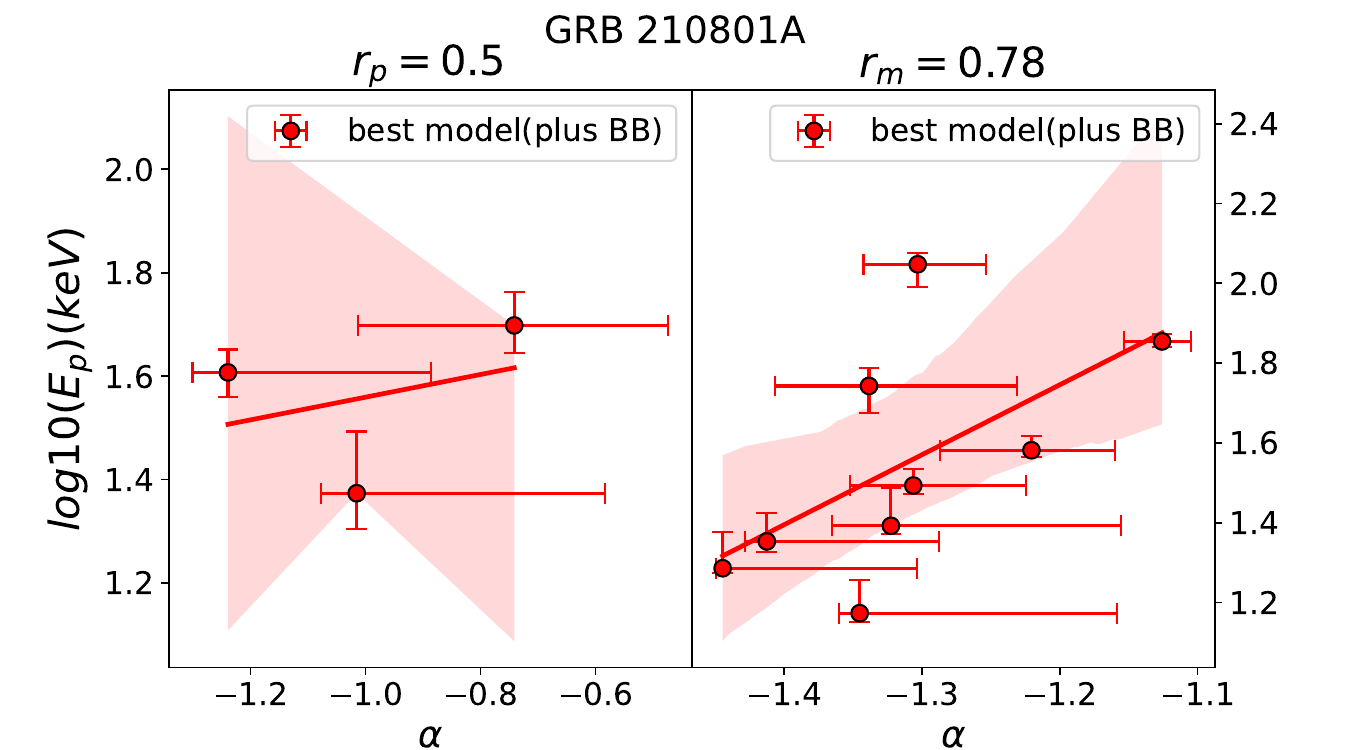}
\includegraphics [width=8.5cm,height=4.5cm]{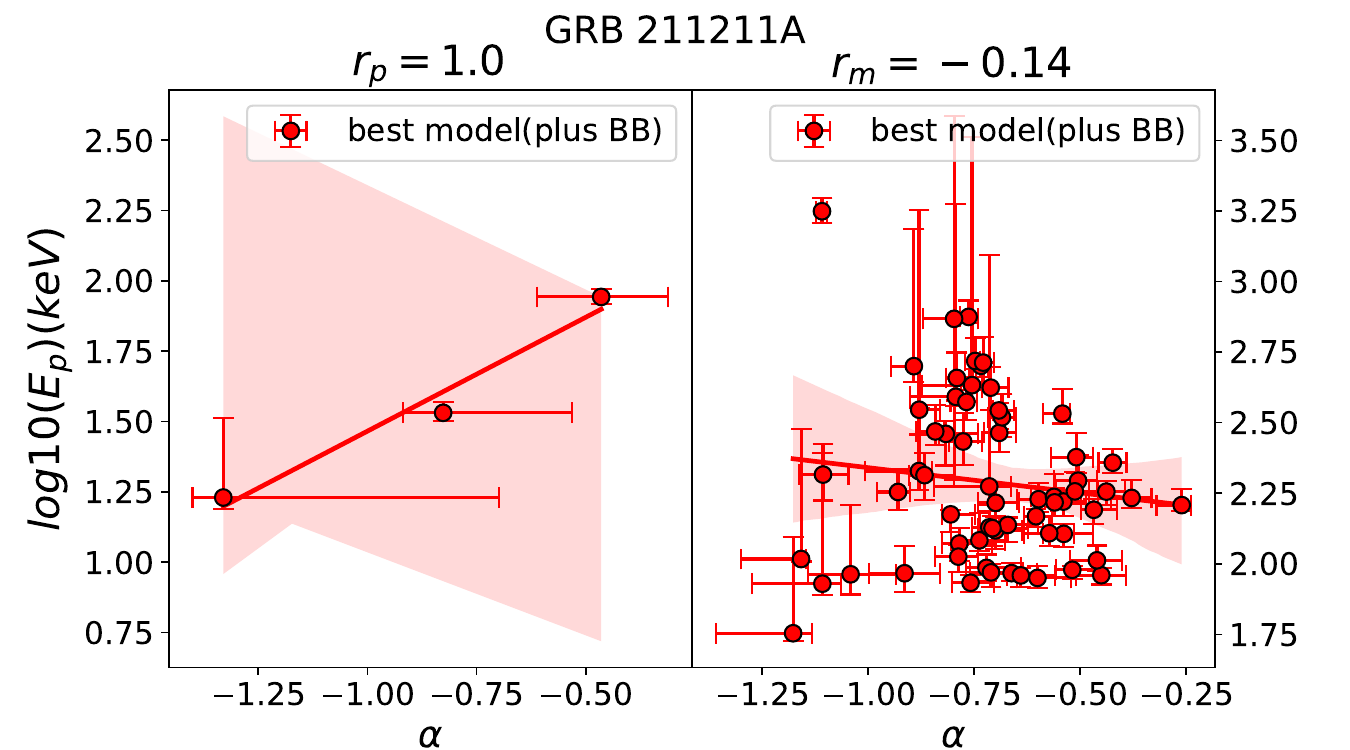}
\includegraphics [width=8.5cm,height=4.5cm]{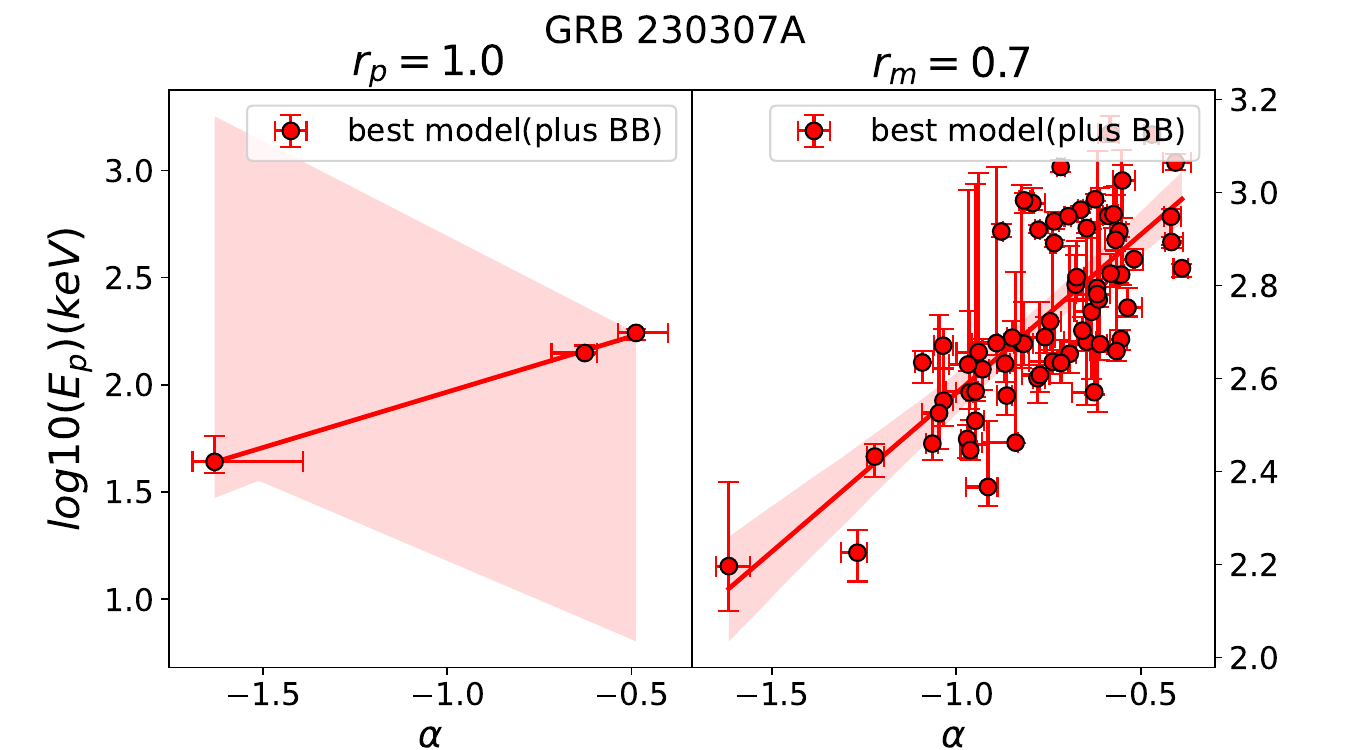}
\includegraphics [width=8.5cm,height=4.5cm]{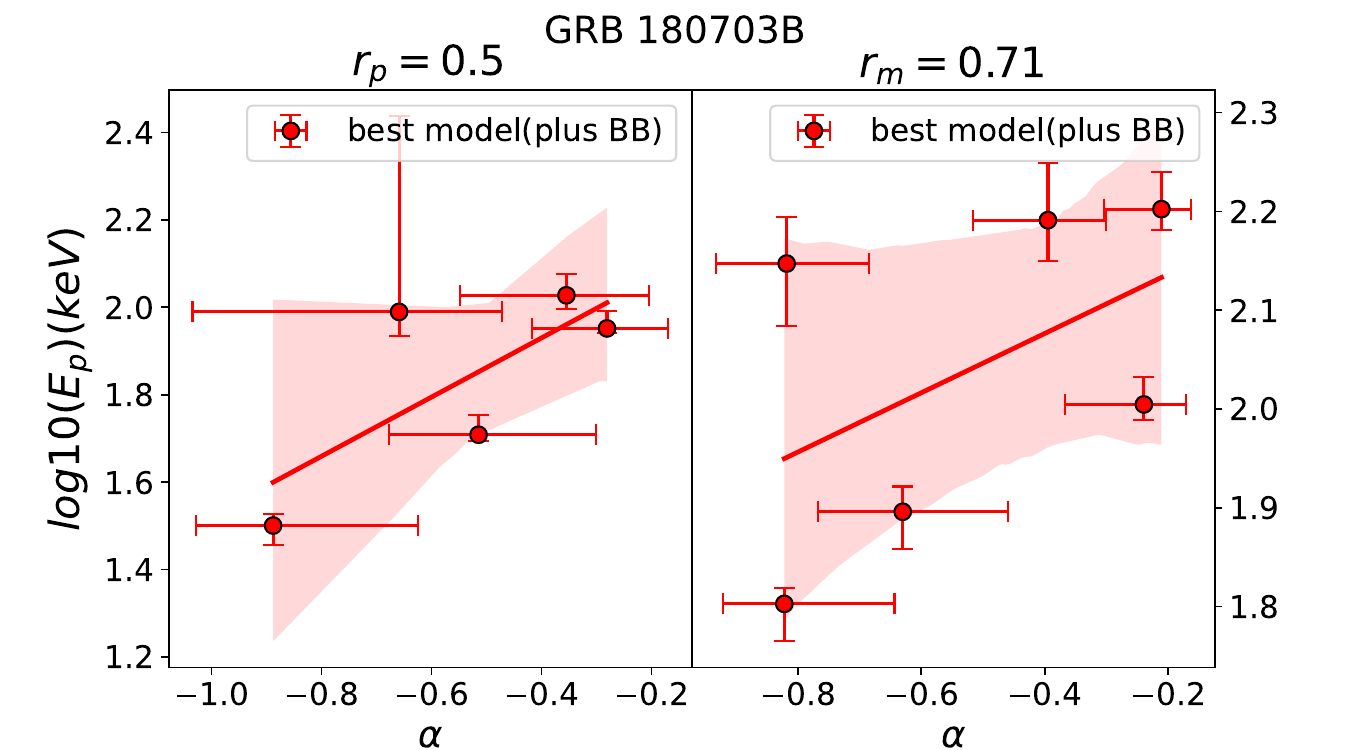}
\includegraphics [width=8.5cm,height=4.5cm]{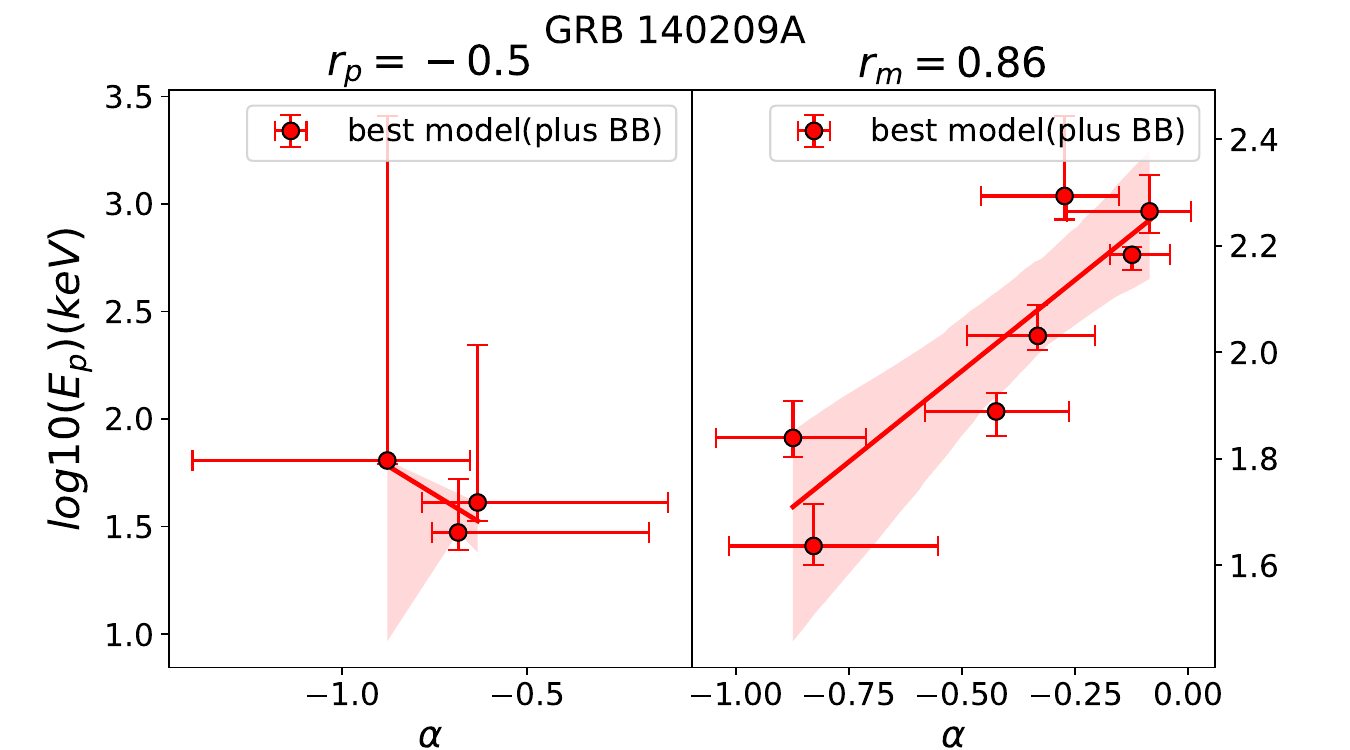}
   \figcaption{Correlations between $E_{p}$ and $\alpha$ fitted with the best model for the precursor and main bursts. All symbols are the same as in Figure \ref{fig 2}. \label{fig C3}}
\end{figure}

\begin{figure}[htbp]
\centering
\includegraphics [width=8.5cm,height=4.5cm]{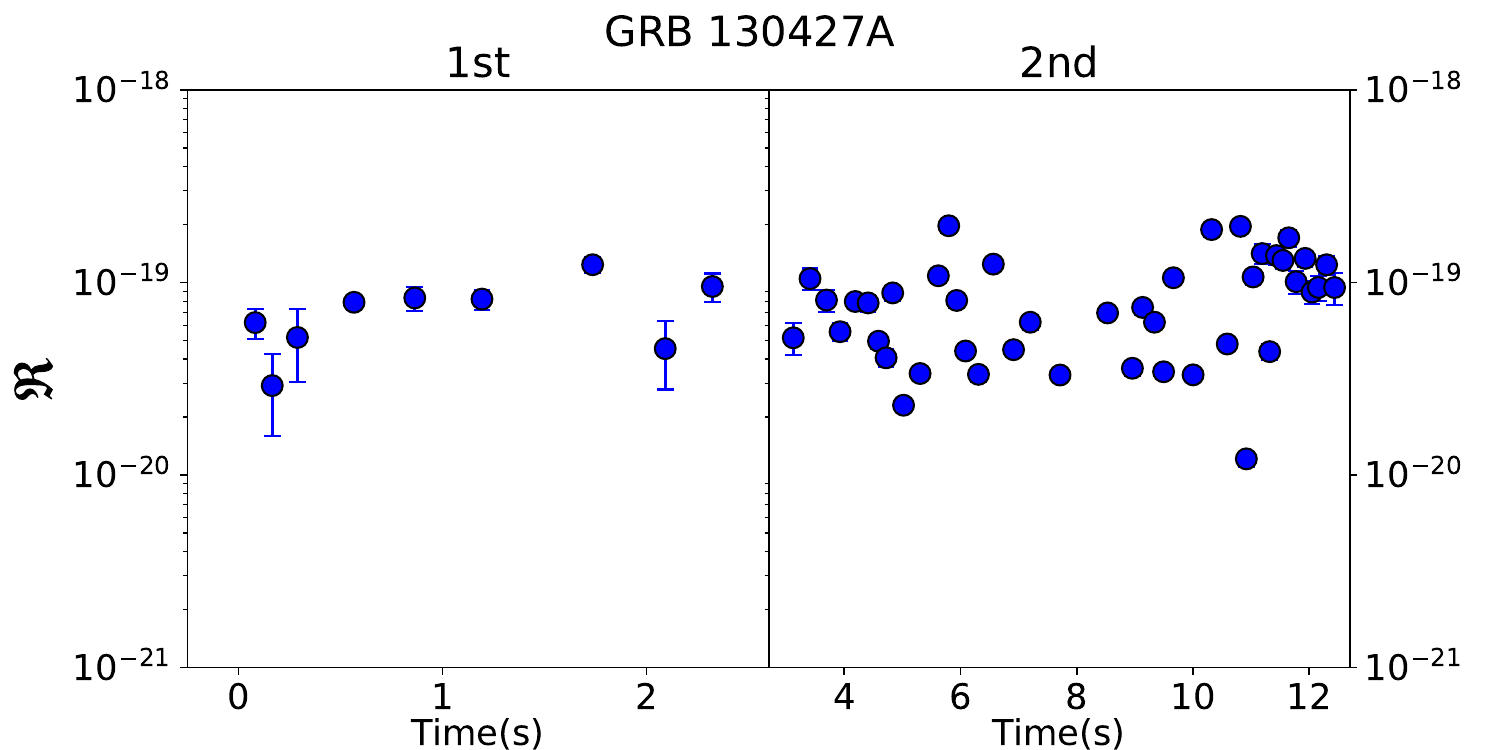}
\includegraphics [width=8.5cm,height=4.5cm]{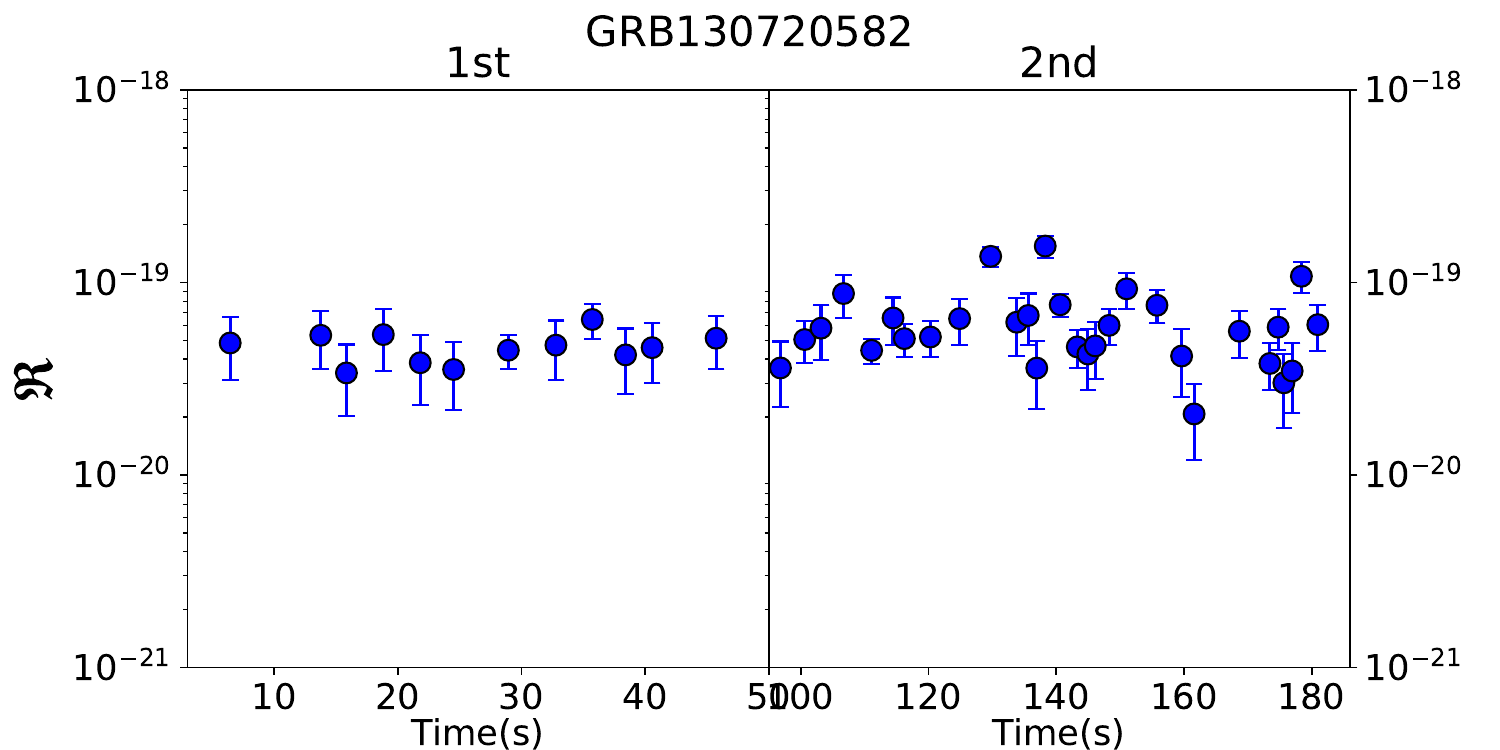}
\includegraphics [width=8.5cm,height=4.5cm]{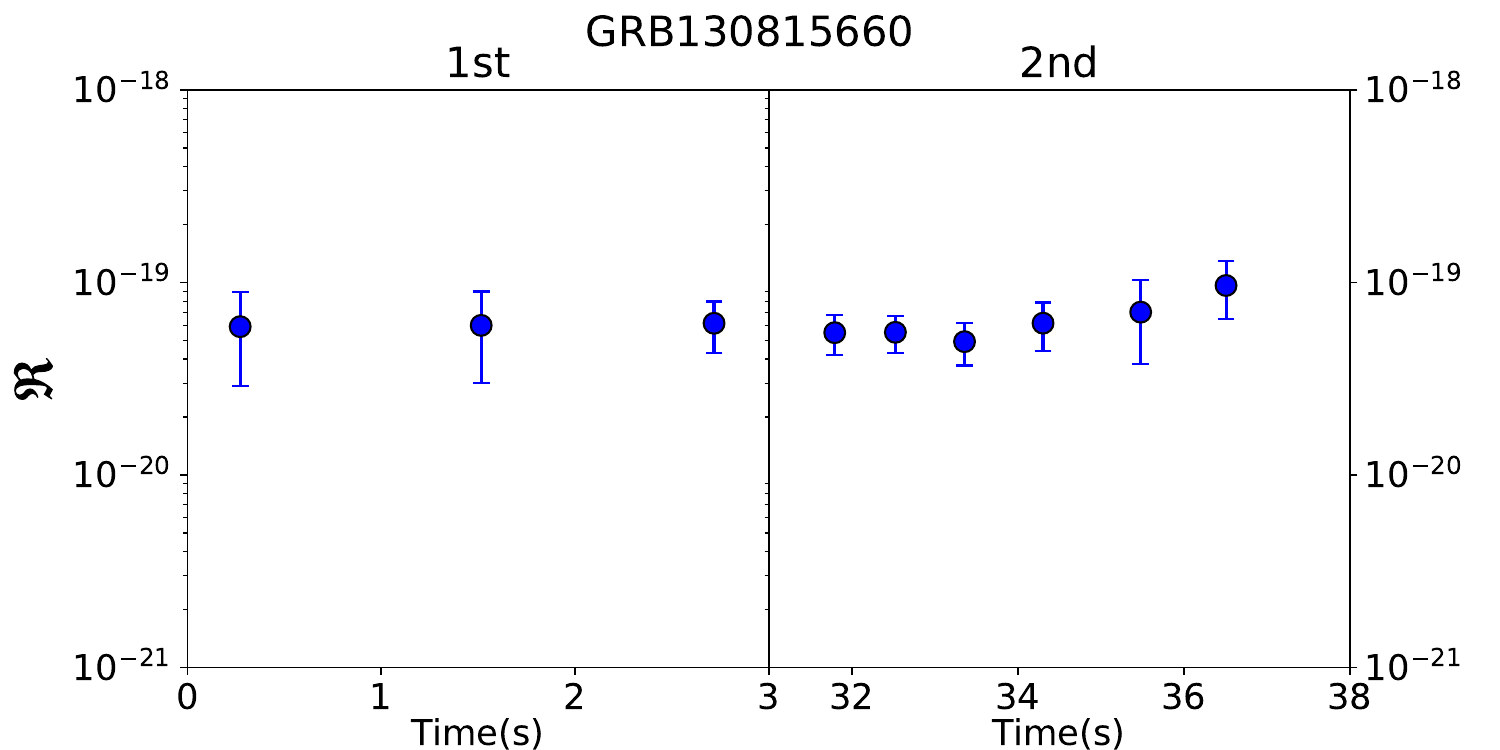}
\includegraphics [width=8.5cm,height=4.5cm]{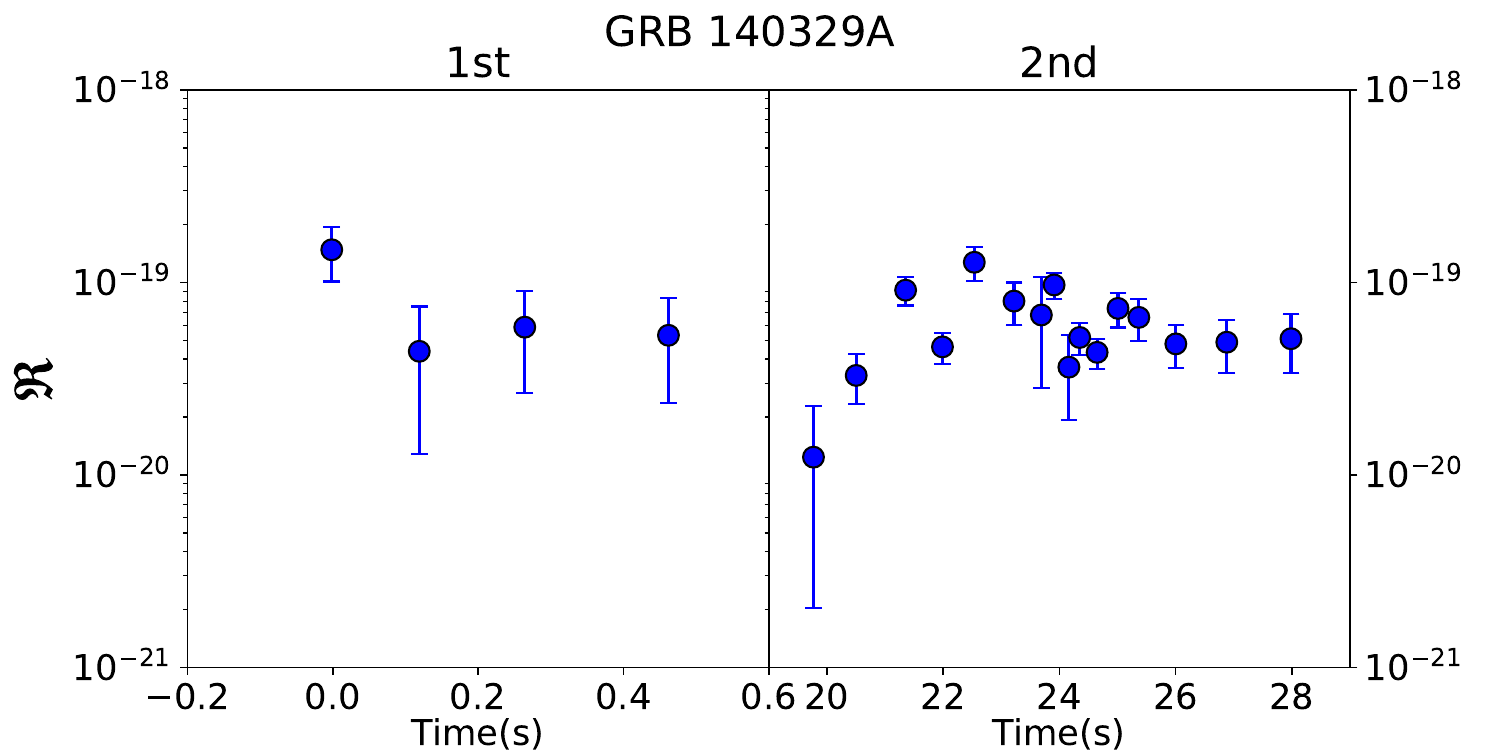}
\includegraphics [width=8.5cm,height=4.5cm]{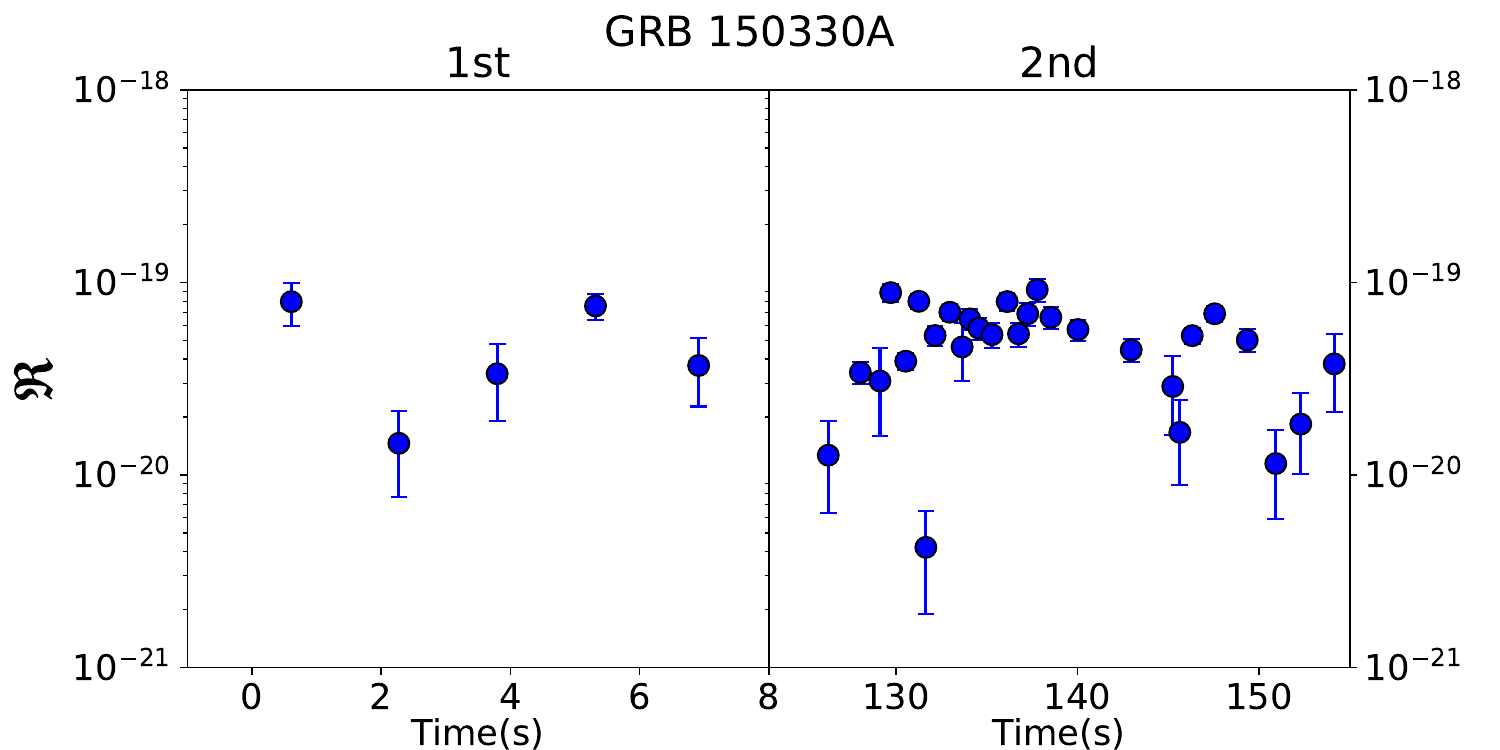}
\includegraphics [width=8.5cm,height=4.5cm]{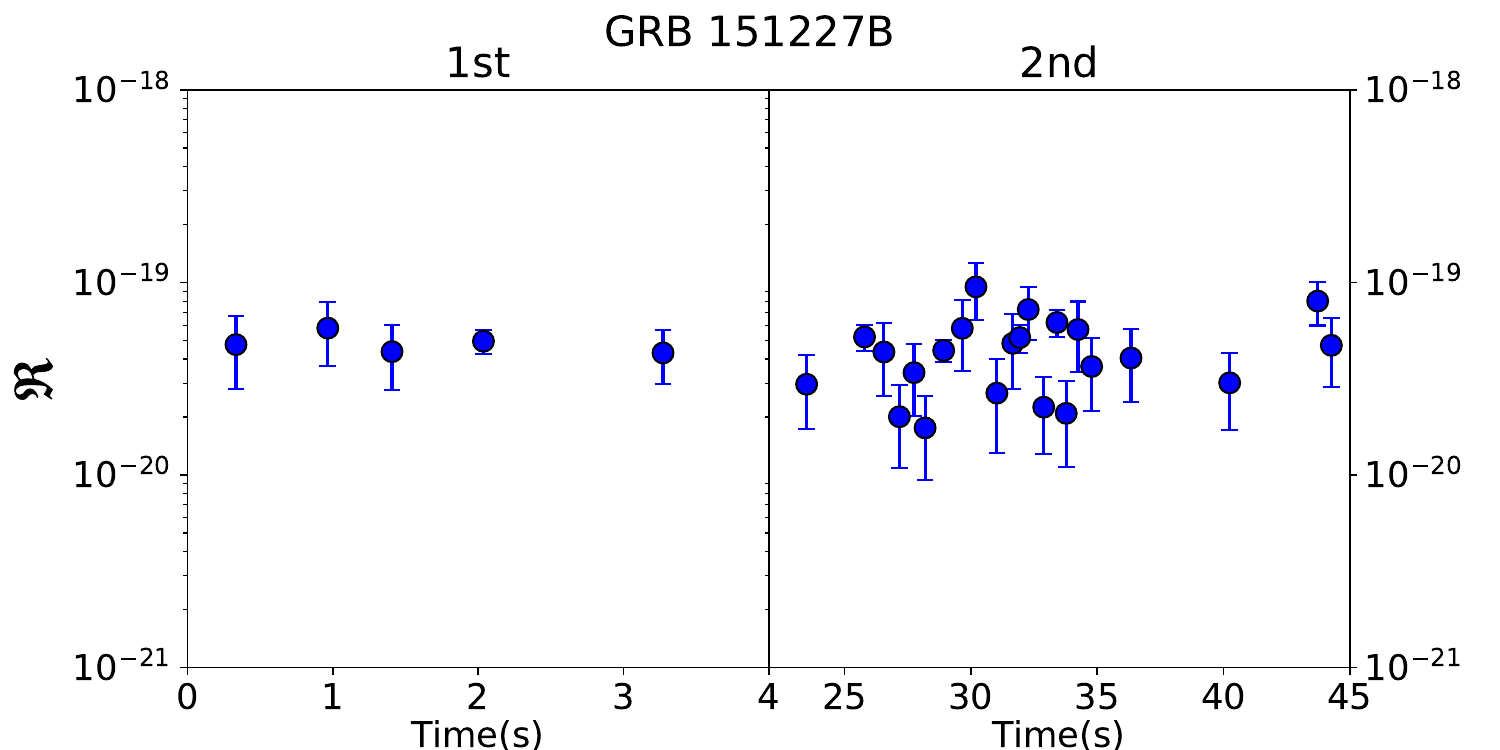}
\includegraphics [width=8.5cm,height=4.5cm]{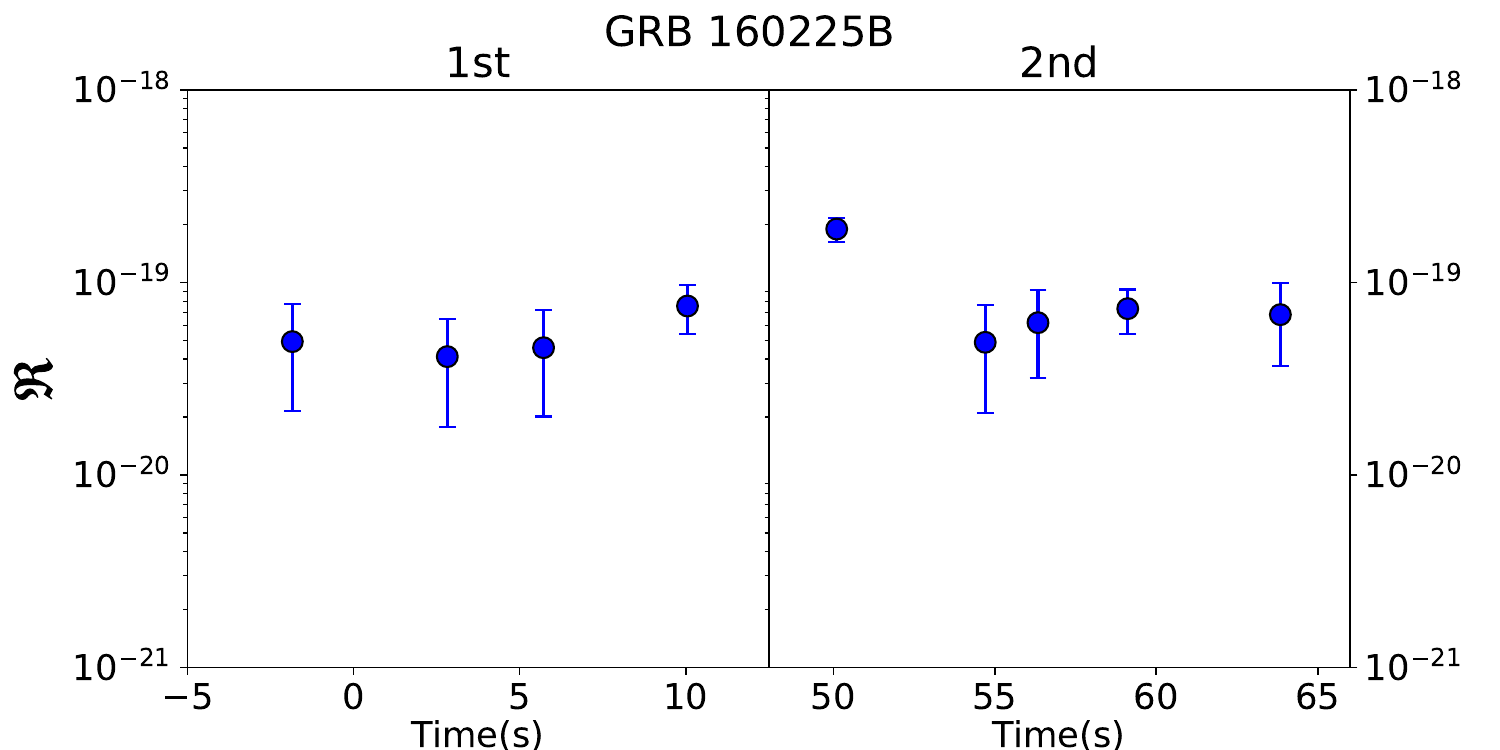}
\includegraphics [width=8.5cm,height=4.5cm]{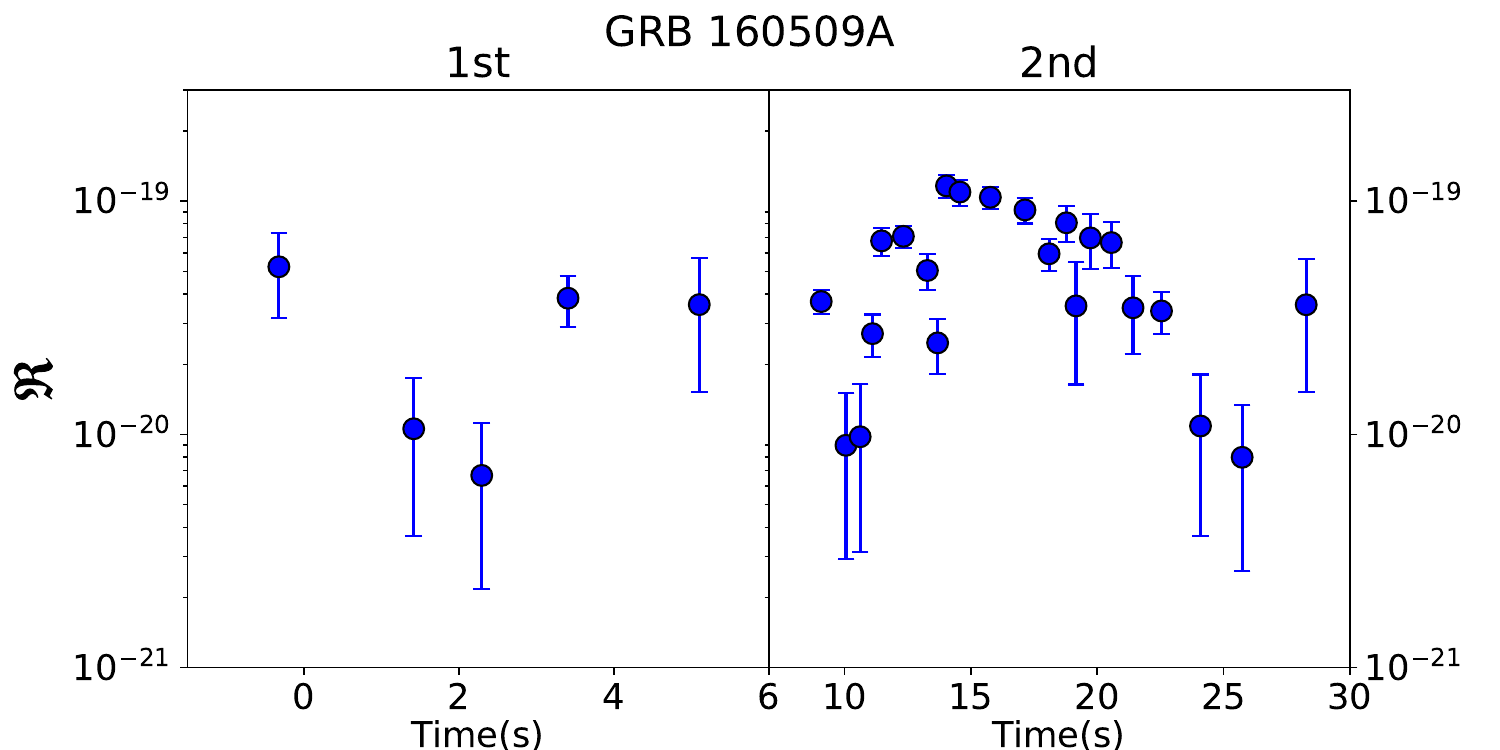}
\includegraphics [width=8.5cm,height=4.5cm]{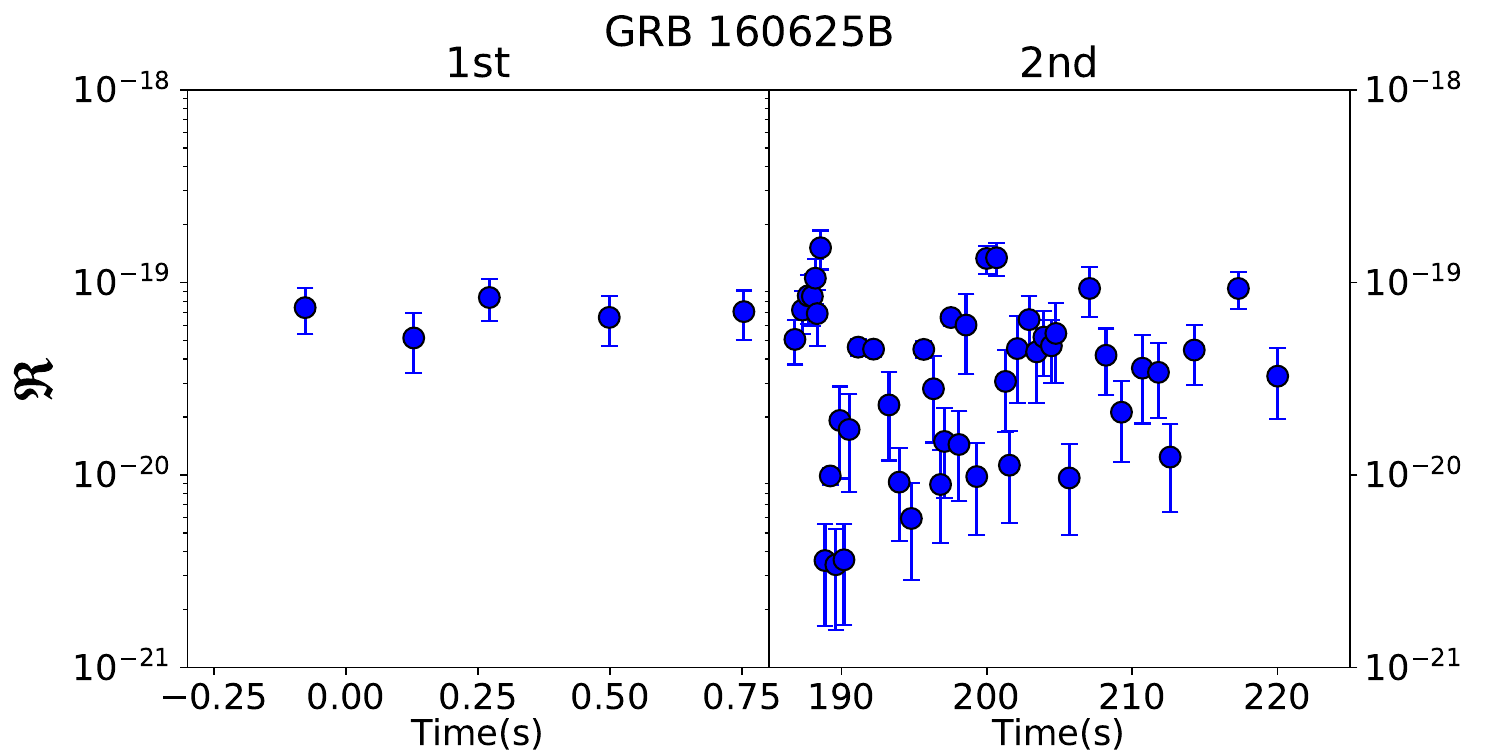}
\includegraphics [width=8.5cm,height=4.5cm]{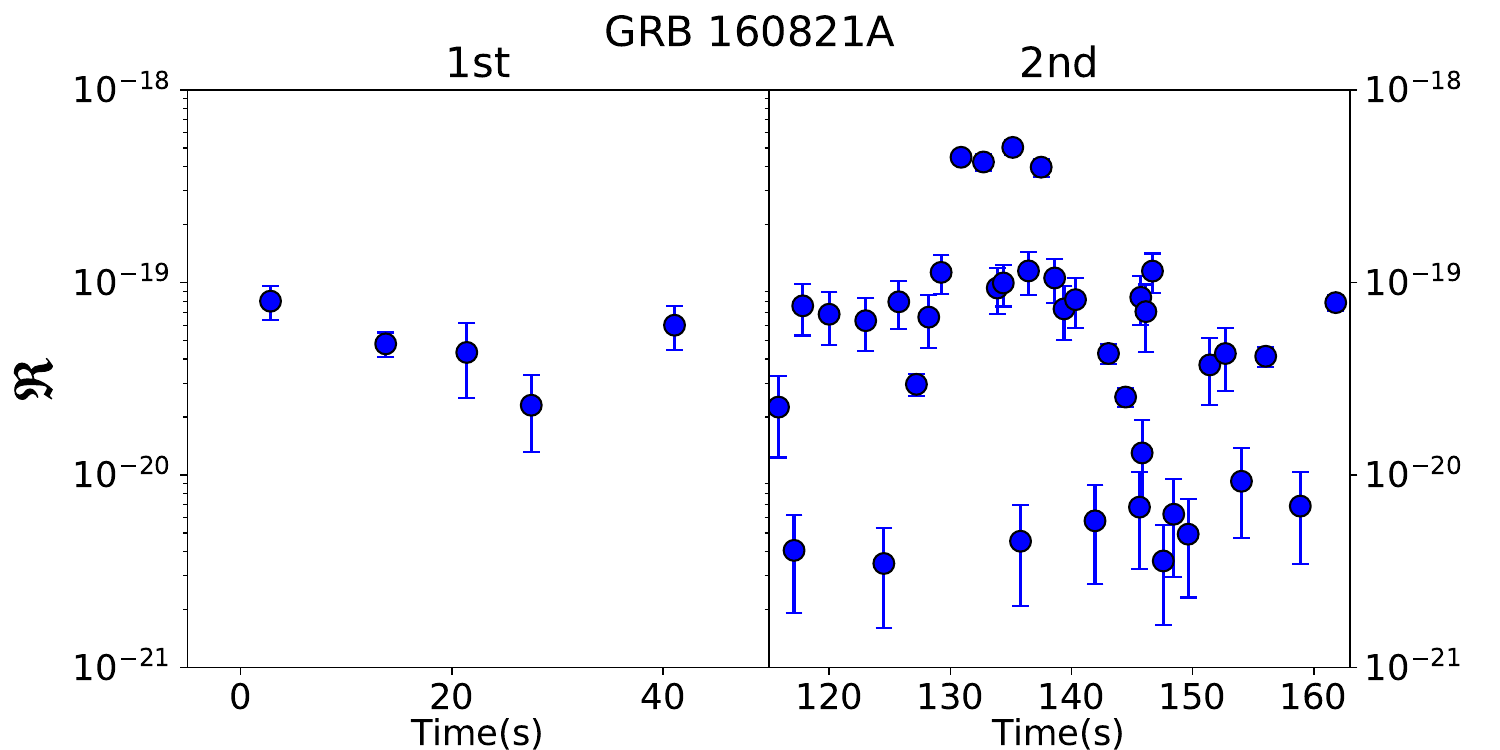}
\end{figure}
\begin{figure}[htbp]
\centering
\includegraphics [width=8.5cm,height=4.5cm]{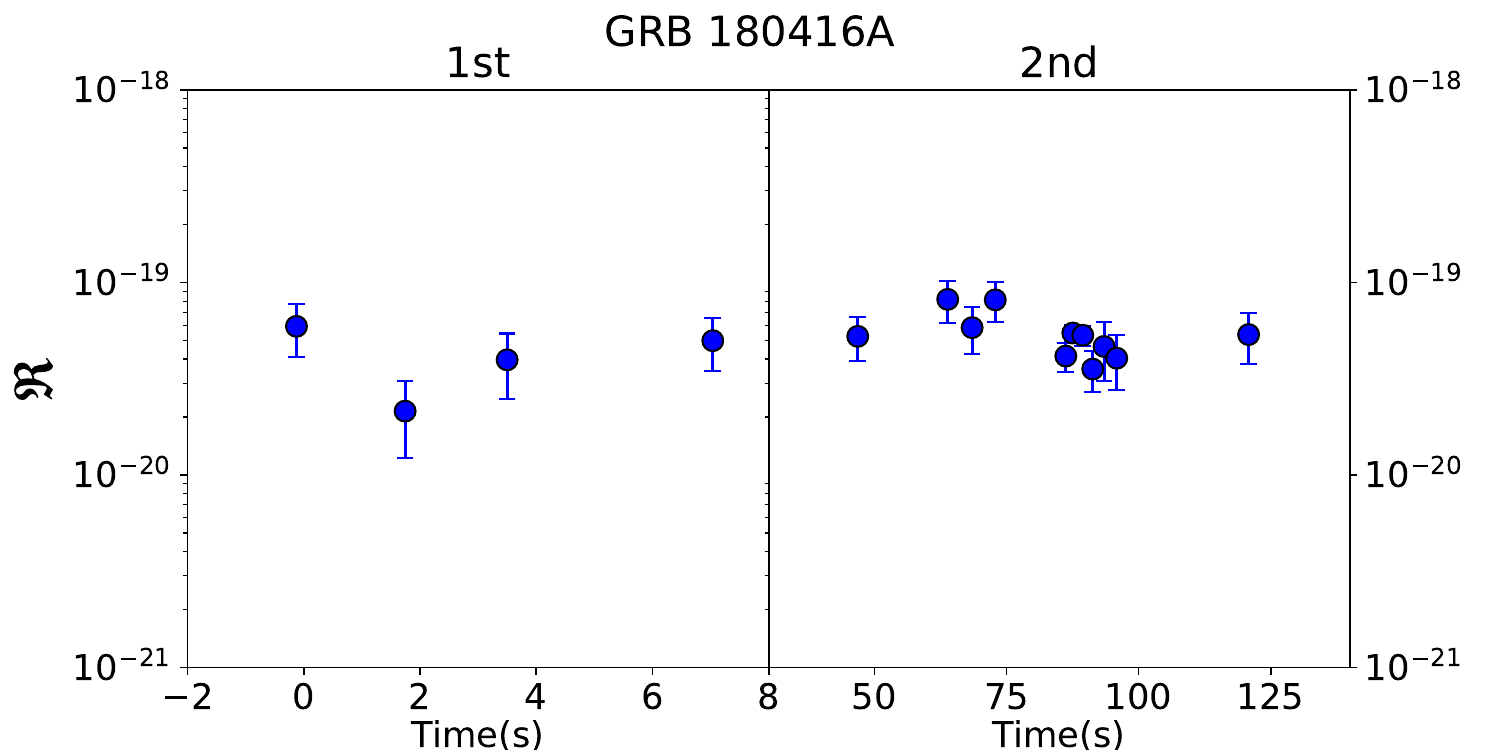}
\includegraphics [width=8.5cm,height=4.5cm]{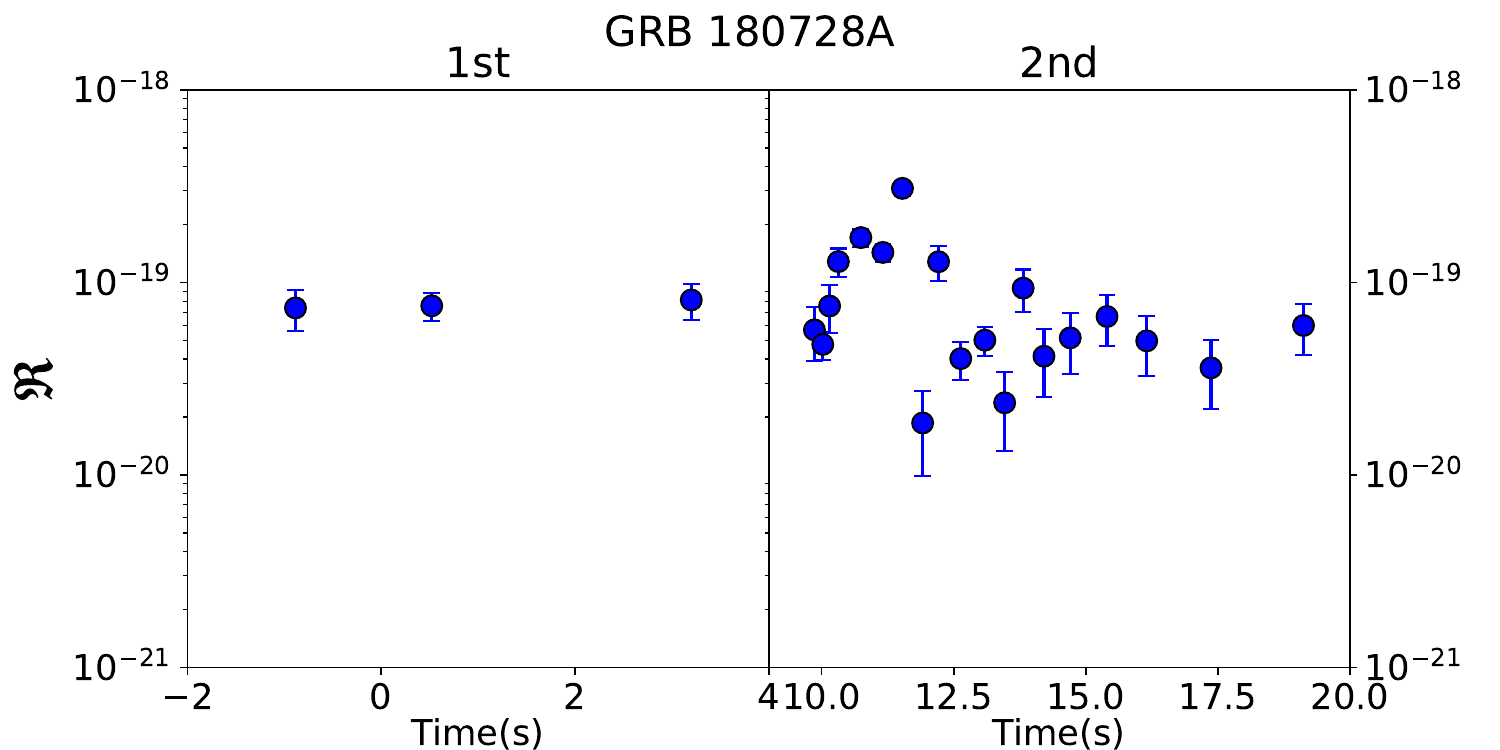}
\includegraphics [width=8.5cm,height=4.5cm]{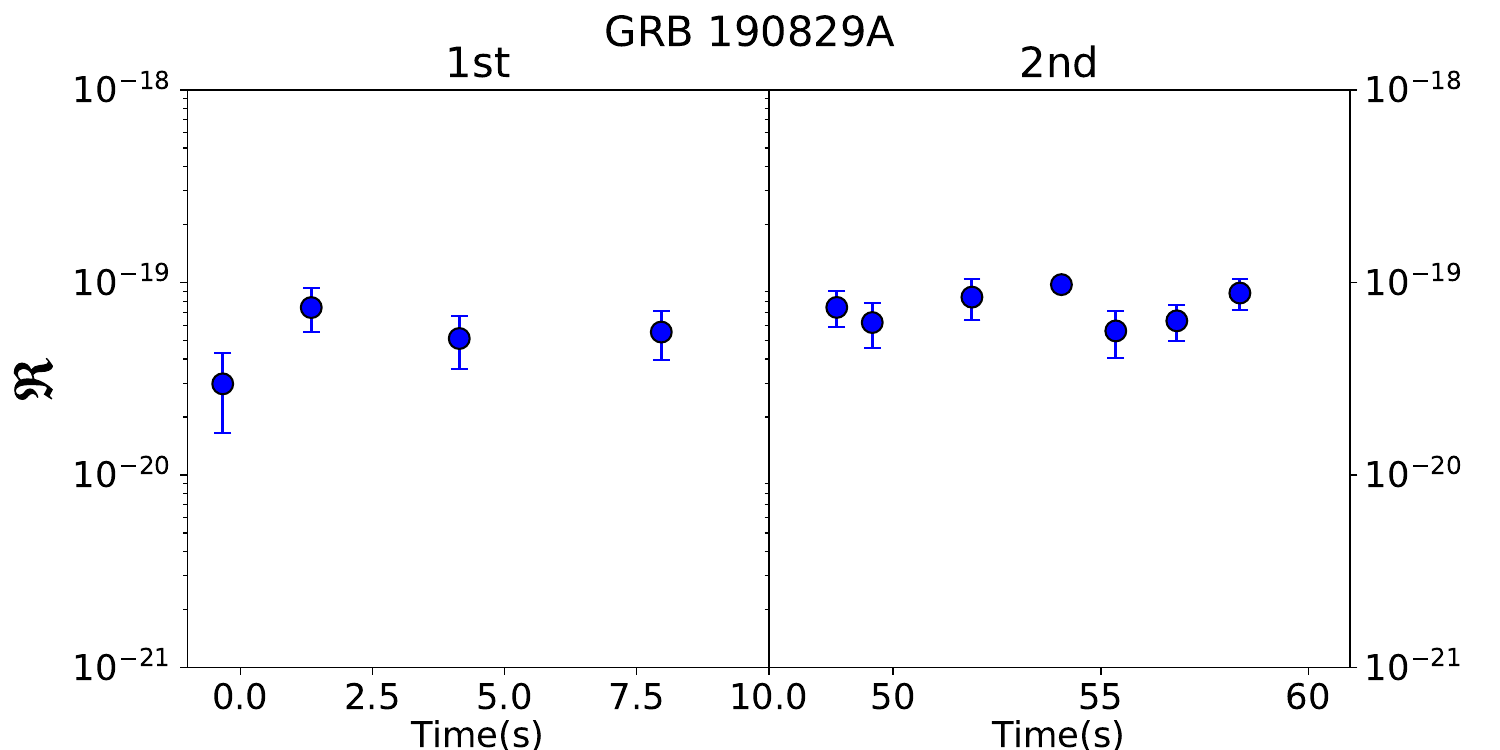}
\includegraphics [width=8.5cm,height=4.5cm]{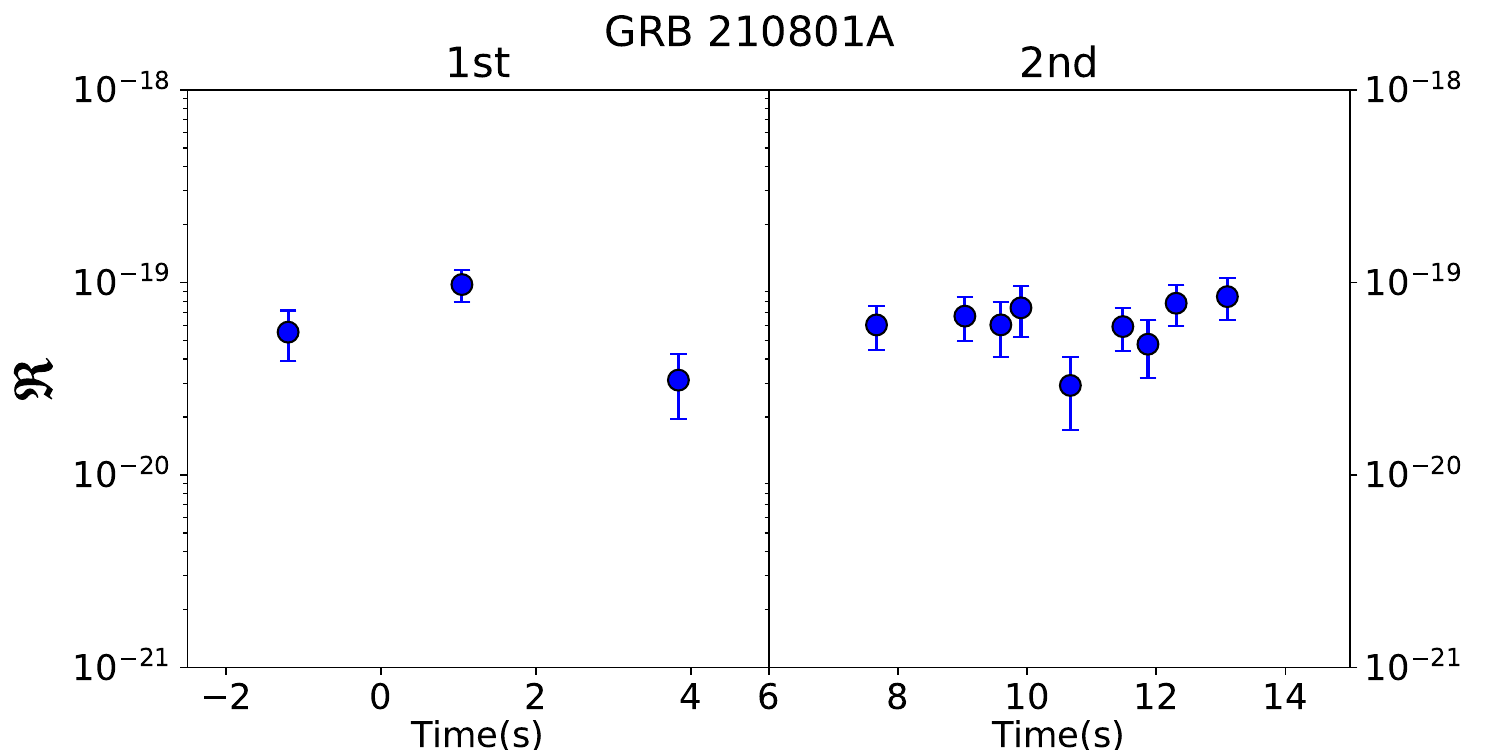}
\includegraphics [width=8.5cm,height=4.5cm]{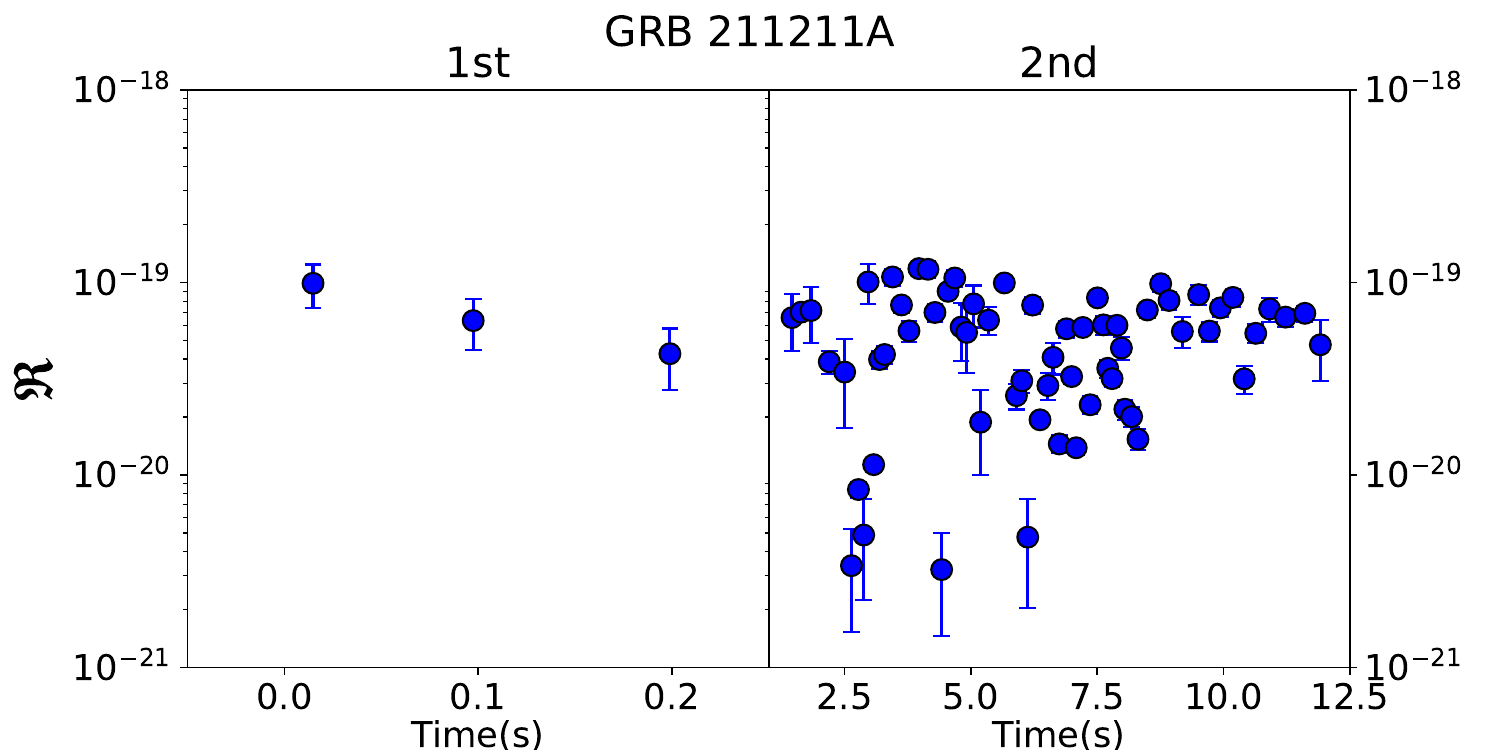}
\includegraphics [width=8.5cm,height=4.5cm]{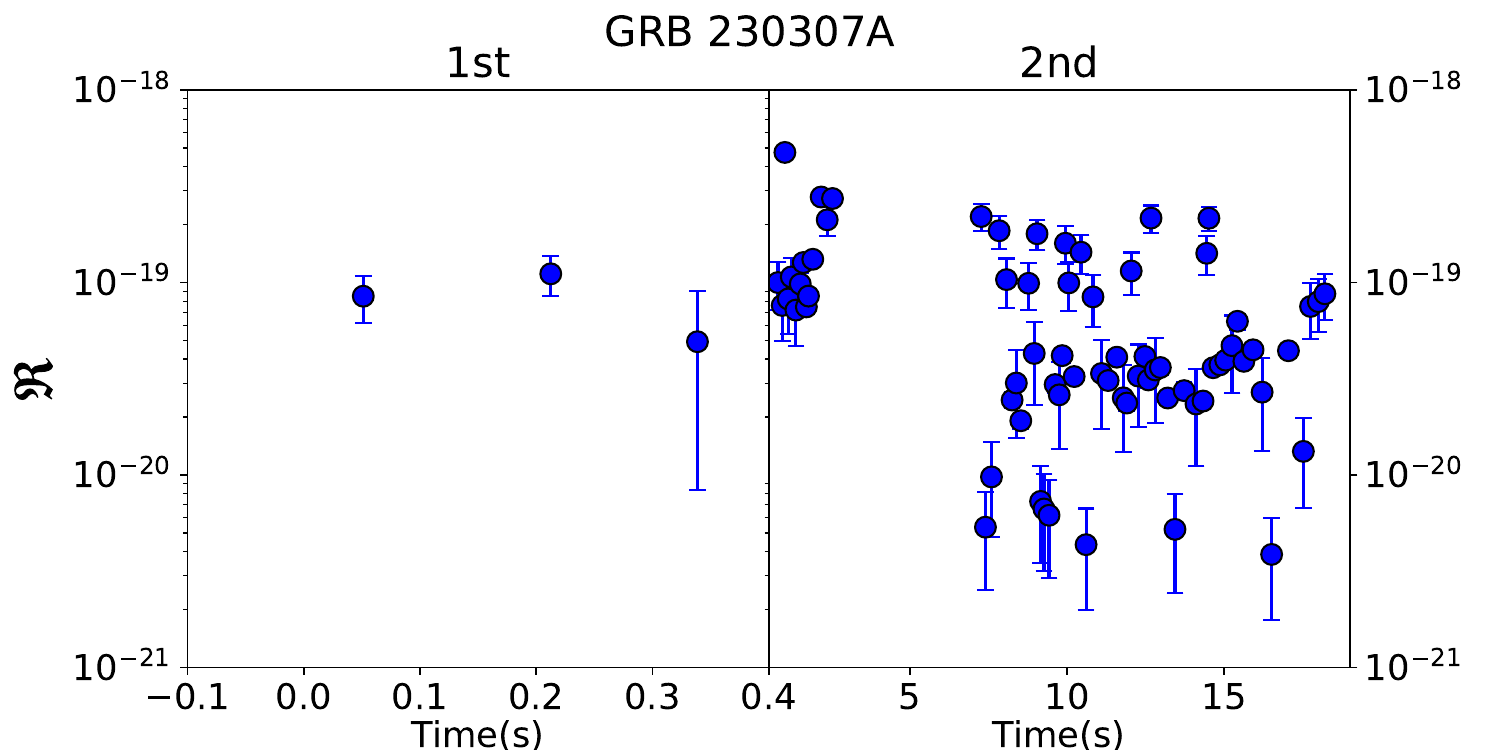}
\includegraphics [width=8.5cm,height=4.5cm]{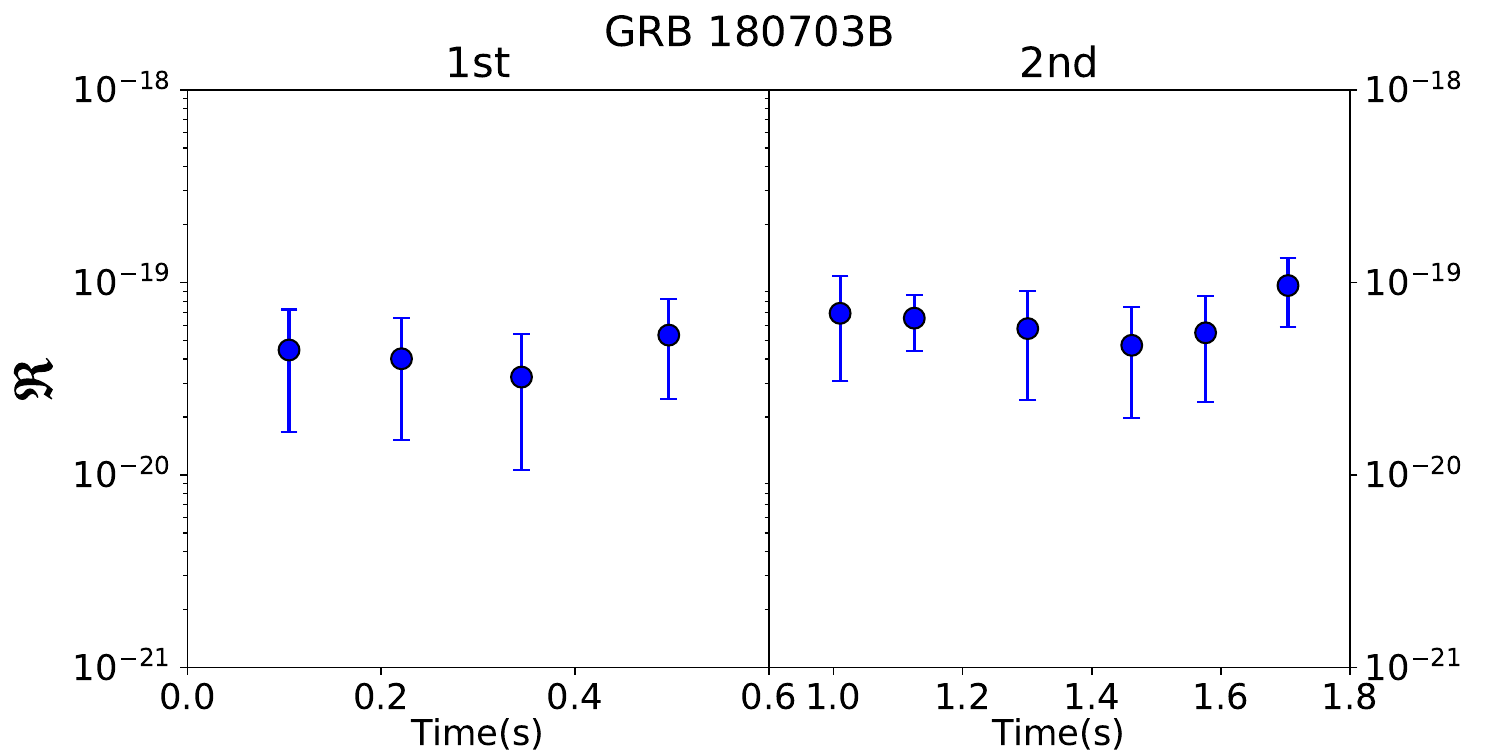}
\includegraphics [width=8.5cm,height=4.5cm]{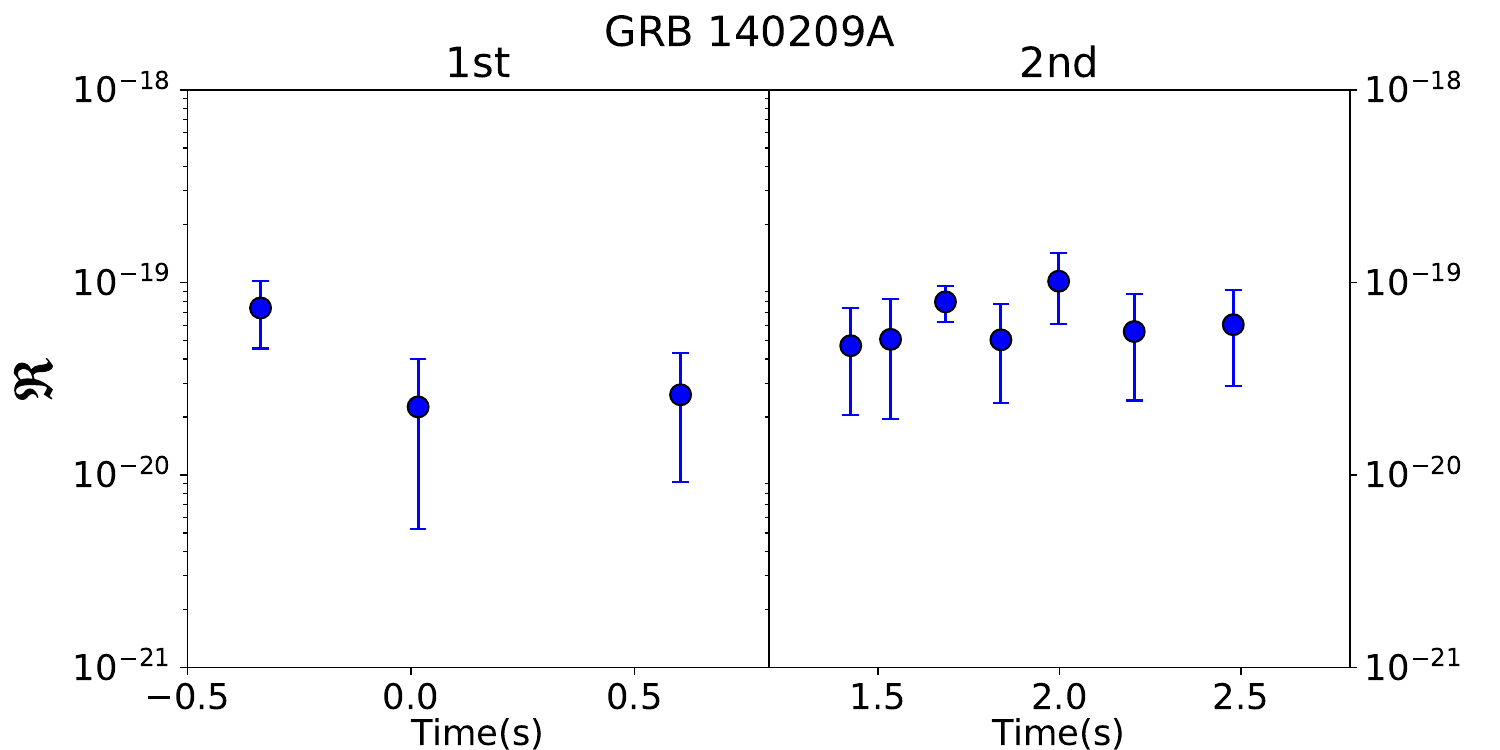}
   \figcaption{The time evolution of $\Re$. ``1st'' denotes precursors, and ``2nd'' denotes main bursts. \label{fig D1}}
\end{figure}

\begin{figure}[htbp]
\centering
\includegraphics [width=8.5cm,height=4.5cm]{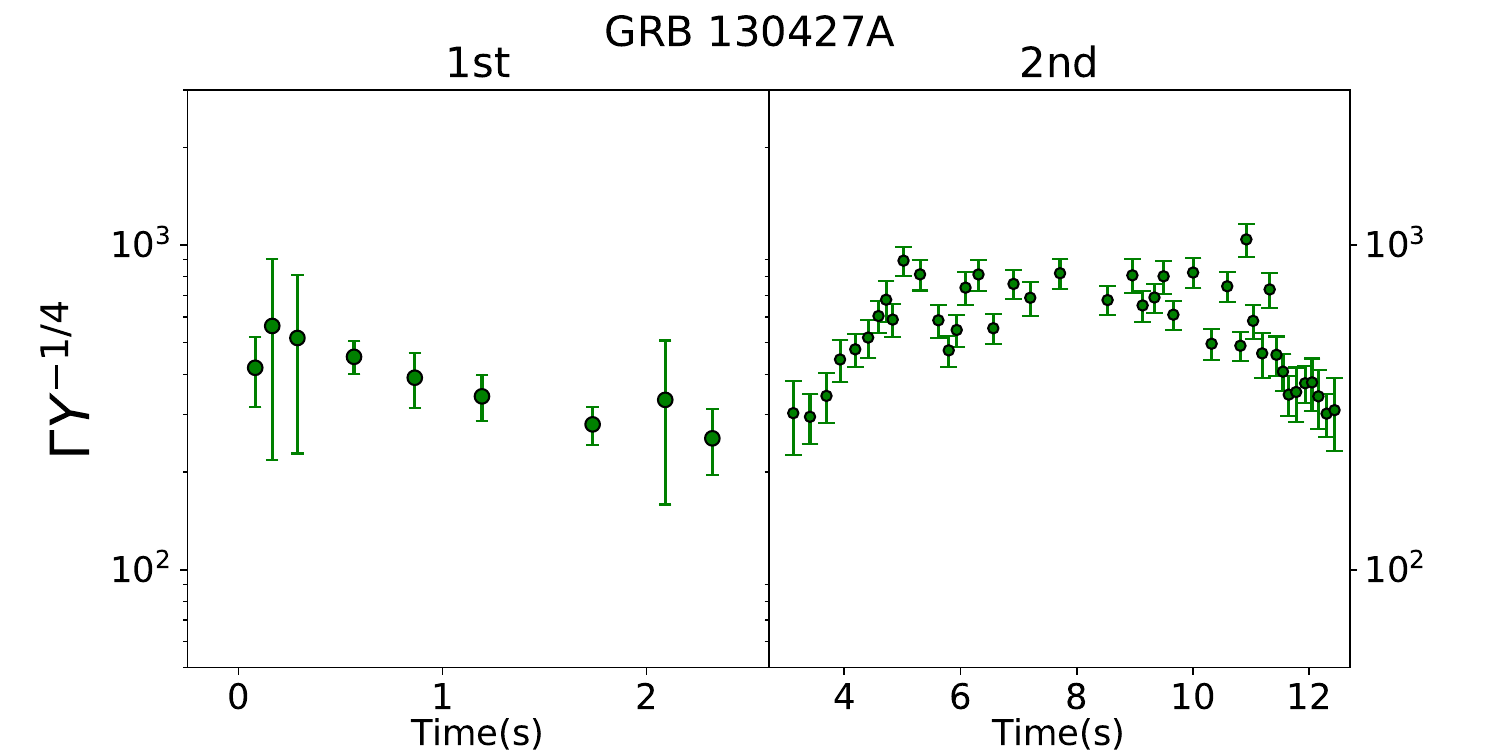}
\includegraphics [width=8.5cm,height=4.5cm]{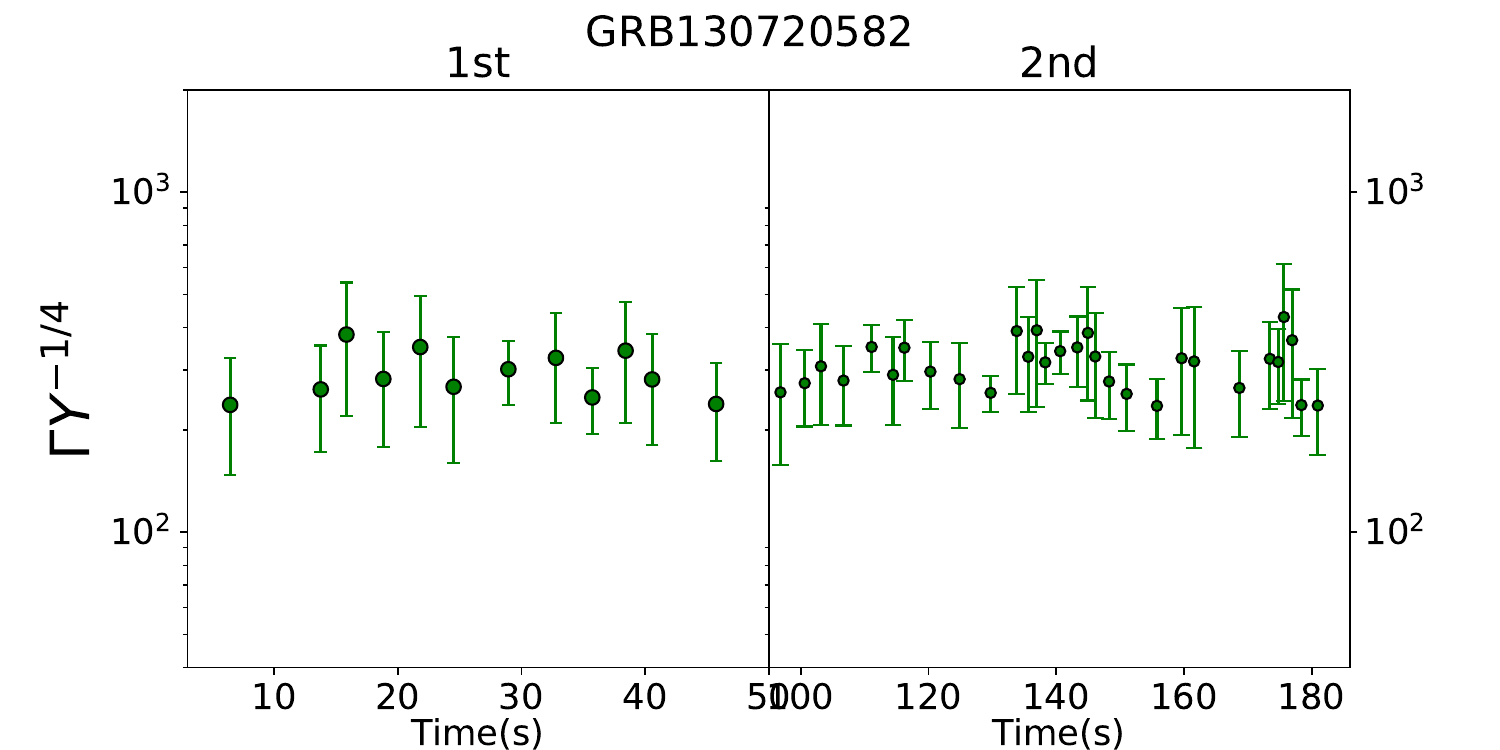}
\includegraphics [width=8.5cm,height=4.5cm]{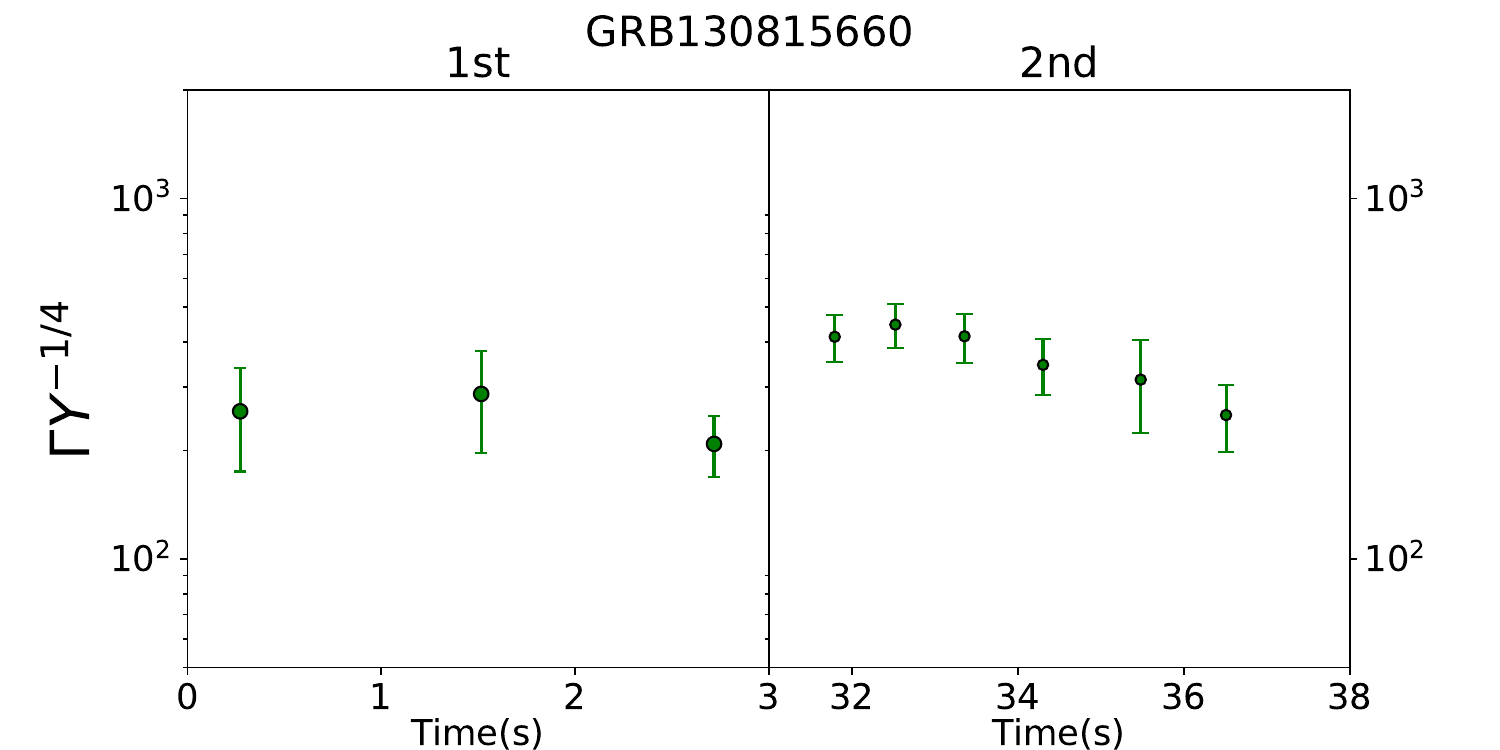}
\includegraphics [width=8.5cm,height=4.5cm]{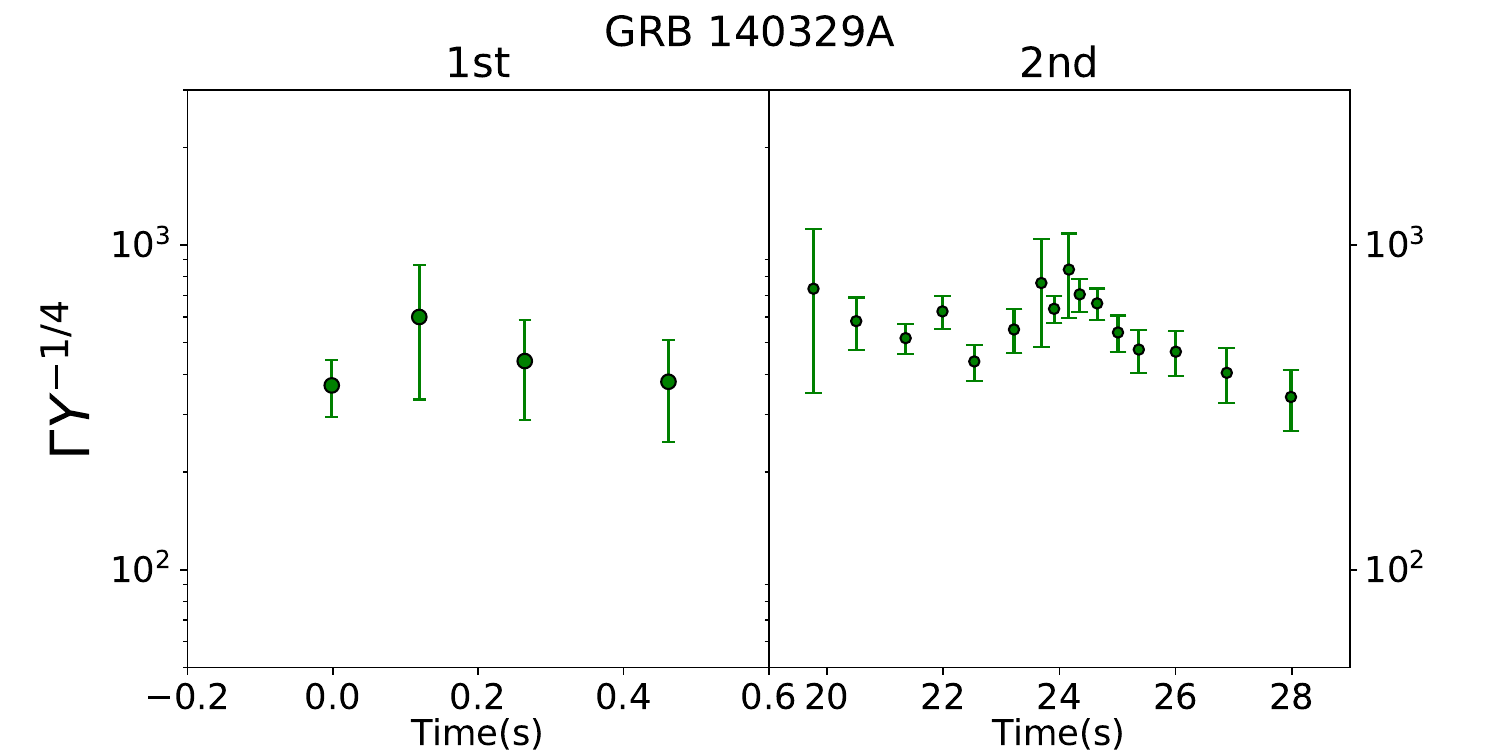}
\includegraphics [width=8.5cm,height=4.5cm]{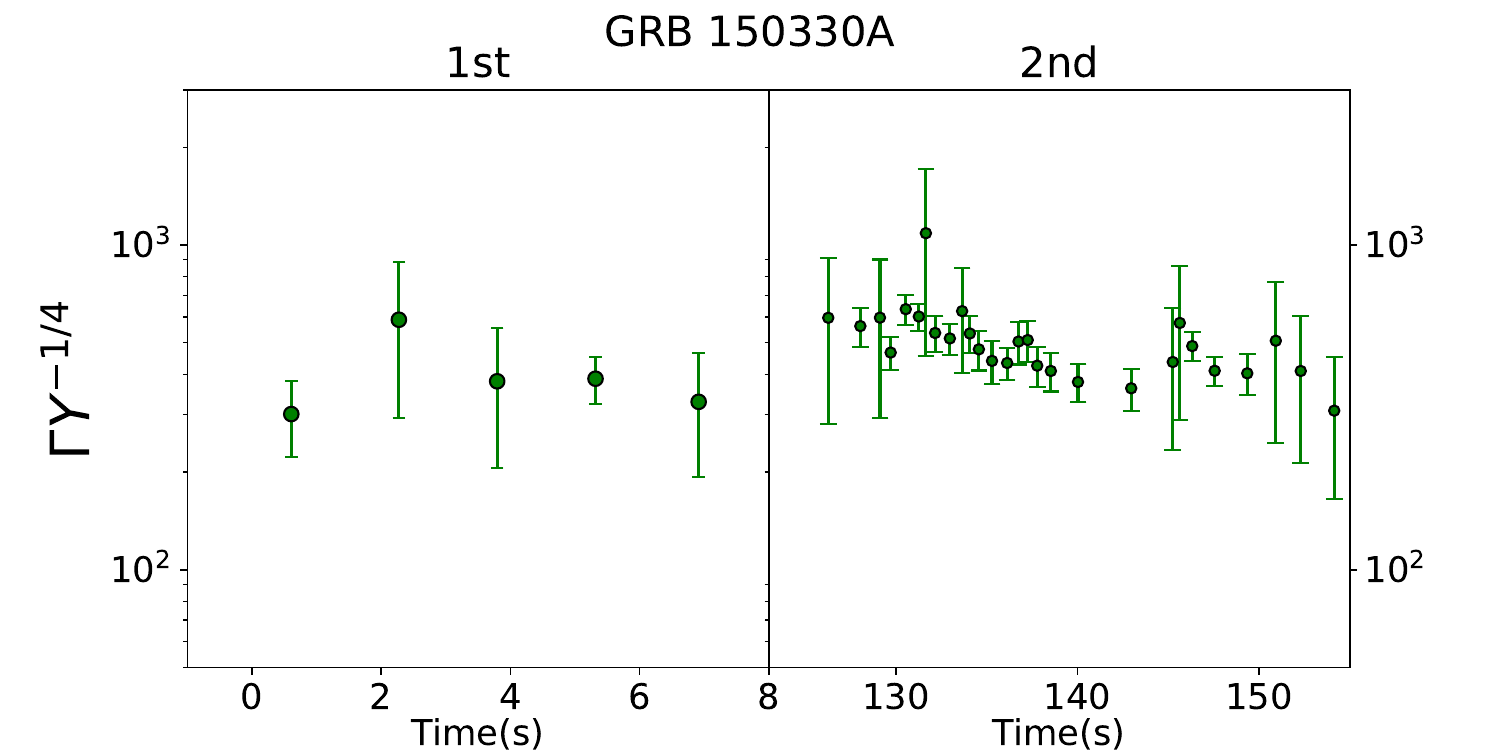}
\includegraphics [width=8.5cm,height=4.5cm]{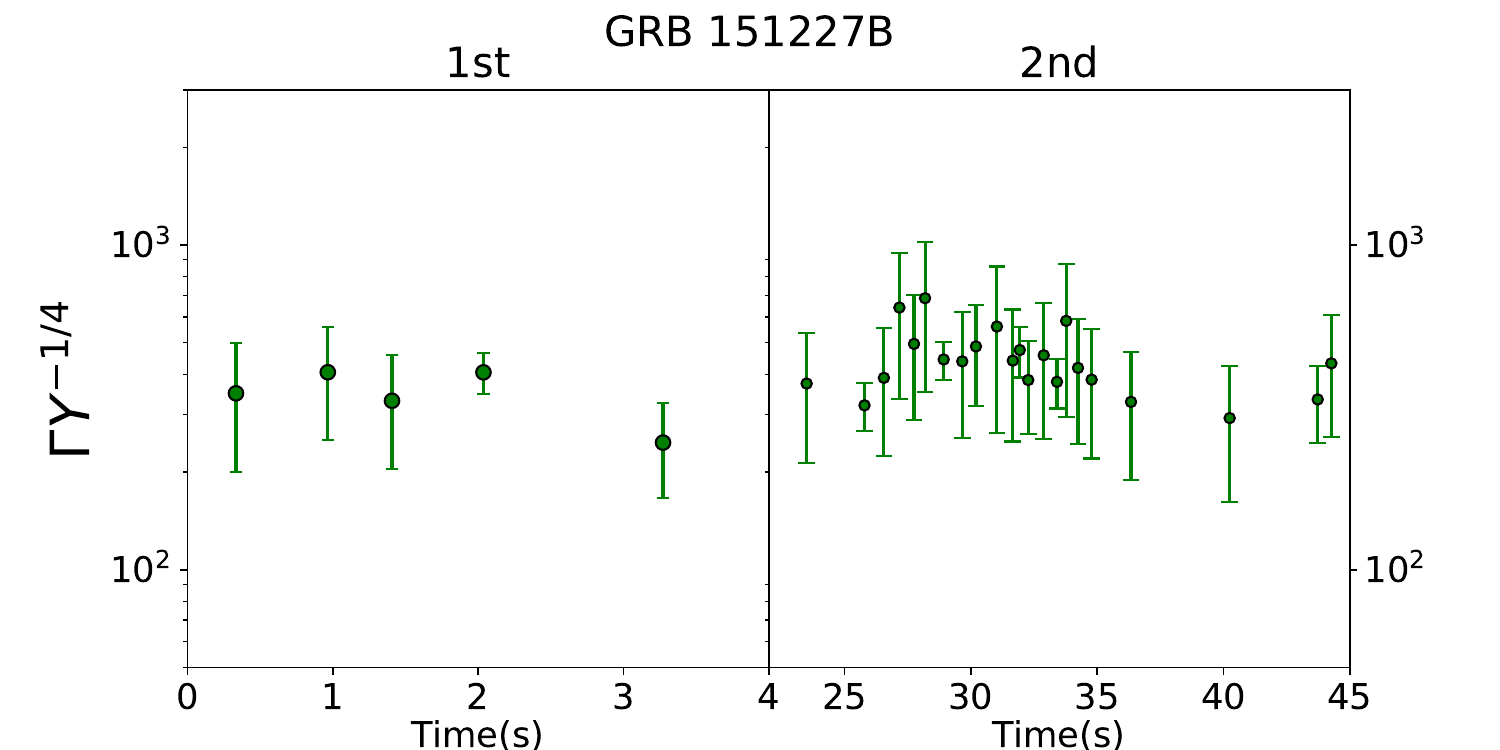}
\includegraphics [width=8.5cm,height=4.5cm]{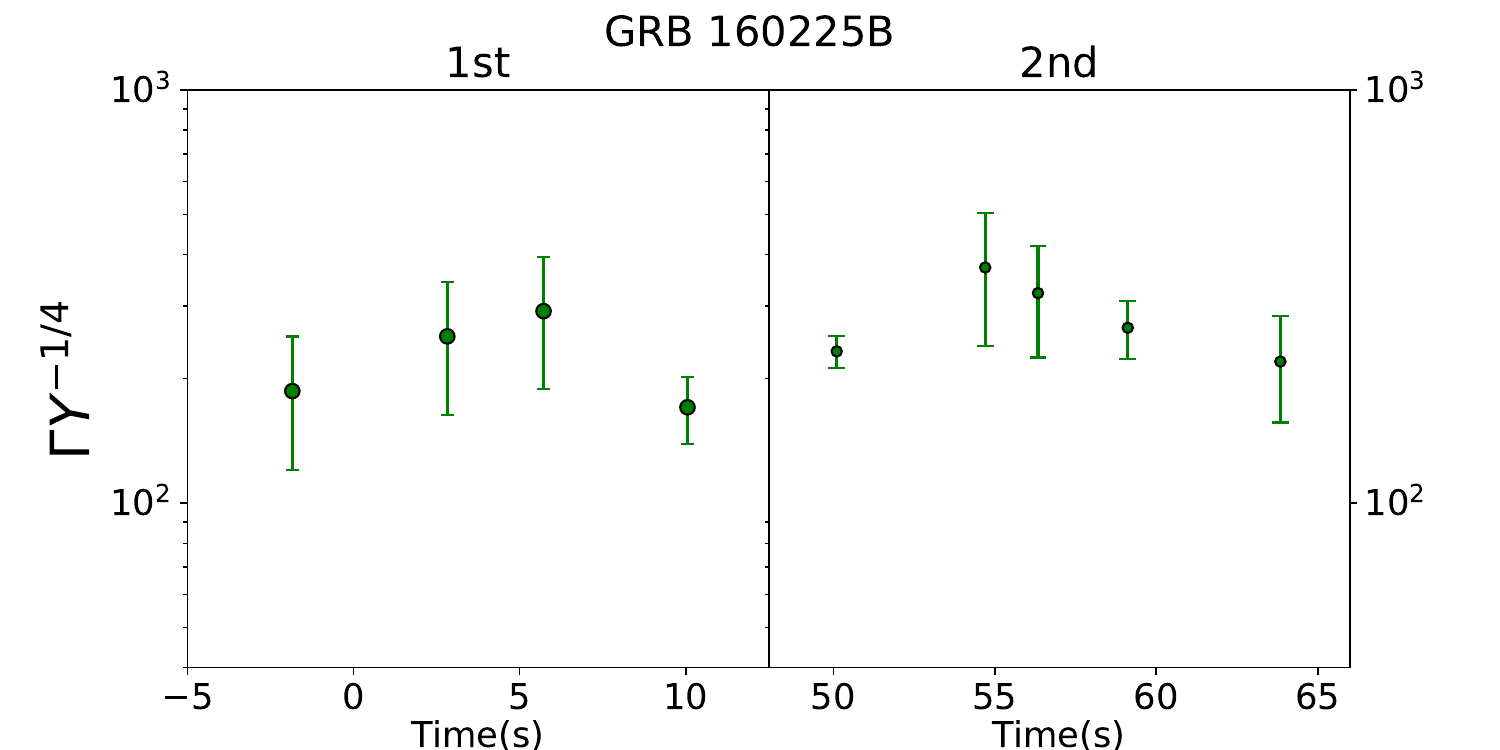}
\includegraphics [width=8.5cm,height=4.5cm]{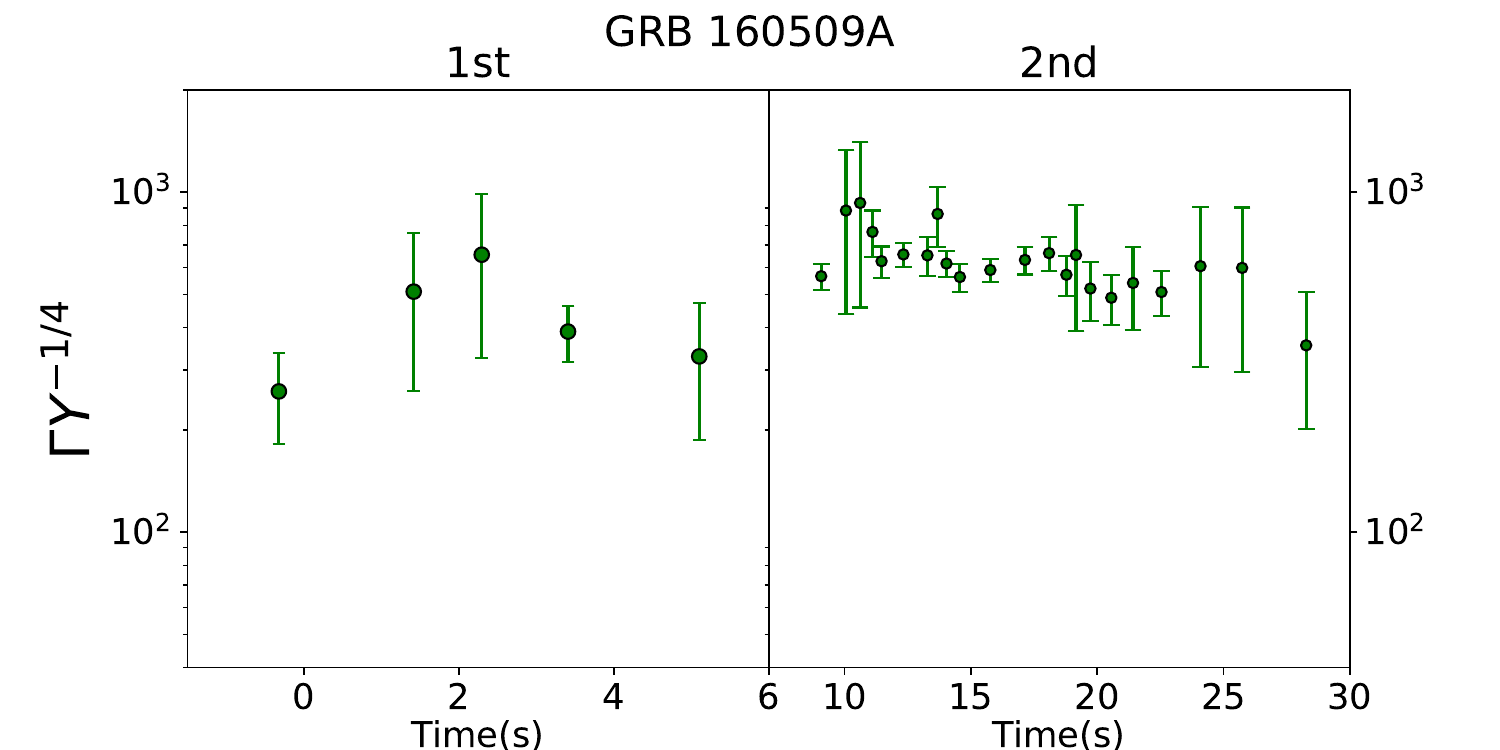}
\includegraphics [width=8.5cm,height=4.5cm]{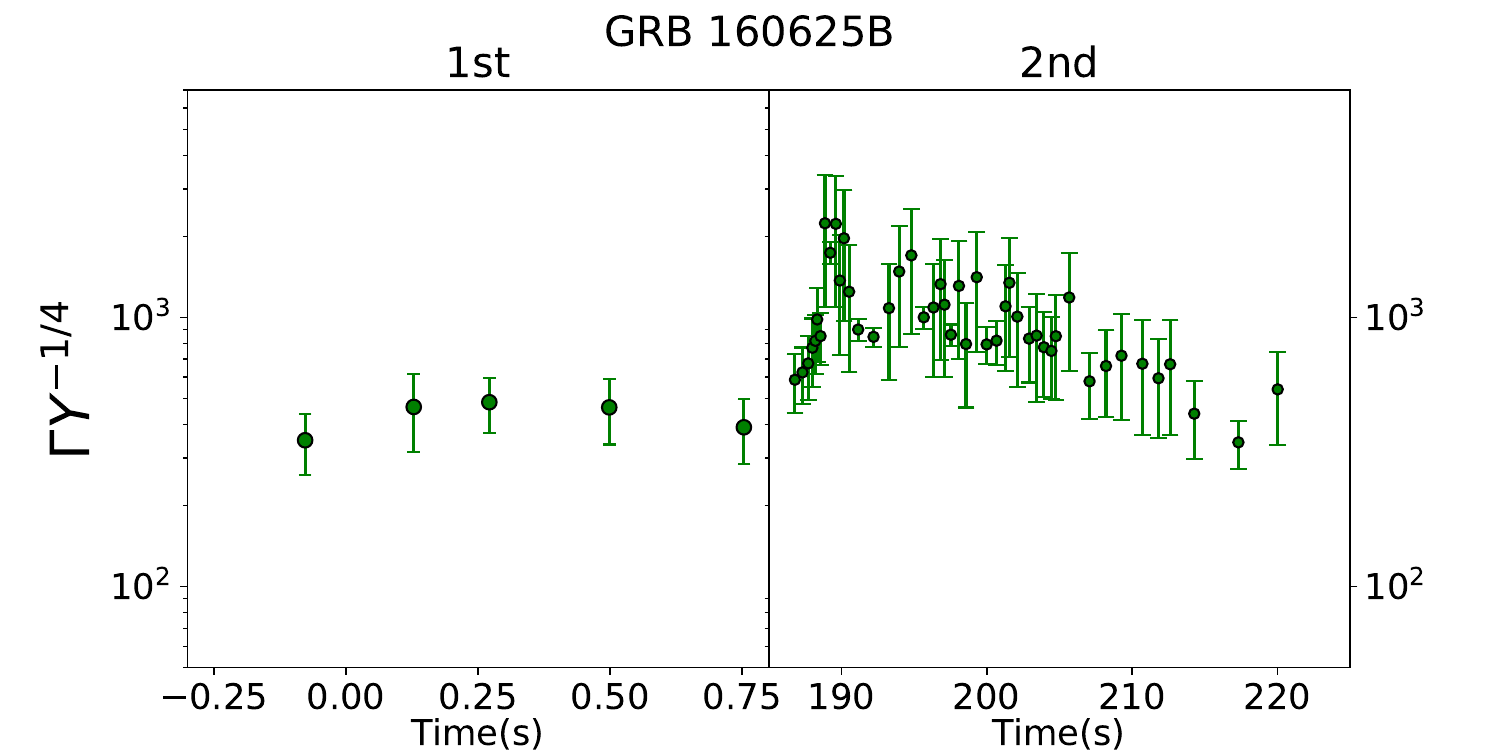}
\includegraphics [width=8.5cm,height=4.5cm]{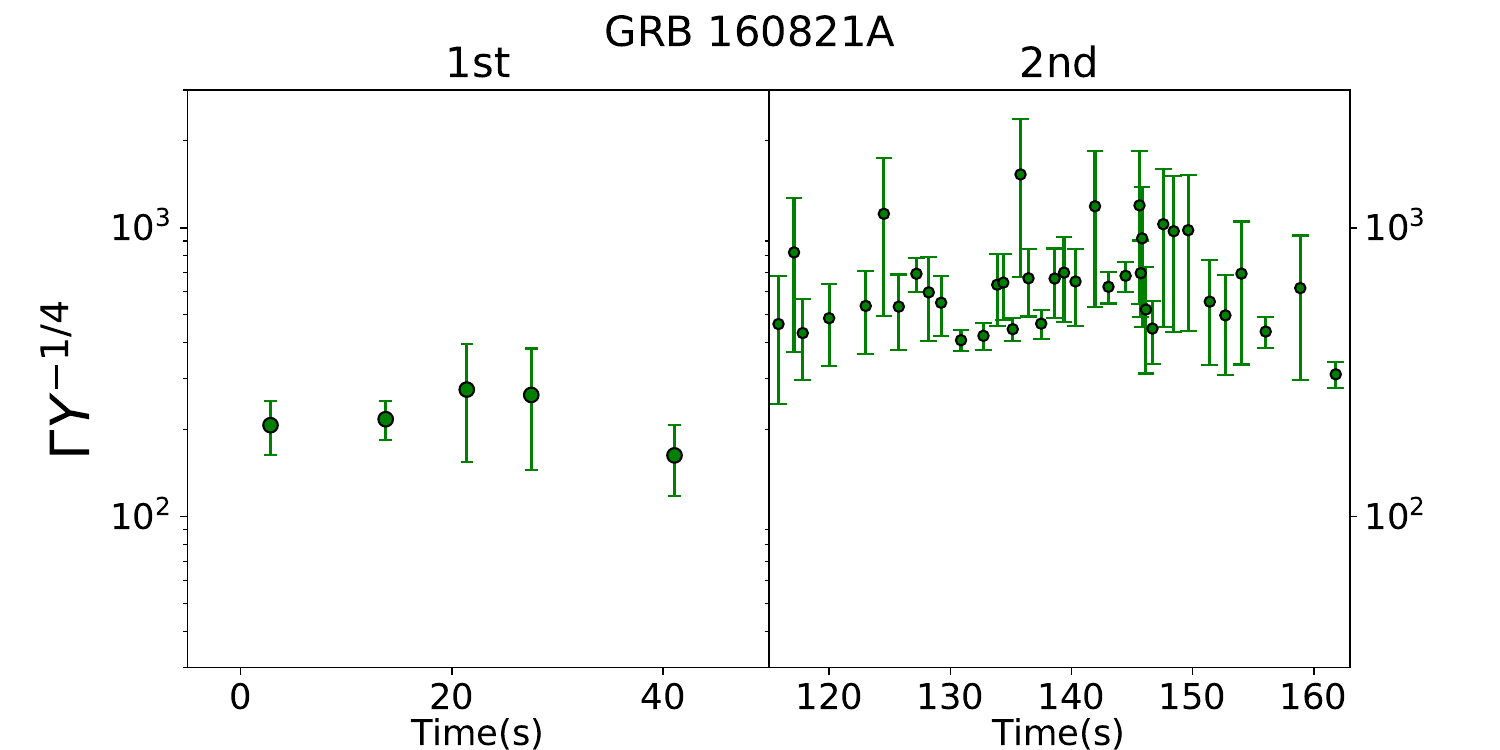}
\end{figure}
\begin{figure}[htbp]
\centering
\includegraphics [width=8.5cm,height=4.5cm]{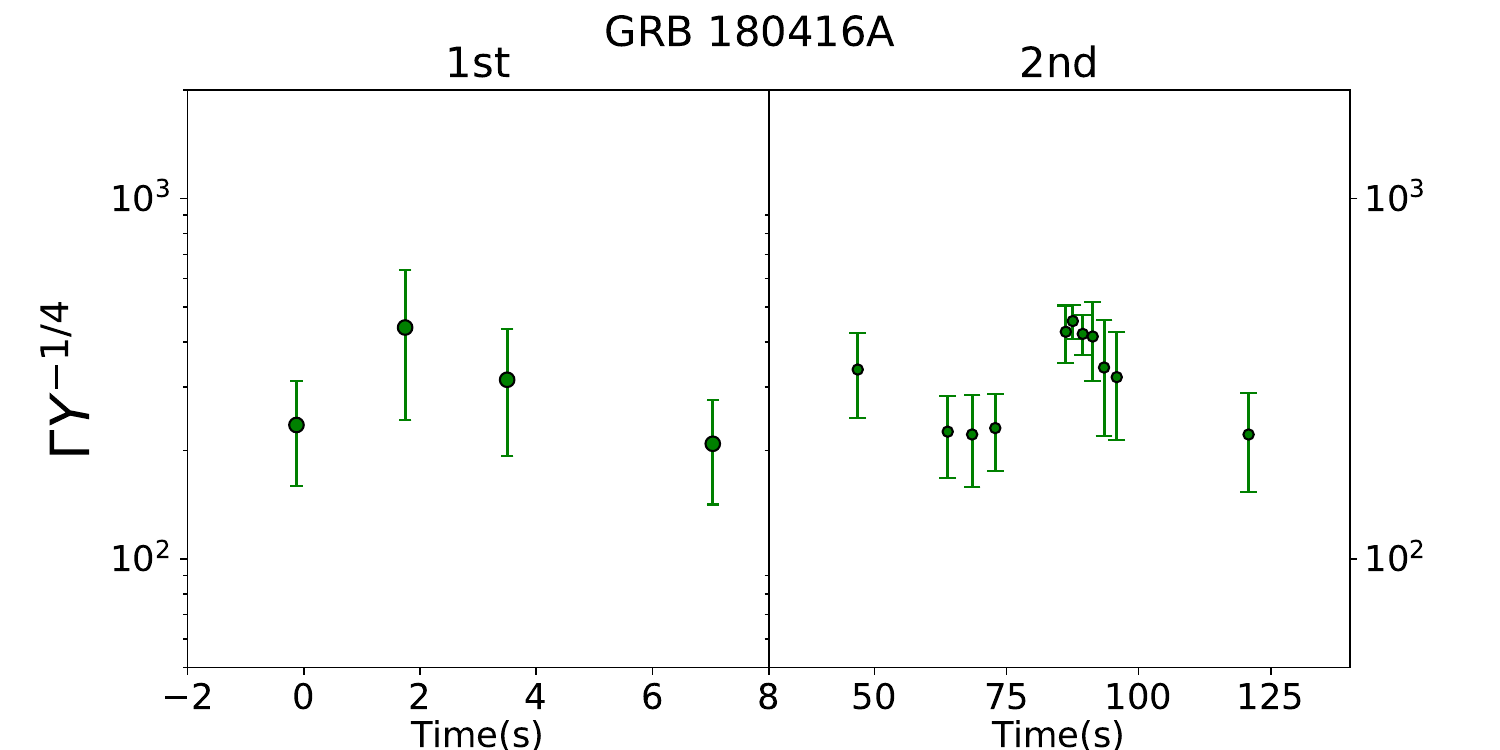}
\includegraphics [width=8.5cm,height=4.5cm]{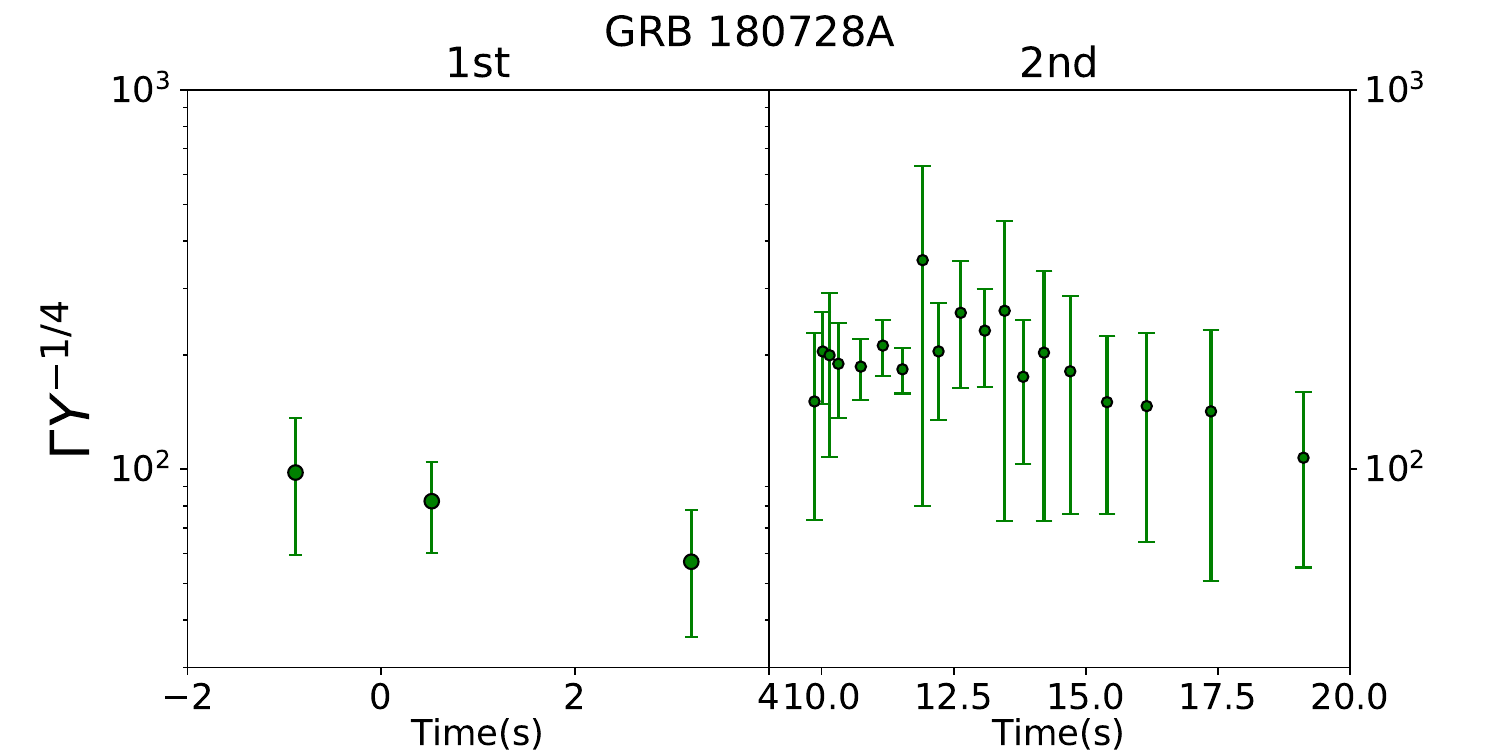}
\includegraphics [width=8.5cm,height=4.5cm]{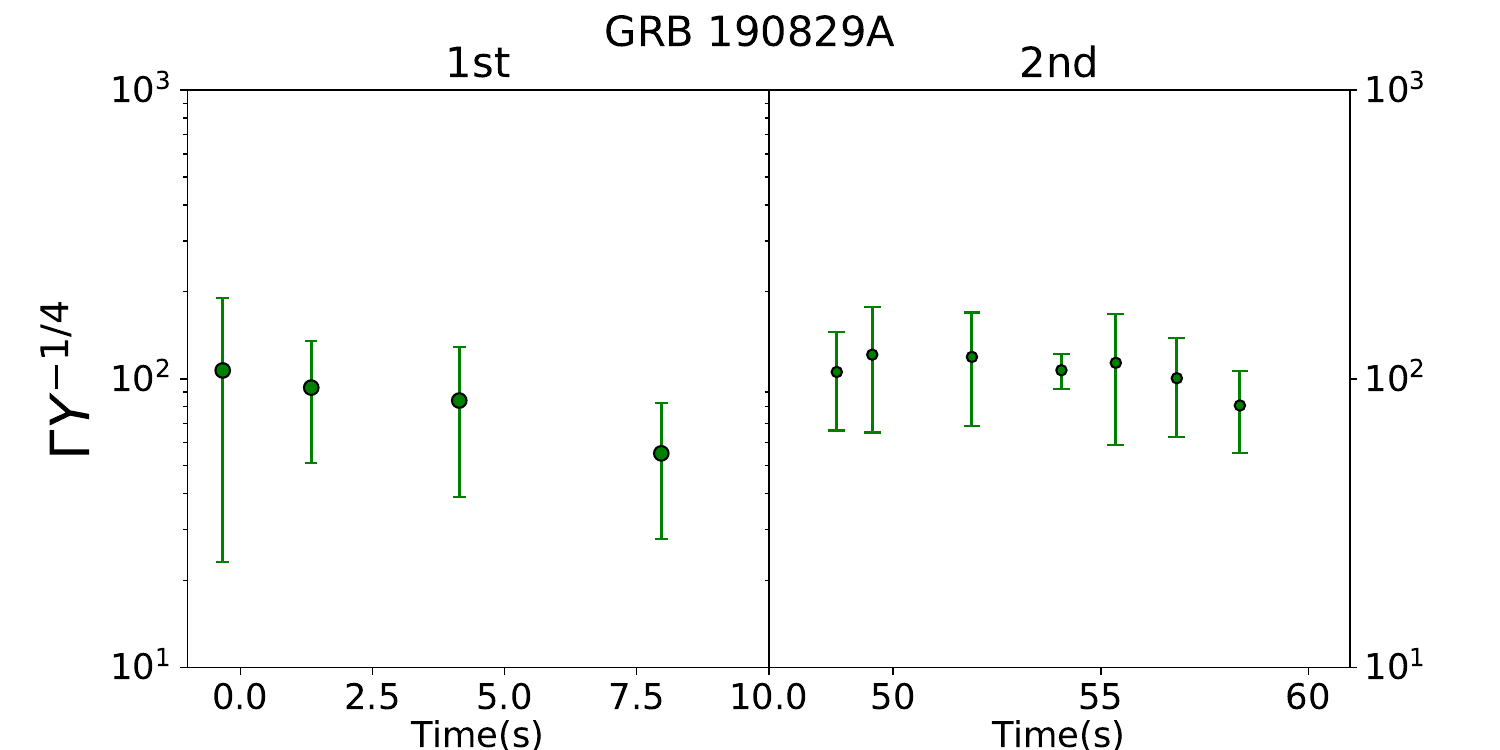}
\includegraphics [width=8.5cm,height=4.5cm]{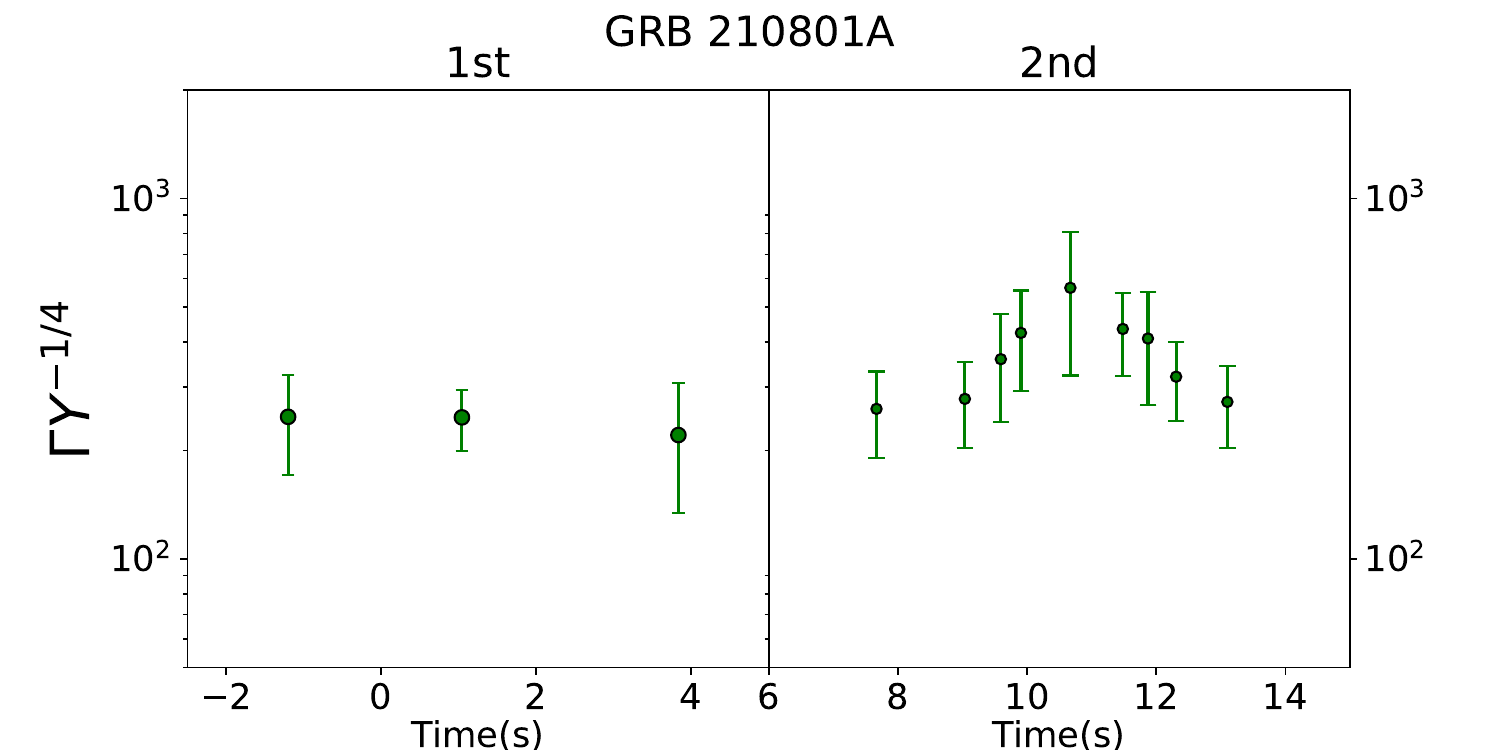}
\includegraphics [width=8.5cm,height=4.5cm]{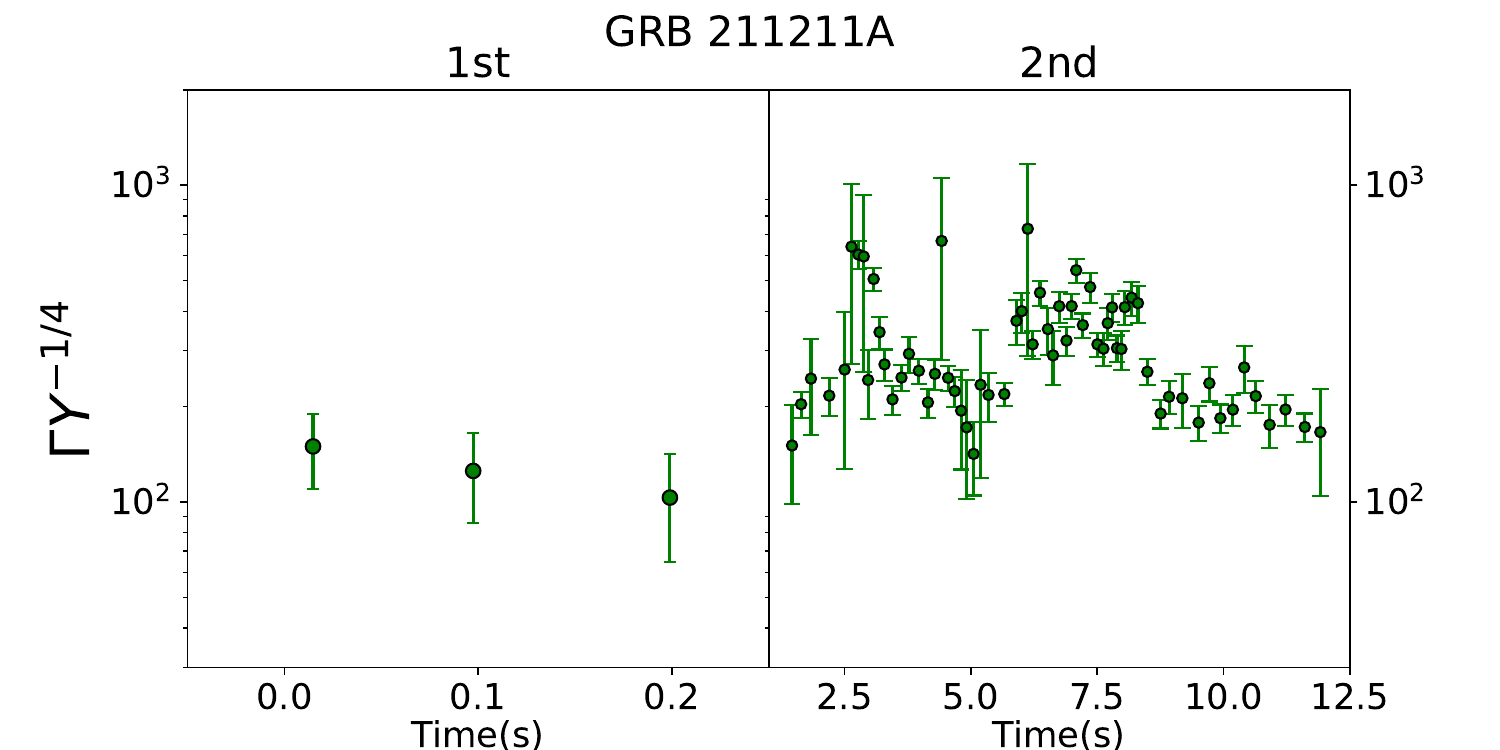}
\includegraphics [width=8.5cm,height=4.5cm]{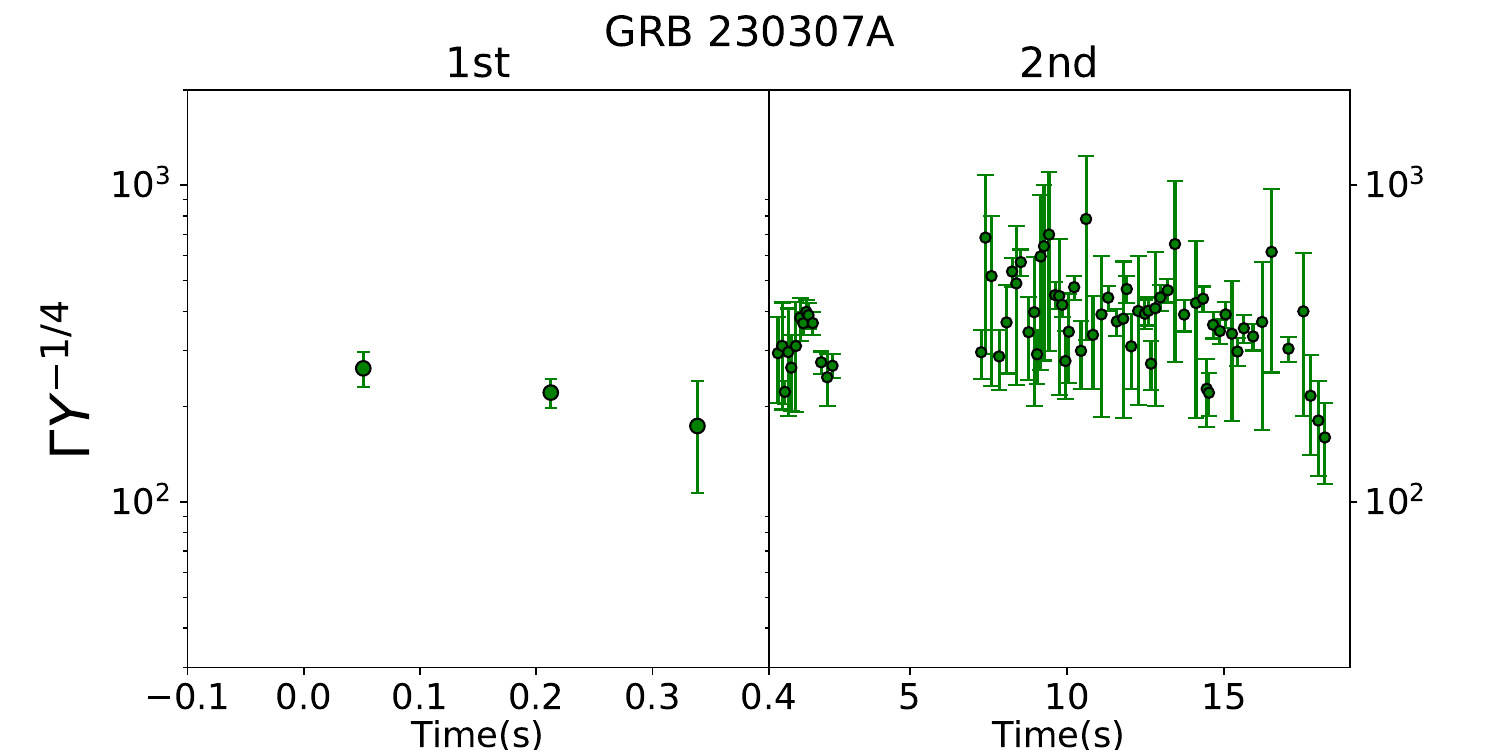}
\includegraphics [width=8.5cm,height=4.5cm]{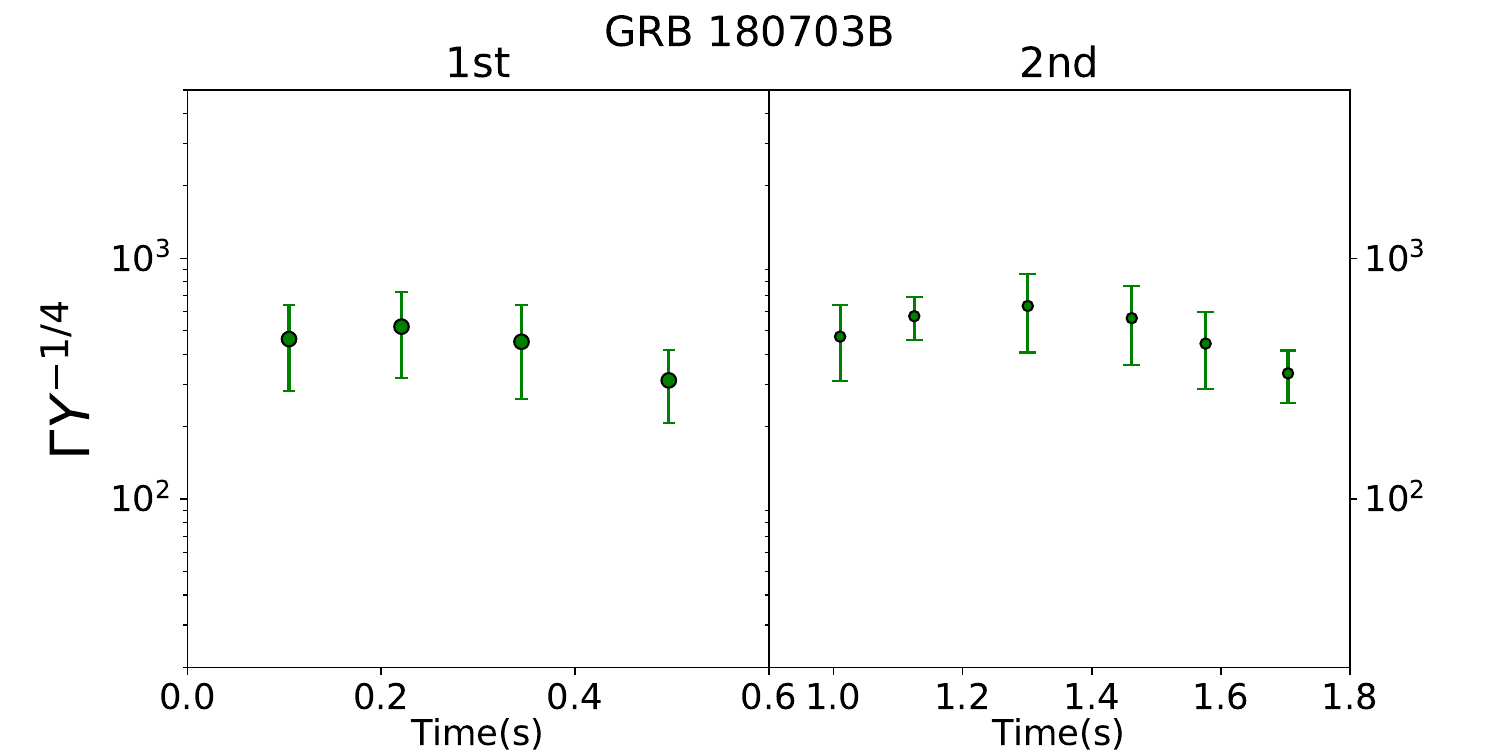}
\includegraphics [width=8.5cm,height=4.5cm]{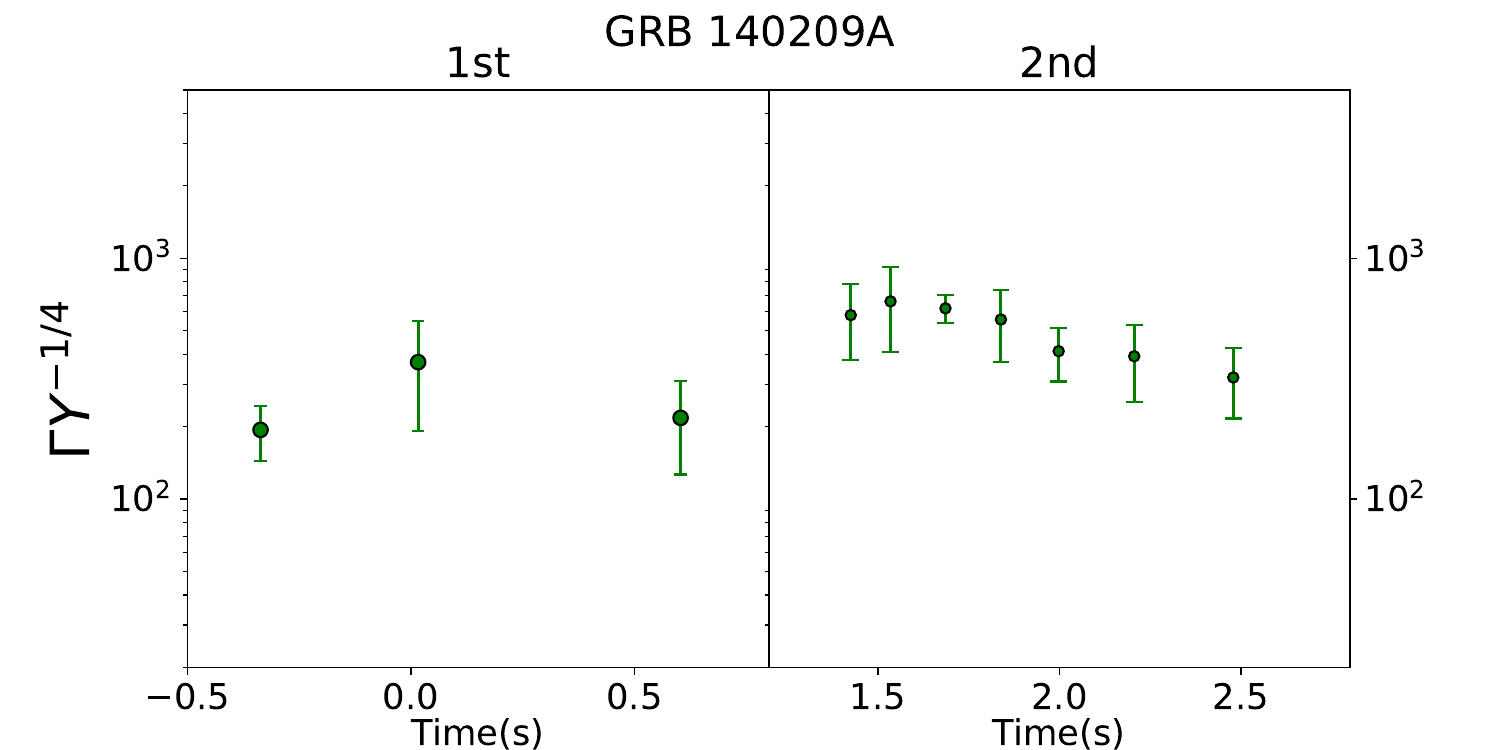}
   \figcaption{The time evolution of  $\Gamma$. ``1st'' denotes precursors, and ``2nd'' denotes main bursts. \label{fig D2}}
\end{figure}

\begin{figure}[htbp]
\centering
\includegraphics [width=8.5cm,height=4.5cm]{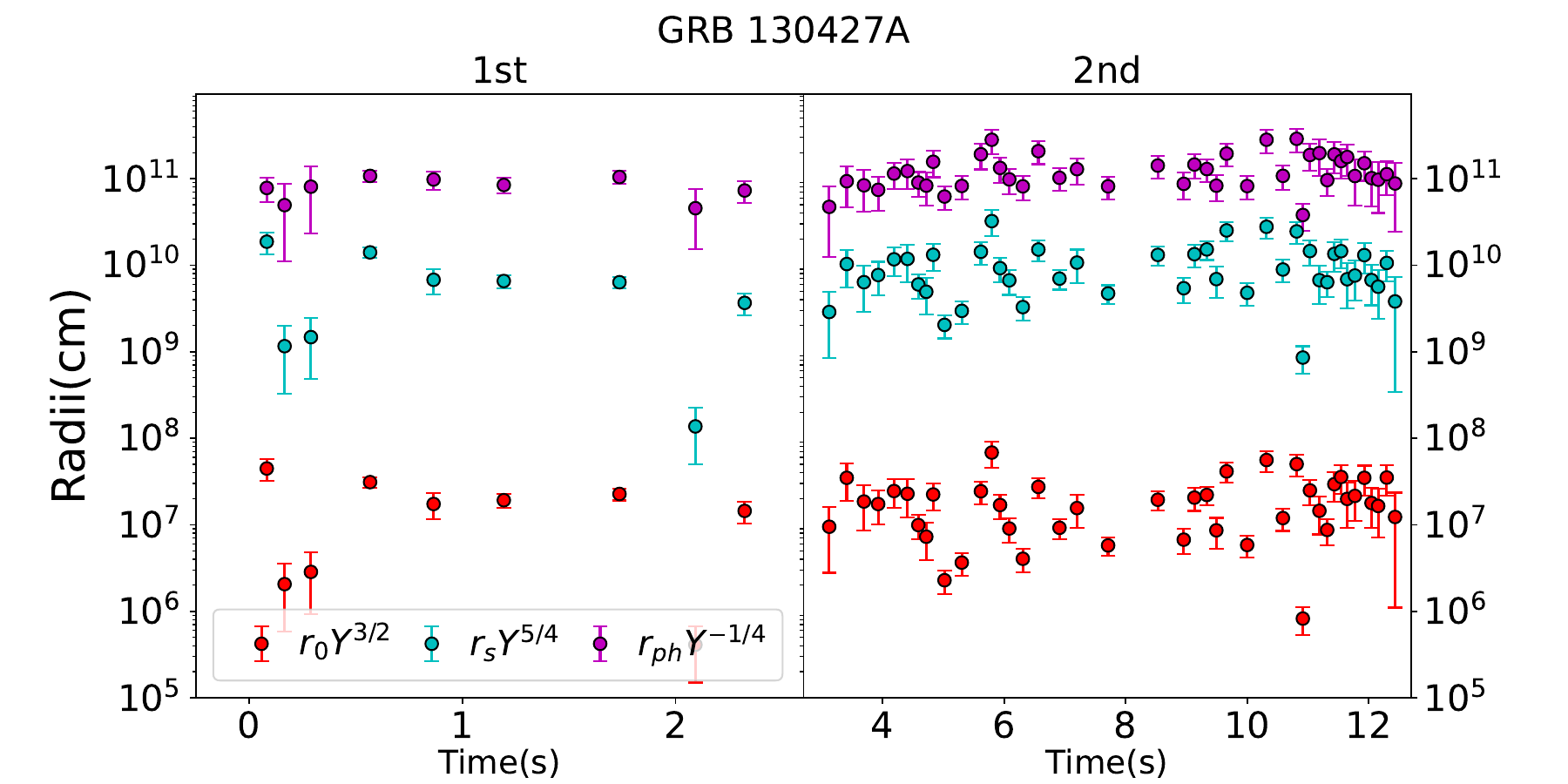}
\includegraphics [width=8.5cm,height=4.5cm]{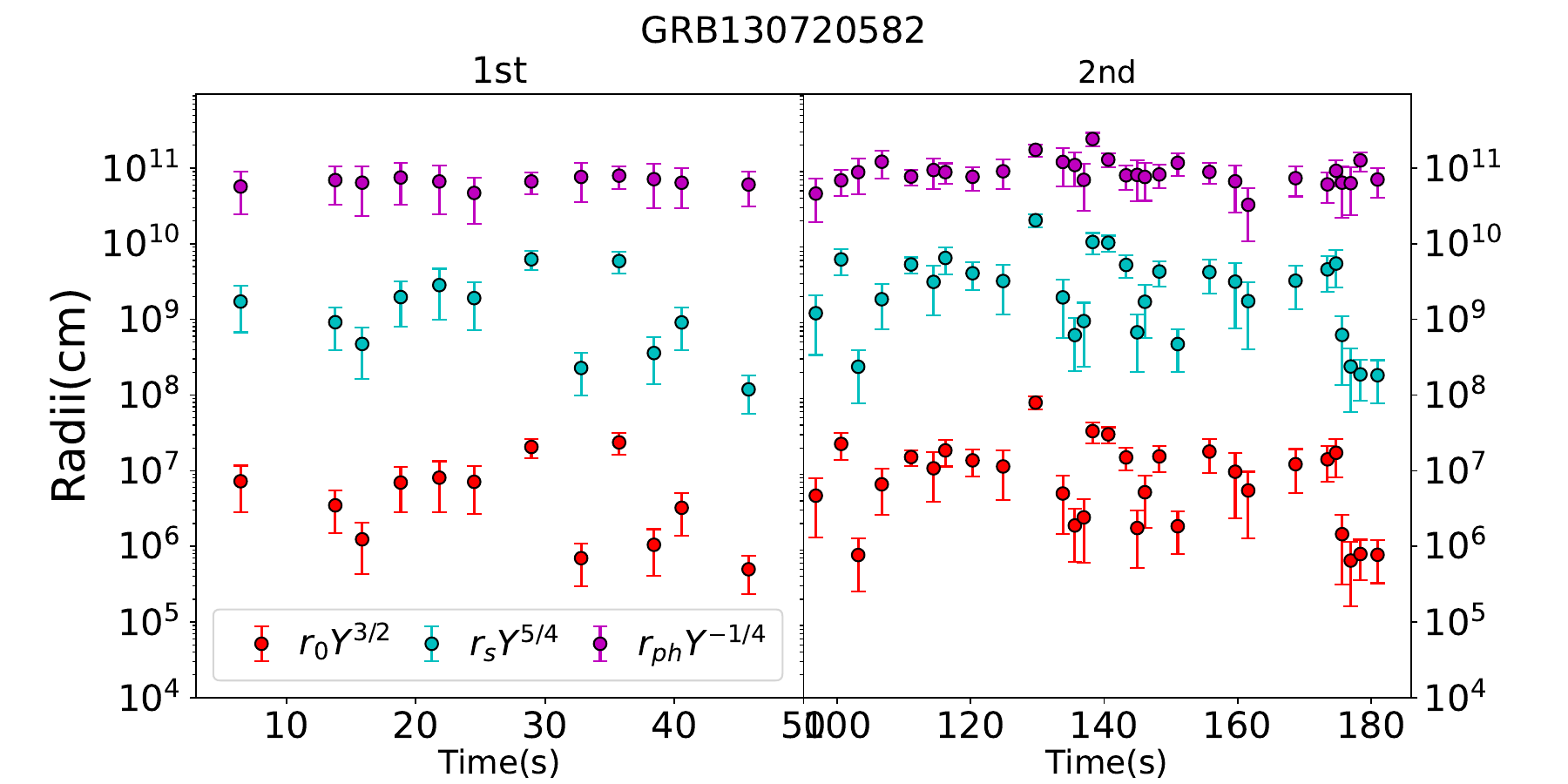}
\includegraphics [width=8.5cm,height=4.5cm]{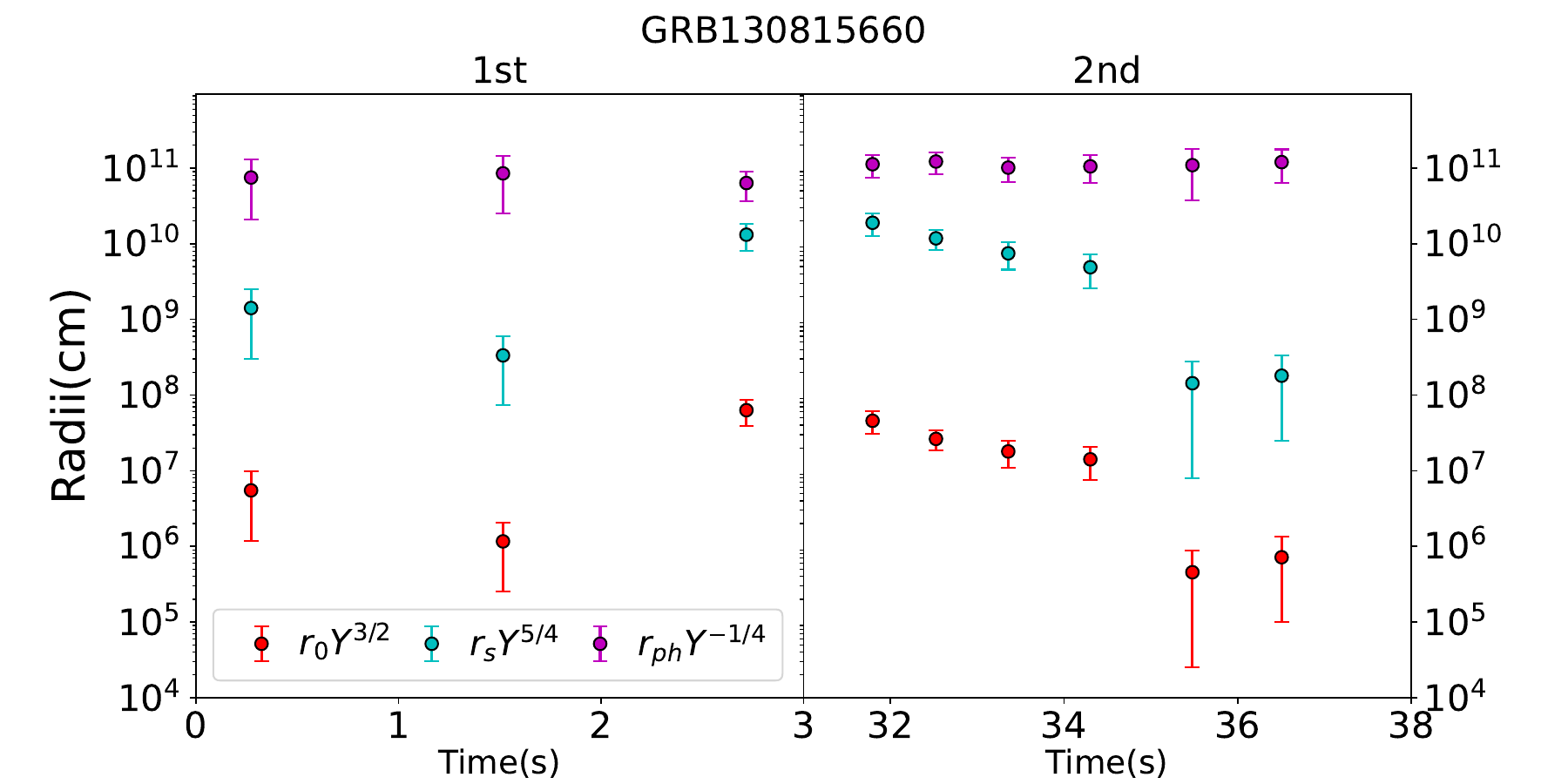}
\includegraphics [width=8.5cm,height=4.5cm]{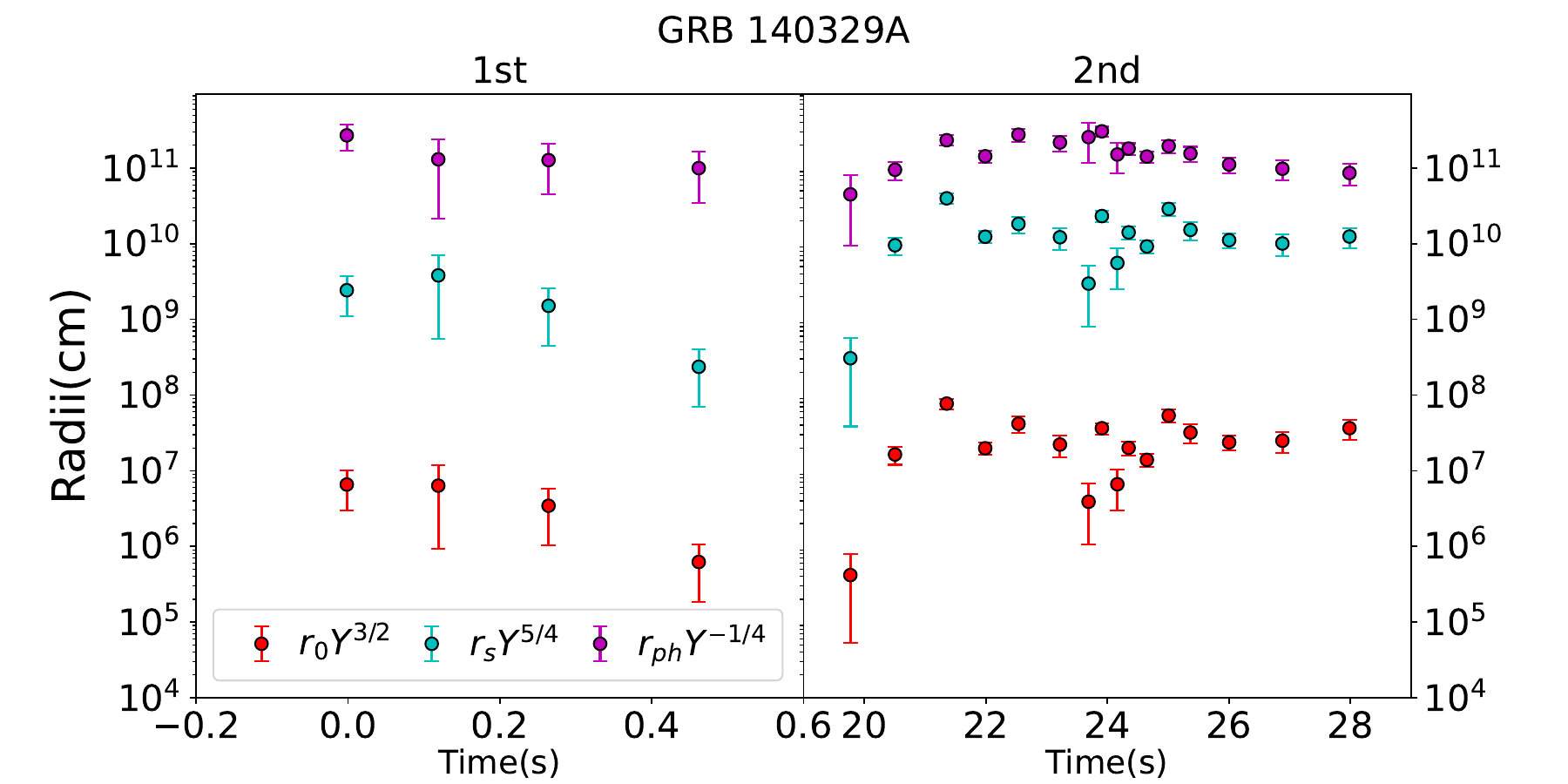}
\includegraphics [width=8.5cm,height=4.5cm]{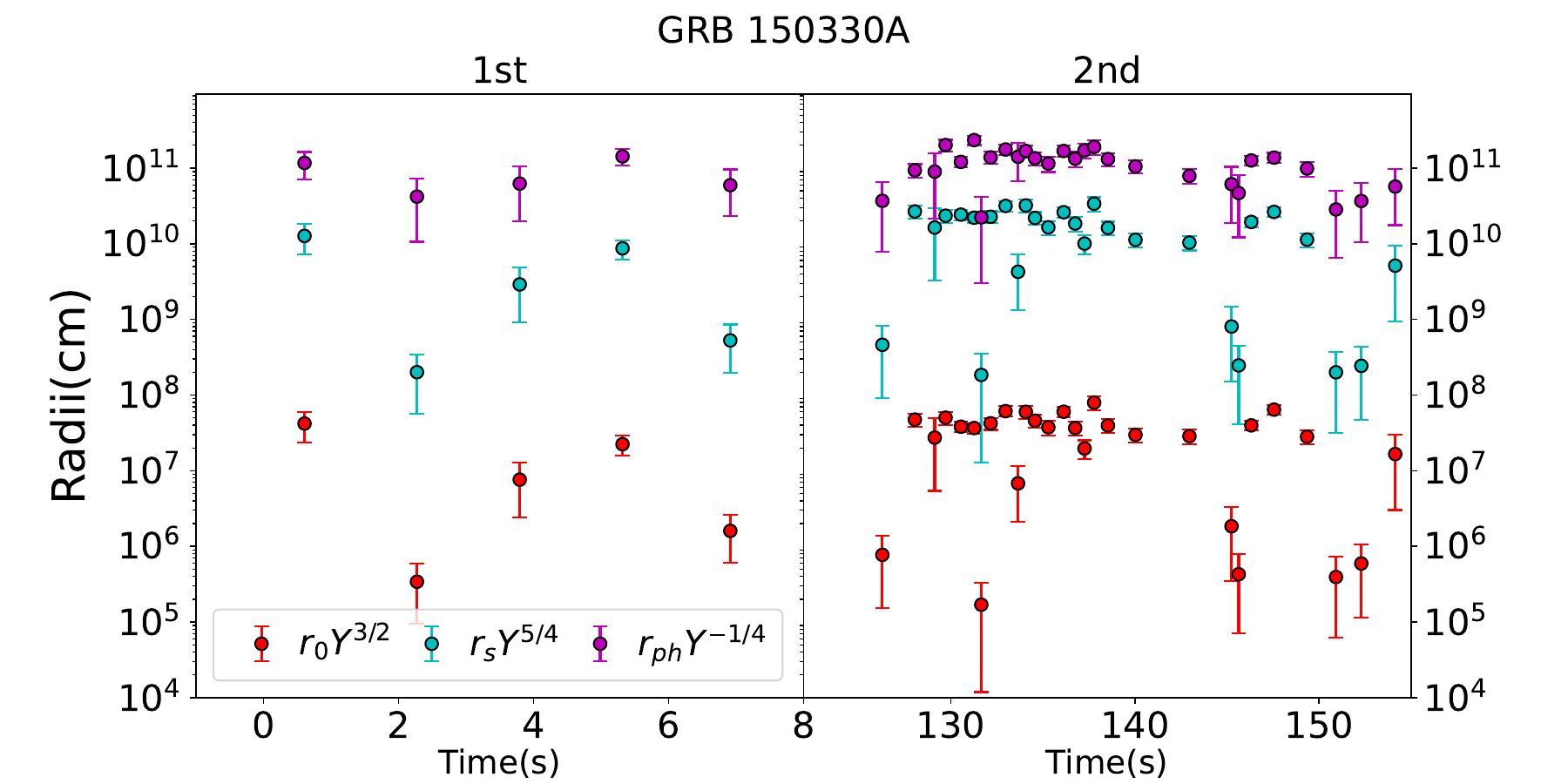}
\includegraphics [width=8.5cm,height=4.5cm]{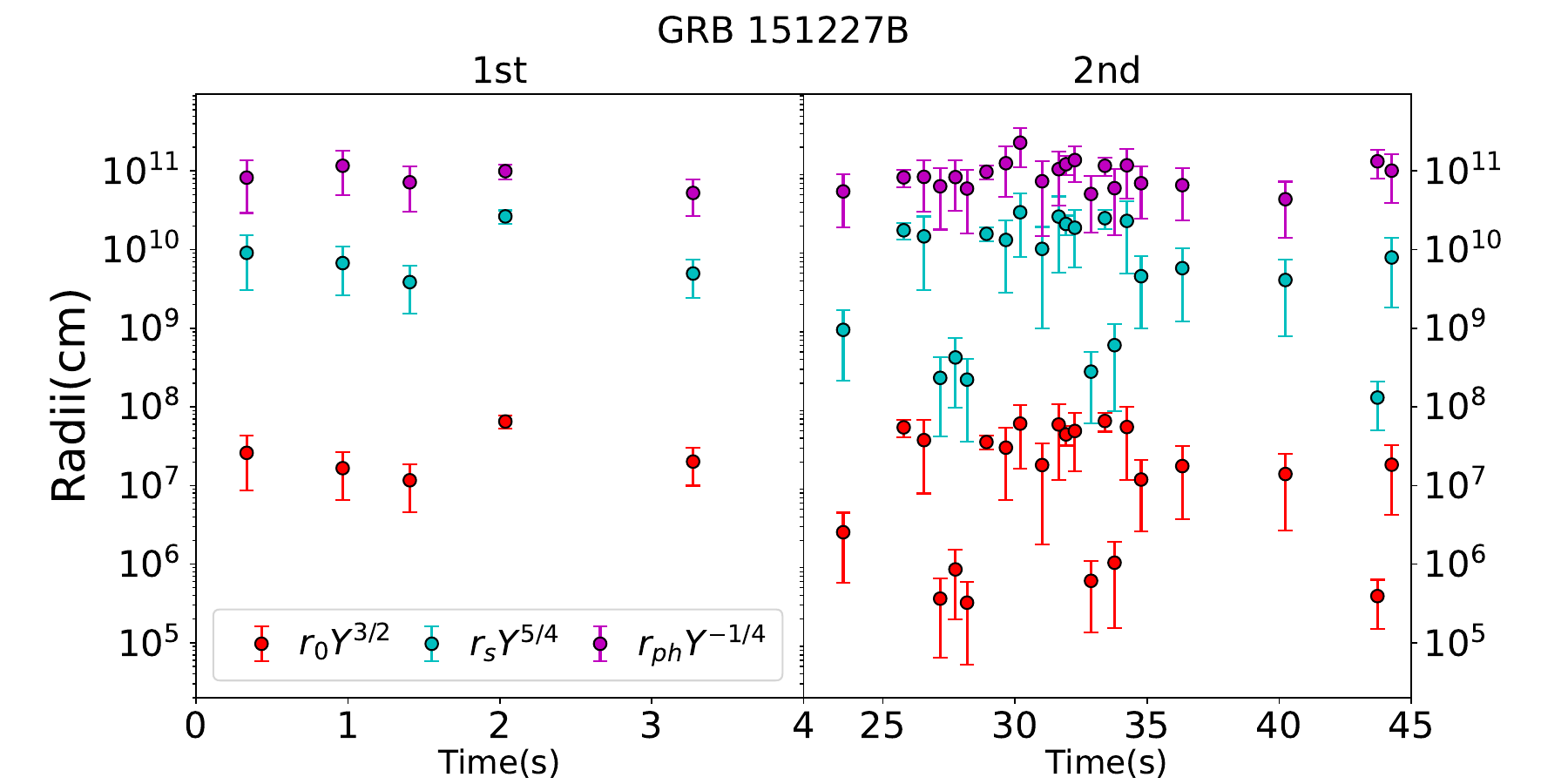}
\includegraphics [width=8.5cm,height=4.5cm]{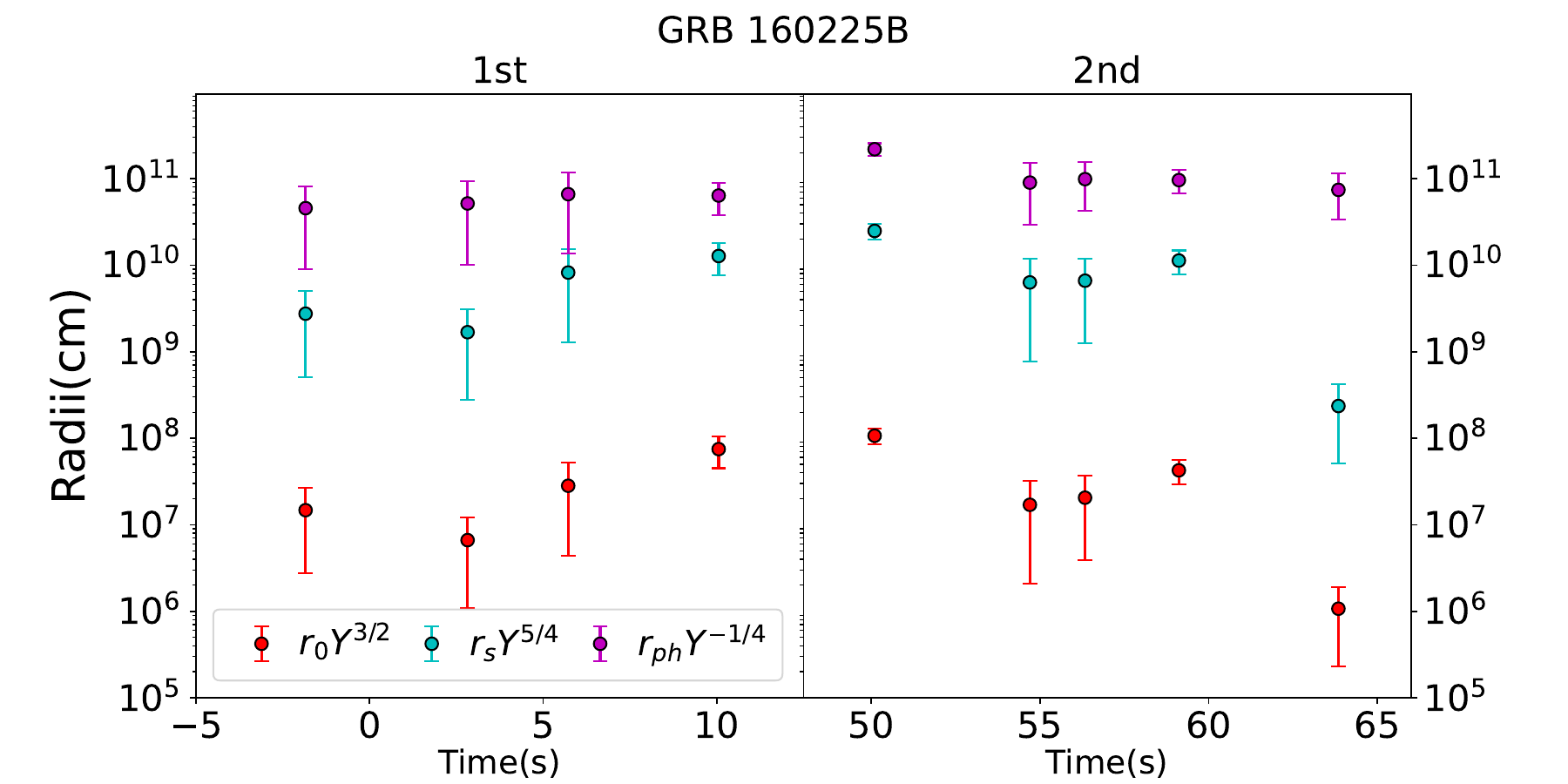}
\includegraphics [width=8.5cm,height=4.5cm]{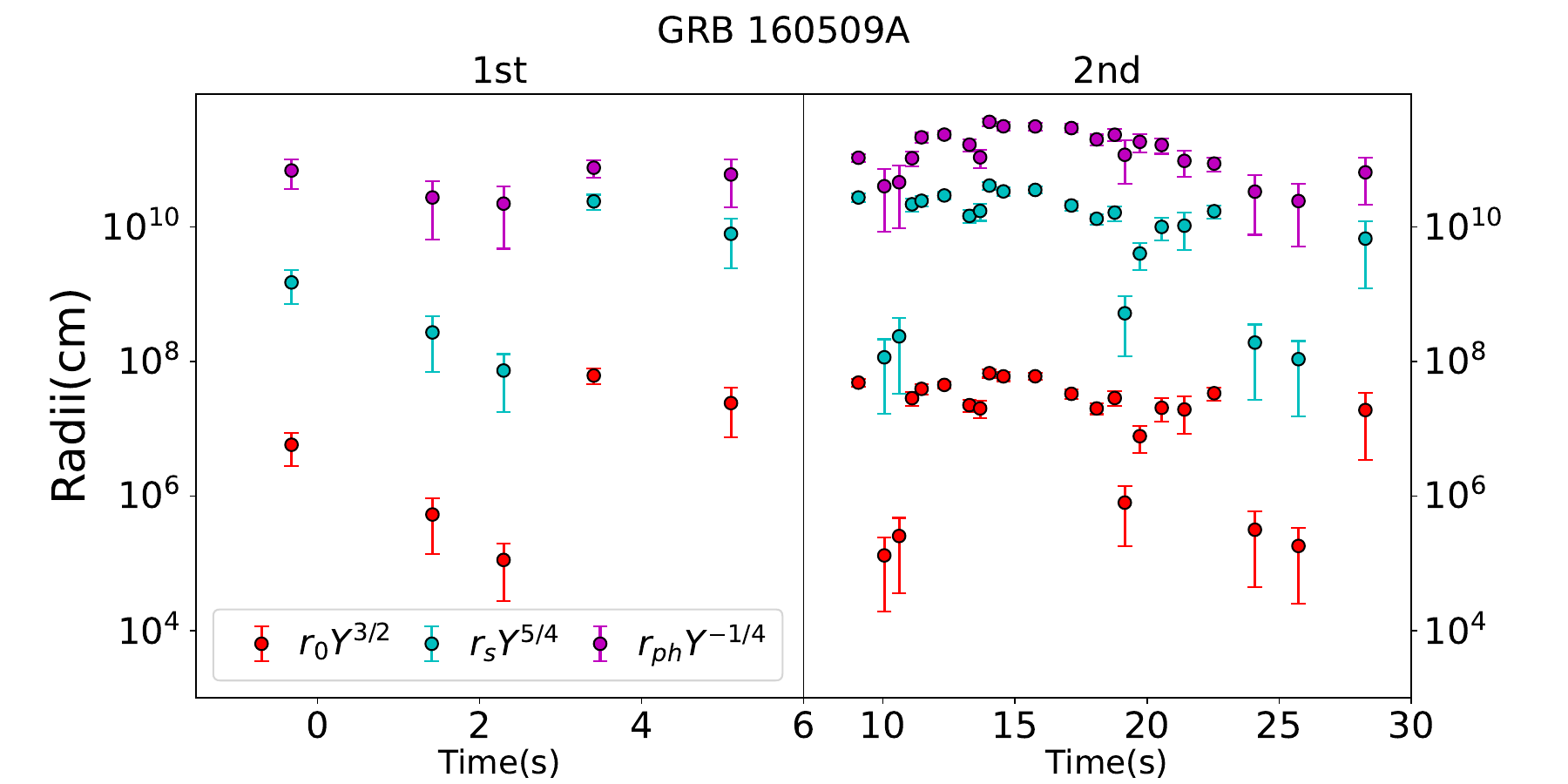}
\includegraphics [width=8.5cm,height=4.5cm]{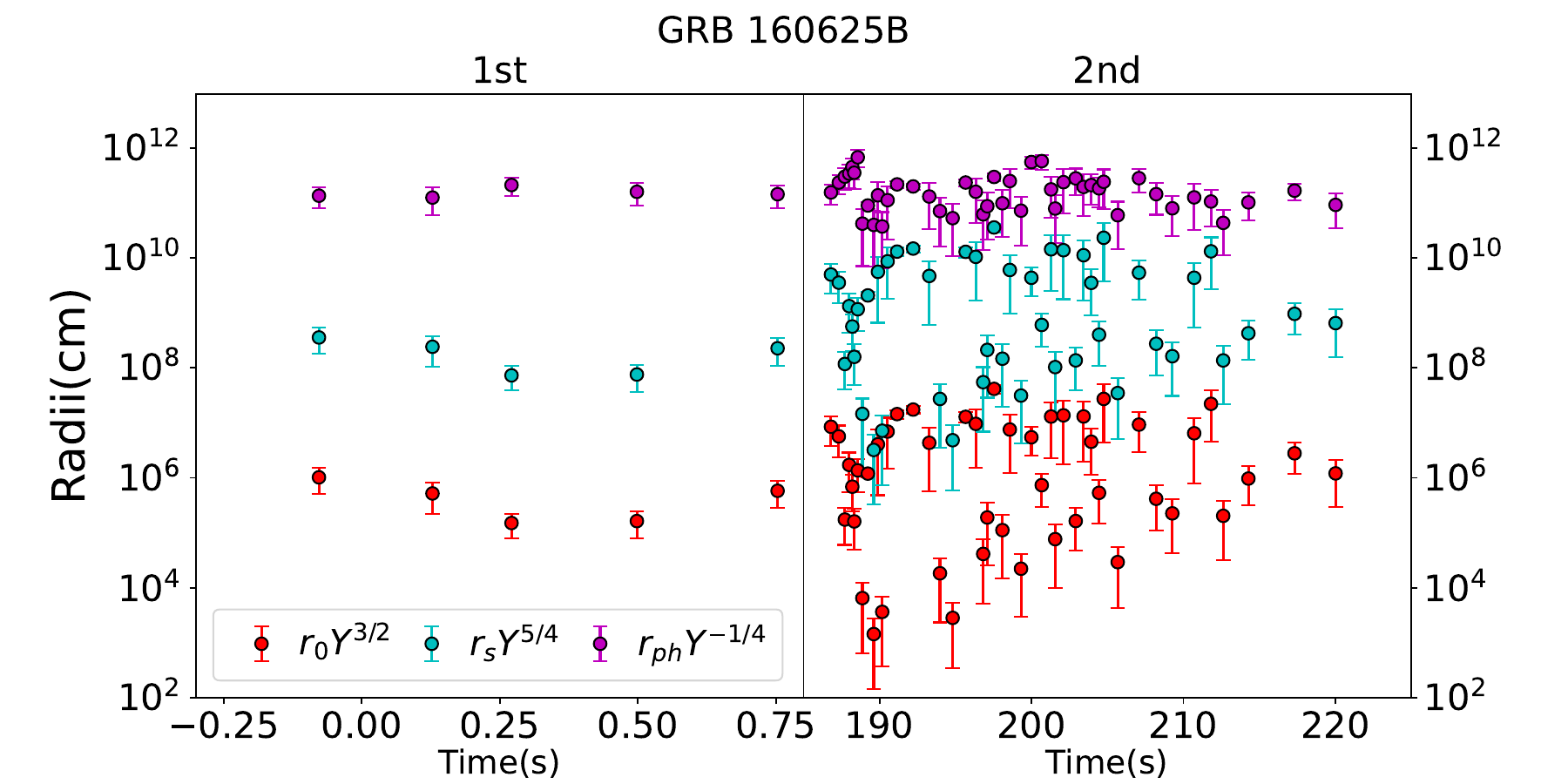}
\includegraphics [width=8.5cm,height=4.5cm]{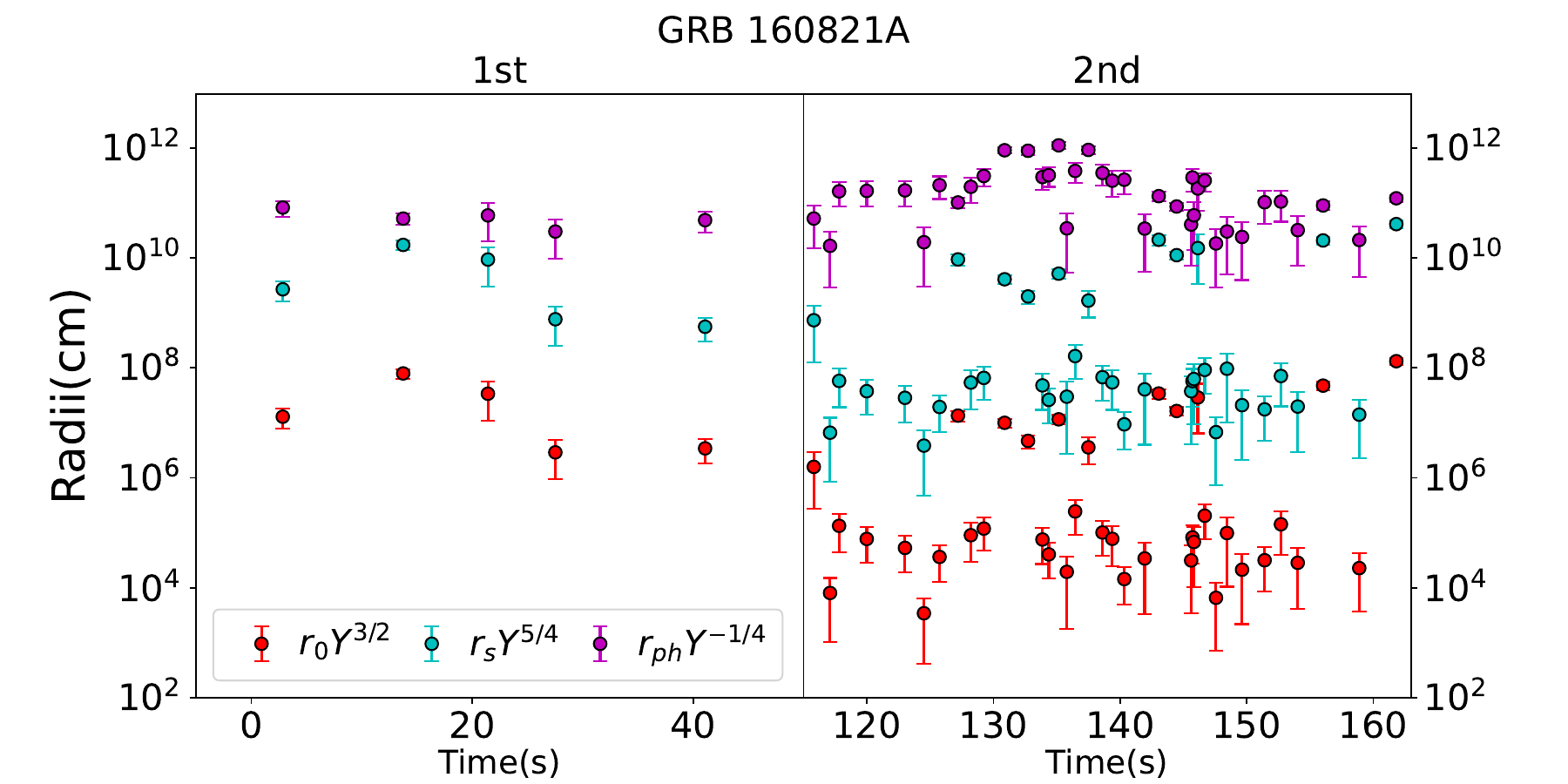}
\end{figure}
\begin{figure}[htbp]
\centering
\includegraphics [width=8.5cm,height=4.5cm]{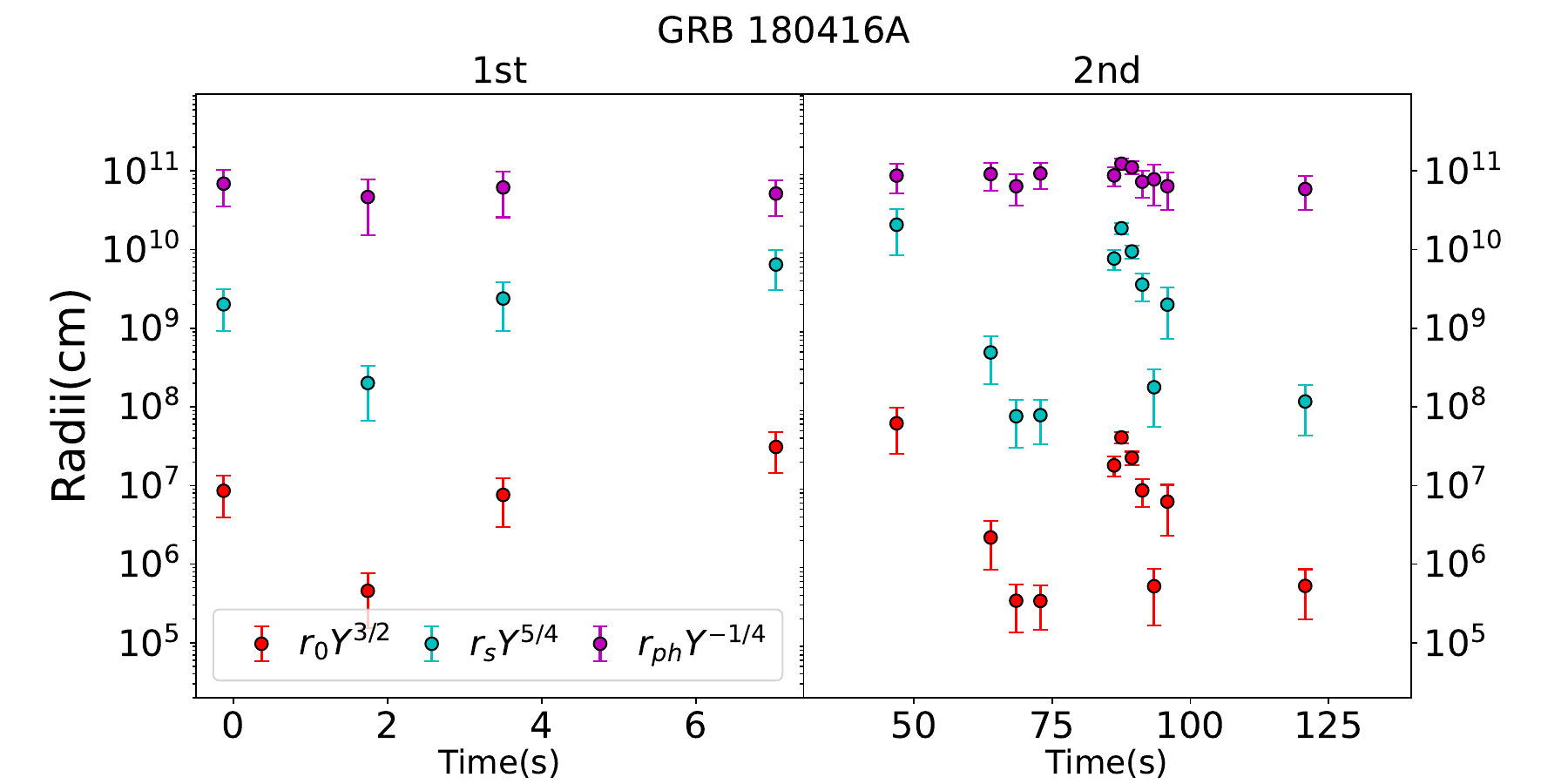}
\includegraphics [width=8.5cm,height=4.5cm]{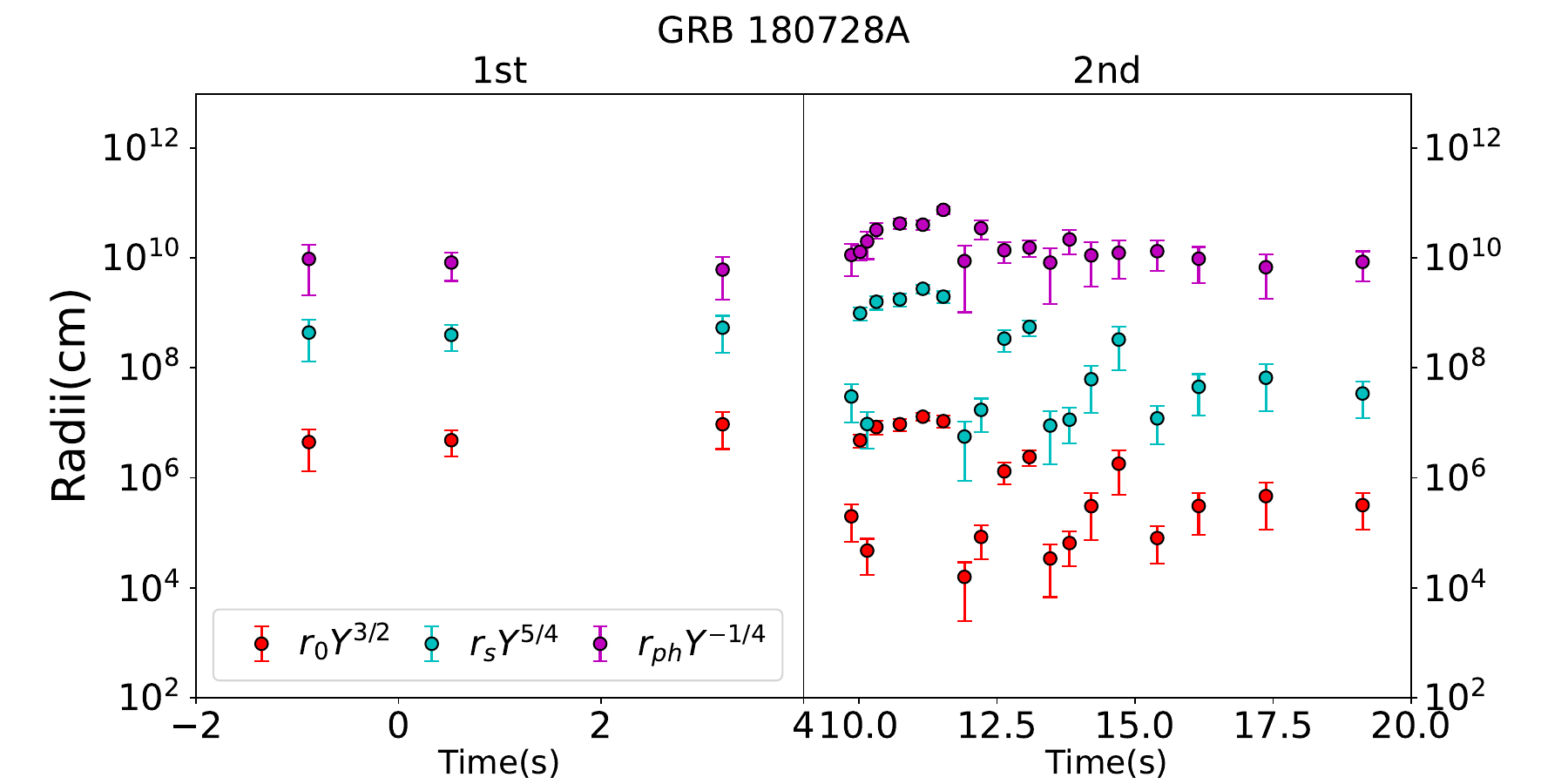}
\includegraphics [width=8.5cm,height=4.5cm]{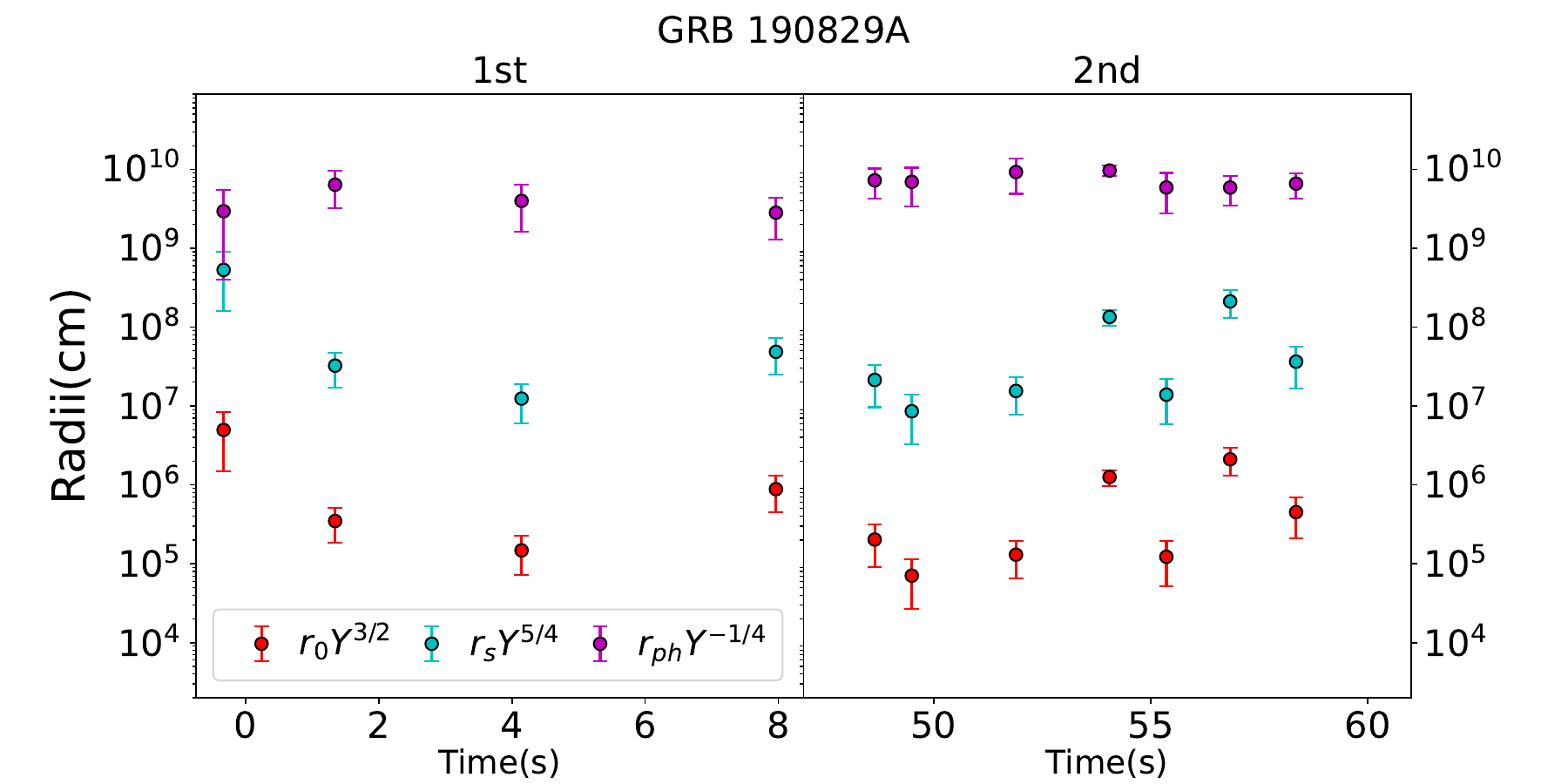}
\includegraphics [width=8.5cm,height=4.5cm]{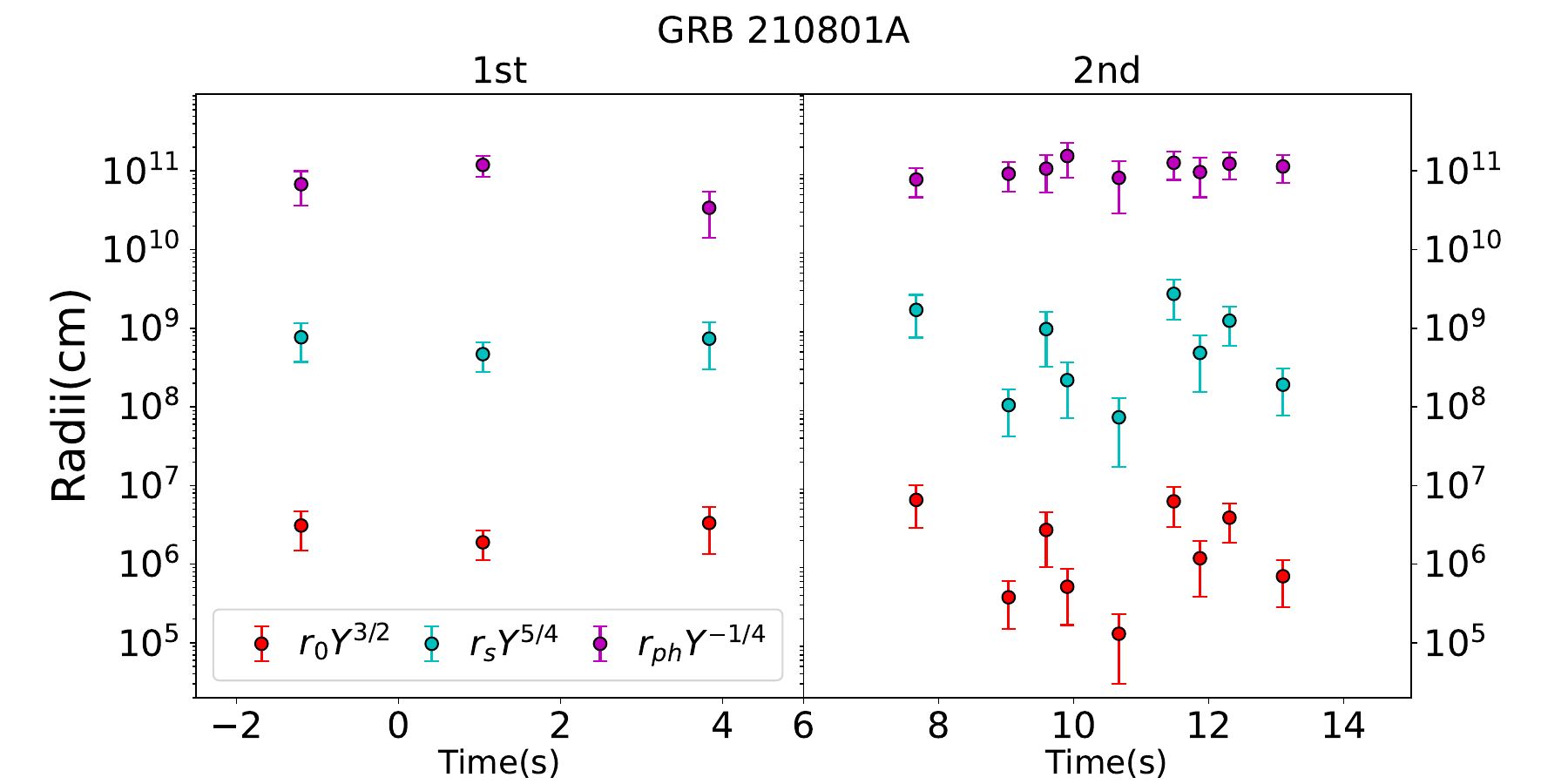}
\includegraphics [width=8.5cm,height=4.5cm]{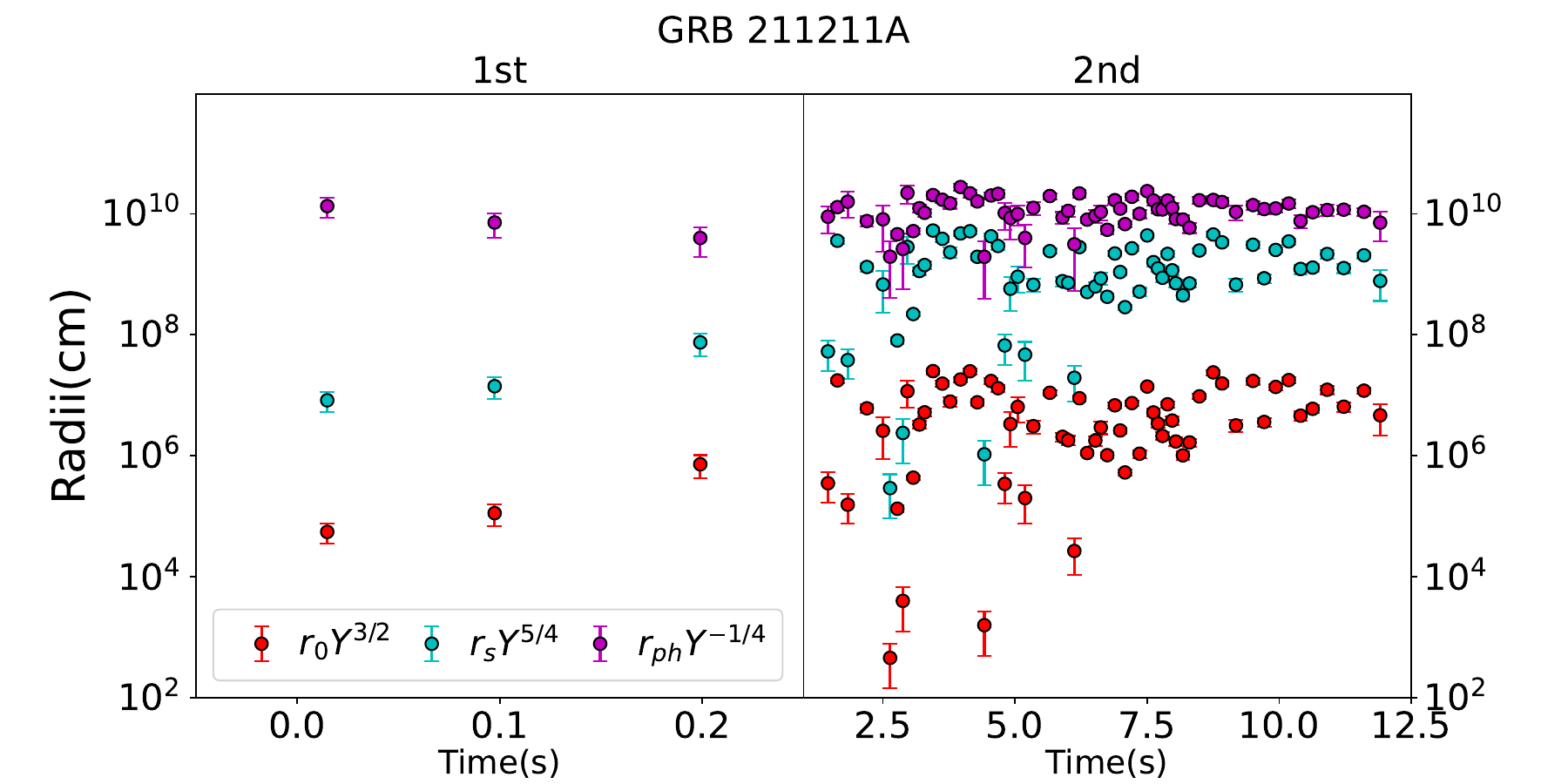}
\includegraphics [width=8.5cm,height=4.5cm]{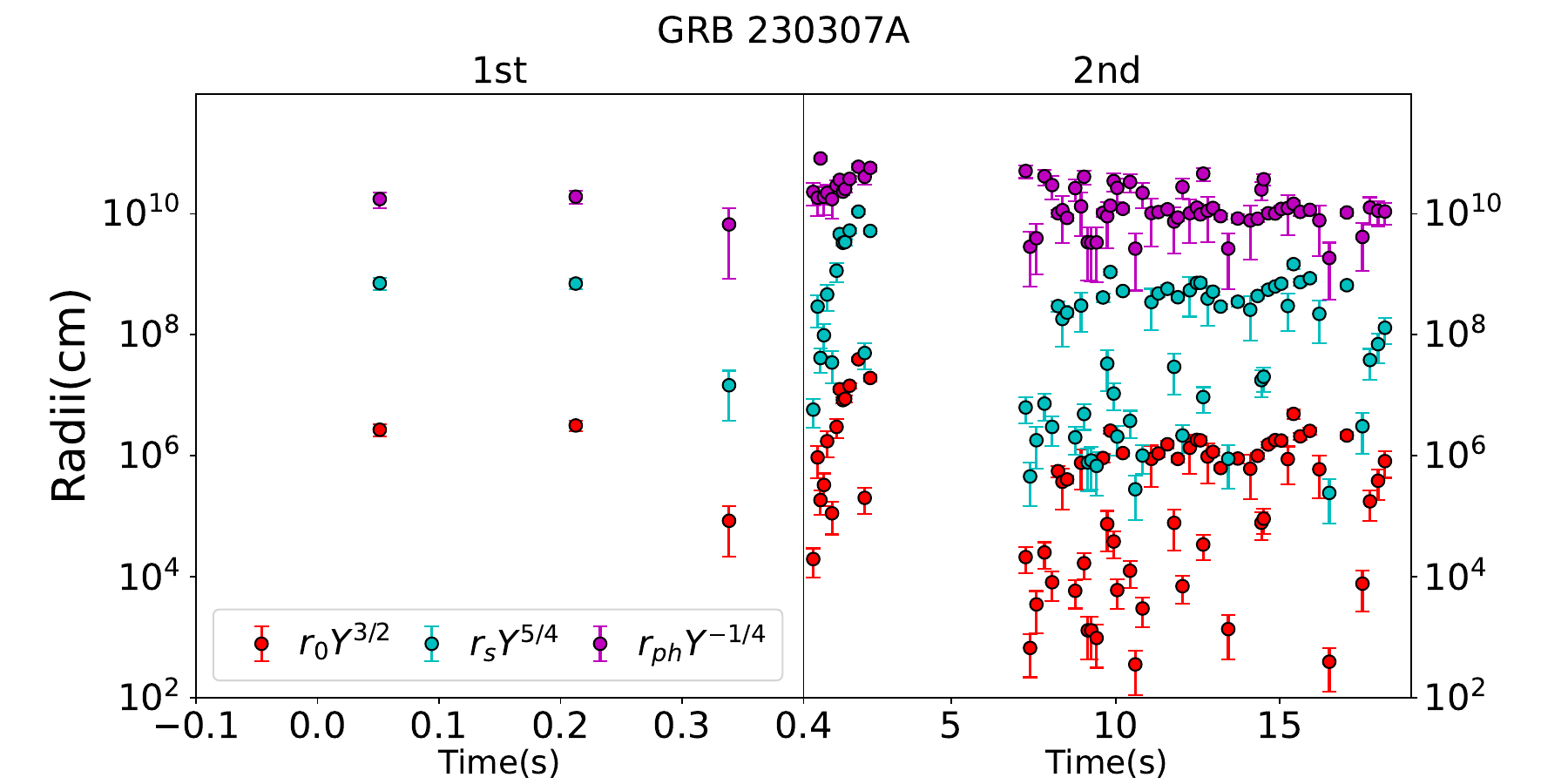}
\includegraphics [width=8.5cm,height=4.5cm]{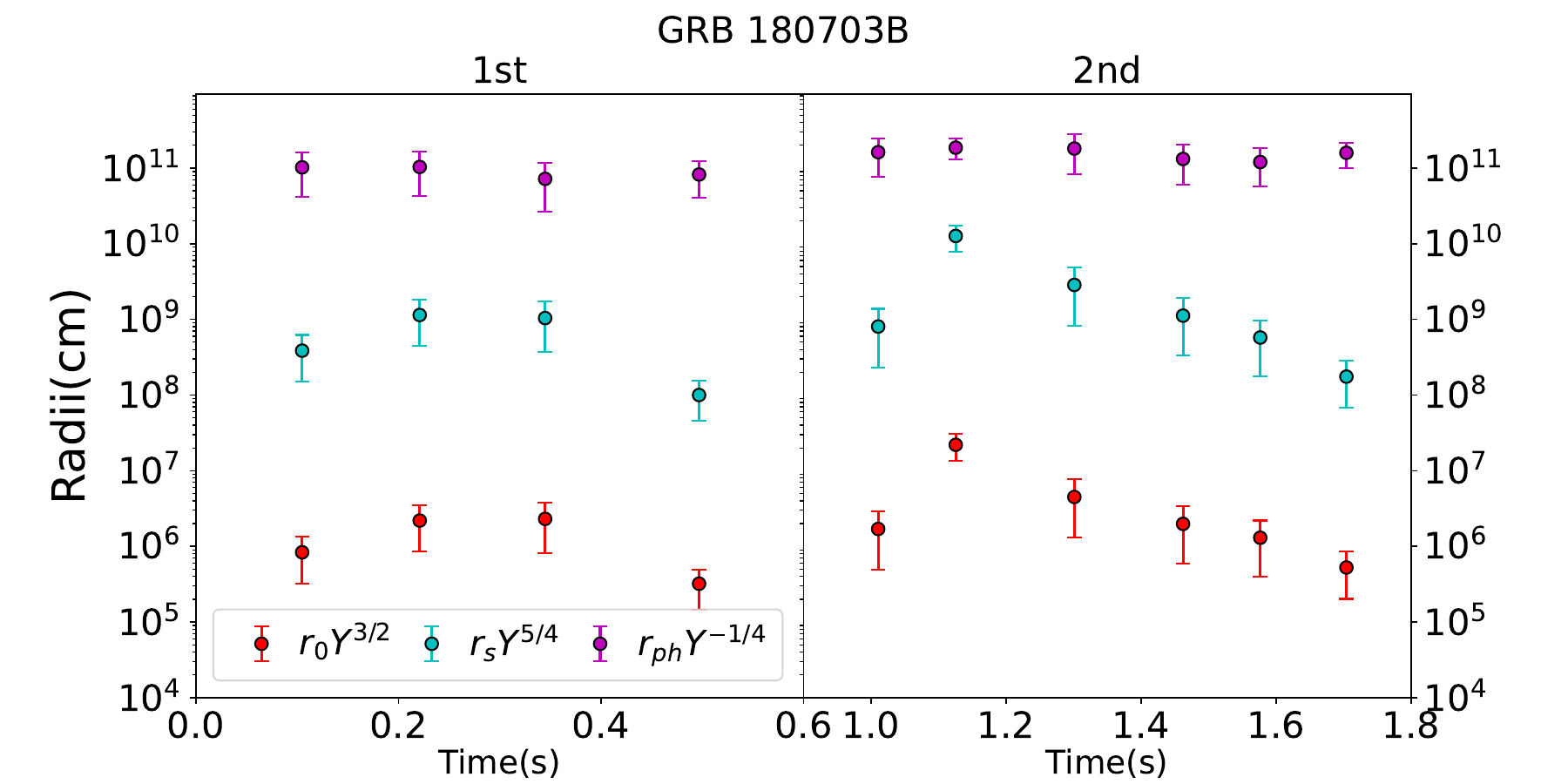}
\includegraphics [width=8.5cm,height=4.5cm]{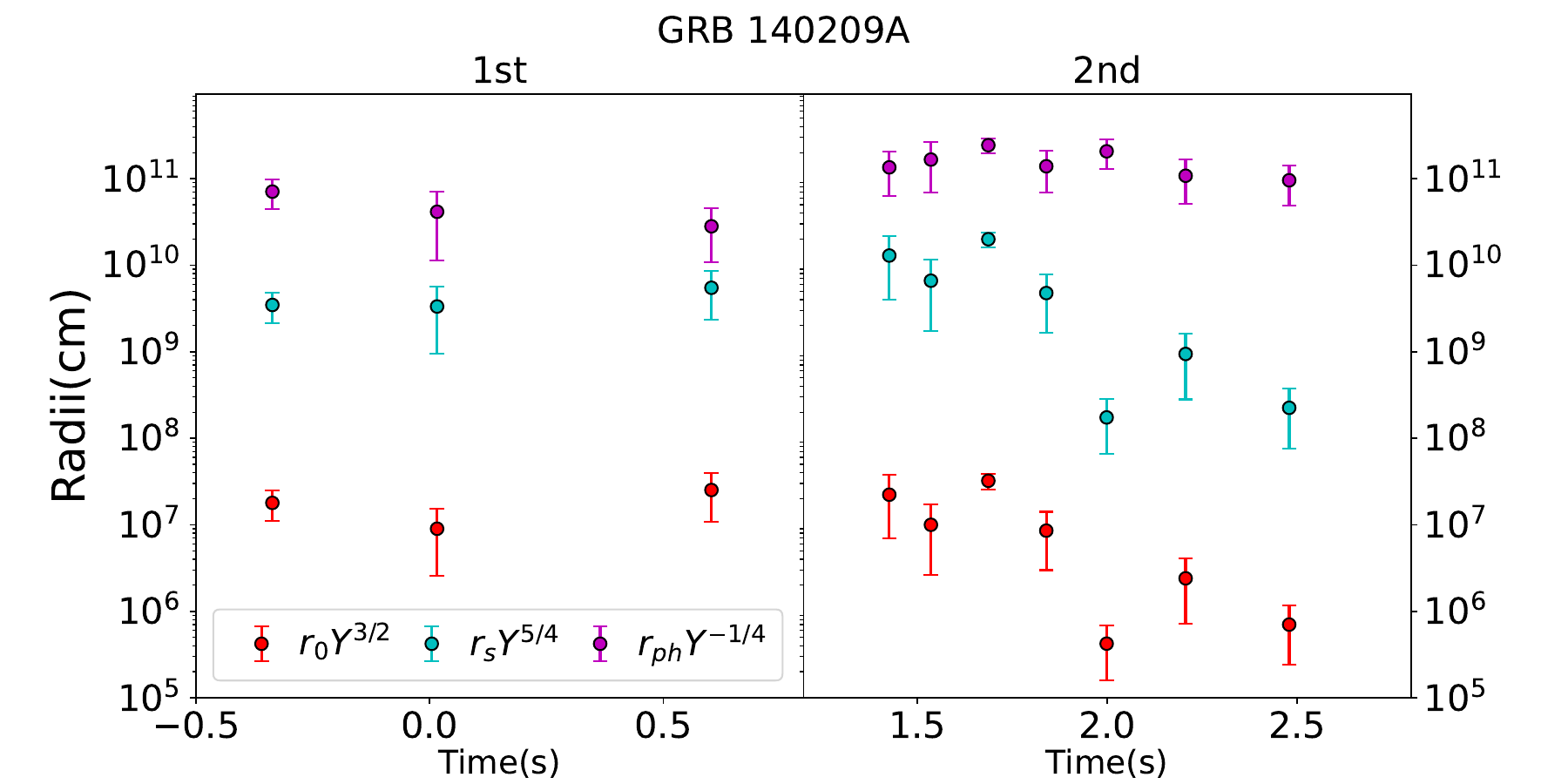}
   \figcaption{The time evolution of Parameter $r_{0}$, $r_{s}$ and $r_{ph}$. ``1st'' denotes precursors, and ``2nd'' denotes main bursts. \label{fig D3}}
\end{figure}

\clearpage
\bibliography{reference}{}
\bibliographystyle{aasjournal}

\end{document}